\PassOptionsToPackage{table}{xcolor}
\documentclass[acmsmall,screen,nonacm]{acmart}

\setcopyright{none}
\renewcommand\footnotetextcopyrightpermission[1]{}
\usepackage{dblfloatfix}
\usepackage{amsmath}
\usepackage[ruled,vlined]{algorithm2e}
\usepackage{mathtools}
\usepackage{url}
\usepackage{multirow}
\usepackage{tabularx}
\usepackage{subcaption} 
\usepackage{color,soul} 
\usepackage{makecell}
\usepackage{pdflscape}
\usepackage{caption}
\usepackage{longtable}

\usepackage{pifont}
\usepackage{textgreek}
\usepackage{array}
\usepackage{url}
\usepackage[autostyle]{csquotes}
\usepackage{rotating}
\usepackage{color,soul}
\AtBeginDocument{%
  }

\acmMonth{xx}

\begin{document}

%%
%% The "title" command has an optional parameter,
%% allowing the author to define a "short title" to be used in page headers.
\title[EDA-Schema-V2: Multimodal Schema, Open Datasets, and Benchmarks for ML in Digital Physical Design]{EDA-Schema-V2: A Multimodal Schema, Open Datasets, and Benchmarks for Machine Learning in Digital Physical Design}
% \thanks{Preprint submitted to arXiv.}
%%
%% The "author" command and its associated commands are used to define
%% the authors and their affiliations.
%% Of note is the shared affiliation of the first two authors, and the
%% "authornote" and "authornotemark" commands
%% used to denote shared contribution to the research.
% \author{Pratik Shrestha}
% % \authornote{Both authors contributed equally to this research.}
% \email{ps937@drexel.edu}
% \orcid{0000-0002-7128-0893}
% \author{G.K.M. Tobin}
% \authornotemark[1]
% \email{webmaster@marysville-ohio.com}
% \affiliation{%
%   \institution{Institute for Clarity in Documentation}
%   \city{Dublin}
%   \state{Ohio}
%   \country{USA}
% }

\author{Pratik Shrestha}
\affiliation{%
  \institution{Drexel University}
  \city{Philadelphia}
  \country{USA}}
\email{ps937@drexel.edu}
\orcid{0000-0002-7128-0893}

\author{Alec Aversa}
\affiliation{%
  \institution{Drexel University}
  \city{Philadelphia}
  \country{USA}}
\email{aja367@drexel.edu}
\orcid{0000-0001-9432-5073}

\author{Ioannis Savidis}
\affiliation{%
  \institution{Drexel University}
  \city{Philadelphia}
  \country{USA}}
\email{is338@drexel.edu}
\orcid{0000-0003-4230-1795}

%%
%% By default, the full list of authors will be used in the page
%% headers. Often, this list is too long, and will overlap
%% other information printed in the page headers. This command allows
%% the author to define a more concise list
%% of authors' names for this purpose.
% \renewcommand{\shortauthors}{Trovato et al.}

%%
%% The abstract is a short summary of the work to be presented in the
%% article.
\begin{abstract}
    The continuous scaling of CMOS technology has significantly increased the complexity of very large-scale integrated (VLSI) circuits, driving growing interest in applying machine learning (ML) algorithms to electronic design automation (EDA). However, the limited availability of open datasets and the absence of standardized data representations present major challenges to the research community. In particular, the lack of interoperability and comparability among ML-based research in digital physical design hinders reproducibility and collaborative advancement.
    This paper introduces \textit{EDA-Schema-V2}, an open and comprehensive multimodal graph and image-based schema designed to address interoperability and comparability by providing a structured framework for the representation and analysis of datasets in digital physical design. The schema includes representations of both physical attributes and quality-of-results (QoR) metrics of circuits across multiple stages of the design flow, including logic synthesis, floorplanning, placement, clock network synthesis, and routing.

    Utilizing the SkyWater 130 nm, Nangate 45 nm, IHP SG13G2 130 nm, and ASAP 7 nm open-source process design kits (PDKs), together with the OpenROAD tool flow, datasets of physical designs of circuits from the IWLS’05 benchmark suite are generated and analyzed. The dataset is comprised of approximately 7,800 design instances spanning 18 benchmark circuits and includes stage-resolved representations from synthesis through detailed routing. The design instances are generated through systematic parameter sweeps over clock period, core utilization, and aspect ratio, while capturing both timing clean and timing violating implementations across technology nodes. The dataset contains over 275 million gates, 75 million nets, and more than 36 million extracted timing paths. Multimodal data is provided in the form of circuit graphs, spatial layout images, and detailed quality of results metrics. 
    The resulting datasets are publicly released to support reproducible ML research and to establish standardized benchmarks for the evaluation of ML-based approaches in digital physical design.
    In addition, twelve representative prediction tasks are identified across technology nodes and benchmark circuits, spanning timing, power, area, and routing metrics.
    A comprehensive baseline analysis of the identified tasks is provided to characterize stage-to-stage predictability across the physical design flow.
    The resulting baseline establishes standardized reference points for benchmarking and evaluating future machine learning (ML) methods.
\end{abstract}

%%
%% The code below is generated by the tool at http://dl.acm.org/ccs.cfm.
%% Please copy and paste the code instead of the example below.
%%
% \begin{CCSXML}
% <ccs2012>
%    <concept>
%        <concept_id>10010583.10010682.10010697</concept_id>
%        <concept_desc>Hardware~Physical design (EDA)</concept_desc>
%        <concept_significance>500</concept_significance>
%        </concept>
%    <concept>
%        <concept_id>10010147.10010257</concept_id>
%        <concept_desc>Computing methodologies~Machine learning</concept_desc>
%        <concept_significance>500</concept_significance>
%        </concept>
%    <concept>
%        <concept_id>10002951.10002952.10002953</concept_id>
%        <concept_desc>Information systems~Database design and models</concept_desc>
%        <concept_significance>500</concept_significance>
%        </concept>
%  </ccs2012>
% \end{CCSXML}


% \received{20 February 2007}
% \received[revised]{12 March 2009}
% \received[accepted]{5 June 2009}

%%
%% This command processes the author and affiliation and title
%% information and builds the first part of the formatted document.
\maketitle

\section{Introduction}
\label{sec:introduction}
With the continued scaling of CMOS technology, the design complexity of integrated circuits (ICs) has increased substantially, challenging traditional design methodologies.
While electronic design automation (EDA) has long relied on heuristic, statistical, and optimization-based techniques that implicitly capture design patterns, such approaches are often rule-based, expert-engineered, and limited in the ability to generalize across designs and technology nodes.
To address the growing complexity, researchers have increasingly applied machine learning (ML) techniques, which enable automated pattern extraction and more scalable, data-driven prediction from large datasets \cite{huang2021machine}.
However, the lack of accessible and curated circuit datasets, along with the lack of a standardized schema to organize and share data, has resulted in significant obstacles.
Variations across technology nodes, the limited availability of open-source resources, time-consuming data preparation, inconsistent preprocessing, and the lack of a unified benchmarking framework all limit the development of consistent and reproducible ML approaches for physical design.

While early initiatives including METRICS2.1 \cite{jung2021metrics2} have advanced openness and standardization in EDA research, the scope of such efforts remains limited. METRICS2.1 introduces standardized metrics for RTL-to-GDSII workflows, but reports only aggregate performance data rather than detailed, stage-wise, or structural representations.
More recent efforts, including CircuitNet \cite{chai2023circuitnet} and CircuitOps \cite{liang2023circuitops}, extend open workflows by integrating machine learning with EDA algorithms.
CircuitNet provides graph and image-based datasets generated using commercial design tools and proprietary process design kits (PDKs).
However, the use of closed-source environments restricts reproducibility and public accessibility. CircuitOps, built entirely on an open-source design flow and open PDKs, targets dataset generation, feature extraction, and ML integration.
The datasets generated by CircuitOps primarily target the post–placement and routing stages.
CircuitOps captures circuit graphs and QoR metrics, but excludes pre-route representations and image modalities.

This paper introduces \textit{EDA-Schema-V2}, an open-source, multi-stage, and multimodal schema and dataset for digital physical design. Generated using the OpenROAD tool flow \cite{ajayi2019openroad} and open process design kits (PDKs), the open datasets enable fully reproducible and standardized machine learning methodologies for physical design automation. Building on the original \textit{EDA-Schema} framework \cite{shrestha2024edaschema}, this work extends beyond graph-based representations to include image modalities and data from multiple stages of the RTL-to-GDSII flow.
The key contributions of this work are summarized as follows:
\begin{itemize}
\item \textbf{Standardized and extensible schema:} A unified, stage-aware data model defining structures, relationships, and properties of digital circuits and corresponding subcomponents.
\item \textbf{Open-source multimodal dataset:} A comprehensive set of physical designs of circuits from the IWLS’05 benchmark suite \cite{albrecht2005iwls} generated using the OpenROAD flow \cite{ajayi2019openroad} and the SkyWater 130 nm, Nangate 45 nm, IHP SG13G2 130 nm, and ASAP 7 nm open-source PDKs.
\item \textbf{Comprehensive dataset analysis:} A quantitative characterization of the dataset demonstrating completeness and data diversity.
% \item \textbf{Baseline ML benchmarks and leaderboard:} A compilation of prediction and optimization tasks and baseline metrics, which are supported by a public leaderboard for transparent and reproducible model comparison.
\item \textbf{Baseline ML benchmark tasks:} A comprehensive profiling of the dataset and standardized metrics for baseline evaluation of representative prediction tasks to support a transparent and reproducible comparison of models.
\end{itemize}

\noindent The formatted dataset of physical designs and the source code of the data model schema are released on GitHub\footnote{A processed dataset of  physical designs along with code to access and represent the data as an EDA-schema data model have been made available
on \url{https://github.com/drexel-ice/EDA-schema}}.

The remainder of this paper is organized as follows.
Background on the design automation flow and the data generated through each stage of the flow is provided in Section~\ref{sec:electronic_design_automation}.
The \textit{EDA-Schema-V2} data model, including the property-graph structure and corresponding entities are described in Section~\ref{sec:eda_datamodel_schema}.
The open datasets, including details on the data-generation process, benchmark circuits, utilized tools, and applied setup parameters are discussed in Section~\ref{sec:open_dataset}, followed by an analysis of the dataset in Section~\ref{sec:dataset_analysis}.
The applicability of the proposed schema and dataset for machine-learning–based EDA tasks, including baseline analysis across multiple design stages and technology nodes, is discussed in Section~\ref{sec:machine_learning}.
Finally, some concluding remarks are provided in Section~\ref{sec:conclusions}.

\section{Electronic Design Automation Flow}
\label{sec:electronic_design_automation}

\begin{figure*}[!h]
    \begin{center}
        \includegraphics[width=0.85\linewidth]{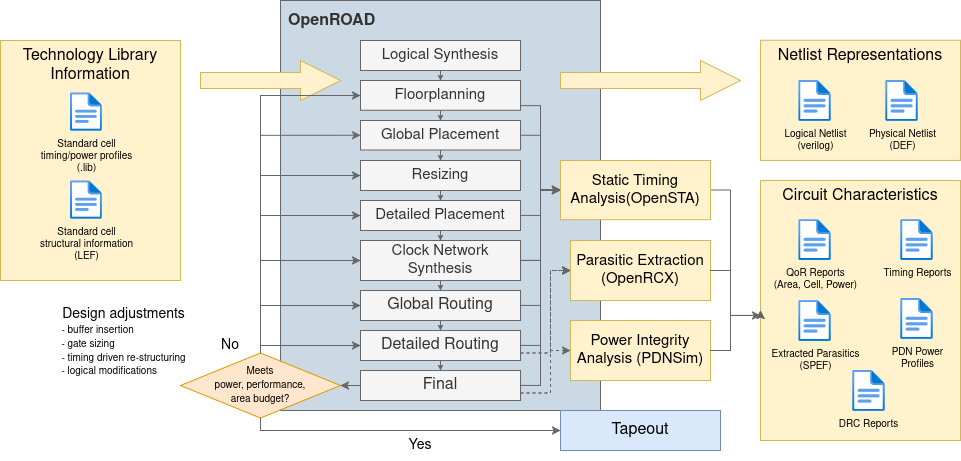}
    \end{center}
    \caption{Block representation of the automated physical design flow.}
    \label{fig:eda_flow}
\end{figure*}

The digital circuit design flow includes a sequence of automated stages to achieve functionally correct and power, performance, and area (PPA) optimized circuit implementations. PDKs provide technology-specific device models and layout rules, which guide EDA tools in generating and optimizing circuit layouts.
This work utilizes the open-source OpenROAD toolchain, an RTL-to-GDSII flow built on open PDKs and standardized file formats.
As shown in Fig.~\ref{fig:eda_flow}, which depicts the design stage structure provided by OpenROAD \cite{ajayi2019openroad}, the flow progresses through logic synthesis, floorplanning, placement (global, resizing, detailed), clock tree synthesis, routing (global and detailed), and final filler cell insertion. Each stage affects design quality, with outputs from the previous stage impacting current stage outcomes.

% \vspace{-0.1in}
% \vspace{-0.05in}

\noindent The completion of each stage generates specific files and reports that describe the logical, physical, and analytical aspects of the design:

\begin{enumerate}
\item \textbf{Logical netlist} is represented by technology-specific Verilog or VHDL files that utilize standard cells defined in the Liberty (.lib) files of a process technology node.
The netlist files describe the logical structure of the circuit. The Liberty files provide detailed timing and electrical characteristics of digital standard cells for a given PDK, which are used to optimize the circuit and verify that design constraints are satisfied.

\item \textbf{Physical netlist} is described using the Library Exchange Format (LEF) and Design Exchange Format (DEF) files.
The LEF and DEF files define technology-specific layout information, including the physical coordinates of input and output pins, standard cell placements, and the path of interconnect routes.

\item \textbf{Quality-of-results (QoR) reports}
summarize key design metrics, including power, area, and timing, at each stage of the physical design flow.
The reports provide generalized, stage-level overviews of design quality, with reports during synthesis providing preliminary estimates while later design stages yield more accurate and detailed results.

\item \textbf{Timing reports}
are generated using the static timing analysis (STA) tool \textbf{OpenSTA} \cite{openroad_opensta}.
The OpenSTA reports provide detailed path-level analysis of signal arrival times, required times, and path delays, which enables precise evaluation of both critical and non-critical timing paths.

\item \textbf{Net parasitics}
are extracted using \textbf{OpenRCX} \cite{openroad_openrcx}. The resulting Standard Parasitic Exchange Format (SPEF) files \cite{hierarchical2012standard} include the resistance and capacitance of nets and devices, which provide an accurate estimate of the post-routed delay and power consumption.

\item \textbf{Power delivery network (PDN) analysis} is performed using \textbf{PDNSim} \cite{openroad_pdnsim}, which evaluates IR-drop (voltage drop due to current flow through resistances) and electromigration (EM) through current density analysis. The results of the EM analysis are stored as gridded CSV data, while heatmaps are used to visualize the IR-drop and current density.

\item \textbf{Design rule checking (DRC)} ensures that the layout adheres to the manufacturing constraints defined by the PDK.
Execution of DRC returns geometric or spacing violations in frontend and backend of the line layers including vias and other layout structures, which ensures that the design is compliant with foundry rules.
\end{enumerate}

% Although our work and open datasets utilizes OpenROAD as an open-source tool, the developed schema is compatible with commercial EDA workflows and proprietary design environments.

\section{EDA DataModel Schema}
\label{sec:eda_datamodel_schema}

\begin{figure}[!h]
    \begin{center}
        \includegraphics[width=0.6\columnwidth]{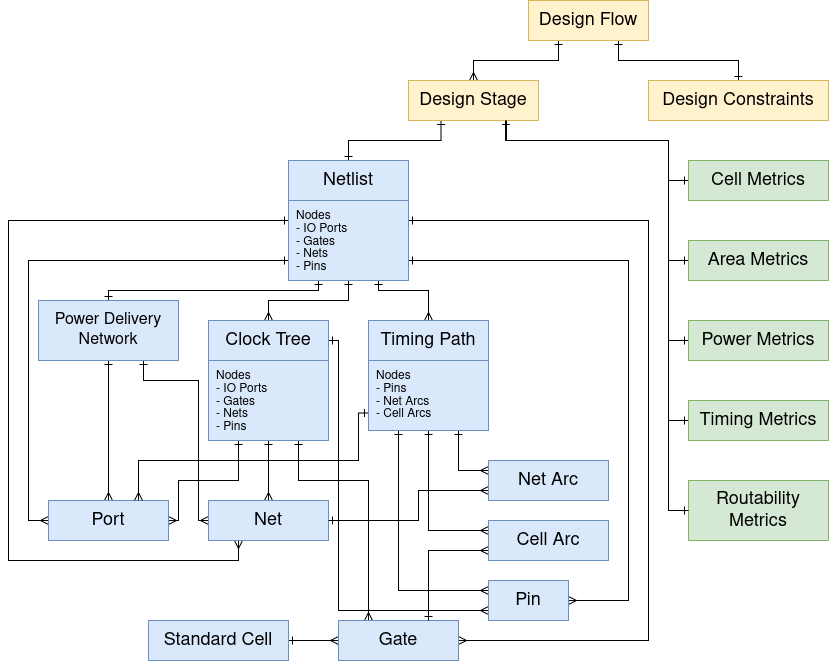}
    \end{center}
    % \vspace{-0.05in}
    \caption{EDA-Schema-V2 entity relationship diagram. Entities in yellow represent the Design Flow Hierarchy (flow, stage, and constraints). Entities in blue correspond to structural design data, while entities in green capture performance and quality-of-results metrics.}
    \label{fig:er_diagram}
\end{figure}

The physical design flow shown in Fig.~\ref{fig:eda_flow} provides two main classes of data that define circuit quality. The first is structural data of the circuit, which is provided in the Liberty (.lib), LEF/DEF, and SPEF files that describe logical connectivity, physical geometry, and parasitic impedance, respectively. The second is analysis data included in power, timing, and area reports, along with reports generated from analysis of the PDN, which include IR-drop and EM profiling. The resulting reports are also used to generate spatial representations of placements and routing, as well as heatmaps of analyzed circuit features.
\textit{EDA-Schema-V2} organizes such artifacts into a unified multimodal schema that models each circuit as a collection of related entities with graph-structured and image-based attributes, forming a taxonomy of design flow components, structural elements, and performance metrics.
Graph relationships between the entities capture logical and physical connectivity among components while image attributes represent spatial patterns and heatmaps of performed analyses.
The entity–relationship structure is shown in Figure~\ref{fig:er_diagram} and the attributes associated with each entity are summarized in Table~\ref{tab:datamodel_features}.
The schema is defined using standard EDA file formats and technology layer conventions to ensure compatibility across different design platforms.
All geometric coordinates are expressed in DEF database units, and technology layers are derived directly from LEF definitions to maintain consistent interpretation across tools and PDKs.
\textbf{Although EDA-schema is instantiated on OpenROAD and open-source PDKs, the schema remains compatible with commercial design tools that utilize the same data exchange standards.
}

The schema defines three types of entities: organizational entities that capture the design flow and the associated netlist and design and performance metrics, structural entities that represent the logical and physical composition of the circuit, and metric entities that store performance and quality-of-results data.
The design flow and metric entities included in EDA-Schema-V2 are described in Section~\ref{subsec:design_flow_and_metrics}, while structural entities including \textit{Netlist}, \textit{Clock Network}, \textit{Power Delivery Network}, and \textit{Timing Path} are detailed in Sections~\ref{subsec:netlist_graph}, \ref{subsec:clock_network_graph}, \ref{subsec:pdn}, and~\ref{subsec:timing_path_graph}, respectively.
Routability features and metrics are discussed in 
Section~\ref{subsec:routability_metrics}.

\subsection{Design Flow and Metrics}
\label{subsec:design_flow_and_metrics}

At the top level, the schema represents an RTL-to-GDSII execution of the design flow through three interdependent entities, \textit{DesignFlow}, \textit{DesignStage}, and \textit{DesignConstraint}, which together define the progression of a circuit design through the various stages of the automated flow.
\textit{DesignFlow} represents high level metadata of the executing flow including the name of the design, the PDK, and the tool versions, along with the sequence of design stages. \textit{DesignConstraint} defines timing and physical area limits, including clock periods, I/O delays, and transition or latency bounds with SDC-style specifications. \textit{DesignStage} captures each stage of the digital design flow, including synthesis, placement, routing, and signoff.
In OpenROAD, the stages are further partitioned into floorplanning (\texttt{floorplan}), global placement (\texttt{global\_place}), resizing (\texttt{place\_resize}), detailed placement (\texttt{detailed\_place}), clock tree synthesis (\texttt{cts}), global routing (\texttt{global\_route}), detailed routing (\texttt{detailed\_route}), and final filler insertion (\texttt{final}).

At each stage, the quality-of-results metrics are grouped with the corresponding metrics entities, which include \textit{Cell Metrics}, \textit{Area Metrics}, \textit{Power Metrics}, \textit{Timing Metrics}, and \textit{Routability Metrics}. The integrated representation of flow, constraints, stages, and metrics provides a consistent and extensible framework for provenance, reproducibility, and performance evaluation across tools and PDKs.

\begin{table}[!h]
    \begin{center}
        % \vspace{-0.05in}
        \caption{List of circuit features represented by the EDA-Schema-V2 datamodel. Each entity is associated with a set of attributes, corresponding data types, and units where applicable. The Stages column indicates the design stages where each attribute is available, which includes floorplan (FP), global placement (GP), placement resize (PR), detailed placement (DP), clock tree synthesis (CTS), global routing (GR), detailed routing (DR), and final layout (F).}
        % \vspace{-0.025in}
        \label{tab:datamodel_features}
        \setlength{\tabcolsep}{3pt}
        \fontsize{5.75}{5.75}\selectfont{
        % \fontsize{6.75}{7.5}\selectfont{
        % \fontsize{6}{7}\selectfont{
            \begin{tabular}{llllll|l|l|l|l|l|}
            \cline{1-5} \cline{7-11}
            \multicolumn{1}{|l|}{\textbf{Entity}}                                                                               & \multicolumn{1}{l|}{\textbf{Feature}}               & \multicolumn{1}{l|}{\textbf{Datatype}}    & \multicolumn{1}{l|}{\textbf{Unit}}            & \multicolumn{1}{l|}{\textbf{Stages}}                             &  & \textbf{Entity}                                                                         & \textbf{Feature}             & \textbf{Datatype} & \textbf{Unit}                   & \textbf{Stages} \\ \cline{1-5} \cline{7-11} 
            \multicolumn{1}{|l|}{}                                                                                              & \multicolumn{1}{l|}{id}                             & \multicolumn{1}{l|}{string}               & \multicolumn{1}{l|}{}                         & \multicolumn{1}{l|}{\cellcolor[HTML]{EFEFEF}}                    &  &                                                                                         & name                         & string            &                                 & FP - F          \\ \cline{2-4} \cline{8-11} 
            \multicolumn{1}{|l|}{}                                                                                              & \multicolumn{1}{l|}{toolchain}                      & \multicolumn{1}{l|}{string}               & \multicolumn{1}{l|}{}                         & \multicolumn{1}{l|}{\cellcolor[HTML]{EFEFEF}}                    &  &                                                                                         & is\_special\_net             & bool              &                                 & FP - F          \\ \cline{2-4} \cline{8-11} 
            \multicolumn{1}{|l|}{}                                                                                              & \multicolumn{1}{l|}{design}                         & \multicolumn{1}{l|}{string}               & \multicolumn{1}{l|}{}                         & \multicolumn{1}{l|}{\cellcolor[HTML]{EFEFEF}}                    &  &                                                                                         & no\_of\_fanouts              & int               &                                 & FP - F          \\ \cline{2-4} \cline{8-11} 
            \multicolumn{1}{|l|}{\multirow{-4}{*}{\textbf{\begin{tabular}[c]{@{}l@{}}Design\\ Flow\end{tabular}}}}              & \multicolumn{1}{l|}{run\_status}                    & \multicolumn{1}{l|}{string}               & \multicolumn{1}{l|}{}                         & \multicolumn{1}{l|}{\cellcolor[HTML]{EFEFEF}}                    &  &                                                                                         & x\_min                       & float             & \textmu m        & GP - F          \\ \cline{1-4} \cline{8-11} 
            \multicolumn{1}{|l|}{}                                                                                              & \multicolumn{1}{l|}{clock\_period}                  & \multicolumn{1}{l|}{float}                & \multicolumn{1}{l|}{ns}                       & \multicolumn{1}{l|}{\cellcolor[HTML]{EFEFEF}}                    &  &                                                                                         & y\_min                       & float             & \textmu m        & GP - F          \\ \cline{2-4} \cline{8-11} 
            \multicolumn{1}{|l|}{}                                                                                              & \multicolumn{1}{l|}{clock\_uncertainty}             & \multicolumn{1}{l|}{float}                & \multicolumn{1}{l|}{ns}                       & \multicolumn{1}{l|}{\cellcolor[HTML]{EFEFEF}}                    &  &                                                                                         & x\_max                       & float             & \textmu m        & GP - F          \\ \cline{2-4} \cline{8-11} 
            \multicolumn{1}{|l|}{}                                                                                              & \multicolumn{1}{l|}{clock\_latency}                 & \multicolumn{1}{l|}{float}                & \multicolumn{1}{l|}{ns}                       & \multicolumn{1}{l|}{\cellcolor[HTML]{EFEFEF}}                    &  &                                                                                         & y\_max                       & float             & \textmu m        & GP - F          \\ \cline{2-4} \cline{8-11} 
            \multicolumn{1}{|l|}{}                                                                                              & \multicolumn{1}{l|}{clock\_transition}              & \multicolumn{1}{l|}{float}                & \multicolumn{1}{l|}{ns}                       & \multicolumn{1}{l|}{\cellcolor[HTML]{EFEFEF}}                    &  &                                                                                         & length                       & float             & \textmu m        & DR - F          \\ \cline{2-4} \cline{8-11} 
            \multicolumn{1}{|l|}{}                                                                                              & \multicolumn{1}{l|}{input\_delay}                   & \multicolumn{1}{l|}{float}                & \multicolumn{1}{l|}{ns}                       & \multicolumn{1}{l|}{\cellcolor[HTML]{EFEFEF}}                    &  &                                                                                         & hpwl                         & float             & \textmu m        & GP - F          \\ \cline{2-4} \cline{8-11} 
            \multicolumn{1}{|l|}{}                                                                                              & \multicolumn{1}{l|}{output\_delay}                  & \multicolumn{1}{l|}{float}                & \multicolumn{1}{l|}{ns}                       & \multicolumn{1}{l|}{\cellcolor[HTML]{EFEFEF}}                    &  &                                                                                         & resistance                   & float             & ohm                             & DR - F          \\ \cline{2-4} \cline{8-11} 
            \multicolumn{1}{|l|}{}                                                                                              & \multicolumn{1}{l|}{aspect\_ratio}                  & \multicolumn{1}{l|}{float}                & \multicolumn{1}{l|}{}                         & \multicolumn{1}{l|}{\cellcolor[HTML]{EFEFEF}}                    &  &                                                                                         & capacitance                  & float             & fF                              & DR - F          \\ \cline{2-4} \cline{8-11} 
            \multicolumn{1}{|l|}{\multirow{-8}{*}{\textbf{Constraints}}}                                                        & \multicolumn{1}{l|}{utilization}                    & \multicolumn{1}{l|}{float}                & \multicolumn{1}{l|}{}                         & \multicolumn{1}{l|}{\cellcolor[HTML]{EFEFEF}}                    &  & \multirow{-12}{*}{\textbf{Net}}                                                         & total\_coupling\_capacitance & float             & fF                              & DR - F          \\ \cline{1-4} \cline{7-11} 
            \multicolumn{1}{|l|}{}                                                                                              & \multicolumn{1}{l|}{name}                           & \multicolumn{1}{l|}{string}               & \multicolumn{1}{l|}{}                         & \multicolumn{1}{l|}{\cellcolor[HTML]{EFEFEF}}                    &  &                                                                                         & name                         & string            &                                 & FP - F          \\ \cline{2-4} \cline{8-11} 
            \multicolumn{1}{|l|}{\multirow{-2}{*}{\textbf{\begin{tabular}[c]{@{}l@{}}Design\\ Stage\end{tabular}}}}             & \multicolumn{1}{l|}{run\_status}                    & \multicolumn{1}{l|}{string}               & \multicolumn{1}{l|}{}                         & \multicolumn{1}{l|}{\multirow{-14}{*}{\cellcolor[HTML]{EFEFEF}}} &  &                                                                                         & direction                    & string            &                                 & FP - F          \\ \cline{1-5} \cline{8-11} 
            \multicolumn{1}{|l|}{}                                                                                              & \multicolumn{1}{l|}{width}                          & \multicolumn{1}{l|}{float}                & \multicolumn{1}{l|}{\textmu m} & \multicolumn{1}{l|}{FP - F}                                      &  &                                                                                         & x\_min                       & float             & \textmu m        & GP - F          \\ \cline{2-5} \cline{8-11} 
            \multicolumn{1}{|l|}{}                                                                                              & \multicolumn{1}{l|}{height}                         & \multicolumn{1}{l|}{float}                & \multicolumn{1}{l|}{\textmu m} & \multicolumn{1}{l|}{FP - F}                                      &  &                                                                                         & y\_min                       & float             & \textmu m        & GP - F          \\ \cline{2-5} \cline{8-11} 
            \multicolumn{1}{|l|}{}                                                                                              & \multicolumn{1}{l|}{no\_of\_inputs}                 & \multicolumn{1}{l|}{int}                  & \multicolumn{1}{l|}{}                         & \multicolumn{1}{l|}{FP - F}                                      &  &                                                                                         & x\_max                       & float             & \textmu m        & GP - F          \\ \cline{2-5} \cline{8-11} 
            \multicolumn{1}{|l|}{}                                                                                              & \multicolumn{1}{l|}{no\_of\_outputs}                & \multicolumn{1}{l|}{int}                  & \multicolumn{1}{l|}{}                         & \multicolumn{1}{l|}{FP - F}                                      &  &                                                                                         & y\_max                       & float             & \textmu m        & GP - F          \\ \cline{2-5} \cline{8-11} 
            \multicolumn{1}{|l|}{}                                                                                              & \multicolumn{1}{l|}{no\_of\_cells}                  & \multicolumn{1}{l|}{int}                  & \multicolumn{1}{l|}{}                         & \multicolumn{1}{l|}{FP - F}                                      &  &                                                                                         & is\_startpoint               & bool              &                                 & FP - F          \\ \cline{2-5} \cline{8-11} 
            \multicolumn{1}{|l|}{}                                                                                              & \multicolumn{1}{l|}{no\_of\_nets}                   & \multicolumn{1}{l|}{int}                  & \multicolumn{1}{l|}{}                         & \multicolumn{1}{l|}{FP - F}                                      &  &                                                                                         & is\_endpoint                 & bool              &                                 & FP - F          \\ \cline{2-5} \cline{8-11} 
            \multicolumn{1}{|l|}{}                                                                                              & \multicolumn{1}{l|}{no\_of\_pins}                   & \multicolumn{1}{l|}{int}                  & \multicolumn{1}{l|}{}                         & \multicolumn{1}{l|}{FP - F}                                      &  &                                                                                         & setup\_rise\_slew            & float             & ns                              & FP - F          \\ \cline{2-5} \cline{8-11} 
            \multicolumn{1}{|l|}{}                                                                                              & \multicolumn{1}{l|}{utilization}                    & \multicolumn{1}{l|}{float}                & \multicolumn{1}{l|}{}                         & \multicolumn{1}{l|}{FP - F}                                      &  &                                                                                         & setup\_fall\_slew            & float             & ns                              & FP - F          \\ \cline{2-5} \cline{8-11} 
            \multicolumn{1}{|l|}{}                                                                                              & \multicolumn{1}{l|}{total\_wirelength}              & \multicolumn{1}{l|}{float}                & \multicolumn{1}{l|}{\textmu m} & \multicolumn{1}{l|}{DR - F}                                      &  &                                                                                         & hold\_rise\_slew             & float             & ns                              & FP - F          \\ \cline{2-5} \cline{8-11} 
            \multicolumn{1}{|l|}{}                                                                                              & \multicolumn{1}{l|}{total\_hpwl}                    & \multicolumn{1}{l|}{float}                & \multicolumn{1}{l|}{\textmu m} & \multicolumn{1}{l|}{FP - F}                                      &  &                                                                                         & hold\_fall\_slew             & float             & ns                              & FP - F          \\ \cline{2-5} \cline{8-11} 
            \multicolumn{1}{|l|}{}                                                                                              & \multicolumn{1}{l|}{cell\_placement}                & \multicolumn{1}{l|}{binary map}           & \multicolumn{1}{l|}{}                         & \multicolumn{1}{l|}{GP - F}                                      &  &                                                                                         & setup\_rise\_slack           & float             & ns                              & FP - F          \\ \cline{2-5} \cline{8-11} 
            \multicolumn{1}{|l|}{}                                                                                              & \multicolumn{1}{l|}{cell\_placement\_combinational} & \multicolumn{1}{l|}{binary map}           & \multicolumn{1}{l|}{}                         & \multicolumn{1}{l|}{GP - F}                                      &  &                                                                                         & setup\_fall\_slack           & float             & ns                              & FP - F          \\ \cline{2-5} \cline{8-11} 
            \multicolumn{1}{|l|}{}                                                                                              & \multicolumn{1}{l|}{cell\_placement\_sequential}    & \multicolumn{1}{l|}{binary map}           & \multicolumn{1}{l|}{}                         & \multicolumn{1}{l|}{GP - F}                                      &  &                                                                                         & hold\_rise\_slack            & float             & ns                              & FP - F          \\ \cline{2-5} \cline{8-11} 
            \multicolumn{1}{|l|}{}                                                                                              & \multicolumn{1}{l|}{cell\_placement\_filler}        & \multicolumn{1}{l|}{binary map}           & \multicolumn{1}{l|}{}                         & \multicolumn{1}{l|}{F}                                           &  &                                                                                         & hold\_fall\_slack            & float             & ns                              & FP - F          \\ \cline{2-5} \cline{8-11} 
            \multicolumn{1}{|l|}{}                                                                                              & \multicolumn{1}{l|}{pin\_placement}                 & \multicolumn{1}{l|}{binary map}           & \multicolumn{1}{l|}{}                         & \multicolumn{1}{l|}{GP - F}                                      &  &                                                                                         & load\_capacitance            & float             & fF                              & FP - F          \\ \cline{2-5} \cline{8-11} 
            \multicolumn{1}{|l|}{}                                                                                              & \multicolumn{1}{l|}{routing}                        & \multicolumn{1}{l|}{binary map}           & \multicolumn{1}{l|}{}                         & \multicolumn{1}{l|}{DR - F}                                      &  & \multirow{-18}{*}{\textbf{Pin}}                                                         & switching\_activity          & float             &                                 & FP - F          \\ \cline{2-5} \cline{7-11} 
            \multicolumn{1}{|l|}{\multirow{-17}{*}{\textbf{Netlist}}}                                                           & \multicolumn{1}{l|}{routing\_by\_metal\_layers}     & \multicolumn{1}{l|}{list{[}binary map{]}} & \multicolumn{1}{l|}{}                         & \multicolumn{1}{l|}{DR - F}                                      &  &                                                                                         & startpoint                   & string            &                                 & FP - F          \\ \cline{1-5} \cline{8-11} 
            \multicolumn{1}{|l|}{}                                                                                              & \multicolumn{1}{l|}{clock\_source}                  & \multicolumn{1}{l|}{string}               & \multicolumn{1}{l|}{}                         & \multicolumn{1}{l|}{FP - F}                                      &  &                                                                                         & endpoint                     & string            &                                 & FP - F          \\ \cline{2-5} \cline{8-11} 
            \multicolumn{1}{|l|}{}                                                                                              & \multicolumn{1}{l|}{no\_of\_buffers}                & \multicolumn{1}{l|}{int}                  & \multicolumn{1}{l|}{}                         & \multicolumn{1}{l|}{CTS - F}                                     &  &                                                                                         & path\_type                   & string            &                                 & FP - F          \\ \cline{2-5} \cline{8-11} 
            \multicolumn{1}{|l|}{}                                                                                              & \multicolumn{1}{l|}{no\_of\_clock\_sinks}           & \multicolumn{1}{l|}{int}                  & \multicolumn{1}{l|}{}                         & \multicolumn{1}{l|}{CTS - F}                                     &  &                                                                                         & arrival\_time                & float             & ns                              & FP - F          \\ \cline{2-5} \cline{8-11} 
            \multicolumn{1}{|l|}{}                                                                                              & \multicolumn{1}{l|}{cell\_placement}                & \multicolumn{1}{l|}{binary map}           & \multicolumn{1}{l|}{}                         & \multicolumn{1}{l|}{CTS - F}                                     &  &                                                                                         & required\_time               & float             & ns                              & FP - F          \\ \cline{2-5} \cline{8-11} 
            \multicolumn{1}{|l|}{}                                                                                              & \multicolumn{1}{l|}{cell\_placement\_combinational} & \multicolumn{1}{l|}{binary map}           & \multicolumn{1}{l|}{}                         & \multicolumn{1}{l|}{CTS - F}                                     &  &                                                                                         & slack                        & float             & ns                              & FP - F          \\ \cline{2-5} \cline{8-11} 
            \multicolumn{1}{|l|}{}                                                                                              & \multicolumn{1}{l|}{cell\_placement\_sequential}    & \multicolumn{1}{l|}{binary map}           & \multicolumn{1}{l|}{}                         & \multicolumn{1}{l|}{CTS - F}                                     &  &                                                                                         & no\_of\_pins                 & int               &                                 & FP - F          \\ \cline{2-5} \cline{8-11} 
            \multicolumn{1}{|l|}{}                                                                                              & \multicolumn{1}{l|}{pin\_placement}                 & \multicolumn{1}{l|}{binary map}           & \multicolumn{1}{l|}{}                         & \multicolumn{1}{l|}{CTS - F}                                     &  & \multirow{-8}{*}{\textbf{\begin{tabular}[c]{@{}l@{}}Timing\\ Path\end{tabular}}}        & is\_critical\_path           & bool              &                                 & FP - F          \\ \cline{2-5} \cline{7-11} 
            \multicolumn{1}{|l|}{}                                                                                              & \multicolumn{1}{l|}{routing}                        & \multicolumn{1}{l|}{binary map}           & \multicolumn{1}{l|}{}                         & \multicolumn{1}{l|}{DR - F}                                      &  &                                                                                         & gate                         & Gate Instance     &                                 & FP - F          \\ \cline{2-5} \cline{8-11} 
            \multicolumn{1}{|l|}{\multirow{-9}{*}{\textbf{\begin{tabular}[c]{@{}l@{}}Clock\\ Tree\end{tabular}}}}               & \multicolumn{1}{l|}{routing\_by\_metal}             & \multicolumn{1}{l|}{list{[}binary map{]}} & \multicolumn{1}{l|}{}                         & \multicolumn{1}{l|}{DR - F}                                      &  &                                                                                         & delay                        & float             & ns                              & FP - F          \\ \cline{1-5} \cline{8-11} 
            \multicolumn{1}{|l|}{}                                                                                              & \multicolumn{1}{l|}{pdn\_routing\_vdd}              & \multicolumn{1}{l|}{binary map}           & \multicolumn{1}{l|}{}                         & \multicolumn{1}{l|}{FP - F}                                      &  &                                                                                         & arrival\_time                & float             & ns                              & FP - F          \\ \cline{2-5} \cline{8-11} 
            \multicolumn{1}{|l|}{}                                                                                              & \multicolumn{1}{l|}{pdn\_routing\_vss}              & \multicolumn{1}{l|}{binary map}           & \multicolumn{1}{l|}{}                         & \multicolumn{1}{l|}{FP - F}                                      &  & \multirow{-4}{*}{\textbf{Cell Arc}}                                                     & slew                         & float             & ns                              & FP - F          \\ \cline{2-5} \cline{7-11} 
            \multicolumn{1}{|l|}{}                                                                                              & \multicolumn{1}{l|}{voltage\_source}                & \multicolumn{1}{l|}{binary map}           & \multicolumn{1}{l|}{}                         & \multicolumn{1}{l|}{FP - F}                                      &  &                                                                                         & net                          & Net               &                                 & FP - F          \\ \cline{2-5} \cline{8-11} 
            \multicolumn{1}{|l|}{}                                                                                              & \multicolumn{1}{l|}{ir\_drop\_vdd}                  & \multicolumn{1}{l|}{scalar map}           & \multicolumn{1}{l|}{mV}                       & \multicolumn{1}{l|}{DR - F}                                      &  &                                                                                         & delay                        & float             & ns                              & FP - F          \\ \cline{2-5} \cline{8-11} 
            \multicolumn{1}{|l|}{}                                                                                              & \multicolumn{1}{l|}{ir\_drop\_vss}                  & \multicolumn{1}{l|}{scalar map}           & \multicolumn{1}{l|}{mV}                       & \multicolumn{1}{l|}{DR - F}                                      &  &                                                                                         & arrival\_time                & float             & ns                              & FP - F          \\ \cline{2-5} \cline{8-11} 
            \multicolumn{1}{|l|}{}                                                                                              & \multicolumn{1}{l|}{em\_vdd}                        & \multicolumn{1}{l|}{scalar map}           & \multicolumn{1}{l|}{mV}                       & \multicolumn{1}{l|}{DR - F}                                      &  &                                                                                         & slew                         & float             & ns                              & FP - F          \\ \cline{2-5} \cline{8-11} 
            \multicolumn{1}{|l|}{\multirow{-7}{*}{\textbf{\begin{tabular}[c]{@{}l@{}}Power\\ Delivery\\ Network\end{tabular}}}} & \multicolumn{1}{l|}{em\_vss}                        & \multicolumn{1}{l|}{scalar map}           & \multicolumn{1}{l|}{mV}                       & \multicolumn{1}{l|}{DR - F}                                      &  & \multirow{-5}{*}{\textbf{Net Arc}}                                                      & capacitance                  & float             & fF                              & FP - F          \\ \cline{1-5} \cline{7-11} 
            \multicolumn{1}{|l|}{}                                                                                              & \multicolumn{1}{l|}{name}                           & \multicolumn{1}{l|}{string}               & \multicolumn{1}{l|}{}                         & \multicolumn{1}{l|}{FP - F}                                      &  &                                                                                         & no\_of\_combinational\_cells & int               &                                 & FP - F          \\ \cline{2-5} \cline{8-11} 
            \multicolumn{1}{|l|}{}                                                                                              & \multicolumn{1}{l|}{direction}                      & \multicolumn{1}{l|}{string}               & \multicolumn{1}{l|}{}                         & \multicolumn{1}{l|}{FP - F}                                      &  &                                                                                         & no\_of\_sequential\_cells    & int               &                                 & FP - F          \\ \cline{2-5} \cline{8-11} 
            \multicolumn{1}{|l|}{}                                                                                              & \multicolumn{1}{l|}{x}                              & \multicolumn{1}{l|}{float}                & \multicolumn{1}{l|}{\textmu m} & \multicolumn{1}{l|}{GP - F}                                      &  &                                                                                         & no\_of\_buffers              & int               &                                 & FP - F          \\ \cline{2-5} \cline{8-11} 
            \multicolumn{1}{|l|}{\multirow{-4}{*}{\textbf{Port}}}                                                               & \multicolumn{1}{l|}{y}                              & \multicolumn{1}{l|}{float}                & \multicolumn{1}{l|}{\textmu m} & \multicolumn{1}{l|}{GP - F}                                      &  &                                                                                         & no\_of\_inverters            & int               &                                 & FP - F          \\ \cline{1-5} \cline{8-11} 
            \multicolumn{1}{|l|}{}                                                                                              & \multicolumn{1}{l|}{name}                           & \multicolumn{1}{l|}{string}               & \multicolumn{1}{l|}{}                         & \multicolumn{1}{l|}{\cellcolor[HTML]{EFEFEF}}                    &  &                                                                                         & no\_of\_fillers              & int               &                                 & FP - F          \\ \cline{2-4} \cline{8-11} 
            \multicolumn{1}{|l|}{}                                                                                              & \multicolumn{1}{l|}{function}                       & \multicolumn{1}{l|}{string}               & \multicolumn{1}{l|}{}                         & \multicolumn{1}{l|}{\cellcolor[HTML]{EFEFEF}}                    &  &                                                                                         & no\_of\_tap\_cells           & int               &                                 & FP - F          \\ \cline{2-4} \cline{8-11} 
            \multicolumn{1}{|l|}{}                                                                                              & \multicolumn{1}{l|}{width}                          & \multicolumn{1}{l|}{float}                & \multicolumn{1}{l|}{\textmu m} & \multicolumn{1}{l|}{\cellcolor[HTML]{EFEFEF}}                    &  &                                                                                         & no\_of\_diodes               & int               &                                 & FP - F          \\ \cline{2-4} \cline{8-11} 
            \multicolumn{1}{|l|}{}                                                                                              & \multicolumn{1}{l|}{height}                         & \multicolumn{1}{l|}{float}                & \multicolumn{1}{l|}{\textmu m} & \multicolumn{1}{l|}{\cellcolor[HTML]{EFEFEF}}                    &  &                                                                                         & no\_of\_macros               & int               &                                 & FP - F          \\ \cline{2-4} \cline{8-11} 
            \multicolumn{1}{|l|}{}                                                                                              & \multicolumn{1}{l|}{no\_of\_input\_pins}            & \multicolumn{1}{l|}{int}                  & \multicolumn{1}{l|}{}                         & \multicolumn{1}{l|}{\cellcolor[HTML]{EFEFEF}}                    &  & \multirow{-9}{*}{\textbf{\begin{tabular}[c]{@{}l@{}}Cell\\ Metrics\end{tabular}}}       & no\_of\_total\_cells         & int               &                                 & FP - F          \\ \cline{2-4} \cline{7-11} 
            \multicolumn{1}{|l|}{}                                                                                              & \multicolumn{1}{l|}{no\_of\_output\_pins}           & \multicolumn{1}{l|}{int}                  & \multicolumn{1}{l|}{}                         & \multicolumn{1}{l|}{\cellcolor[HTML]{EFEFEF}}                    &  &                                                                                         & combinational\_cell\_area    & float             & \textmu m$^{-2}$ & FP - F          \\ \cline{2-4} \cline{8-11} 
            \multicolumn{1}{|l|}{}                                                                                              & \multicolumn{1}{l|}{is\_sequential}                 & \multicolumn{1}{l|}{bool}                 & \multicolumn{1}{l|}{}                         & \multicolumn{1}{l|}{\cellcolor[HTML]{EFEFEF}}                    &  &                                                                                         & sequential\_cell\_area       & float             & \textmu m$^{-2}$ & FP - F          \\ \cline{2-4} \cline{8-11} 
            \multicolumn{1}{|l|}{}                                                                                              & \multicolumn{1}{l|}{is\_inverter}                   & \multicolumn{1}{l|}{bool}                 & \multicolumn{1}{l|}{}                         & \multicolumn{1}{l|}{\cellcolor[HTML]{EFEFEF}}                    &  &                                                                                         & buffer\_area                 & float             & \textmu m$^{-2}$ & FP - F          \\ \cline{2-4} \cline{8-11} 
            \multicolumn{1}{|l|}{}                                                                                              & \multicolumn{1}{l|}{is\_buffer}                     & \multicolumn{1}{l|}{bool}                 & \multicolumn{1}{l|}{}                         & \multicolumn{1}{l|}{\cellcolor[HTML]{EFEFEF}}                    &  &                                                                                         & inverter\_area               & float             & \textmu m$^{-2}$ & FP - F          \\ \cline{2-4} \cline{8-11} 
            \multicolumn{1}{|l|}{}                                                                                              & \multicolumn{1}{l|}{is\_filler}                     & \multicolumn{1}{l|}{bool}                 & \multicolumn{1}{l|}{}                         & \multicolumn{1}{l|}{\cellcolor[HTML]{EFEFEF}}                    &  &                                                                                         & filler\_area                 & float             & \textmu m$^{-2}$ & FP - F          \\ \cline{2-4} \cline{8-11} 
            \multicolumn{1}{|l|}{}                                                                                              & \multicolumn{1}{l|}{is\_diode}                      & \multicolumn{1}{l|}{bool}                 & \multicolumn{1}{l|}{}                         & \multicolumn{1}{l|}{\cellcolor[HTML]{EFEFEF}}                    &  &                                                                                         & tap\_cell\_area              & float             & \textmu m$^{-2}$ & FP - F          \\ \cline{2-4} \cline{8-11} 
            \multicolumn{1}{|l|}{}                                                                                              & \multicolumn{1}{l|}{drive\_strength}                & \multicolumn{1}{l|}{float}                & \multicolumn{1}{l|}{}                         & \multicolumn{1}{l|}{\cellcolor[HTML]{EFEFEF}}                    &  &                                                                                         & diode\_area                  & float             & \textmu m$^{-2}$ & FP - F          \\ \cline{2-4} \cline{8-11} 
            \multicolumn{1}{|l|}{}                                                                                              & \multicolumn{1}{l|}{input\_capacitance\_min}        & \multicolumn{1}{l|}{float}                & \multicolumn{1}{l|}{fF}                       & \multicolumn{1}{l|}{\cellcolor[HTML]{EFEFEF}}                    &  &                                                                                         & macro\_area                  & float             & \textmu m$^{-2}$ & FP - F          \\ \cline{2-4} \cline{8-11} 
            \multicolumn{1}{|l|}{}                                                                                              & \multicolumn{1}{l|}{input\_capacitance\_max}        & \multicolumn{1}{l|}{float}                & \multicolumn{1}{l|}{fF}                       & \multicolumn{1}{l|}{\cellcolor[HTML]{EFEFEF}}                    &  &                                                                                         & cell\_area                   & float             & \textmu m$^{-2}$ & FP - F          \\ \cline{2-4} \cline{8-11} 
            \multicolumn{1}{|l|}{}                                                                                              & \multicolumn{1}{l|}{output\_capacitance\_min}       & \multicolumn{1}{l|}{float}                & \multicolumn{1}{l|}{fF}                       & \multicolumn{1}{l|}{\cellcolor[HTML]{EFEFEF}}                    &  & \multirow{-10}{*}{\textbf{\begin{tabular}[c]{@{}l@{}}Area\\ Metrics\end{tabular}}}      & total\_area                  & float             & \textmu m$^{-2}$ & FP - F          \\ \cline{2-4} \cline{7-11} 
            \multicolumn{1}{|l|}{}                                                                                              & \multicolumn{1}{l|}{output\_capacitance\_max}       & \multicolumn{1}{l|}{float}                & \multicolumn{1}{l|}{fF}                       & \multicolumn{1}{l|}{\cellcolor[HTML]{EFEFEF}}                    &  &                                                                                         & combinational\_power         & float             & \textmu W        & FP - F          \\ \cline{2-4} \cline{8-11} 
            \multicolumn{1}{|l|}{}                                                                                              & \multicolumn{1}{l|}{leakage\_power\_min}            & \multicolumn{1}{l|}{float}                & \multicolumn{1}{l|}{\textmu W} & \multicolumn{1}{l|}{\cellcolor[HTML]{EFEFEF}}                    &  &                                                                                         & sequential\_power            & float             & \textmu W        & FP - F          \\ \cline{2-4} \cline{8-11} 
            \multicolumn{1}{|l|}{\multirow{-18}{*}{\textbf{\begin{tabular}[c]{@{}l@{}}Standard\\ Cell\end{tabular}}}}           & \multicolumn{1}{l|}{leakage\_power\_max}            & \multicolumn{1}{l|}{float}                & \multicolumn{1}{l|}{\textmu W} & \multicolumn{1}{l|}{\multirow{-18}{*}{\cellcolor[HTML]{EFEFEF}}} &  &                                                                                         & macro\_power                 & float             & \textmu W        & FP - F          \\ \cline{1-5} \cline{8-11} 
            \multicolumn{1}{|l|}{}                                                                                              & \multicolumn{1}{l|}{name}                           & \multicolumn{1}{l|}{string}               & \multicolumn{1}{l|}{}                         & \multicolumn{1}{l|}{FP - F}                                      &  &                                                                                         & internal\_power              & float             & \textmu W        & FP - F          \\ \cline{2-5} \cline{8-11} 
            \multicolumn{1}{|l|}{}                                                                                              & \multicolumn{1}{l|}{standard\_cell}                 & \multicolumn{1}{l|}{StandardCell}         & \multicolumn{1}{l|}{}                         & \multicolumn{1}{l|}{FP - F}                                      &  &                                                                                         & switching\_power             & float             & \textmu W        & FP - F          \\ \cline{2-5} \cline{8-11} 
            \multicolumn{1}{|l|}{}                                                                                              & \multicolumn{1}{l|}{x\_min}                         & \multicolumn{1}{l|}{float}                & \multicolumn{1}{l|}{\textmu m} & \multicolumn{1}{l|}{GP - F}                                      &  &                                                                                         & leakage\_power               & float             & \textmu W        & FP - F          \\ \cline{2-5} \cline{8-11} 
            \multicolumn{1}{|l|}{}                                                                                              & \multicolumn{1}{l|}{y\_min}                         & \multicolumn{1}{l|}{float}                & \multicolumn{1}{l|}{\textmu m} & \multicolumn{1}{l|}{GP - F}                                      &  & \multirow{-7}{*}{\textbf{\begin{tabular}[c]{@{}l@{}}Power\\ Metrics\end{tabular}}}      & total\_power                 & float             & \textmu W        & FP - F          \\ \cline{2-5} \cline{7-11} 
            \multicolumn{1}{|l|}{}                                                                                              & \multicolumn{1}{l|}{x\_max}                         & \multicolumn{1}{l|}{float}                & \multicolumn{1}{l|}{\textmu m} & \multicolumn{1}{l|}{GP - F}                                      &  &                                                                                         & total\_negative\_slack       & float             & ns                              & FP - F          \\ \cline{2-5} \cline{8-11} 
            \multicolumn{1}{|l|}{}                                                                                              & \multicolumn{1}{l|}{y\_max}                         & \multicolumn{1}{l|}{float}                & \multicolumn{1}{l|}{\textmu m} & \multicolumn{1}{l|}{GP - F}                                      &  &                                                                                         & worst\_slack                 & float             & ns                              & FP - F          \\ \cline{2-5} \cline{8-11} 
            \multicolumn{1}{|l|}{}                                                                                              & \multicolumn{1}{l|}{no\_of\_inputs}                 & \multicolumn{1}{l|}{int}                  & \multicolumn{1}{l|}{}                         & \multicolumn{1}{l|}{FP - F}                                      &  &                                                                                         & worst\_arrival\_time         & float             & ns                              & FP - F          \\ \cline{2-5} \cline{8-11} 
            \multicolumn{1}{|l|}{}                                                                                              & \multicolumn{1}{l|}{no\_of\_outputs}                & \multicolumn{1}{l|}{int}                  & \multicolumn{1}{l|}{}                         & \multicolumn{1}{l|}{FP - F}                                      &  &                                                                                         & worst\_required\_time        & float             & ns                              & FP - F          \\ \cline{2-5} \cline{8-11} 
            \multicolumn{1}{|l|}{}                                                                                              & \multicolumn{1}{l|}{internal\_power}                & \multicolumn{1}{l|}{float}                & \multicolumn{1}{l|}{\textmu W} & \multicolumn{1}{l|}{FP - F}                                      &  &                                                                                         & critical\_path\_startpoint   & string            &                                 & FP - F          \\ \cline{2-5} \cline{8-11} 
            \multicolumn{1}{|l|}{}                                                                                              & \multicolumn{1}{l|}{switching\_power}               & \multicolumn{1}{l|}{float}                & \multicolumn{1}{l|}{\textmu W} & \multicolumn{1}{l|}{FP - F}                                      &  &                                                                                         & critical\_path\_endpoint     & string            &                                 & FP - F          \\ \cline{2-5} \cline{8-11} 
            \multicolumn{1}{|l|}{}                                                                                              & \multicolumn{1}{l|}{leakage\_power}                 & \multicolumn{1}{l|}{float}                & \multicolumn{1}{l|}{\textmu W} & \multicolumn{1}{l|}{FP - F}                                      &  &                                                                                         & no\_of\_endpoints            & int               &                                 & FP - F          \\ \cline{2-5} \cline{8-11} 
            \multicolumn{1}{|l|}{}                                                                                              & \multicolumn{1}{l|}{total\_power}                   & \multicolumn{1}{l|}{float}                & \multicolumn{1}{l|}{\textmu W} & \multicolumn{1}{l|}{FP - F}                                      &  & \multirow{-8}{*}{\textbf{\begin{tabular}[c]{@{}l@{}}Timing\\ Metrics\end{tabular}}}     & no\_of\_violating\_endpoints & int               &                                 & FP - F          \\ \cline{2-5} \cline{7-11} 
            \multicolumn{1}{|l|}{}                                                                                              & \multicolumn{1}{l|}{ir\_drop\_vdd}                  & \multicolumn{1}{l|}{float}                & \multicolumn{1}{l|}{mV}                       & \multicolumn{1}{l|}{DR - F}                                      &  &                                                                                         & rudy\_net                    & scalar map        &                                 & DR - F          \\ \cline{2-5} \cline{8-11} 
            \multicolumn{1}{|l|}{\multirow{-14}{*}{\textbf{Gate}}}                                                              & \multicolumn{1}{l|}{ir\_drop\_vss}                  & \multicolumn{1}{l|}{float}                & \multicolumn{1}{l|}{mV}                       & \multicolumn{1}{l|}{DR - F}                                      &  &                                                                                         & rudy\_net\_long              & scalar map        &                                 & DR - F          \\ \cline{1-5} \cline{8-11} 
                                                                                                                                &                                                     &                                           &                                               &                                                                  &  &                                                                                         & rudy\_net\_short             & scalar map        &                                 & DR - F          \\ \cline{8-11} 
                                                                                                                                &                                                     &                                           &                                               &                                                                  &  & \multirow{-4}{*}{\textbf{\begin{tabular}[c]{@{}l@{}}Routability\\ Metric\end{tabular}}} & rudy\_pin                    & scalar map        &                                 & DR - F          \\ \cline{7-11} 
            \end{tabular}
    }
    \end{center}
    \vspace{-0.2in}
\end{table}

\subsection{Netlist}
\label{subsec:netlist_graph}

\begin{figure}[!h]
    \begin{center}
        \includegraphics[width=0.6\columnwidth]{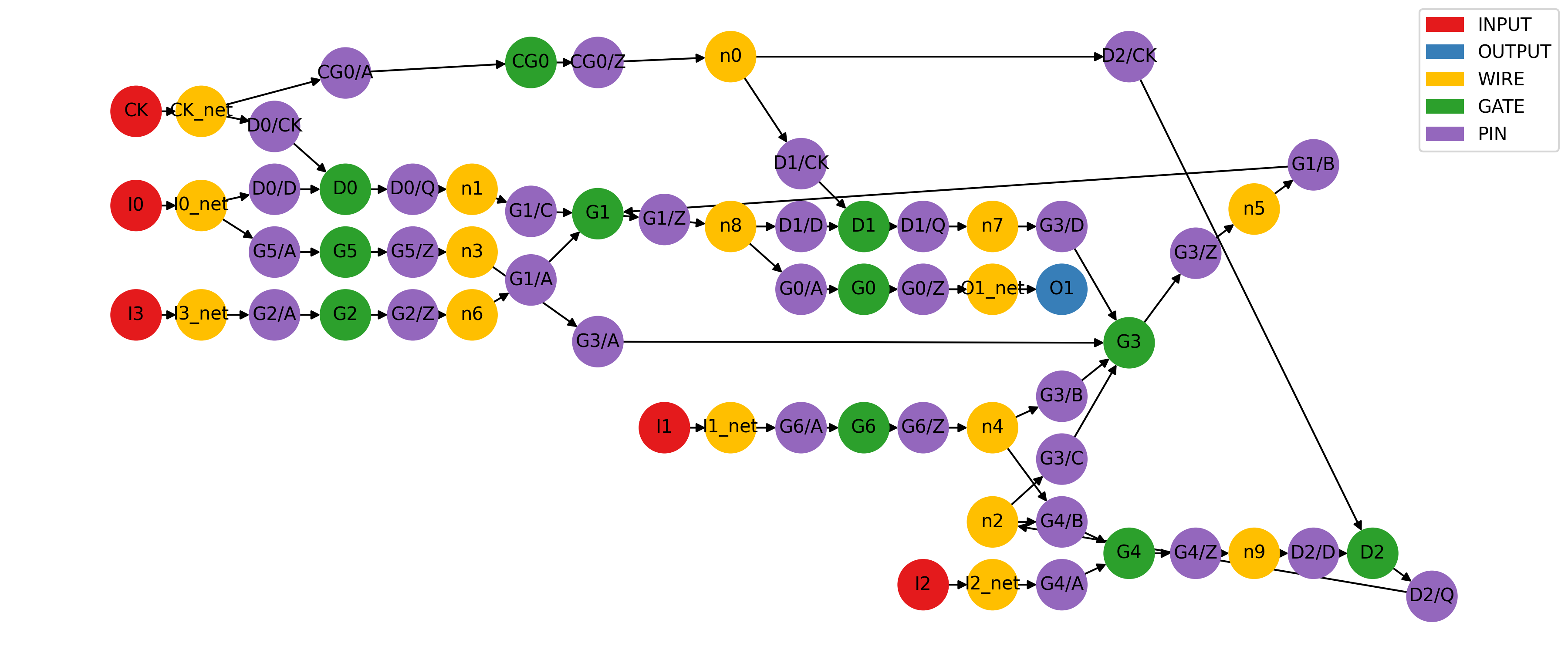}
    \end{center}
    % \end{subfigure}
    \caption{Sample circuit represented as a netlist graph $NLG$.}
    \label{fig:netlist_graph}
    % \vspace{-0.05in}
\end{figure}

LEF and DEF files provide structural information on logic gates, input/output ports, standard cells, coordinates of the placed gates and macros, and routing connectivity. The information is captured in the netlist entity, which represents the complete logical and physical composition of the circuit as a heterogeneous graph described by $NLG = (V, E)$, where nodes $v \in V$ correspond to gates ($G$), pins ($P$), nets ($N$), and I/O ports ($IO$), and edges $e \in E$ define the logical and physical connections between nodes. An example of a circuit and the corresponding netlist graph is shown in Fig.~\ref{fig:netlist_graph}.

Gate entities represent placed instances of standard cells and serve as the primary functional components of the netlist graph. Each gate entity is associated with a technology-specific standard-cell type, with cell geometry and library information derived from LEF files.
Pin entities represent the input and output terminals of gate entities and capture timing and electrical-related attributes including slew, slack, and load capacitances, obtained from static timing analysis (STA) reports and SPEF files.
I/O port entities represent the external interface of the circuit or circuit-block and specify signal direction and physical location.
For net entities, structural attributes are extracted from DEF files, while parameters relating to parasitic impedance are obtained from SPEF files, together modeling physical routing and electrical parasitics.
Prior to placement, net entities represent only logical connections among gate pins.
After placement and routing, the physical geometry of each net is established, which results in semi-fixed or fixed net locations and enables the computation of metrics including half-perimeter wire length (HPWL) after placement and total true wire length after routing.

The netlist entity also maintains relationships with primary analysis entities. The PDN entity models the spatial and electrical properties of the power grid, while clock tree and timing path entities describe clock propagation and timing dependencies, respectively.
The relationships between the netlist and analysis entities allow the schema to represent both circuit structure and performance behavior consistently across the design hierarchy.

\begin{figure}[!h]
    \begin{center}
    \begin{subfigure}[t]{0.16\columnwidth}
        \begin{center}
            \fbox{\includegraphics[width=\columnwidth]{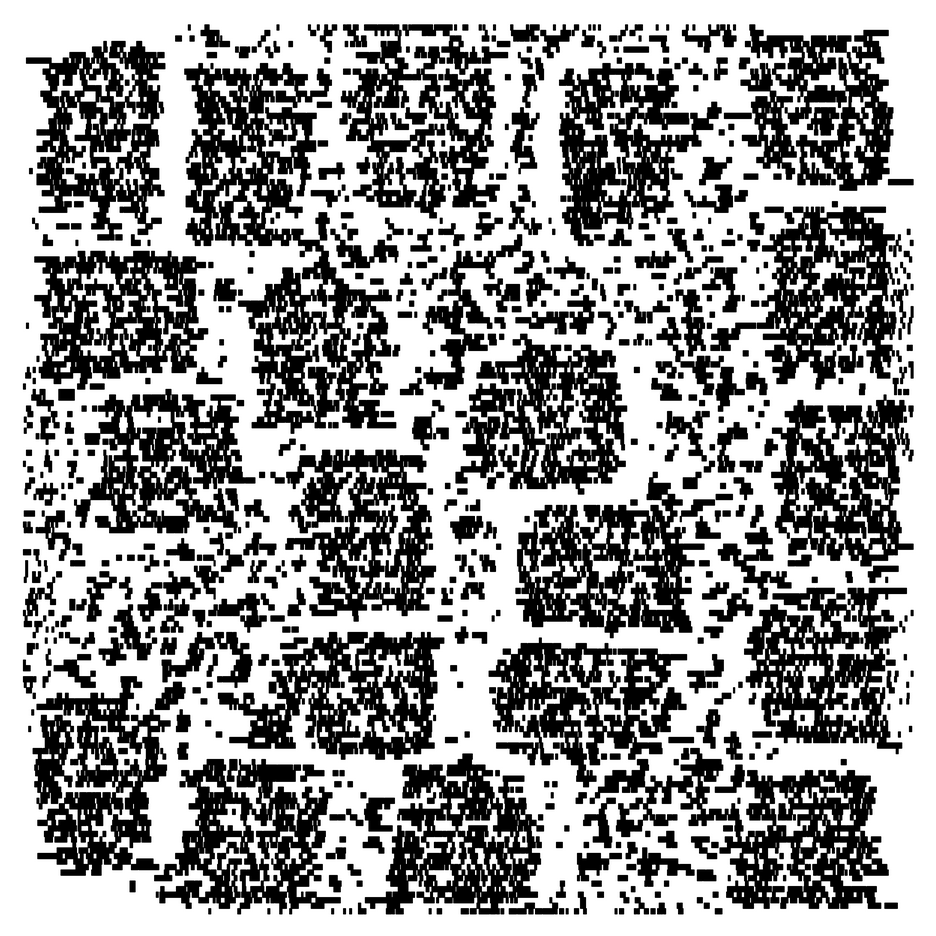}}
            \caption{}
            % \label{fig:}
        \end{center}
    \end{subfigure}
    %\hfill
    \begin{subfigure}[t]{0.16\columnwidth}
        \begin{center}
            \fbox{\includegraphics[width=\columnwidth]{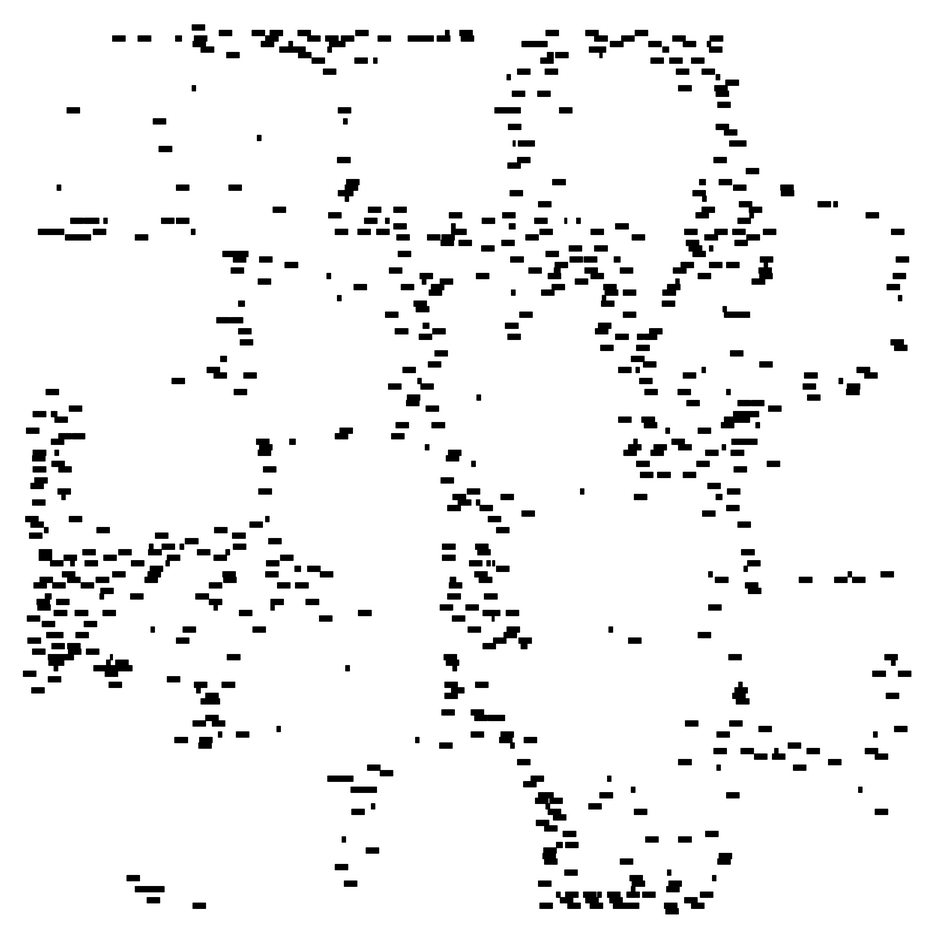}}
            \caption{}
            % \label{fig:}
        \end{center}
    \end{subfigure}
    \begin{subfigure}[t]{0.16\columnwidth}
        \begin{center}
            \fbox{\includegraphics[width=\columnwidth]{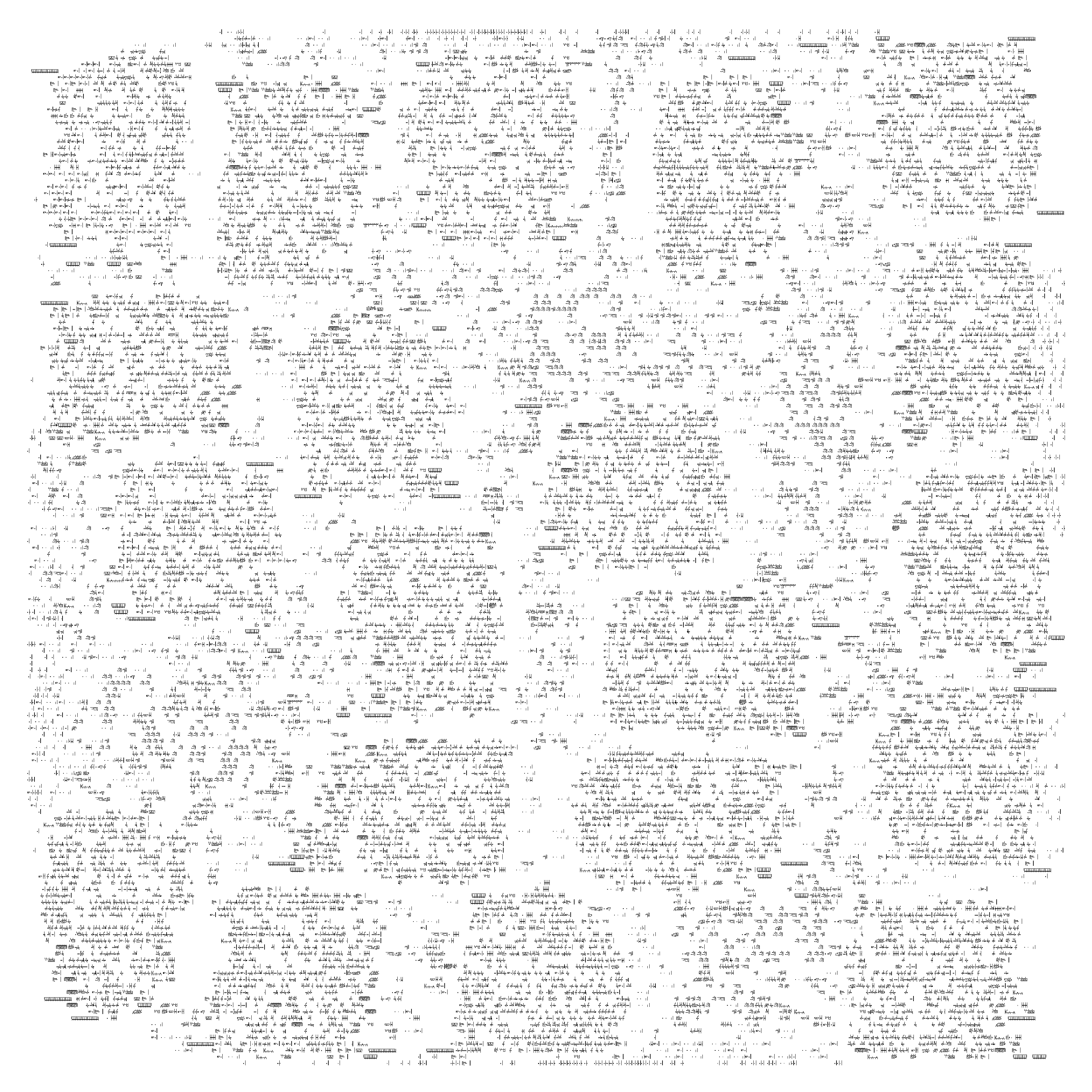}}
            \caption{}
            % \label{fig:}
        \end{center}
    \end{subfigure}
    \begin{subfigure}[t]{0.16\columnwidth}
        \begin{center}
            \fbox{\includegraphics[width=\columnwidth]{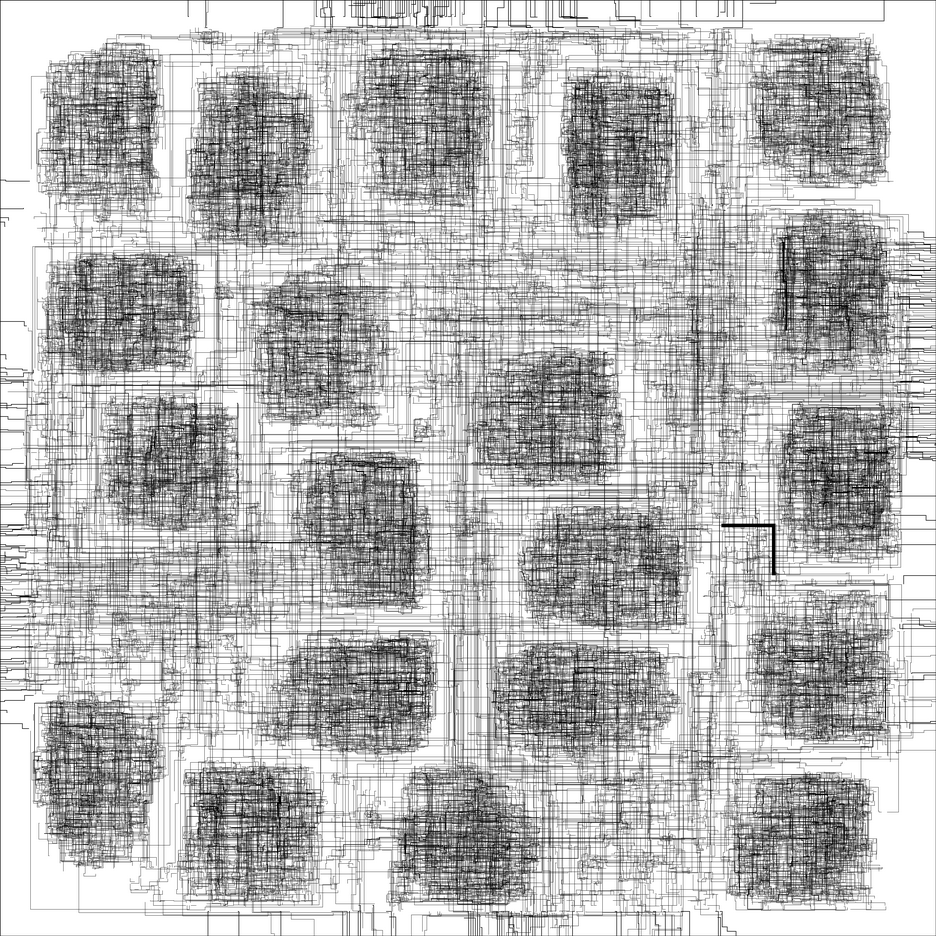}}
            \caption{}
            % \label{fig:}
        \end{center}
    \end{subfigure}
    \begin{subfigure}[t]{0.16\columnwidth}
        \begin{center}
            \fbox{\includegraphics[width=\columnwidth]{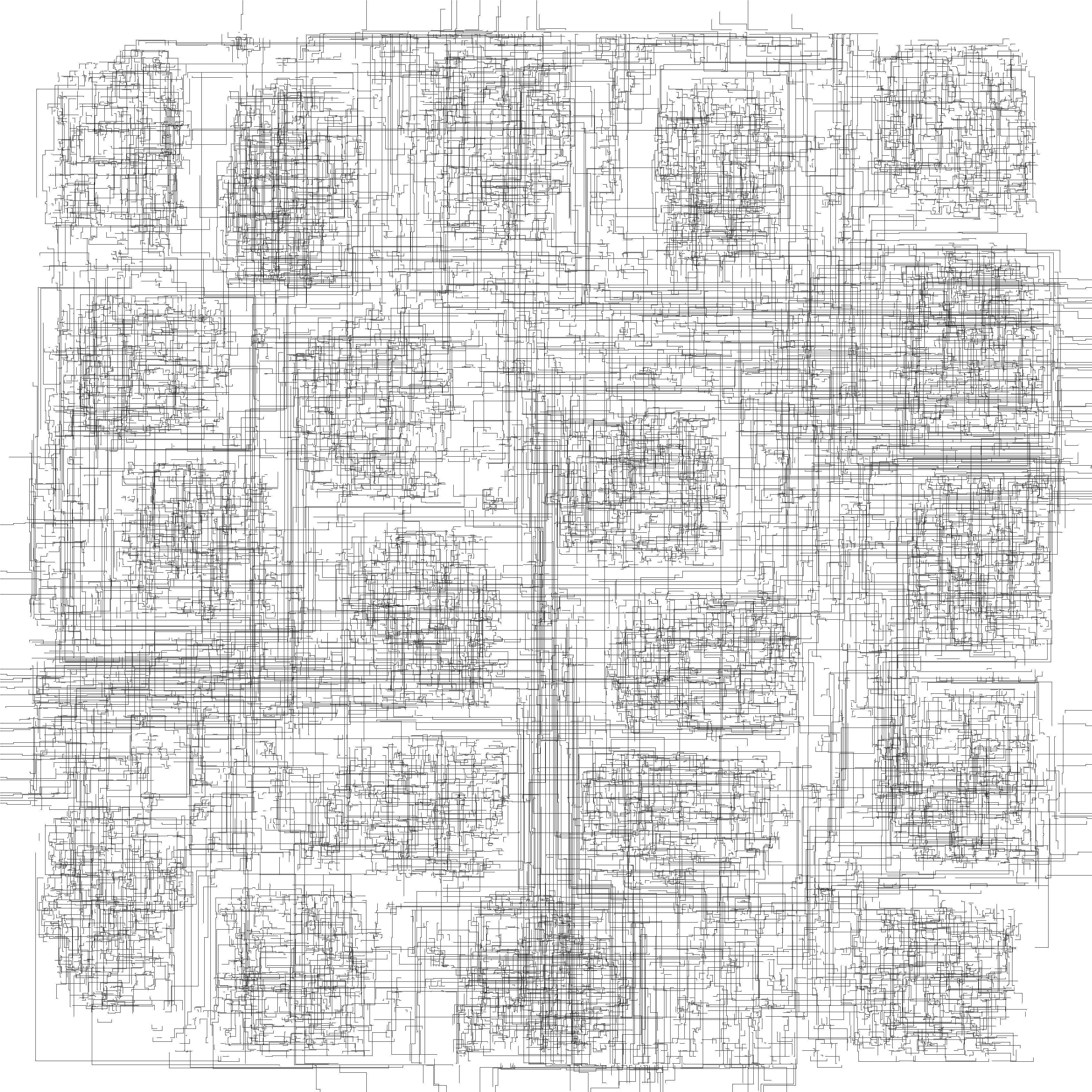}}
            \caption{}
            % \label{fig:}
        \end{center}
    \end{subfigure}
    \begin{subfigure}[t]{0.16\columnwidth}
        \begin{center}
            \fbox{\includegraphics[width=\columnwidth]{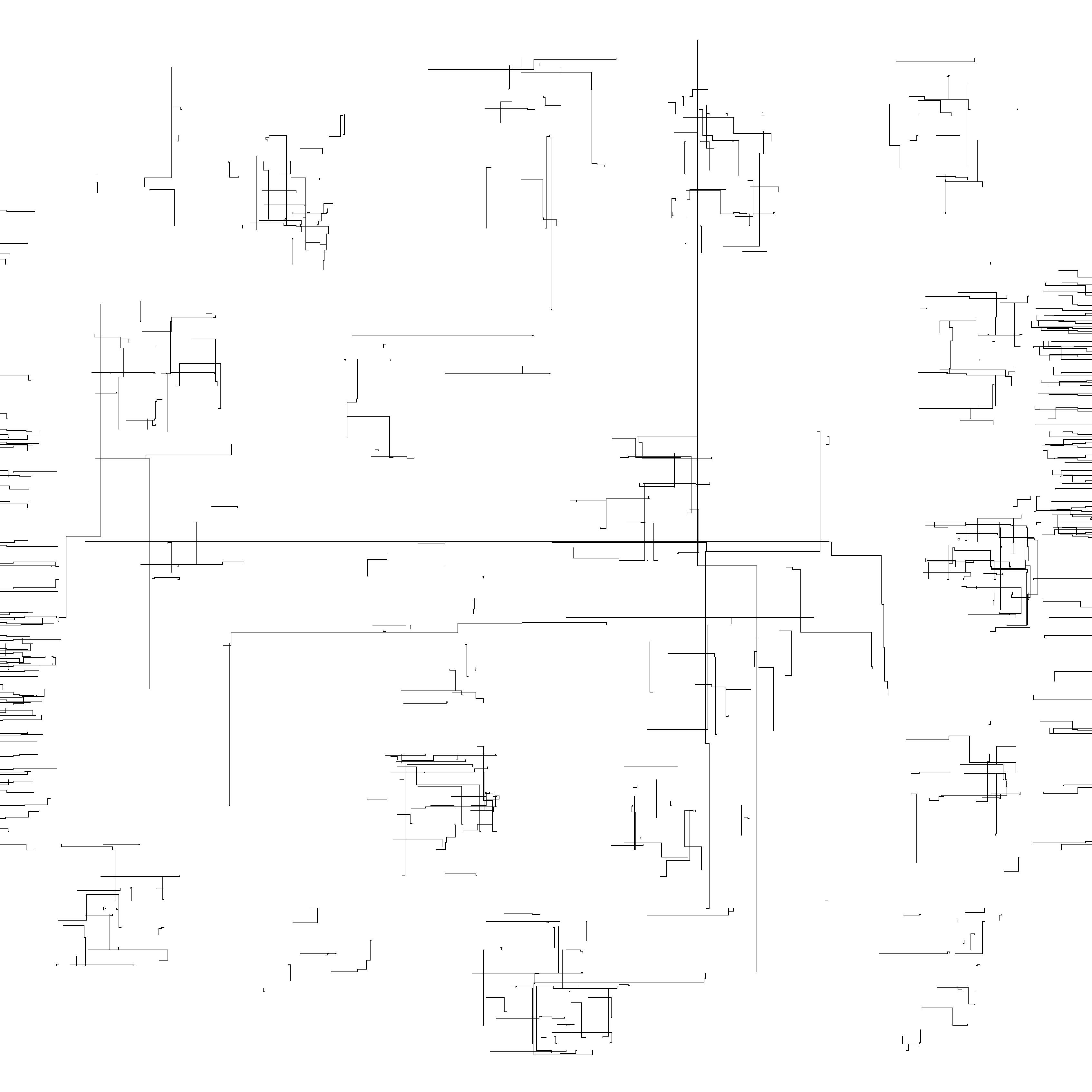}}
            \caption{}
            % \label{fig:}
        \end{center}
    \end{subfigure}
    \caption{Image attributes associated with the netlist entity for the \textit{ac97\_ctrl} circuit implemented in the Nangate 45 nm technology node, with each image represented as binary spatial maps capturing key physical aspects of the circuit layout at the final design stage. Subfigures depict (a) combinational cell placement, (b) sequential cell placement, (c) pin locations, (d) overall routing, (e) metal~1 routing, and (f) metal~5 routing.}
    \label{fig:netlist_images}
    \end{center}
\end{figure}

In addition to graph-based representations, the netlist entity includes image attributes that capture spatial views of the physical layout as two-dimensional binary maps.
The images provide visualization of cell placement (overall, combinational, sequential, and filler), pin locations, and detailed routing, including routing on each metal layer.
The image attributes associated with the netlist entity for the \textit{ac97\_ctrl} circuit implemented in the Nangate 45 nm (NG45) technology node are illustrated in Fig.~\ref{fig:netlist_images}.
Each pixel of the images corresponds to a fixed physical region of the layout, represented as a square with side lengths equal to the minimum width of metal~1 for the given PDK. Defining each pixel length with respect to the width of metal 1 ensures consistent physical scaling across technology nodes. The image resolution is defined as
\begin{equation}
(\text{Resolution}_x, \text{Resolution}_y) =
\left(\frac{L}{w_{M1}}, \frac{W}{w_{M1}}\right),
\label{eqn:resolution}
\end{equation}
where $L$ and $W$ denote the length and width of the circuit layout, respectively, and $w_{M1}$ denotes the minimum width of metal~1.

\subsection{Clock Networks}
\label{subsec:clock_network_graph}

\begin{figure}[!h]
    \begin{center}
    %\hfill
    \begin{subfigure}[t]{0.32\columnwidth}
        \begin{center}
            \fbox{\includegraphics[width=0.485\columnwidth]{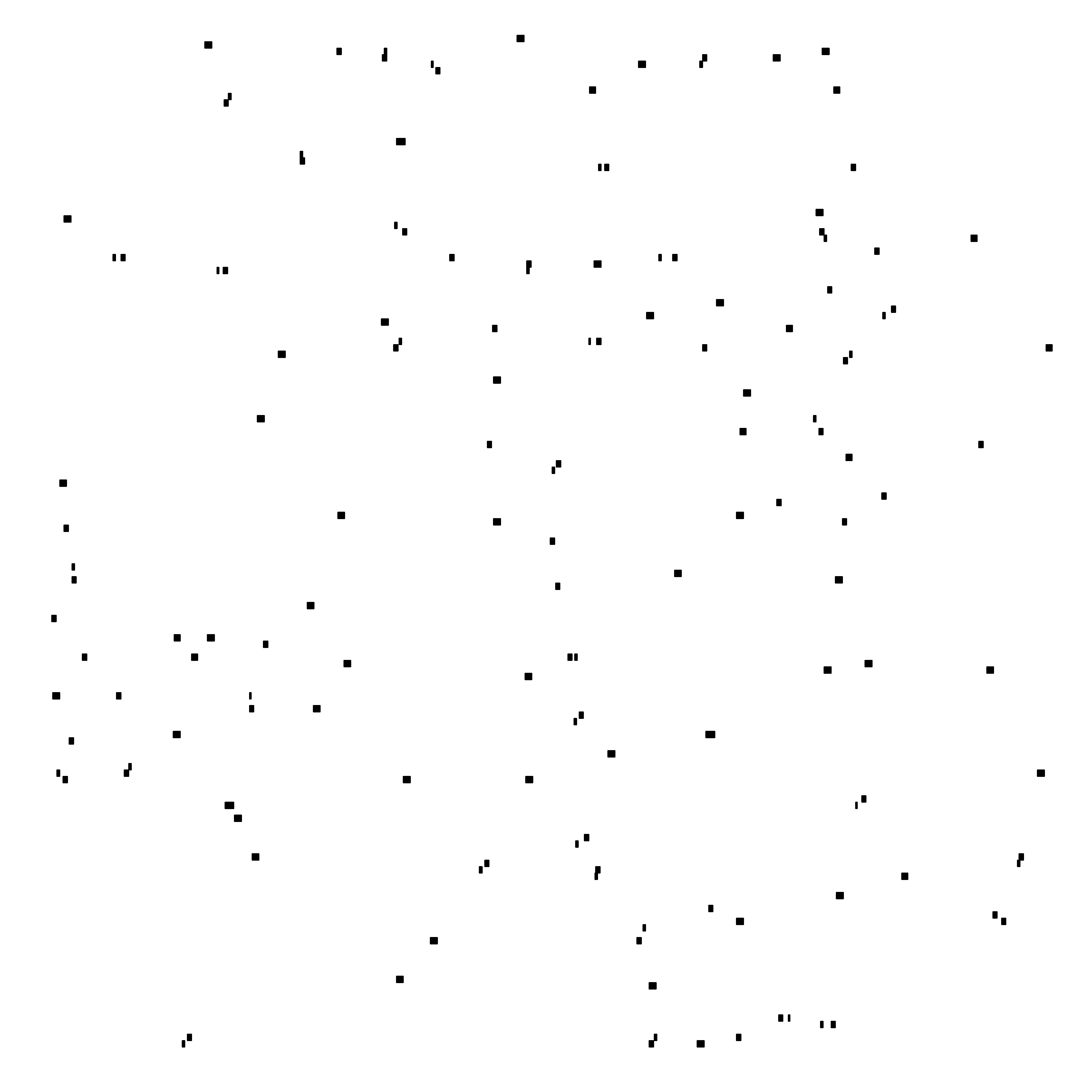}}
            \caption{}
        \end{center}
    \end{subfigure}
    \begin{subfigure}[t]{0.32\columnwidth}
        \begin{center}
            \fbox{\includegraphics[width=0.485\columnwidth]{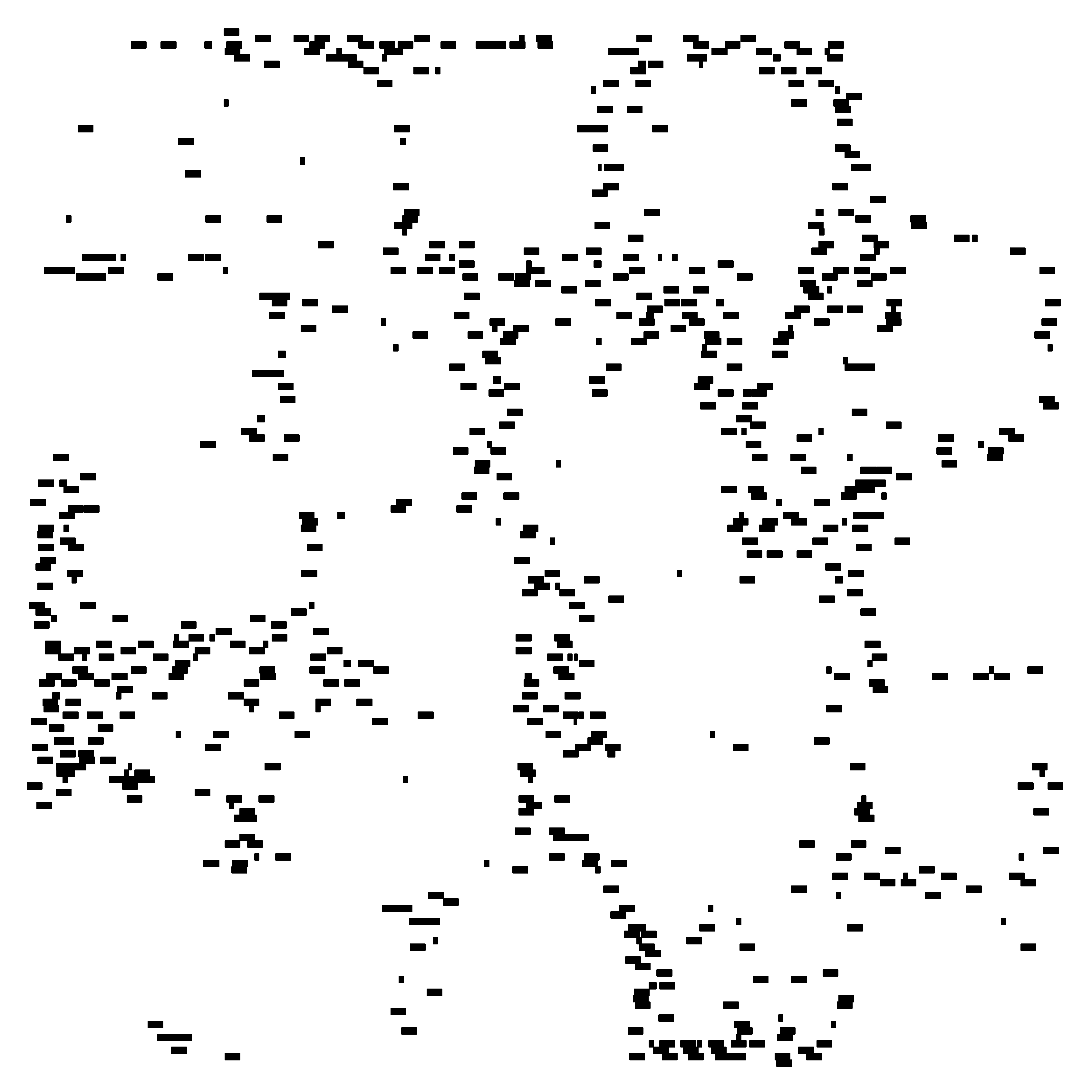}}
            \caption{}
        \end{center}
    \end{subfigure}
    \begin{subfigure}[t]{0.32\columnwidth}
        \begin{center}
            \fbox{\includegraphics[width=0.485\columnwidth]{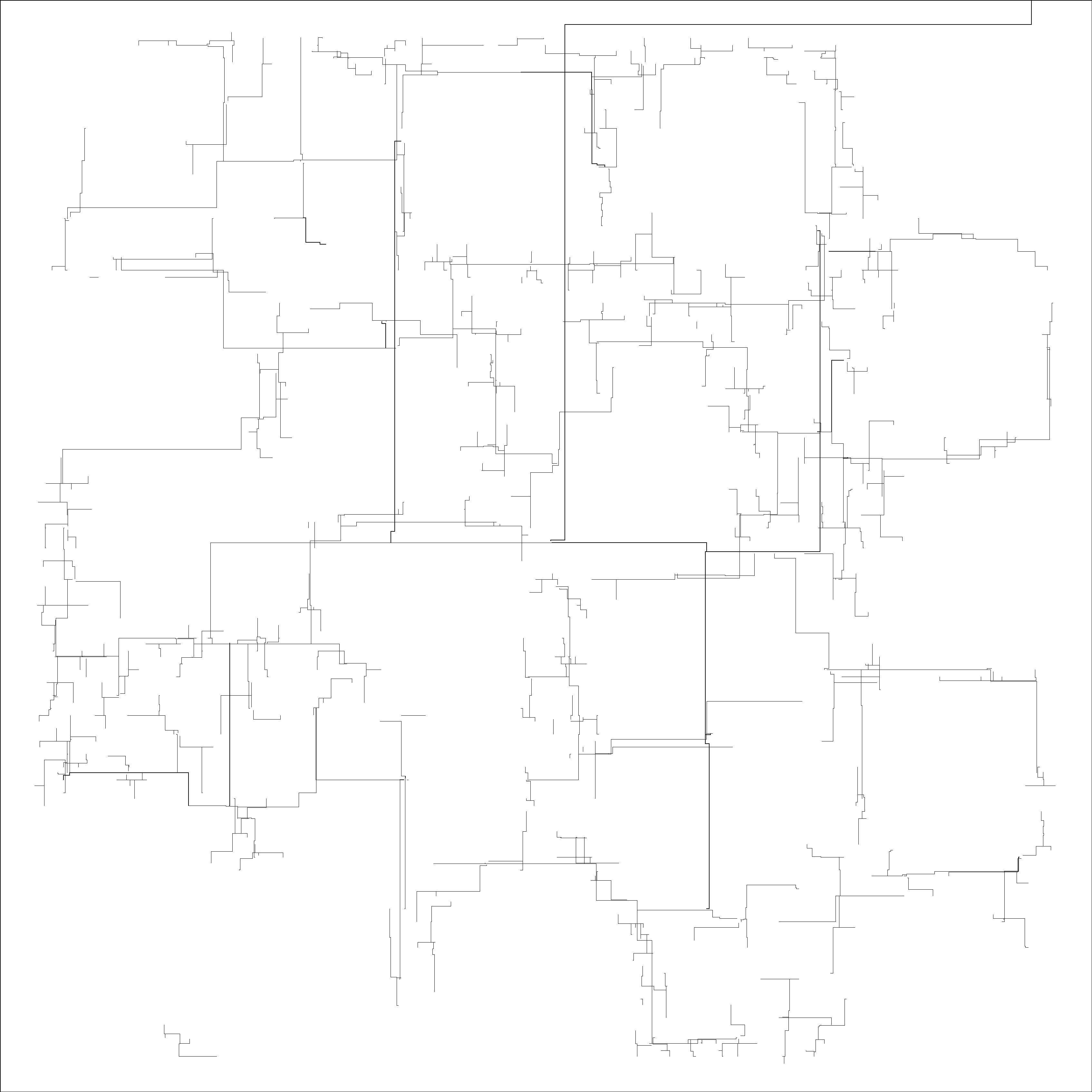}}
            \caption{}
        \end{center}
    \end{subfigure}
    \caption{Binary spatial maps associated with the clock network entity for the \textit{ac97\_ctrl} circuit implemented in the Nangate 45 nm technology node after completing the final design stage. Subfigures depict (a) clock buffer placement, (b) sequential cell placement, and (c) clock routes.}
    \label{fig:clock_network_images}
    \end{center}
    % \vspace{-0.2in}
\end{figure}

Clock networks are substructures derived from the netlist and capture the distribution of the clock signal from source node to all sequential elements. Clock network graphs ($CNG$), similar in structure to netlist graphs, are composed of nodes $v \in V$ representing input/output ports ($IO$), gates ($G$), pins ($P$), and nets ($N$), and edges $e \in E$ denoting logical and physical connections between components.
In addition to the graph representation, the spatial layout of the clock tree is stored as a binary mask of the completed placement and routing of the clock network, as shown in Fig.~\ref{fig:clock_network_images}. The images capture clock buffer locations, locations of sequential elements, and clock routing across multiple metal layers. All clock network image attributes utilize the same resolution as defined by (\ref{eqn:resolution}), which ensures consistent physical scaling and alignment with netlist image representations.

\subsection{Power Delivery Networks}
\label{subsec:pdn}

The PDN entity represents the physical and electrical network responsible for distributing power across the integrated circuit (IC). The PDN entity captures the structure of the power grid through metal routing information of both the supply and ground networks (VDD and VSS), with geometry and metal-layer assignments extracted from the DEF and LEF technology files.
Results from electrical analysis, including metrics evaluating IR-drop and electromigration (EM), are obtained from PDNSim within the OpenROAD flow.

The PDN entity includes two classes of images: images of physical routing and heatmaps of analyzed circuit data. Images of physical routing are represented as two-dimensional binary maps that capture VDD metal routing, VSS metal routing, and locations of power sources, and follow the same spatial resolution as defined by (\ref{eqn:resolution}), which ensures consistent physical alignment with other layout-level image entities. In contrast, heatmaps of IR-drop and EM are stored as scalar-valued maps that represent spatially aggregated electrical quantities. For image entities stored as scalar maps, each pixel corresponds to a square physical region with side length equal to $k w_{M1}$, where $w_{M1}$ denotes the minimum width of metal~1 and $k$ is a downsampling factor that controls the level of spatial aggregation. The resolution, with the downsampling factor accounted for, is defined by
\begin{equation}
(\text{Resolution}_x^{\text{scalar}}, \text{Resolution}_y^{\text{scalar}}) =
\left(\frac{L}{kw_{M1}}, \frac{W}{kw_{M1}}\right),
\label{eqn:scalar_resolution}
\end{equation}
where $L$ and $W$ denote the length and width of the circuit layout, respectively. The coarser resolution reflects the sampling granularity provided by the analysis performed by PDNSim while preserving spatial agreement with the underlying physical circuit.

\vspace{0.05in}
\begin{figure}[!h]
    \begin{center}
    \begin{subfigure}[t]{0.195\linewidth}
        \begin{center}
            \fbox{\includegraphics[width=0.81\columnwidth]{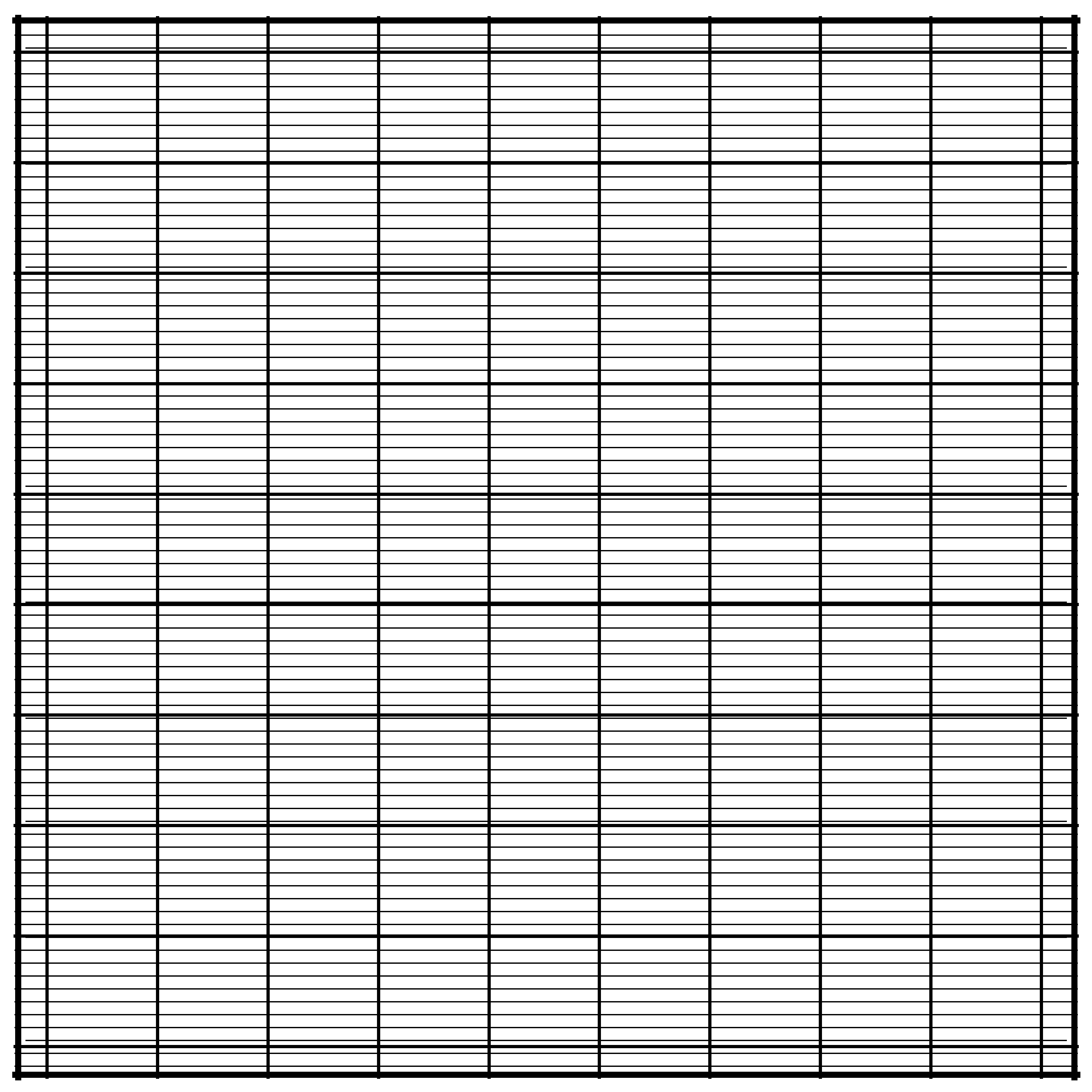}}
            \caption{}
            \label{fig:}
        \end{center}
    \end{subfigure}
    \begin{subfigure}[t]{0.195\linewidth}
        \begin{center}
            \fbox{\includegraphics[width=0.81\columnwidth]{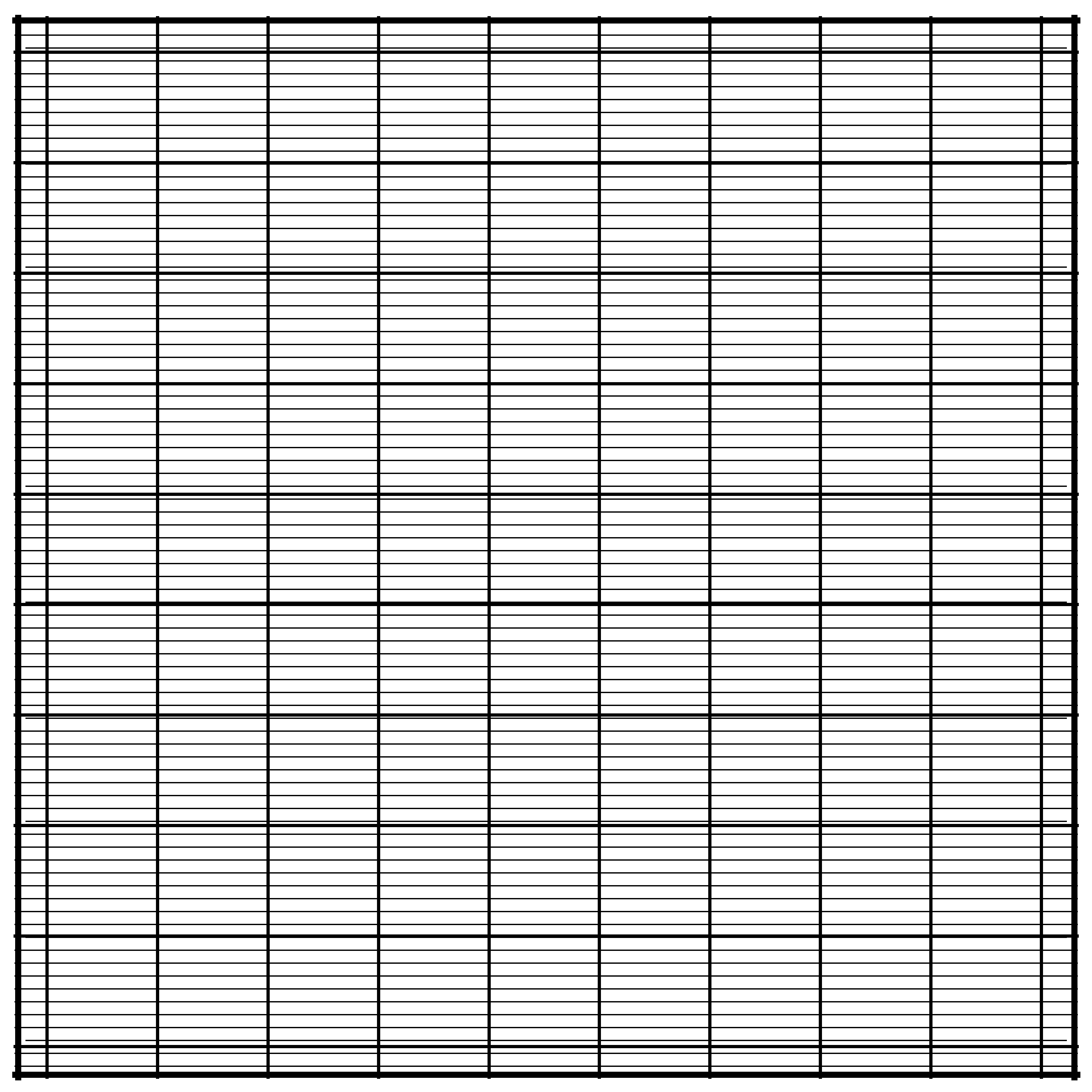}}
            \caption{}
            \label{fig:}
        \end{center}
    \end{subfigure}
    \begin{subfigure}[t]{0.195\linewidth}
        \begin{center}
            \fbox{\includegraphics[width=0.81\columnwidth]{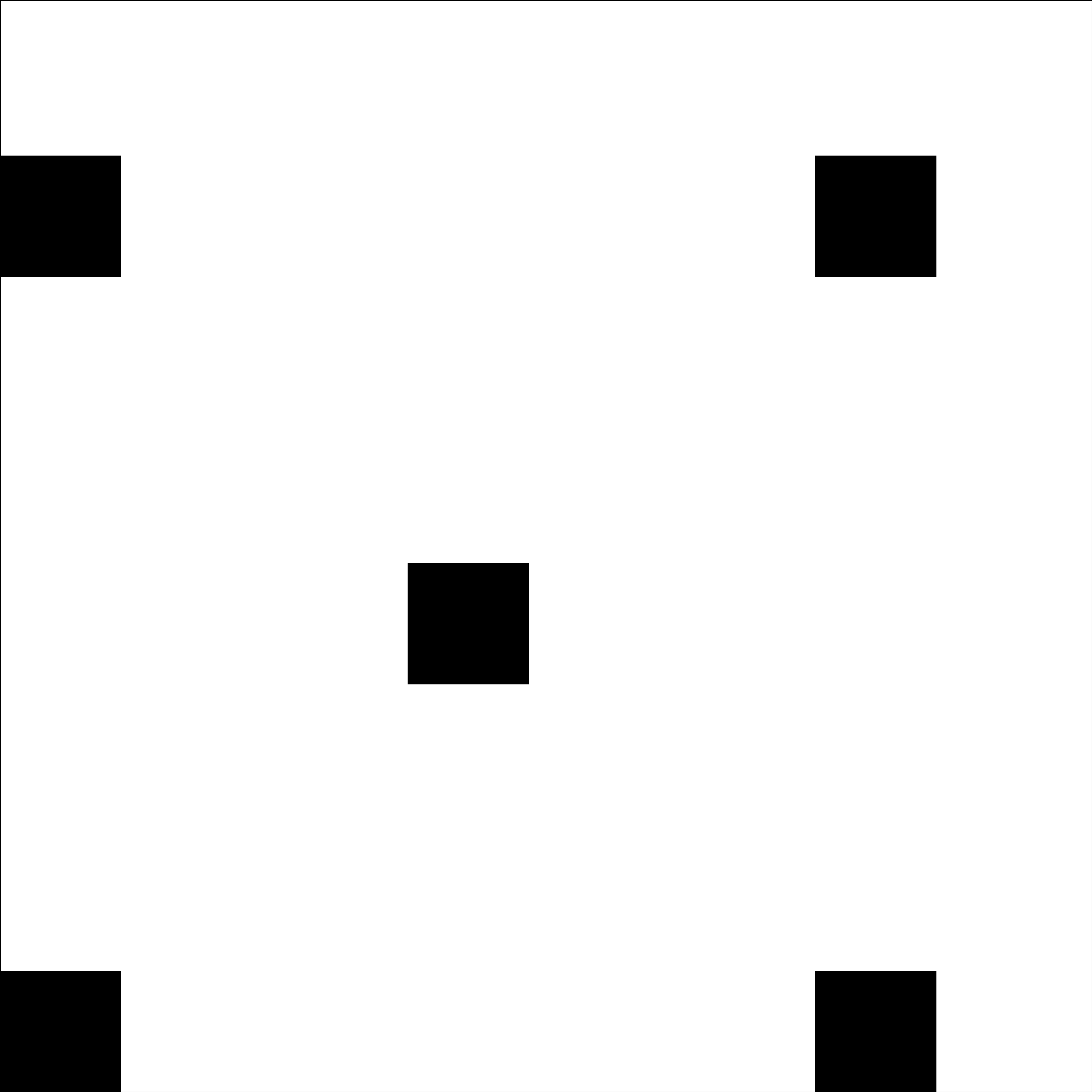}}
            \caption{}
            \label{fig:}
        \end{center}
    \end{subfigure}
    \begin{subfigure}[t]{0.195\linewidth}
        \begin{center}
            \fbox{\includegraphics[width=0.81\columnwidth]{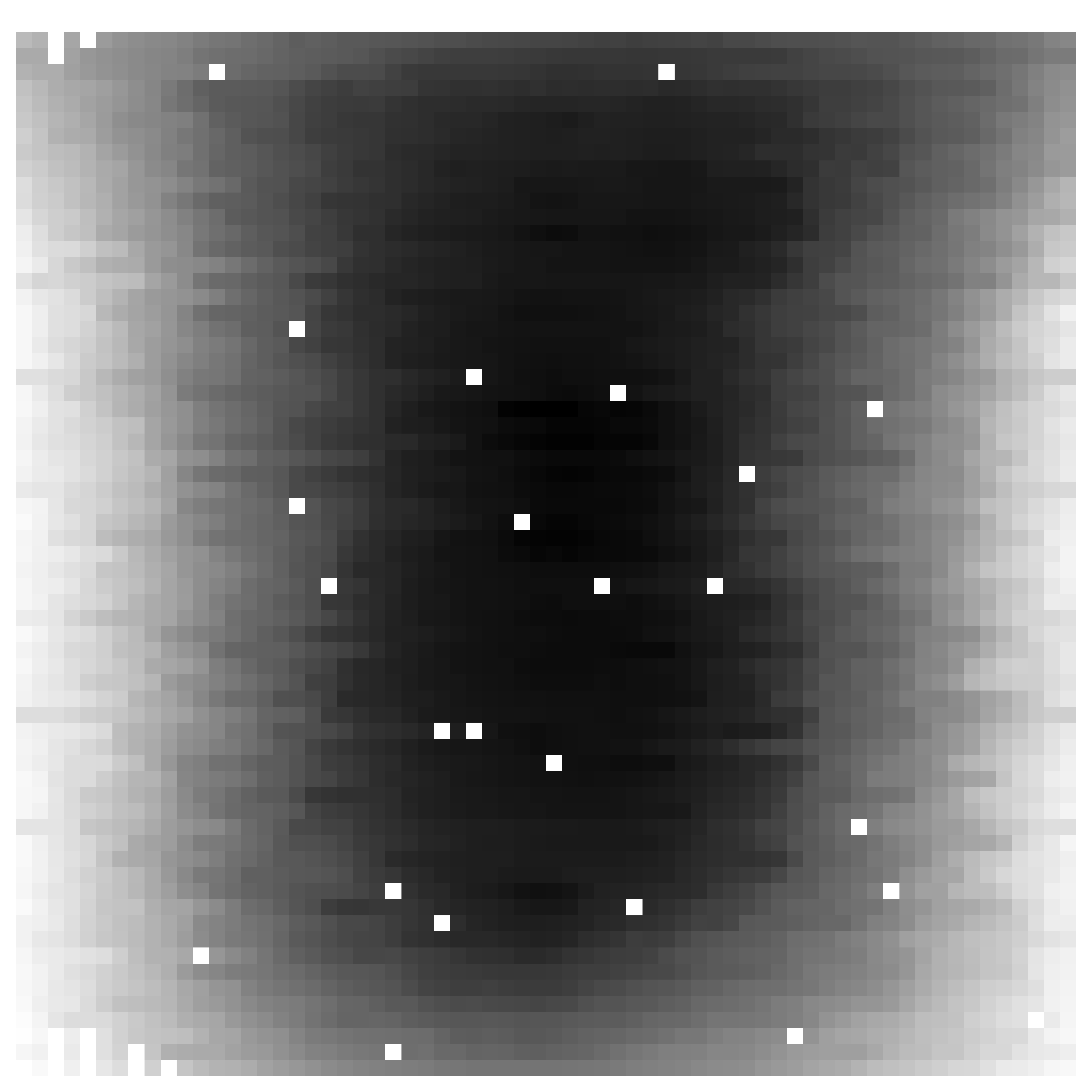}}
            \caption{}
            \label{fig:}
        \end{center}
    \end{subfigure}
    \begin{subfigure}[t]{0.195\linewidth}
        \begin{center}
            \fbox{\includegraphics[width=0.81\columnwidth]{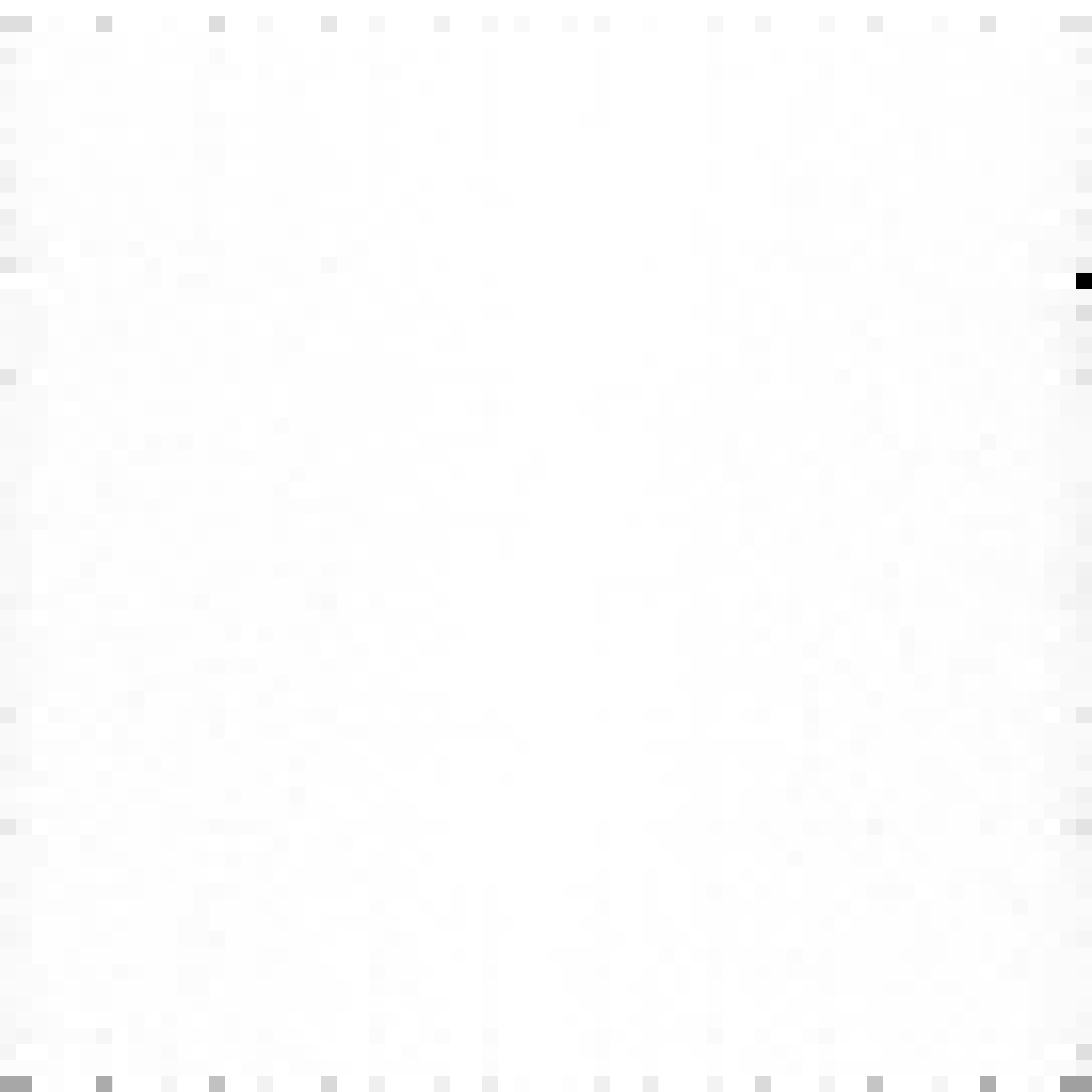}}
            \caption{}
            \label{fig:}
        \end{center}
    \end{subfigure}
    \caption{Image attributes associated with the power delivery network (PDN) entity after the completion of the final design stage for the \textit{ac97\_ctrl} circuit implemented using the Nangate 45 nm technology node. Subfigures depict (a) VSS routing represented as a binary spatial map, (b) VDD routing represented as a binary spatial map, (c) power source locations for PDNsim represented as a binary spatial map, (d) IR-drop represented as a scalar heatmap, and (e) electromigration (EM) represented as a scalar heatmap. For scalar maps, each pixel corresponds to an aggregated physical region of size $k \times w_{M1}$ by $k \times w_{M1}$, where $k=50$ and $w_{M1}$ denotes the minimum width of metal~1.}
    \label{fig:pdn_heatmaps}
    \end{center}
    \vspace{-0.1in}
\end{figure}

\subsection{Timing Paths}
\label{subsec:timing_path_graph}

\begin{figure}[!h]
    % \vspace{-0.05in}
    \begin{center}
        \includegraphics[width=0.45\columnwidth]{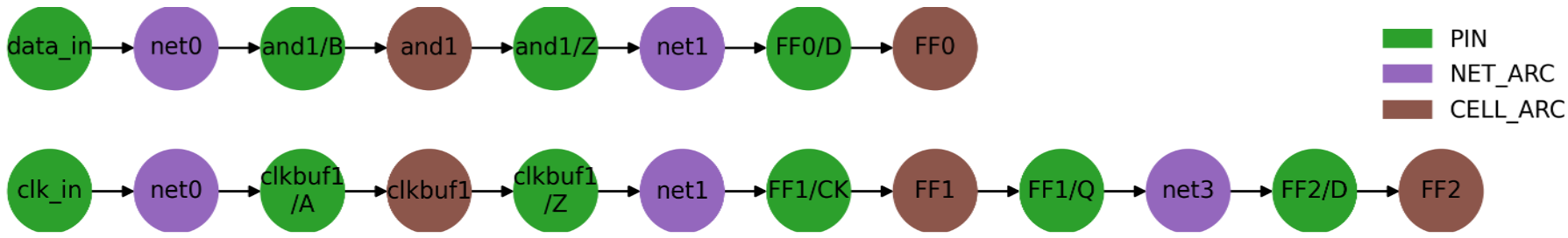}
    \end{center}
    \caption{Timing path graphs $TPG$ extracted from the netlist graph $NLG$ of the circuit shown in Fig. \ref{fig:netlist_graph}.}
    \label{fig:timing_path_graph}
    \vspace{-0.2in}
\end{figure}

The timing path entity represents individual signal paths extracted from static timing analysis (STA).
Each timing path is modeled as a directed graph represented by \( TPG = (V, E) \), where nodes \( v \in V \) correspond to pins (\(P\)), I/O ports (\(IO\)), cell timing arcs (\(GA\)), and net timing arcs (\(NA\)), and edges \( e \in E \) represent signal propagation between two successive elements on the timing path.
Cell timing arcs model delay and slew propagation through logic elements between an input pin and an output pin of a gate, capturing the gate delay and the output transition produced by the cell. Net timing arcs capture interconnect propagation between a driving pin and one or more receiving pins, modeling both the interconnect delay and the resulting slew at the sink pin after accounting for the $RC$ delay of the wire.
The dichotomy between cell and net timing arcs reflects the structure of STA reports, in which  contributions from the gates and interconnects are evaluated and reported as distinct components of a timing path.
A subset of the timing paths annotated with STA-derived data for the circuit shown in Figure~\ref{fig:netlist_graph} are depicted in Figure~\ref{fig:timing_path_graph}.

\subsection{Routability Metrics}
\label{subsec:routability_metrics}

Routability metrics quantify routing demand and congestion characteristics that affect the ability of a tool to complete routing without violations. The \textit{RoutabilityMetrics} entity captures scalar-valued spatial maps derived from Rectangular Uniform wire DensitY (RUDY)~\cite{spindler2007fast} scores, which provide an estimate of routing demand by uniformly distributing net bounding-box contributions across the layout.
For a net $n$ with bounding-box dimensions $(w_n, h_n)$, the net-based and pin-based RUDY contributions to a spatial tile $(i,j)$ are defined as, respectively,

\begin{equation}
\begin{aligned}
\mathrm{RUDY}_n(i,j) &= \frac{w_n + h_n}{w_n \cdot h_n} \cdot A_{n,i,j}\text{, and}
\end{aligned}
\end{equation}
\begin{equation}
\begin{aligned}
\mathrm{RUDY}_{\text{pin}}(i,j) &= \frac{w_n + h_n}{w_n \cdot h_n},
\end{aligned}
\end{equation}
where $A_{n,i,j}$ denotes the area of overlap between the bounding box of net $n$ and tile $(i,j)$.
Each tile $(i,j)$ corresponds to a spatial region of size $k w_{M1} \times k w_{M1}$ as defined by the scalar-map resolution given by (\ref{eqn:scalar_resolution}).
Values for each tile $(i,j)$ are obtained by summing contributions from all overlapping nets or pins within the tile.
Additional RUDY-based variants, namely long-range and short-range RUDY, separate routing demand based on whether nets span multiple tiles, which enables analysis of routing demand across different spatial scales. Long-range and short-range RUDY are defined as, respectively,

\begin{equation}
\mathrm{RUDY}_{\text{long}}(i,j) = \sum_{n \in \mathcal{N}_{\text{long}}} \frac{w_n + h_n}{w_n \cdot h_n} \cdot A_{n,i,j},
\end{equation}

\begin{equation}
\mathrm{RUDY}_{\text{short}}(i,j) = \sum_{n \in \mathcal{N}_{\text{short}}} \frac{w_n + h_n}{w_n \cdot h_n} \cdot A_{n,i,j}.
\end{equation}
where, $\mathcal{N}_{\text{long}}$ and $\mathcal{N}_{\text{short}}$ denote the sets of long-range and short-range nets, respectively, where long-range nets span at least two tiles and short-range nets are confined to a single tile.

% The schema includes net-based and pin-based RUDY maps, along with long-range and short-range variants that separate contributions based on whether nets span multiple tiles, which enables analysis of routing demand across different spatial scales.

\begin{figure}[!h]
    \begin{center}
    \begin{subfigure}[t]{0.245\linewidth}
        \begin{center}
            \fbox{\includegraphics[width=0.667\columnwidth]{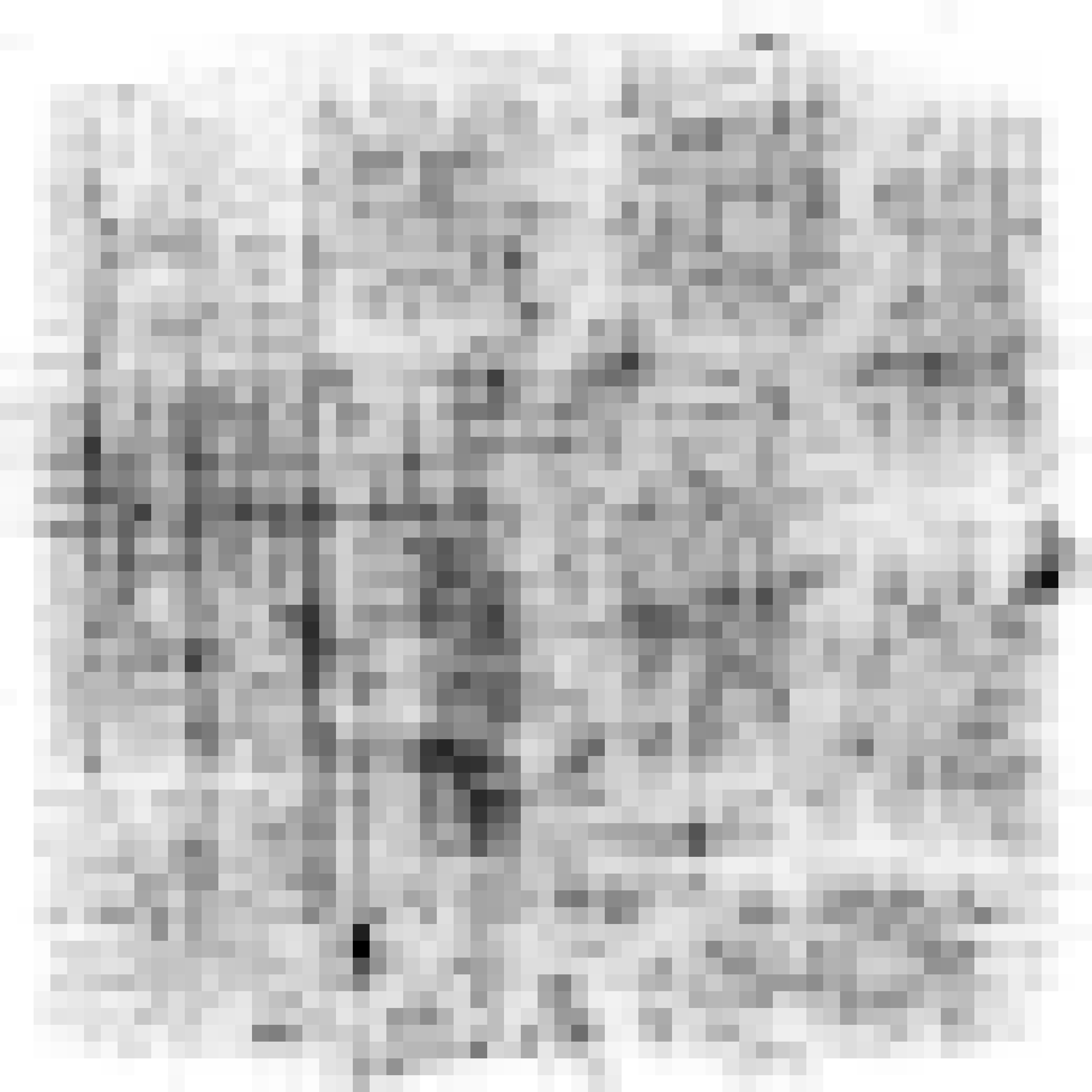}}
            \caption{}
            \label{fig:}
        \end{center}
    \end{subfigure}
    \begin{subfigure}[t]{0.245\linewidth}
        \begin{center}
            \fbox{\includegraphics[width=0.667\columnwidth]{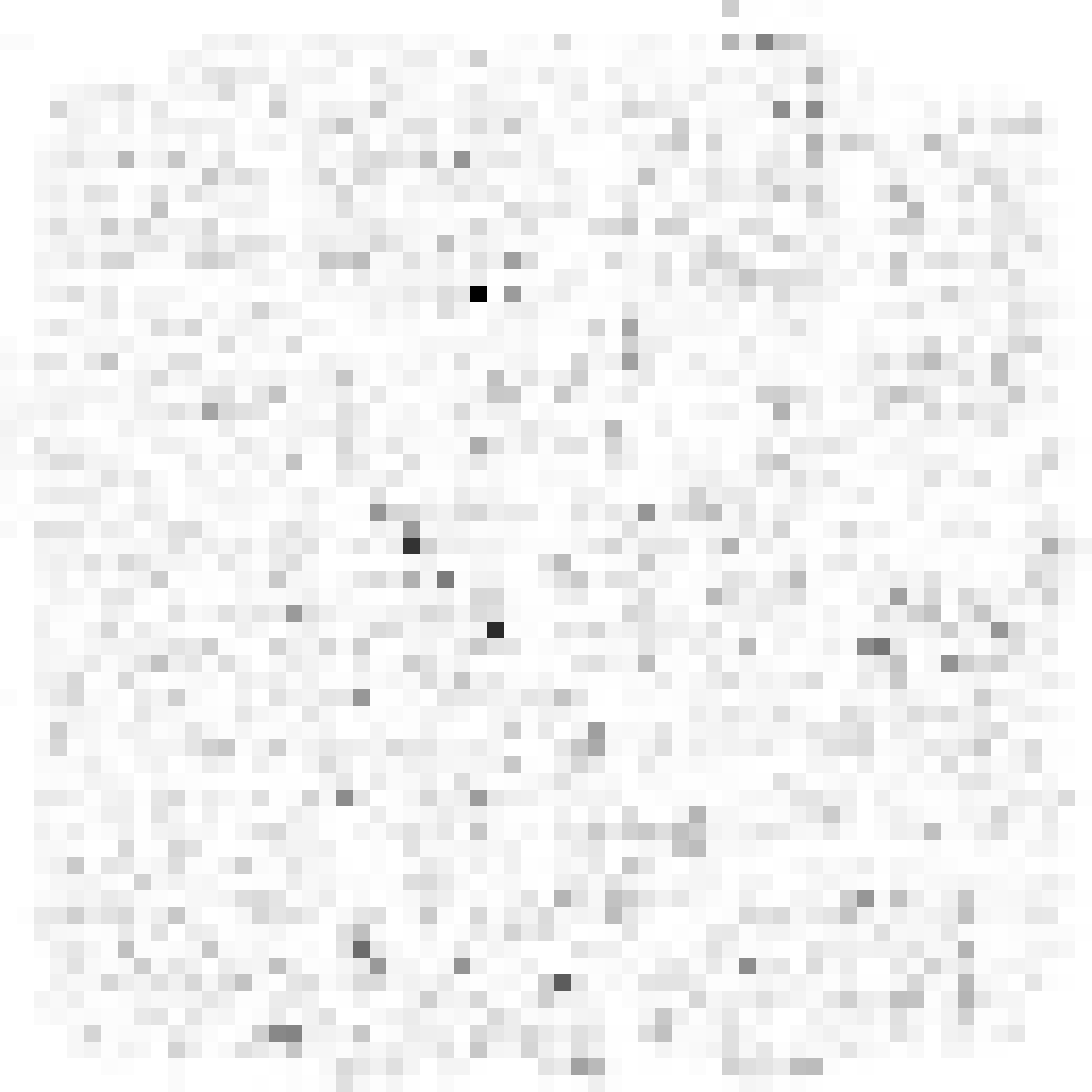}}
            \caption{}
            \label{fig:}
        \end{center}
    \end{subfigure}
    \begin{subfigure}[t]{0.245\linewidth}
        \begin{center}
            \fbox{\includegraphics[width=0.667\columnwidth]{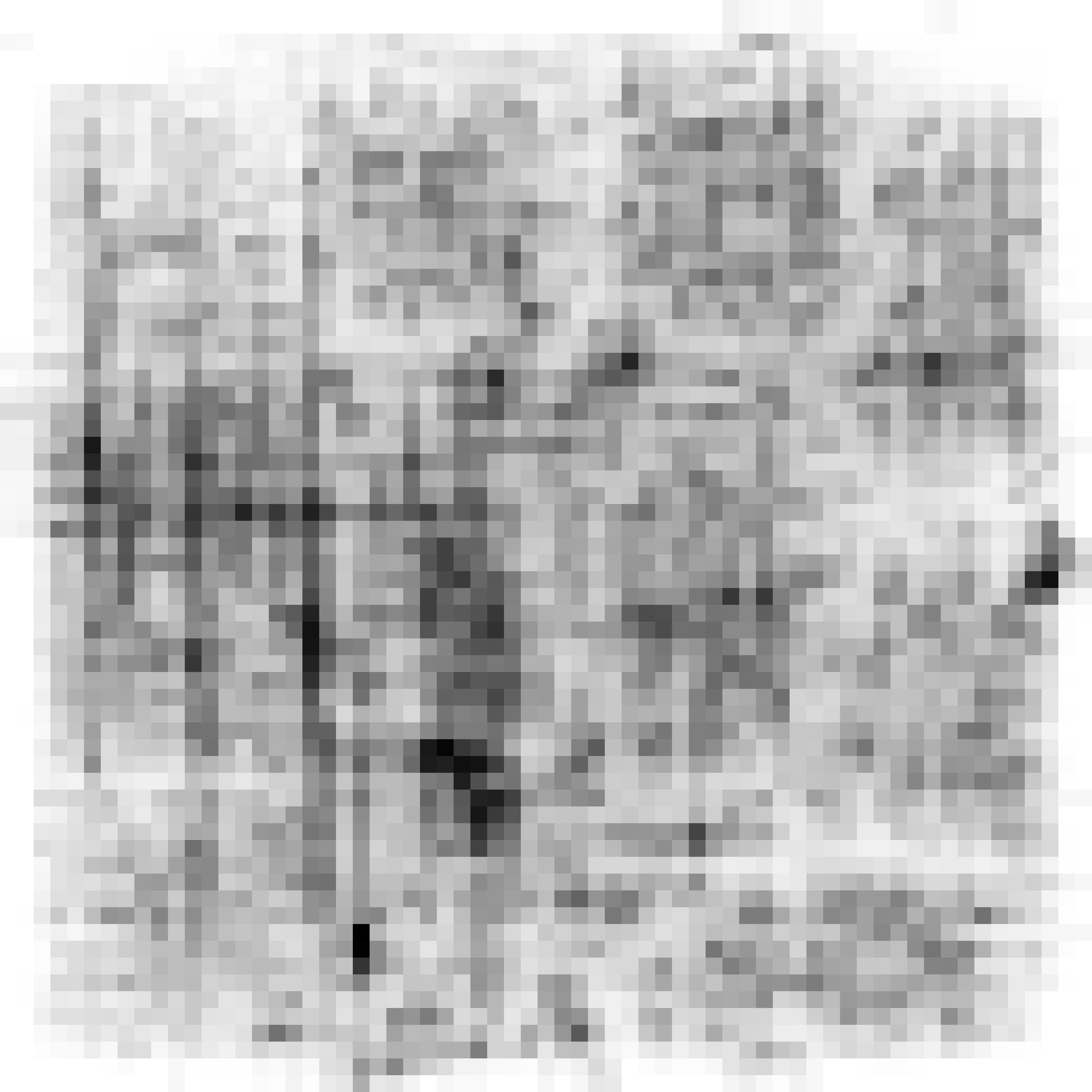}}
            \caption{}
            \label{fig:}
        \end{center}
    \end{subfigure}
    \begin{subfigure}[t]{0.245\linewidth}
        \begin{center}
            \fbox{\includegraphics[width=0.667\columnwidth]{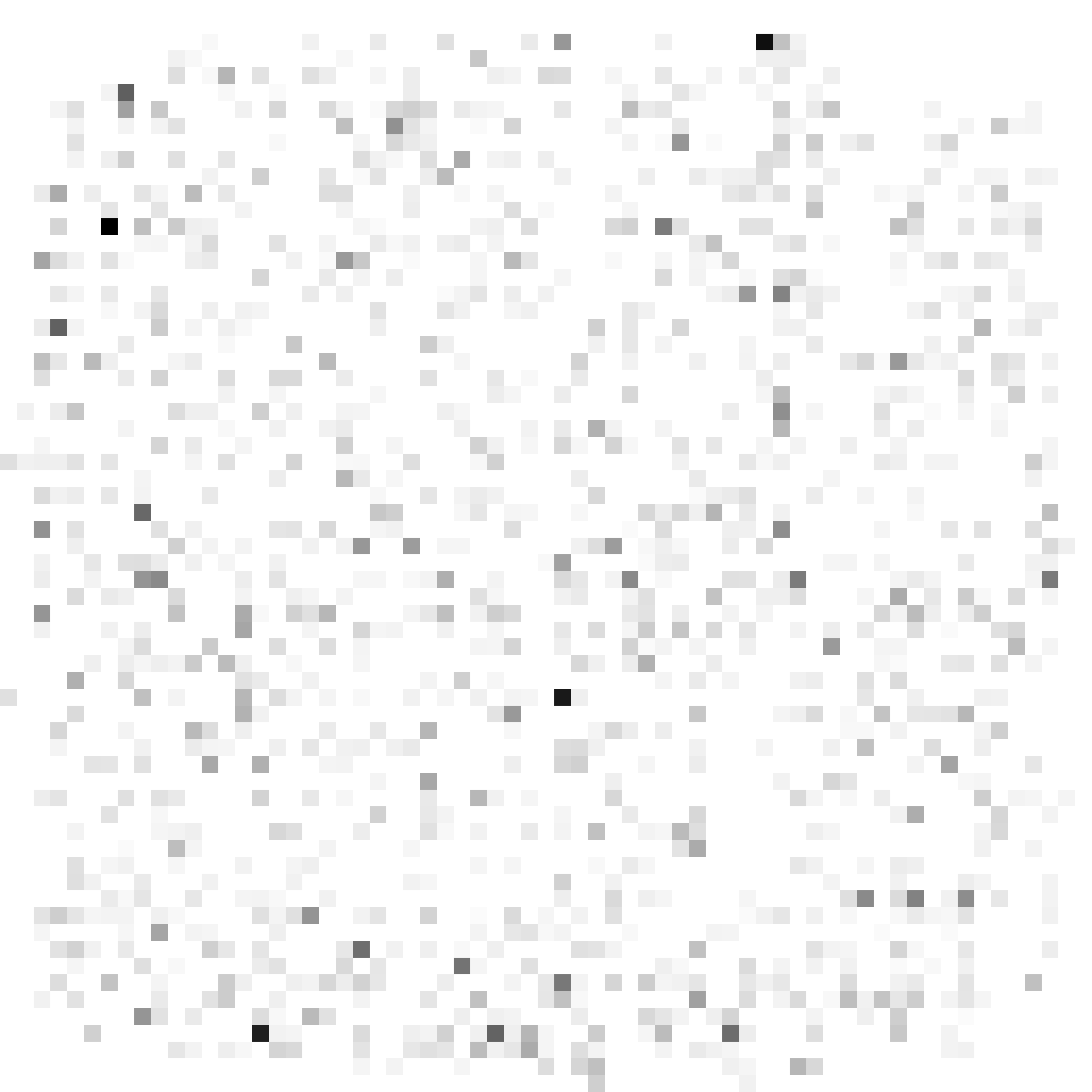}}
            \caption{}
            \label{fig:}
        \end{center}
    \end{subfigure}
    
    \caption{Routability metrics represented as scalar-valued RUDY maps at the final design stage for the \textit{ac97\_ctrl} circuit implemented using the Nangate 45 nm technology node. Subfigures depict (a) net-based RUDY, (b) pin-based RUDY, (c) long-range net-based RUDY, and (d) short-range net-based RUDY. All maps are generated using a scalar-map resolution with aggregation factor $k$ set to 50, where each pixel corresponds to a $50 \times w_{M1}$ width by $50 \times w_{M1}$ width region of the layout.}
    \label{fig:routability_metrics}
    \end{center}
    \vspace{-0.1in}
\end{figure}

All RUDY-based attributes are stored in EDA-Schema-V2 as two-dimensional scalar maps using the scalar-map resolution defined by (\ref{eqn:scalar_resolution}).
RUDY-based routability maps for the \textit{ac97\_ctrl} circuit implemented in the nangate 45 nm technology node are shown in Fig.~\ref{fig:routability_metrics}.

\section{Open Dataset}
\label{sec:open_dataset}

Publication or open profiling of datasets generated using commercial EDA tools and proprietary PDKs is often limited by terms of use restrictions.
Generating open data by utilizing open-source tools and publicly available PDKs and benchmark circuits allows for the unrestricted sharing of the generated data, which enables broader research and collaboration.
In this section, open datasets of physical designs of public benchmark circuits generated across four open-source PDKs are described.
The benchmark circuits and technology nodes used to construct the dataset are presented in Section~\ref{subsec:benchmarks_pdks}.
The design constraints used to generate baseline physical implementations are described in Section~\ref{subsec:bp_bf_definition}.
An expansion of baseline physical designs is performed by varying key physical design parameters, as described in Section~\ref{subsec:dataset_expansion}.

\subsection{Benchmarks and Technology Nodes}
\label{subsec:benchmarks_pdks}

The dataset is based on the IWLS’05 sequential benchmark suite with circuits and corresponding number of inputs, outputs, and registers as summarized in Table~\ref{tab:benchmarks}. The IWLS05 suite offers diverse circuit types with varying functional and structural characteristics, which enables a broad evaluation of physical design properties.
To capture variability across technology nodes, the dataset is generated with four publicly available PDKs:
\begin{itemize}
    \item \textbf{Nangate 45 nm (NG45) \cite{nangate45}}: An academic PDK that bridges modern process challenges with accessible design flows. NG45 contains 10 metal routing layers.
    \item \textbf{Skywater 130 nm (SKY130) \cite{google_sky130pdk}}: A mature process node widely used in academic and open-source EDA research. SKY130 contains 5 metal routing layers.
    \item \textbf{IHP SG13G2 130 nm (IHP130) \cite{ihp130}}: A SiGe BiCMOS PDK that supports research in RF, analog, and mixed-signal circuit design. IHP130 contains 7 metal routing layers.
    \item \textbf{ASAP 7 nm (ASAP7) \cite{asap7}}: A predictive process node that enables exploration and analysis of advanced scaling challenges. ASAP7 contains 9 metal routing layers.
\end{itemize}
Circuits complete physical design using the open-source \textbf{OpenROAD} toolchain.

% Please add the following required packages to your document preamble:
% \usepackage{multirow}
\begin{table}[!h]
    \begin{center}
        % \vspace{-0.05in}
        \caption{Characteristics of the circuit of the IWLS’05 benchmark suite used with EDA-Schema-V2. The listed attributes are derived from pre-synthesis RTL (Verilog) descriptions.}
        % \vspace{-0.05in}
        \label{tab:benchmarks}
        \fontsize{8}{10}\selectfont{
        % \resizebox{0.64\columnwidth}{!}{
            \begin{tabular}{|l|r|r|r|}
            \hline
            \textbf{Circuits}    & \multicolumn{1}{l|}{\textbf{No. of inputs}} & \multicolumn{1}{l|}{\textbf{No. of outputs}} & \multicolumn{1}{l|}{\textbf{No. of Registers}} \\ \hline
            \textbf{ac97\_ctrl}  & 84                                          & 48                                           & 2211                                           \\ \hline
            \textbf{aes\_core}   & 259                                         & 129                                          & 562                                            \\ \hline
            \textbf{des3\_area}  & 240                                         & 64                                           & 64                                             \\ \hline
            \textbf{ethernet}    & 96                                          & 115                                          & 10544                                          \\ \hline
            \textbf{i2c}         & 19                                          & 14                                           & 129                                            \\ \hline
            \textbf{jpeg}        & 20                                          & 27                                           & 4383                                           \\ \hline
            \textbf{mem\_ctrl}   & 115                                         & 152                                          & 1083                                           \\ \hline
            \textbf{pci}         & 162                                         & 207                                          & 3220                                           \\ \hline
            \textbf{sasc}        & 16                                          & 12                                           & 118                                            \\ \hline
            \textbf{simple\_spi} & 16                                          & 12                                           & 131                                            \\ \hline
            \textbf{spi}         & 47                                          & 45                                           & 229                                            \\ \hline
            \textbf{ss\_pcm}     & 19                                          & 9                                            & 87                                             \\ \hline
            \textbf{systemcaes}  & 260                                         & 129                                          & 670                                            \\ \hline
            \textbf{systemcdes}  & 132                                         & 65                                           & 190                                            \\ \hline
            \textbf{tv80}        & 14                                          & 32                                           & 361                                            \\ \hline
            \textbf{usb\_funct}  & 128                                         & 121                                          & 1740                                           \\ \hline
            \textbf{usb\_phy}    & 15                                          & 18                                           & 108                                            \\ \hline
            \textbf{wb\_dma}     & 217                                         & 215                                          & 521                                            \\ \hline
            \end{tabular}
        }
    \end{center}
    \vspace{-0.15in}
\end{table}

\subsection{Barely-Pass and Barely-Fail Operating Points}
\label{subsec:bp_bf_definition}

Dataset generation begins by identifying two critical operating points for each circuit: the \textbf{Barely Pass (BP)} point, where the generated circuit just meets timing constraints, and the \textbf{Barely Fail (BF)} point, where the generated circuit slightly violates timing constraints. 
The two classifications of BP and BF are defined based on the \textbf{Slack to Clock Period Ratio (SCPR)}, which is given as
\begin{equation}
    \text{SCPR} = \frac{\text{Worst Slack}}{\text{Clock Period}} \times 100\%
\end{equation}
\noindent Designs are categorized using the following thresholds:
\begin{itemize}
\item \textbf{Barely Fail:} SCPR $\in$ (–10\%, 0\%), which indicates slight negative slack.
\item \textbf{Barely Pass:} SCPR $\in$ (0\%, +10\%), which indicates slight positive slack.
\end{itemize}
Initial BP and BF points are identified for all PDKs using default configurations of the OpenROAD Flow Scripts (ORFS), where the aspect ratio is fixed to 1.0 and the placement density is assumed uniform.
Core utilization is set to 40\% for ASAP7 and NG45, while SKY130 and IHP130 are set to a lower utilization of 30\% to account for the smaller number of metal layers available in each PDK.

For constraints on the clock network, the transition time of the clock signal for each PDK is defined as 12.5\% of the fastest clock period of any of the IWLS'05 benchmark circuits analyzed, which is that of the \textit{ss\_pcm} circuit. 
Constraining the transition time to the clock period of the fastest circuit provides a consistent yet technology-aware baseline across the dataset.
Input and output delays are fixed to 20\% of the target clock period, clock latency is fixed to 1\% of the target clock period but capped to a maximum of 50~ps, and clock uncertainty is set to 5\% of the clock period with a maximum of 250~ps.

Similarly, the configuration of the parameters of the PDN for the barely pass and barely fail design points is defined per PDK using technology-aware properties. Each configuration of the stripes and rings of the PDN is based on minimum width and spacing constraints of the metal layers used to form the PDN, with exceptions made to maintain symmetry in the power network. For all PDKs, alternating, interdigitated VDD and VSS stripes are placed on intermediate routing layers, while core rings are formed around the placement boundary using semi-global or global metal layers.
The width of the vertical and horizontal metal stripes are set to five times the minimum width of the corresponding metal layer. An exception is applied to the SKY130 PDK, where the width of the horizontal stripe is adjusted to the nearest allowable multiple of the minimum width of the given metal layer that best matches the width of the vertical stripe due to the large disparity between the minimum widths of the metals used for the stripes. Ring widths are set to three times the minimum width of the metal layer with the largest minimum width. An exception is applied to the ASAP7 PDK as design rules restrict the allowable metal widths to every other odd integer multiple of the minimum width, which results in ring widths that are set to the maximum permitted value for each layer. The spacing between power and ground stripes and rings is set to the minimum allowed spacing between two metals on the same layer as specified in the LEF files. An exception is applied to the SKY130 PDK, where the spacing between vertical stripes and rings is set to five times the minimum allowed value to more closely match the spacing of the horizontal stripe and ring. Finally, the pitch between interdigitated stripe pairs is set to twenty times the rail-rail pitch of the standard cells. 
Detailed PDN parameters for each PDK are listed in   Table~\ref{tab:pdn_config}.

For the analysis of IR-drop and electromigration (EM) analysis using PDNSim, voltage sources are placed at the origin of the layout, which is set to the lower left corner boundary of the circuit, and at regularly spaced intervals of twice the power-strap pitch in both dimensions.
Using the defined configurations and constraints for general circuit parameters, the clock network, and the power network, the corresponding Barely-Pass (BP) and Barely-Fail (BF) operating points are determined for each circuit. The resulting initial BP and BF operating points are listed in Table~\ref{tab:initial_bp_bf}.

\begin{table}[!h]
\centering
\caption{Configuration of PDN parameters across PDKs. All values are in micrometers (\textmu m) unless otherwise noted. (V/H) denotes the vertical and horizontal PDN metal layers, respectively.}
\label{tab:pdn_config}
\fontsize{8}{10}\selectfont{
% \resizebox{0.5\columnwidth}{!}{
    \begin{tabular}{|lllll|}
    \hline
    \multicolumn{1}{|l|}{\textbf{PDN Metric}}     & \multicolumn{1}{l|}{\textbf{NG45}} & \multicolumn{1}{l|}{\textbf{SKY130}} & \multicolumn{1}{l|}{\textbf{IHP130}} & \textbf{ASAP7} \\ \hline
    \multicolumn{1}{|l|}{Core Offset (µm)}        & \multicolumn{1}{l|}{5.6}           & \multicolumn{1}{l|}{22.5}            & \multicolumn{1}{l|}{28}              & 1.8            \\ \hline
    \multicolumn{1}{|l|}{Rail to Rail Pitch (µm)} & \multicolumn{1}{l|}{1.4}           & \multicolumn{1}{l|}{2.72}            & \multicolumn{1}{l|}{3.78}            & 0.27           \\ \hline
    \multicolumn{5}{|c|}{\cellcolor[HTML]{C0C0C0}{\color[HTML]{000000} \textbf{Stripes}}}                                                                                             \\ \hline
    \multicolumn{1}{|l|}{Metal Layers (V/H)}      & \multicolumn{1}{l|}{M4 / M5}       & \multicolumn{1}{l|}{M4 / M5}         & \multicolumn{1}{l|}{M3 / M4}         & M5 / M6        \\ \hline
    \multicolumn{1}{|l|}{Min. metal width (µm)}   & \multicolumn{1}{l|}{0.14 / 0.14}   & \multicolumn{1}{l|}{0.3 / 1.6}       & \multicolumn{1}{l|}{0.2 / 0.2}       & 0.024 / 0.032  \\ \hline
    \multicolumn{1}{|l|}{Width (µm)}              & \multicolumn{1}{l|}{0.7 / 0.7}     & \multicolumn{1}{l|}{1.5 / 1.6}       & \multicolumn{1}{l|}{1.0 / 1.0}       & 0.12 / 0.16    \\ \hline
    \multicolumn{1}{|l|}{Spacing (µm)}            & \multicolumn{1}{l|}{0.14 / 0.14}   & \multicolumn{1}{l|}{1.5 / 1.6}       & \multicolumn{1}{l|}{0.21 / 0.21}     & 0.072 / 0.072  \\ \hline
    \multicolumn{1}{|l|}{Pitch (µm)}              & \multicolumn{1}{l|}{28.0 / 28.0}   & \multicolumn{1}{l|}{54.4 / 54.4}     & \multicolumn{1}{l|}{75.6 / 75.6}     & 5.4 / 5.4      \\ \hline
    \multicolumn{5}{|c|}{\cellcolor[HTML]{C0C0C0}\textbf{Rings}}                                                                                                                      \\ \hline
    \multicolumn{1}{|l|}{Metal Layers (V/H)}      & \multicolumn{1}{l|}{M8 / M7}       & \multicolumn{1}{l|}{M4 / M5}         & \multicolumn{1}{l|}{TM2 / TM1}       & M7 / M8        \\ \hline
    \multicolumn{1}{|l|}{Min. metal width (µm)}   & \multicolumn{1}{l|}{0.4 / 0.4}     & \multicolumn{1}{l|}{0.3 / 1.6}       & \multicolumn{1}{l|}{2.0 / 2.0}       & 0.032 / 0.04   \\ \hline
    \multicolumn{1}{|l|}{Width (µm)}              & \multicolumn{1}{l|}{1.2 / 1.2}     & \multicolumn{1}{l|}{4.8 / 4.8}       & \multicolumn{1}{l|}{6.0 / 6.0}       & 0.288 / 0.288  \\ \hline
    \multicolumn{1}{|l|}{Spacing (µm)}            & \multicolumn{1}{l|}{0.4 / 0.4}     & \multicolumn{1}{l|}{1.5 / 1.6}       & \multicolumn{1}{l|}{2.0 / 2.0}       & 0.072 / 0.072  \\ \hline
    \end{tabular}
}
\end{table}
\vspace{-0.2in}

% \vspace{-0.1in}
\begin{table}[h!]
\centering
\caption{Initial barely pass and barely fail target clock periods (TCP), worst slack (WS), and calculated slack to clock period ratio (SCPR) for each circuit across the four PDKs. All clock periods and worst slacks are in ns.}
\label{tab:initial_bp_bf}
\resizebox{\columnwidth}{!}{
    \begin{tabular}{|l|c|lll|lll|c|lll|lll|}
    \hline
    \multicolumn{1}{|c|}{\multirow{2}{*}{\textbf{Circuit}}} & \multirow{2}{*}{\textbf{PDK}}                                                            & \multicolumn{3}{c|}{\textbf{Barely Pass}}                                                                 & \multicolumn{3}{c|}{\textbf{Barely Fail}}                                                                 & \multirow{2}{*}{\textbf{PDK}}                                                            & \multicolumn{3}{c|}{\textbf{Barely Pass}}                                                                 & \multicolumn{3}{c|}{\textbf{Barely Fail}}                                                                 \\ \cline{3-8} \cline{10-15} 
    \multicolumn{1}{|c|}{}                                  &                                                                                          & \multicolumn{1}{c|}{\textbf{TCP}} & \multicolumn{1}{c|}{\textbf{WS}} & \multicolumn{1}{c|}{\textbf{SCPR}} & \multicolumn{1}{c|}{\textbf{TCP}} & \multicolumn{1}{c|}{\textbf{WS}} & \multicolumn{1}{c|}{\textbf{SCPR}} &                                                                                          & \multicolumn{1}{c|}{\textbf{TCP}} & \multicolumn{1}{c|}{\textbf{WS}} & \multicolumn{1}{c|}{\textbf{SCPR}} & \multicolumn{1}{c|}{\textbf{TCP}} & \multicolumn{1}{c|}{\textbf{WS}} & \multicolumn{1}{c|}{\textbf{SCPR}} \\ \hline
    \textbf{ac97\_ctrl}                                     & \multirow{18}{*}{\textbf{\begin{tabular}[c]{@{}c@{}}N\\ G\\ 4\\ 5\end{tabular}}}         & \multicolumn{1}{l|}{0.6}          & \multicolumn{1}{l|}{0.0078}      & 1.30\%                             & \multicolumn{1}{l|}{0.5}          & \multicolumn{1}{l|}{-0.0425}     & -8.50\%                            & \multirow{18}{*}{\textbf{\begin{tabular}[c]{@{}c@{}}A\\ S\\ A\\ P\\ 7\end{tabular}}}     & \multicolumn{1}{l|}{0.375}        & \multicolumn{1}{l|}{0.0030}      & 0.80\%                             & \multicolumn{1}{l|}{0.35}         & \multicolumn{1}{l|}{-0.0066}     & -1.90\%                            \\ \cline{1-1} \cline{3-8} \cline{10-15} 
    \textbf{aes\_core}                                      &                                                                                          & \multicolumn{1}{l|}{1.5}          & \multicolumn{1}{l|}{0.1158}      & 7.72\%                             & \multicolumn{1}{l|}{1.25}         & \multicolumn{1}{l|}{-0.0412}     & -3.30\%                            &                                                                                          & \multicolumn{1}{l|}{0.75}         & \multicolumn{1}{l|}{0.0074}      & 0.99\%                             & \multicolumn{1}{l|}{0.725}        & \multicolumn{1}{l|}{-0.0105}     & -1.45\%                            \\ \cline{1-1} \cline{3-8} \cline{10-15} 
    \textbf{des3\_area}                                     &                                                                                          & \multicolumn{1}{l|}{2.5}          & \multicolumn{1}{l|}{0.0006}      & 0.02\%                             & \multicolumn{1}{l|}{2.25}         & \multicolumn{1}{l|}{-0.0411}     & -1.83\%                            &                                                                                          & \multicolumn{1}{l|}{1.225}        & \multicolumn{1}{l|}{0.0014}      & 0.11\%                             & \multicolumn{1}{l|}{1.2}          & \multicolumn{1}{l|}{-0.0042}     & -0.35\%                            \\ \cline{1-1} \cline{3-8} \cline{10-15} 
    \textbf{ethernet}                                       &                                                                                          & \multicolumn{1}{l|}{1.65}         & \multicolumn{1}{l|}{0.0204}      & 1.24\%                             & \multicolumn{1}{l|}{1.5}          & \multicolumn{1}{l|}{-0.022}      & -1.47\%                            &                                                                                          & \multicolumn{1}{l|}{0.75}         & \multicolumn{1}{l|}{0.0166}      & 2.22\%                             & \multicolumn{1}{l|}{0.725}        & \multicolumn{1}{l|}{-0.0404}     & -5.57\%                            \\ \cline{1-1} \cline{3-8} \cline{10-15} 
    \textbf{i2c}                                            &                                                                                          & \multicolumn{1}{l|}{0.65}         & \multicolumn{1}{l|}{0.0103}      & 1.58\%                             & \multicolumn{1}{l|}{0.55}         & \multicolumn{1}{l|}{-0.0154}     & -2.79\%                            &                                                                                          & \multicolumn{1}{l|}{0.375}        & \multicolumn{1}{l|}{0.0042}      & 1.13\%                             & \multicolumn{1}{l|}{0.35}         & \multicolumn{1}{l|}{-0.0079}     & -2.27\%                            \\ \cline{1-1} \cline{3-8} \cline{10-15} 
    \textbf{jpeg}                                           &                                                                                          & \multicolumn{1}{l|}{1.8}          & \multicolumn{1}{l|}{0.0483}      & 2.68\%                             & \multicolumn{1}{l|}{1.6}          & \multicolumn{1}{l|}{-0.0117}     & -0.73\%                            &                                                                                          & \multicolumn{1}{l|}{1.125}        & \multicolumn{1}{l|}{0.0343}      & 3.05\%                             & \multicolumn{1}{l|}{1.1}          & \multicolumn{1}{l|}{-0.0234}     & -2.13\%                            \\ \cline{1-1} \cline{3-8} \cline{10-15} 
    \textbf{mem\_ctrl}                                      &                                                                                          & \multicolumn{1}{l|}{1.75}         & \multicolumn{1}{l|}{0.0202}      & 1.15\%                             & \multicolumn{1}{l|}{1.25}         & \multicolumn{1}{l|}{-0.0558}     & -4.46\%                            &                                                                                          & \multicolumn{1}{l|}{0.875}        & \multicolumn{1}{l|}{0.0006}      & 0.07\%                             & \multicolumn{1}{l|}{0.85}         & \multicolumn{1}{l|}{-0.0043}     & -0.51\%                            \\ \cline{1-1} \cline{3-8} \cline{10-15} 
    \textbf{pci}                                            &                                                                                          & \multicolumn{1}{l|}{1}            & \multicolumn{1}{l|}{0.0398}      & 3.98\%                             & \multicolumn{1}{l|}{0.75}         & \multicolumn{1}{l|}{-0.0328}     & -4.37\%                            &                                                                                          & \multicolumn{1}{l|}{0.825}        & \multicolumn{1}{l|}{0.0378}      & 4.58\%                             & \multicolumn{1}{l|}{0.81}         & \multicolumn{1}{l|}{-0.0257}     & -3.17\%                            \\ \cline{1-1} \cline{3-8} \cline{10-15} 
    \textbf{sasc}                                           &                                                                                          & \multicolumn{1}{l|}{0.6}          & \multicolumn{1}{l|}{0.0252}      & 4.19\%                             & \multicolumn{1}{l|}{0.5}          & \multicolumn{1}{l|}{-0.0392}     & -7.84\%                            &                                                                                          & \multicolumn{1}{l|}{0.4}          & \multicolumn{1}{l|}{0.0177}      & 4.43\%                             & \multicolumn{1}{l|}{0.375}        & \multicolumn{1}{l|}{-0.0031}     & -0.82\%                            \\ \cline{1-1} \cline{3-8} \cline{10-15} 
    \textbf{simple\_spi}                                    &                                                                                          & \multicolumn{1}{l|}{0.55}         & \multicolumn{1}{l|}{0.0258}      & 4.69\%                             & \multicolumn{1}{l|}{0.45}         & \multicolumn{1}{l|}{-0.0421}     & -9.35\%                            &                                                                                          & \multicolumn{1}{l|}{0.325}        & \multicolumn{1}{l|}{0.0039}      & 1.20\%                             & \multicolumn{1}{l|}{0.3}          & \multicolumn{1}{l|}{-0.0052}     & -1.73\%                            \\ \cline{1-1} \cline{3-8} \cline{10-15} 
    \textbf{spi}                                            &                                                                                          & \multicolumn{1}{l|}{1.5}          & \multicolumn{1}{l|}{0.0595}      & 3.96\%                             & \multicolumn{1}{l|}{1.25}         & \multicolumn{1}{l|}{-0.0142}     & -1.14\%                            &                                                                                          & \multicolumn{1}{l|}{0.775}        & \multicolumn{1}{l|}{0.0047}      & 0.61\%                             & \multicolumn{1}{l|}{0.75}         & \multicolumn{1}{l|}{-0.0012}     & -0.16\%                            \\ \cline{1-1} \cline{3-8} \cline{10-15} 
    \textbf{ss\_pcm}                                        &                                                                                          & \multicolumn{1}{l|}{0.38}         & \multicolumn{1}{l|}{0.0111}      & 2.92\%                             & \multicolumn{1}{l|}{0.325}        & \multicolumn{1}{l|}{-0.0133}     & -4.08\%                            &                                                                                          & \multicolumn{1}{l|}{0.275}        & \multicolumn{1}{l|}{0.0044}      & 1.58\%                             & \multicolumn{1}{l|}{0.25}         & \multicolumn{1}{l|}{-0.0110}     & -4.40\%                            \\ \cline{1-1} \cline{3-8} \cline{10-15} 
    \textbf{systemcaes}                                     &                                                                                          & \multicolumn{1}{l|}{1.75}         & \multicolumn{1}{l|}{0.0809}      & 4.62\%                             & \multicolumn{1}{l|}{1.5}          & \multicolumn{1}{l|}{-0.0064}     & -0.43\%                            &                                                                                          & \multicolumn{1}{l|}{0.85}         & \multicolumn{1}{l|}{0.0059}      & 0.70\%                             & \multicolumn{1}{l|}{0.825}        & \multicolumn{1}{l|}{-0.0031}     & -0.37\%                            \\ \cline{1-1} \cline{3-8} \cline{10-15} 
    \textbf{systemcdes}                                     &                                                                                          & \multicolumn{1}{l|}{1.45}         & \multicolumn{1}{l|}{0.1219}      & 8.41\%                             & \multicolumn{1}{l|}{1.25}         & \multicolumn{1}{l|}{-0.0143}     & -1.15\%                            &                                                                                          & \multicolumn{1}{l|}{0.75}         & \multicolumn{1}{l|}{0.0066}      & 0.88\%                             & \multicolumn{1}{l|}{0.725}        & \multicolumn{1}{l|}{-0.0062}     & -0.86\%                            \\ \cline{1-1} \cline{3-8} \cline{10-15} 
    \textbf{tv80}                                           &                                                                                          & \multicolumn{1}{l|}{2.25}         & \multicolumn{1}{l|}{0.094}       & 4.18\%                             & \multicolumn{1}{l|}{1.9}          & \multicolumn{1}{l|}{-0.0028}     & -0.15\%                            &                                                                                          & \multicolumn{1}{l|}{1.2}          & \multicolumn{1}{l|}{0.0058}      & 0.48\%                             & \multicolumn{1}{l|}{1.175}        & \multicolumn{1}{l|}{-0.0217}     & -1.85\%                            \\ \cline{1-1} \cline{3-8} \cline{10-15} 
    \textbf{usb\_funct}                                     &                                                                                          & \multicolumn{1}{l|}{1}            & \multicolumn{1}{l|}{0.0035}      & 0.35\%                             & \multicolumn{1}{l|}{0.8}          & \multicolumn{1}{l|}{-0.0068}     & -0.85\%                            &                                                                                          & \multicolumn{1}{l|}{0.425}        & \multicolumn{1}{l|}{0.0041}      & 0.97\%                             & \multicolumn{1}{l|}{0.4}          & \multicolumn{1}{l|}{-0.0010}     & -0.25\%                            \\ \cline{1-1} \cline{3-8} \cline{10-15} 
    \textbf{usb\_phy}                                       &                                                                                          & \multicolumn{1}{l|}{0.45}         & \multicolumn{1}{l|}{0.0068}      & 1.51\%                             & \multicolumn{1}{l|}{0.35}         & \multicolumn{1}{l|}{-0.0193}     & -5.52\%                            &                                                                                          & \multicolumn{1}{l|}{0.3}          & \multicolumn{1}{l|}{0.0040}      & 1.32\%                             & \multicolumn{1}{l|}{0.275}        & \multicolumn{1}{l|}{-0.0066}     & -2.41\%                            \\ \cline{1-1} \cline{3-8} \cline{10-15} 
    \textbf{wb\_dma}                                        &                                                                                          & \multicolumn{1}{l|}{1.25}         & \multicolumn{1}{l|}{0.023}       & 1.84\%                             & \multicolumn{1}{l|}{0.9}          & \multicolumn{1}{l|}{-0.0683}     & -7.59\%                            &                                                                                          & \multicolumn{1}{l|}{0.7}          & \multicolumn{1}{l|}{0.0063}      & 0.90\%                             & \multicolumn{1}{l|}{0.675}        & \multicolumn{1}{l|}{-0.0003}     & -0.04\%                            \\ \hline
    \textbf{ac97\_ctrl}                                     & \multirow{18}{*}{\textbf{\begin{tabular}[c]{@{}c@{}}S\\ K\\ Y\\ 1\\ 3\\ 0\end{tabular}}} & \multicolumn{1}{l|}{2.5}          & \multicolumn{1}{l|}{0.0233}      & 0.93\%                             & \multicolumn{1}{l|}{2.25}         & \multicolumn{1}{l|}{-0.0341}     & -1.52\%                            & \multirow{18}{*}{\textbf{\begin{tabular}[c]{@{}c@{}}I\\ H\\ P\\ 1\\ 3\\ 0\end{tabular}}} & \multicolumn{1}{l|}{2.35}         & \multicolumn{1}{l|}{0.0415}      & 1.77\%                             & \multicolumn{1}{l|}{2.3}          & \multicolumn{1}{l|}{-0.0612}     & -2.66\%                            \\ \cline{1-1} \cline{3-8} \cline{10-15} 
    \textbf{aes\_core}                                      &                                                                                          & \multicolumn{1}{l|}{4.75}         & \multicolumn{1}{l|}{0.0723}      & 1.52\%                             & \multicolumn{1}{l|}{4.5}          & \multicolumn{1}{l|}{-0.0889}     & -1.98\%                            &                                                                                          & \multicolumn{1}{l|}{4.25}         & \multicolumn{1}{l|}{0.1046}      & 2.46\%                             & \multicolumn{1}{l|}{4}            & \multicolumn{1}{l|}{-0.3553}     & -8.88\%                            \\ \cline{1-1} \cline{3-8} \cline{10-15} 
    \textbf{des3\_area}                                     &                                                                                          & \multicolumn{1}{l|}{7.75}         & \multicolumn{1}{l|}{0.0138}      & 0.18\%                             & \multicolumn{1}{l|}{7.5}          & \multicolumn{1}{l|}{-0.2478}     & -3.30\%                            &                                                                                          & \multicolumn{1}{l|}{8.5}          & \multicolumn{1}{l|}{0.2}         & 2.35\%                             & \multicolumn{1}{l|}{8.25}         & \multicolumn{1}{l|}{-0.0561}     & -0.68\%                            \\ \cline{1-1} \cline{3-8} \cline{10-15} 
    \textbf{ethernet}                                       &                                                                                          & \multicolumn{1}{l|}{5.25}         & \multicolumn{1}{l|}{0.0856}      & 1.63\%                             & \multicolumn{1}{l|}{5}            & \multicolumn{1}{l|}{-0.1807}     & -3.61\%                            &                                                                                          & \multicolumn{1}{l|}{5.5}          & \multicolumn{1}{l|}{0.1467}      & 2.67\%                             & \multicolumn{1}{l|}{5.25}         & \multicolumn{1}{l|}{-0.1679}     & -3.20\%                            \\ \cline{1-1} \cline{3-8} \cline{10-15} 
    \textbf{i2c}                                            &                                                                                          & \multicolumn{1}{l|}{2.75}         & \multicolumn{1}{l|}{0.0772}      & 2.81\%                             & \multicolumn{1}{l|}{2.5}          & \multicolumn{1}{l|}{-0.1254}     & -5.02\%                            &                                                                                          & \multicolumn{1}{l|}{2.25}         & \multicolumn{1}{l|}{0.0197}      & 0.88\%                             & \multicolumn{1}{l|}{2}            & \multicolumn{1}{l|}{-0.0857}     & -4.29\%                            \\ \cline{1-1} \cline{3-8} \cline{10-15} 
    \textbf{jpeg}                                           &                                                                                          & \multicolumn{1}{l|}{6}            & \multicolumn{1}{l|}{0.0228}      & 0.38\%                             & \multicolumn{1}{l|}{5.75}         & \multicolumn{1}{l|}{-0.372}      & -6.47\%                            &                                                                                          & \multicolumn{1}{l|}{9.5}          & \multicolumn{1}{l|}{0.375}       & 3.95\%                             & \multicolumn{1}{l|}{9.25}         & \multicolumn{1}{l|}{-0.1166}     & -1.26\%                            \\ \cline{1-1} \cline{3-8} \cline{10-15} 
    \textbf{mem\_ctrl}                                      &                                                                                          & \multicolumn{1}{l|}{6.25}         & \multicolumn{1}{l|}{0.0185}      & 0.30\%                             & \multicolumn{1}{l|}{6}            & \multicolumn{1}{l|}{-0.1917}     & -3.20\%                            &                                                                                          & \multicolumn{1}{l|}{5.25}         & \multicolumn{1}{l|}{0.256}       & 4.88\%                             & \multicolumn{1}{l|}{5}            & \multicolumn{1}{l|}{-0.1356}     & -2.71\%                            \\ \cline{1-1} \cline{3-8} \cline{10-15} 
    \textbf{pci}                                            &                                                                                          & \multicolumn{1}{l|}{3.75}         & \multicolumn{1}{l|}{0.0861}      & 2.30\%                             & \multicolumn{1}{l|}{3.5}          & \multicolumn{1}{l|}{-0.2461}     & -7.03\%                            &                                                                                          & \multicolumn{1}{l|}{3.5}          & \multicolumn{1}{l|}{0.2581}      & 7.37\%                             & \multicolumn{1}{l|}{3.25}         & \multicolumn{1}{l|}{-0.0176}     & -0.54\%                            \\ \cline{1-1} \cline{3-8} \cline{10-15} 
    \textbf{sasc}                                           &                                                                                          & \multicolumn{1}{l|}{2.25}         & \multicolumn{1}{l|}{0.0785}      & 3.49\%                             & \multicolumn{1}{l|}{2}            & \multicolumn{1}{l|}{-0.0914}     & -4.57\%                            &                                                                                          & \multicolumn{1}{l|}{2}            & \multicolumn{1}{l|}{0.0537}      & 2.69\%                             & \multicolumn{1}{l|}{1.75}         & \multicolumn{1}{l|}{-0.0443}     & -2.53\%                            \\ \cline{1-1} \cline{3-8} \cline{10-15} 
    \textbf{simple\_spi}                                    &                                                                                          & \multicolumn{1}{l|}{2.25}         & \multicolumn{1}{l|}{0.0828}      & 3.68\%                             & \multicolumn{1}{l|}{1.9}          & \multicolumn{1}{l|}{-0.1071}     & -5.64\%                            &                                                                                          & \multicolumn{1}{l|}{1.75}         & \multicolumn{1}{l|}{0.0709}      & 4.05\%                             & \multicolumn{1}{l|}{1.5}          & \multicolumn{1}{l|}{-0.0196}     & -1.31\%                            \\ \cline{1-1} \cline{3-8} \cline{10-15} 
    \textbf{spi}                                            &                                                                                          & \multicolumn{1}{l|}{4.75}         & \multicolumn{1}{l|}{0.1341}      & 2.82\%                             & \multicolumn{1}{l|}{4.5}          & \multicolumn{1}{l|}{-0.1181}     & -2.62\%                            &                                                                                          & \multicolumn{1}{l|}{4.75}         & \multicolumn{1}{l|}{0.1467}      & 3.09\%                             & \multicolumn{1}{l|}{4.5}          & \multicolumn{1}{l|}{-0.0779}     & -1.73\%                            \\ \cline{1-1} \cline{3-8} \cline{10-15} 
    \textbf{ss\_pcm}                                        &                                                                                          & \multicolumn{1}{l|}{1.5}          & \multicolumn{1}{l|}{0.0594}      & 3.96\%                             & \multicolumn{1}{l|}{1.35}         & \multicolumn{1}{l|}{-0.0174}     & -1.29\%                            &                                                                                          & \multicolumn{1}{l|}{1.4}          & \multicolumn{1}{l|}{0.1014}      & 7.24\%                             & \multicolumn{1}{l|}{1.2}          & \multicolumn{1}{l|}{-0.0859}     & -7.16\%                            \\ \cline{1-1} \cline{3-8} \cline{10-15} 
    \textbf{systemcaes}                                     &                                                                                          & \multicolumn{1}{l|}{5.75}         & \multicolumn{1}{l|}{0.1246}      & 2.17\%                             & \multicolumn{1}{l|}{5.5}          & \multicolumn{1}{l|}{-0.1404}     & -2.55\%                            &                                                                                          & \multicolumn{1}{l|}{5.5}          & \multicolumn{1}{l|}{0.0415}      & 0.75\%                             & \multicolumn{1}{l|}{5.25}         & \multicolumn{1}{l|}{-0.3941}     & -7.51\%                            \\ \cline{1-1} \cline{3-8} \cline{10-15} 
    \textbf{systemcdes}                                     &                                                                                          & \multicolumn{1}{l|}{5}            & \multicolumn{1}{l|}{0.1677}      & 3.35\%                             & \multicolumn{1}{l|}{4.8}          & \multicolumn{1}{l|}{-0.0352}     & -0.73\%                            &                                                                                          & \multicolumn{1}{l|}{4}            & \multicolumn{1}{l|}{0.2955}      & 7.39\%                             & \multicolumn{1}{l|}{3.75}         & \multicolumn{1}{l|}{-0.0634}     & -1.69\%                            \\ \cline{1-1} \cline{3-8} \cline{10-15} 
    \textbf{tv80}                                           &                                                                                          & \multicolumn{1}{l|}{8.25}         & \multicolumn{1}{l|}{0.0655}      & 0.79\%                             & \multicolumn{1}{l|}{8}            & \multicolumn{1}{l|}{-0.2718}     & -3.40\%                            &                                                                                          & \multicolumn{1}{l|}{7.5}          & \multicolumn{1}{l|}{0.2298}      & 3.06\%                             & \multicolumn{1}{l|}{7.25}         & \multicolumn{1}{l|}{-0.1341}     & -1.85\%                            \\ \cline{1-1} \cline{3-8} \cline{10-15} 
    \textbf{usb\_funct}                                     &                                                                                          & \multicolumn{1}{l|}{2.75}         & \multicolumn{1}{l|}{0.0229}      & 0.83\%                             & \multicolumn{1}{l|}{2.5}          & \multicolumn{1}{l|}{-0.0448}     & -1.79\%                            &                                                                                          & \multicolumn{1}{l|}{3.25}         & \multicolumn{1}{l|}{0.1478}      & 4.55\%                             & \multicolumn{1}{l|}{3}            & \multicolumn{1}{l|}{-0.1618}     & -5.39\%                            \\ \cline{1-1} \cline{3-8} \cline{10-15} 
    \textbf{usb\_phy}                                       &                                                                                          & \multicolumn{1}{l|}{1.75}         & \multicolumn{1}{l|}{0.1531}      & 8.75\%                             & \multicolumn{1}{l|}{1.5}          & \multicolumn{1}{l|}{-0.0402}     & -2.68\%                            &                                                                                          & \multicolumn{1}{l|}{1.4}          & \multicolumn{1}{l|}{0.0876}      & 6.26\%                             & \multicolumn{1}{l|}{1.25}         & \multicolumn{1}{l|}{-0.036}      & -2.88\%                            \\ \cline{1-1} \cline{3-8} \cline{10-15} 
    \textbf{wb\_dma}                                        &                                                                                          & \multicolumn{1}{l|}{4.25}         & \multicolumn{1}{l|}{0.0602}      & 1.42\%                             & \multicolumn{1}{l|}{4}            & \multicolumn{1}{l|}{-0.0645}     & -1.61\%                            &                                                                                          & \multicolumn{1}{l|}{4.5}          & \multicolumn{1}{l|}{0.1333}      & 2.96\%                             & \multicolumn{1}{l|}{4.25}         & \multicolumn{1}{l|}{-0.0114}     & -0.27\%                            \\ \hline
    \end{tabular}
}
\end{table}
\vspace{-0.2in}

\subsection{Dataset Expansion Through Parameter Sweeps}
\label{subsec:dataset_expansion}

After identifying the clock periods for the IWLS benchmark circuits considered under the barely-pass (BP) and barely-fail (BF) design scenarios, additional baseline physical circuit designs are generated through a systematic sweep of parameters.
For each benchmark circuit, multiple layouts are generated by varying a selected set of physical design parameters, including the target clock period, the core aspect ratio, and the core utilization, while all other settings are held constant.
Configurations with relaxed clock periods relative to the BP point are denoted as BP$+$ , while configurations with tighter clock periods relative to the BF point are denoted as BF$-$.
The sweeps provide a sample of design points around the baseline operating conditions, producing a diverse set of layouts that maintain timing characteristics and reflect realistic physical design trade-offs.
The complete set of constraints and parameter values used to generate the dataset is listed in Table~\ref{tab:dataset_parameters}.

The resulting raw outputs (i.e DEF files, SPEF files, reports, etc) produced by the OpenROAD flow are processed and mapped into the EDA-Schema-V2 representation, which enables consistent multimodal analysis across circuits, design stages, and technology nodes. For scalar map representations, including IR-drop, electromigration (EM), and routability (RUDY) metrics, the downsampling factor $k$ in (\ref{eqn:scalar_resolution}) is fixed to 50, which ensures consistent spatial aggregation for computational efficiency across all designs.

% \input{tables/dataset_parameters_openROAD_default}
% Please add the following required packages to your document preamble:
% \usepackage{multirow}
\begin{table}[!h]
    \begin{center}
    % \vspace{0.05in}
    % \vspace{-0.05in}
    \caption{Constraints and parameters utilized to generate the dataset.}
    \label{tab:dataset_parameters}
    \fontsize{8}{10}\selectfont{
        \begin{tabular}{|l|llll|}
        \hline
        \textbf{Parameters}                                                                                                                & \multicolumn{4}{l|}{\textbf{Values or Ranges}}                                                                                                                                                \\ \hline
        Target clock periods (ns)                                                                                                          & \multicolumn{4}{l|}{\{0.8 × BF, BF, BP, 1.2 × BP\}}                                                                                                                                           \\ \hline
        Aspect ratio                                                                                                                       & \multicolumn{4}{l|}{\{0.5, 1, 1.5\}}                                                                                                                                                          \\ \hline
                                                                                                                                           & \multicolumn{1}{l|}{\cellcolor[HTML]{EFEFEF}NG45} & \multicolumn{1}{l|}{\cellcolor[HTML]{EFEFEF}SKY130} & \multicolumn{1}{l|}{\cellcolor[HTML]{EFEFEF}IHP130} & \cellcolor[HTML]{EFEFEF}ASAP7 \\ \cline{2-5} 
        \multirow{-2}{*}{Core Utilization}                                                                                                 & \multicolumn{1}{l|}{\{0.3, 0.4, 0.5\}}            & \multicolumn{1}{l|}{\{0.2, 0.3, 0.4\}}              & \multicolumn{1}{l|}{\{0.2, 0.3, 0.4\}}              & \{0.3, 0.4, 0.5\}             \\ \hline
                                                                                                                                           & \multicolumn{1}{l|}{\cellcolor[HTML]{EFEFEF}NG45} & \multicolumn{1}{l|}{\cellcolor[HTML]{EFEFEF}SKY130} & \multicolumn{1}{l|}{\cellcolor[HTML]{EFEFEF}IHP130} & \cellcolor[HTML]{EFEFEF}ASAP7 \\ \cline{2-5} 
        \multirow{-2}{*}{Core Margin (μm)}                                                                                                 & \multicolumn{1}{l|}{5.72}                         & \multicolumn{1}{l|}{22.4}                           & \multicolumn{1}{l|}{22.72}                          & 2.29                          \\ \hline
        Placement Density                                                                                                                  & \multicolumn{4}{l|}{Uniform, 1.25 × Uniform, 1.5 × Uniform}                                                                                                                                   \\ \hline
        Input/Output Delay (ns)                                                                                                            & \multicolumn{4}{l|}{20\% of Target clock period (TCP)}                                                                                                                                        \\ \hline
        Clock Latency (ns)                                                                                                                 & \multicolumn{4}{l|}{1\% of TCP (capped at 50 ps)}                                                                                                                                             \\ \hline
                                                                                                                                           & \multicolumn{1}{l|}{\cellcolor[HTML]{EFEFEF}NG45} & \multicolumn{1}{l|}{\cellcolor[HTML]{EFEFEF}SKY130} & \multicolumn{1}{l|}{\cellcolor[HTML]{EFEFEF}IHP130} & \cellcolor[HTML]{EFEFEF}ASAP7 \\ \cline{2-5} 
        \multirow{-2}{*}{\begin{tabular}[c]{@{}l@{}}Clock Transition (ns) (12.5\% of TCP\\ of the fastest circuit for a PDK)\end{tabular}} & \multicolumn{1}{l|}{0.05}                         & \multicolumn{1}{l|}{0.1875}                         & \multicolumn{1}{l|}{0.1875}                         & 0.02375                       \\ \hline
        Clock Uncertainty (ns)                                                                                                             & \multicolumn{4}{l|}{5\% of TCP (capped at 50 ps)}                                                                                                                                             \\ \hline
        \end{tabular}
    }
    \end{center}
    % \vspace{-0.05in}
\end{table}

\section{Dataset Analysis}
\label{sec:dataset_analysis}

In this section, an in-depth analysis of the generated dataset is provided.
The objective is to characterize key aspects of the dataset while also providing insights into the composition of the generated circuits.
A high-level overview of the dataset is presented in Section~\ref{subsec:dataset_overview}, with the summary of the size, composition, and coverage of the data provided.
A quantitative analysis of the generated designs, including the timing, power, and area distributions across technology nodes is provided in Section~\ref{subsec:qor_analysis}.
The effect of design and flow parameters including clock period, core utilization, aspect ratio, and placement density on physical and performance metrics is analyzed in Section~\ref{subsec:parameter_sensitivity_analysis}.
The change in the metric scores across different physical design stages is described in Section~\ref{subsec:inter_stage_analysis}.

\subsection{Dataset Overview}
\label{subsec:dataset_overview}

The composition of the dataset across all technology nodes, including the number of design instances, structural elements, timing outcomes, and the numerical range of the final-stage quality-of-results for each PDK are listed in Table~\ref{tab:dataset_summary}.
For each technology node, the dataset contains physical implementations of 18 benchmark circuits evaluated across 108 parameter configurations, yielding 1,944 design instances per PDK.
Across the four supported PDKs (NG45, SKY130, IHP130, and ASAP7), a total of 7,776 design instances are generated spanning all major stages of the digital physical design flow.
A small subset of flows did not complete the global routing stage due to routing congestion and failure, which resulted in missing data for global routing, detailed routing, and the final design stages for 25 design instances (3 in NG45, 17 in SKY130, 3 in IHP130, and 2 in ASAP7). However, data from all preceding stages, up to and including clock tree synthesis (CTS) remain available for the incomplete instances.

The generated dataset includes multimodal artifacts such as circuit graphs, images of spatial layouts, and detailed quality-of-results reports. Structural and metric data is stored in a columnar Parquet format that supports efficient analytical queries and scalable processing, while spatial layout representations are stored as NumPy arrays for compact storage and fast loading in numerical and machine learning workflows. The size  of the dataset for the 1,944 design instances per PDK is approximately 69\,GB for NG45, 78\,GB for SKY130, 92\,GB for IHP130, and 81\,GB for ASAP7.

In aggregate, across all technology nodes, the dataset is comprised of over 275 million gates (82 million excluding filler cells) and 75 million nets, along with more than 36 million timing paths, reflecting a broad range of circuit sizes, architectural complexities, and operating conditions.
Final-stage timing outcomes indicate that both timing-clean and timing-violating designs are well represented across all PDKs, with approximately 36.7 million timing-clean paths and 1.02 million timing-violating paths in aggregate across the dataset, which reflects a balanced coverage of designs that pass and fail timing.
The properties of the dataset demonstrate that an extensive and diverse coverage is provided of realistic physical designs across multiple process technologies, which enables a robust analysis of performance, power, and area behavior, as well as timing, routing, and physical design characteristics under varied design conditions.

% \vspace{-0.05in}
\begin{table}[h!]
\centering
\caption{Dataset composition and final-stage quality metrics across technology nodes. Quality metrics are reported as minimum to maximum values across all final-stage design instances.}
\label{tab:dataset_summary}
\fontsize{8}{10}\selectfont{
% \resizebox{0.5\columnwidth}{!}{
    \begin{tabular}{|lllll|}
    \hline
    \multicolumn{1}{|c|}{}                                              & \multicolumn{4}{c|}{\textbf{PDK}}                                                                                                                                                                                                                                                                                                       \\ \cline{2-5} 
    \multicolumn{1}{|c|}{\multirow{-2}{*}{\textbf{}}}                   & \multicolumn{1}{l|}{\textbf{NG45}}                                                   & \multicolumn{1}{l|}{\textbf{SKY130}}                                                  & \multicolumn{1}{l|}{\textbf{IHP130}}                                                   & \textbf{ASAP7}                                                  \\ \hline
    \multicolumn{5}{|c|}{\cellcolor[HTML]{EFEFEF}\textbf{Dataset Composition}}                                                                                                                                                                                                                                                                                                                                    \\ \hline
    \multicolumn{1}{|l|}{\textbf{Total gates}}                          & \multicolumn{1}{l|}{39136357}                                                        & \multicolumn{1}{l|}{74270175}                                                         & \multicolumn{1}{l|}{108913101}                                                         & 53204206                                                        \\ \hline
    \multicolumn{1}{|l|}{\textbf{Total gates (excluding filler cells)}} & \multicolumn{1}{l|}{17579912}                                                        & \multicolumn{1}{l|}{21359517}                                                         & \multicolumn{1}{l|}{21027098}                                                          & 22055351                                                        \\ \hline
    \multicolumn{1}{|l|}{\textbf{Total nets}}                           & \multicolumn{1}{l|}{17793857}                                                        & \multicolumn{1}{l|}{14838301}                                                         & \multicolumn{1}{l|}{20241099}                                                          & 23086812                                                        \\ \hline
    \multicolumn{1}{|l|}{\textbf{Total pins}}                           & \multicolumn{1}{l|}{61430010}                                                        & \multicolumn{1}{l|}{50825641}                                                         & \multicolumn{1}{l|}{69281157}                                                          & 73797773                                                        \\ \hline
    \multicolumn{1}{|l|}{\textbf{Total timing paths}}                   & \multicolumn{1}{l|}{7959170}                                                         & \multicolumn{1}{l|}{10877094}                                                         & \multicolumn{1}{l|}{10147368}                                                          & 8223683                                                         \\ \hline
    \multicolumn{1}{|l|}{\textbf{Timing clean designs}}                 & \multicolumn{1}{l|}{7707993}                                                         & \multicolumn{1}{l|}{10525557}                                                         & \multicolumn{1}{l|}{9981539}                                                           & 7968189                                                         \\ \hline
    \multicolumn{1}{|l|}{\textbf{Timing violating designs}}             & \multicolumn{1}{l|}{251177}                                                          & \multicolumn{1}{l|}{351537}                                                           & \multicolumn{1}{l|}{165829}                                                            & 255494                                                          \\ \hline
    \multicolumn{5}{|c|}{\cellcolor[HTML]{EFEFEF}\textbf{Final stage Quality-of-Results}}                                                                                                                                                                                                                                                                                                                         \\ \hline
    \multicolumn{1}{|l|}{\textbf{Total area (µm2)}}                     & \multicolumn{1}{l|}{723 to 100505}                                                   & \multicolumn{1}{l|}{3576 to 513155}                                                   & \multicolumn{1}{l|}{16056 to 5592904}                                                  & 58 to 9333                                                      \\ \hline
    \multicolumn{1}{|l|}{\textbf{Total power (µW)}}                     & \multicolumn{1}{l|}{2230 to 456000}                                                  & \multicolumn{1}{l|}{3690 to 700000}                                                   & \multicolumn{1}{l|}{3410 to 383000}                                                    & 654 to 115000                                                   \\ \hline
    \multicolumn{1}{|l|}{\textbf{Worst slack (ns)}}                     & \multicolumn{1}{l|}{-0.3474 to 0.5806}                                               & \multicolumn{1}{l|}{-4.792 to 0.5328}                                                 & \multicolumn{1}{l|}{-1.555 to 2.102}                                                   & -0.4188 to 0.2555                                               \\ \hline
    \multicolumn{1}{|l|}{\textbf{Total negative slack (ns)}}            & \multicolumn{1}{l|}{-463.9339 to 0}                                                  & \multicolumn{1}{l|}{-3045.1636 to 0}                                                  & \multicolumn{1}{l|}{-1672.6881 to 0}                                                   & -434.9989 to 0                                                  \\ \hline
    \multicolumn{1}{|l|}{\textbf{Violating endpoints}}                  & \multicolumn{1}{l|}{0 to 4034}                                                       & \multicolumn{1}{l|}{0 to 9917}                                                        & \multicolumn{1}{l|}{0 to 3910}                                                         & 0 to 12123                                                      \\ \hline
    \multicolumn{1}{|l|}{\textbf{Total wirelength (µm)}}                & \multicolumn{1}{l|}{\begin{tabular}[c]{@{}l@{}}2326.59 to\\ 900314.625\end{tabular}} & \multicolumn{1}{l|}{\begin{tabular}[c]{@{}l@{}}5639.58 to\\ 2137809.055\end{tabular}} & \multicolumn{1}{l|}{\begin{tabular}[c]{@{}l@{}}10166.155 to\\ 3621758.23\end{tabular}} & \begin{tabular}[c]{@{}l@{}}751.681 to\\ 591701.235\end{tabular} \\ \hline
    \end{tabular}
}
\end{table}
% \vspace{-0.05in}

\begin{figure}[!b]
    \begin{center}
        \begin{subfigure}[!h]{0.245\linewidth}
            \includegraphics[width=\columnwidth]{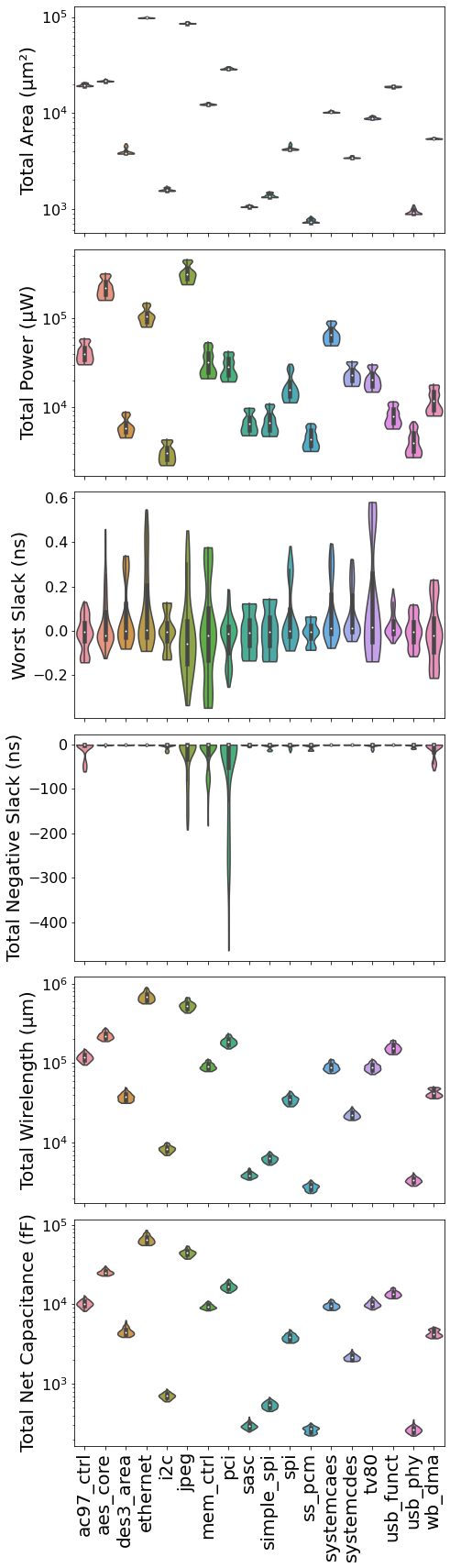}
            \caption{}
            \label{fig:dataset_ppa}
        \end{subfigure}
        \begin{subfigure}[!h]{0.245\linewidth}
            \includegraphics[width=\columnwidth]{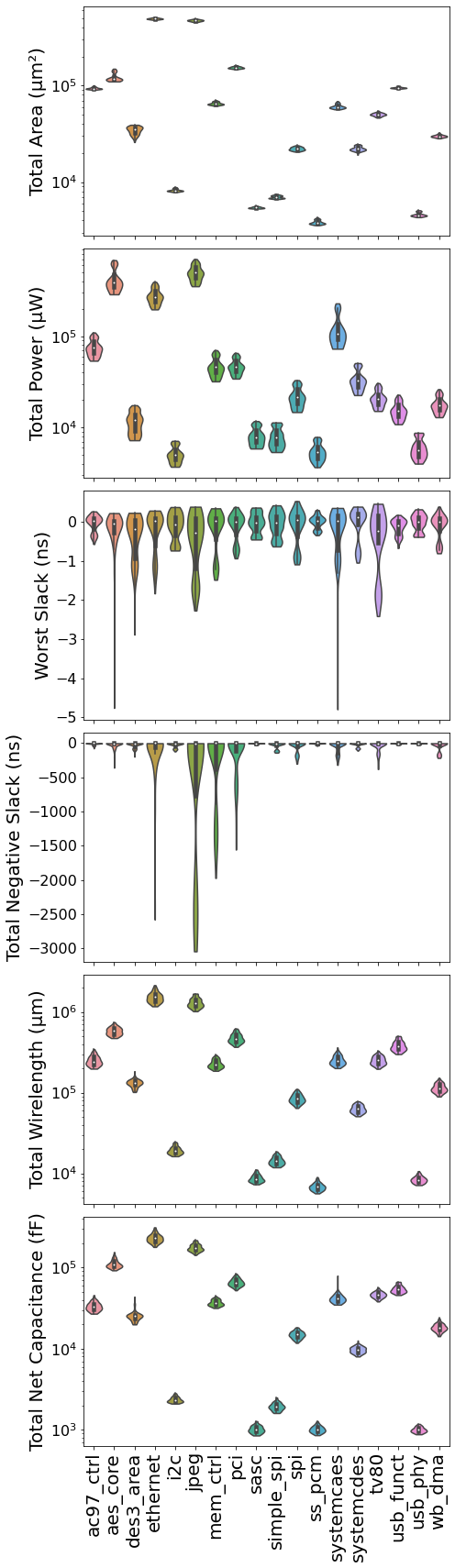}
            \caption{}
            \label{fig:dataset_ppa}
        \end{subfigure}
        \begin{subfigure}[!h]{0.245\linewidth}
            \includegraphics[width=\columnwidth]{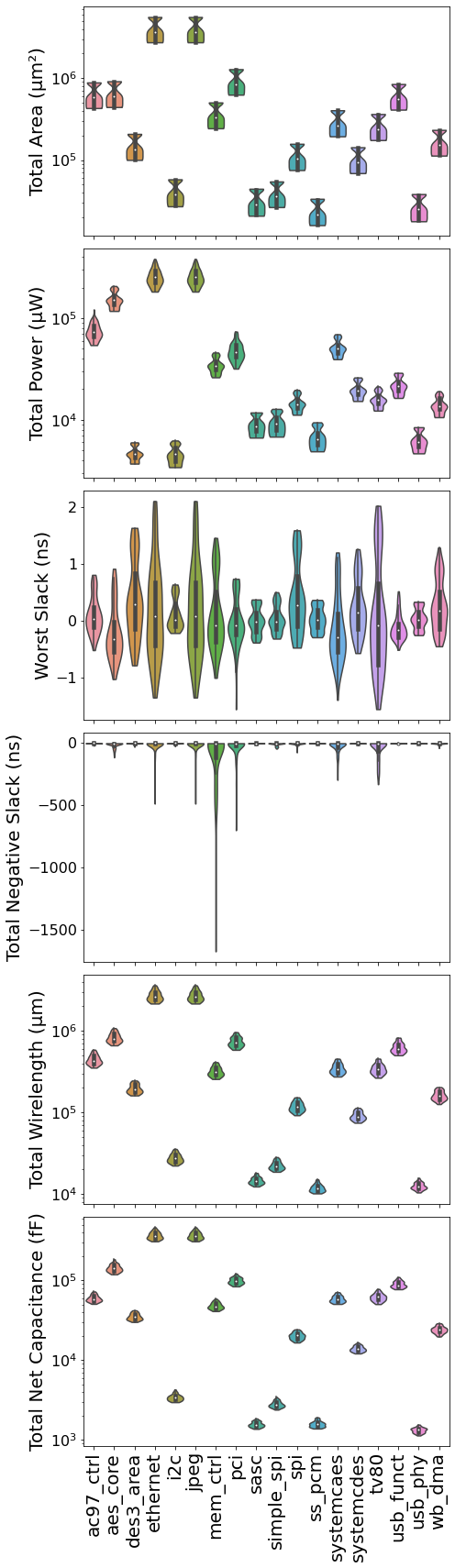}
            \caption{}
            \label{fig:dataset_ppa}
        \end{subfigure}
        \begin{subfigure}[!h]{0.245\linewidth}
            \includegraphics[width=\columnwidth]{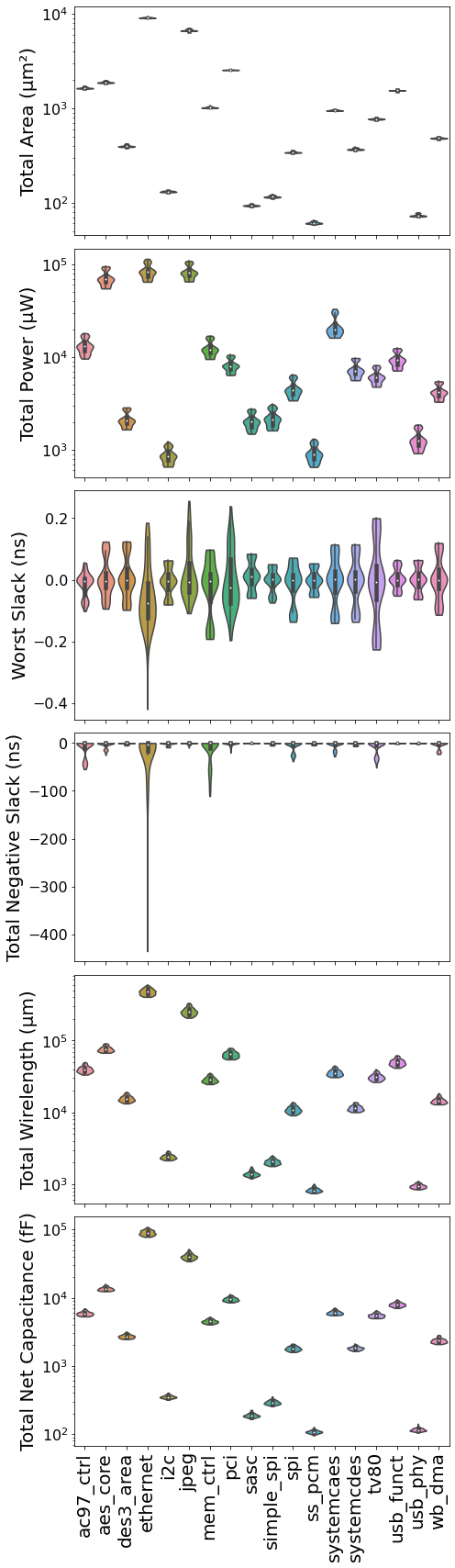}
            \caption{}
            \label{fig:dataset_ppa}
        \end{subfigure}
    \end{center}
    \caption{Distribution of post-routing quality-of-results (QoR) metrics across benchmark circuits for each technology node.
    Box plots characterize (from top to bottom) total area, total power, worst slack (WNS), total negative slack (TNS), total routed wirelength, and total capacitance.
    Total area, total power, total routed wirelength, and total capacitance are shown on a logarithmic scale. Subfigures (a)–(d) correspond to NG45, SKY130, IHP130, and ASAP7, respectively.}
\end{figure}

\subsection{Quality-of-Results (QoR) Analysis}
\label{subsec:qor_analysis}

An analysis of the quality-of-results (QoR) after completion of the final stage of the physical design flow provides characterization of the dataset by examining circuit-level power, performance, and area metrics. The objective of the analysis is to assess the diversity and coverage of the dataset across benchmark circuits and technology nodes, and to verify that the generated designs span a range of physically meaningful operating conditions.
Final-stage QoR metrics that are analyzed include total area, total power, worst slack (WNS), total negative slack (TNS), number of violating timing endpoints, and total routed wirelength. The observed range of each metric and for each technology node is listed in Table~\ref{tab:dataset_summary}, while per-design distributions across all four PDKs using box and violin plots are provided in Fig.~\ref{fig:dataset_ppa}.

The distributions shown in Fig.~\ref{fig:dataset_ppa} indicate an achieved target level of variation in the QoR metrics of the final design stage across the benchmark circuits. The dataset includes both small and large circuits with metric distributions spanning multiple orders of magnitude, as indicated by the area violin plot in Fig.~\ref{fig:dataset_ppa}. Some circuits occupy only tens to thousands of square microns with relatively short routed interconnect, while other circuits occupy several million square microns and exhibit extremely large wirelengths, as indicated by the range of values listed in Table~\ref{tab:dataset_summary}. The observed distributions indicate that the dataset spans a wide range of sizes and physical complexities across benchmark circuits and technology nodes.

\begin{figure}[!h]
    % \vspace{-0.05in}
    \begin{center}
        \begin{subfigure}[!h]{0.49\columnwidth}
            \includegraphics[width=0.95\columnwidth]{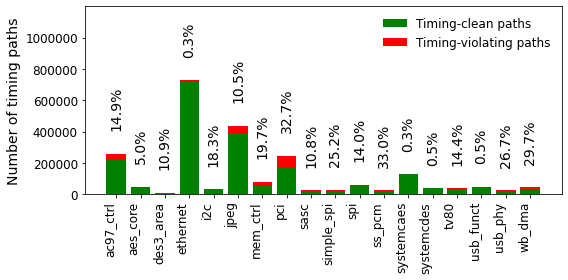}
            % \label{full_dataset_ppa}
            \caption{}
        \end{subfigure}
        \begin{subfigure}[!h]{0.49\columnwidth}
            \includegraphics[width=0.95\columnwidth]{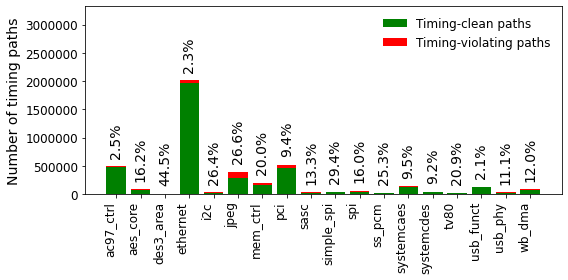}
            \caption{}
            % \label{fig:full_dataset_ppa}
        \end{subfigure}
        \begin{subfigure}[!h]{0.49\columnwidth}
            \includegraphics[width=0.95\columnwidth]{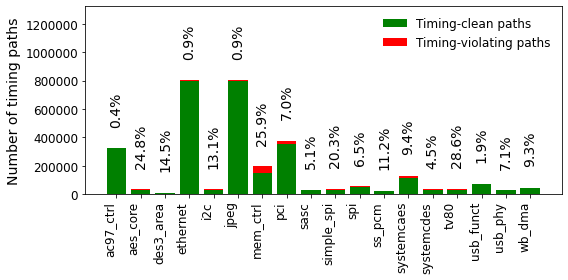}
            \caption{}
            % \label{fig:full_dataset_ppa}
        \end{subfigure}
        \begin{subfigure}[!h]{0.49\columnwidth}
            \includegraphics[width=0.95\columnwidth]{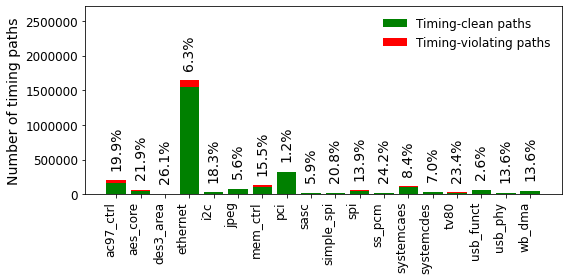}
            \caption{}
            % \label{fig:full_dataset_ppa}
        \end{subfigure}
    \end{center}
    % \vspace{-0.05in}
    \caption{Analysis of timing-clean and timing-violating path distributions after completion of the final design stage. Subfigures (a)–(d) correspond to NG45, SKY130, IHP130, and ASAP7, respectively.}
    \label{fig:timing_paths_summary}
    % \vspace{-0.05in}
\end{figure}

While many generated designs cluster within relatively narrow ranges of values, a smaller subset of circuits produce extreme values that dominate the overall distribution of data, particularly for area, power, and wirelength.
Across all technology nodes \textit{ethernet} and \textit{jpeg} consistently appear at the upper extremes of the area and wirelength distributions.
Both \textit{ethernet} and \textit{jpeg} are larger circuits, include more interconnects, and provide behavior that is consistent across PDKs despite a difference in absolute scale.
Across technology nodes, the relative order of circuits in terms of area and wirelength remains consistent, indicating that architectural characteristics dominate absolute scaling effects introduced by the PDK.

The timing behavior after completion of the final design stage is summarized in Fig.~\ref{fig:timing_paths_summary}, which reports results from analysis of timing-clean and timing-violating path counts per circuit along with the corresponding percentage of paths that include timing violations.
The results indicate that the dataset captures both fully timing-clean designs and designs that exhibit non-zero timing violations, and that the fraction of violating paths varies across circuits and technologies.
The inclusion of timing paths that both meet and violate the timing budget provides a comprehensive range of timing behaviors, which enhances the usefulness of the dataset for timing analysis and modeling.

Overall, the analysis of the final-stage QoR indicates that the dataset covers a wide range of physical design characteristics across circuits and technologies.
The observed variation in circuit size, power, and timing behavior confirms that the dataset includes both simple and complex designs and provides sufficient data diversity to study power, performance, area, timing closure, and routability trade-offs through all stages of the physical design flow.

\subsection{Parameter Sensitivity Analysis}
\label{subsec:parameter_sensitivity_analysis}

An evaluation of the effects of design parameters (clock period, core aspect ratio, core utilization, and placement density) on key quality metrics (total area, total power, worst slack, total negative slack, total wirelength, and net capacitance) is performed.
The characterization of the effects of the parameters on the metrics is utilized to validate the inclusion of each parameter as a component of the dataset.

\begin{figure}[!h]
    % \vspace{-0.05in}
    \begin{center}
        \includegraphics[width=\columnwidth]{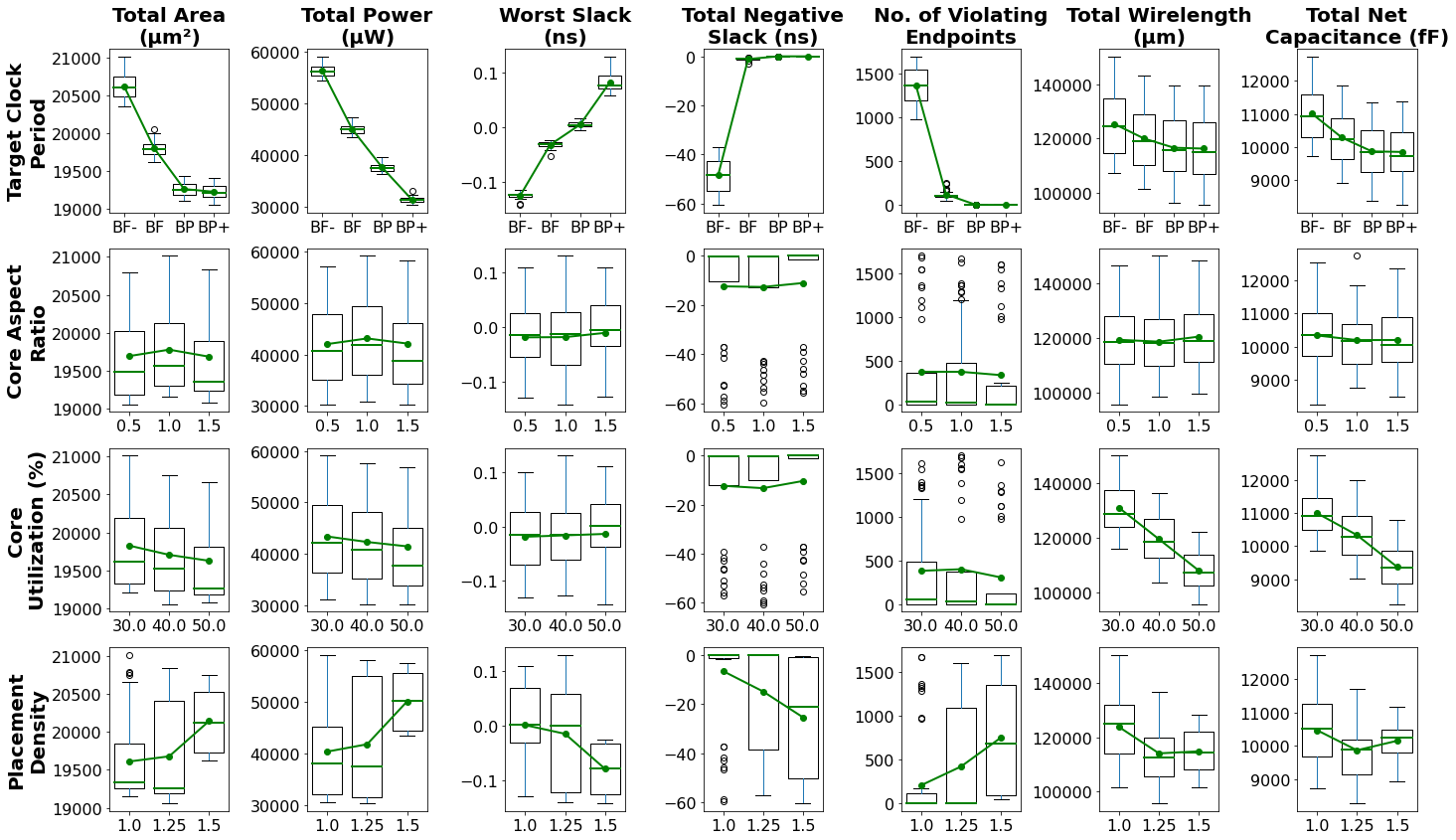}
    \end{center}
    % \vspace{-0.05in}
    \caption{Distribution of final-stage quality-of-results metrics for total area, total power consumption, worst-case slack, total negative slack, number of endpoints that violate slack, total wirelength, and total net capacitance for the \textit{ac97\_ctrl} circuit with variations in clock period, core aspect ratio, core utilization, and placement density using the NG45 technology node.}
    \label{fig:parameter_sensitivity_analysis}
    % \vspace{-0.1in}
    % \vspace{-0.05in}
\end{figure}

Distributions of the quality metrics as a function of design parameters for the \textit{ac97\_ctrl} circuit physically designed in the NG45 technology node are shown in Fig.~\ref{fig:parameter_sensitivity_analysis}.
Results indicate that changes in clock period, core aspect ratio, core utilization, and placement density lead to observable variations in the evaluated quality metrics.
For each parameter, boxplots capture the variability across runs, while the overlaid trend lines mark the mean value of the corresponding metric as a function of the design parameter.
Modifications in design parameters lead to consistent shifts in mean values and changes in the variance of the distribution, which indicates a systematic effect on the post-routing quality of results.
In particular, variations in clock period result in pronounced changes in timing-related metrics, while core utilization and placement density introduce increased dispersion reflecting effects due to congestion and routability.
The observed variation in mean values and distributions confirms the effect of each considered design parameter on the evaluated metrics and motivates the inclusion of each parameter and corresponding metric in the dataset.

To quantitatively characterize the relationship between design parameters and quality of results metrics across the dataset, a correlation analysis is performed using all circuits included in EDA-Schema-V2.
For each circuit implemented in a given PDK, the Pearson correlation coefficient ($r$) is computed between the design parameters used to generate the circuit and the corresponding quality metrics values observed across the parameterized design instances.
The Pearson correlation coefficient ($r$) is computed for each parameter-metric pair to determine the strength and direction of the corresponding linear relationships.
The Pearson correlation coefficient is given by
\begin{equation}
r = \frac{\sum_{i=1}^{n} (x_i - \bar{x})(y_i - \bar{y})}{\sqrt{\sum_{i=1}^{n} (x_i - \bar{x})^2} \sqrt{\sum_{i=1}^{n} (y_i - \bar{y})^2}}
\end{equation}
\noindent where $x_i$ and $y_i$ are the values of the two variables being compared, $\bar{x}$ and $\bar{y}$ are the mean values of each variable, and $n$ is the number of observations.

\begin{figure}[!h]
    % \vspace{-0.05in}
    \includegraphics[width=\columnwidth]{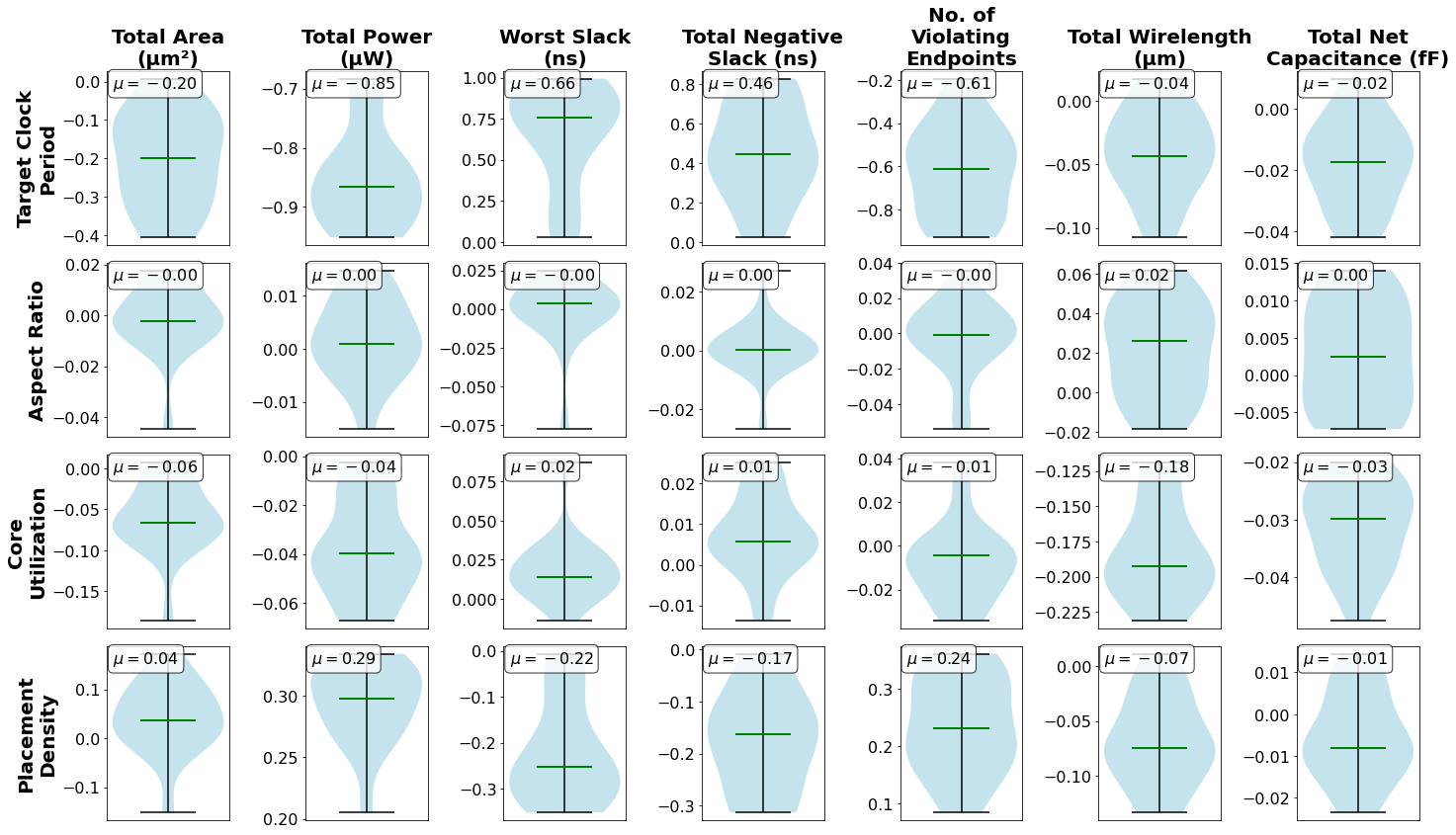}
    \caption{Distribution of Pearson correlation coefficients between quality-of-results metrics and design parameters across all circuits in the NG45 dataset. The correlation coefficient are shown with violin plots for each parameter–metric pair.}
    \label{fig:parameter_correlation_violins}
    % \vspace{-0.05in}
\end{figure}

The correlation between design parameters and quality metrics across all circuits generated in the NG45 technology node is shown in Fig.~\ref{fig:parameter_correlation_violins}.
Violin plots of the distributions of the Pearson correlation coefficients for each parameter–metric pair are provided, which captures the overall shape and spread of correlated parameter-metric pairs across circuits.
Analysis of the distributions demonstrates that the effect of design parameters on quality metrics is consistently observed across the dataset and is not limited to a single circuit.
Many parameter–metric pairs exhibit large correlations, often approaching $\pm1$, which indicates a strong linear relationship between design parameters and post-routing quality metrics across a broad set of circuits.
Clock period is strongly correlated with timing-related metrics and total power, with average correlation magnitudes greater than $\pm 0.4$.
Clock period also exhibits measurable correlation with total area, with average correlation magnitudes greater than $\pm 0.1$.
Similarly, placement density exhibits measurable correlations ($\geq \pm 0.1$) with timing metrics and total power, while core utilization exhibits measurable correlation with total wirelength.
The remaining parameter–metric pairs exhibit weak correlations ($\leq \pm 0.1$), which, while small in magnitude, still indicate minor but consistent variation across circuits.
The observed relationships between design parameters and QoR metrics are consistent across different technology nodes, as demonstrated from analysis provided in Appendix~\ref{sec:appendix_param_sensitivity}.
Overall, the results from analysis of the Pearson correlation coefficients confirm that parameterization systematically influences quality of results across the dataset and captures meaningful, parameter-induced variation rather than isolated circuit behavior.

\subsection{Inter-Design Stage Analysis}
\label{subsec:inter_stage_analysis}

Similar to the parameter sensitivity analysis, inter-design stage analysis provides an evaluation of quality metrics at each stage of the physical design flow, characterizing the effect of each design stage on circuit performance. The analysis characterizes the evolution of total area, total power consumption, total wirelength, total net capacitance, and timing-related metrics as the circuit progresses through successive stages of the physical design flow.
The analysis also quantifies the contribution of individual stages to the final quality of results.

\begin{figure}[!h]
    \centering

    \begin{subfigure}{0.49\columnwidth}
        \centering
        \includegraphics[width=0.8\linewidth]{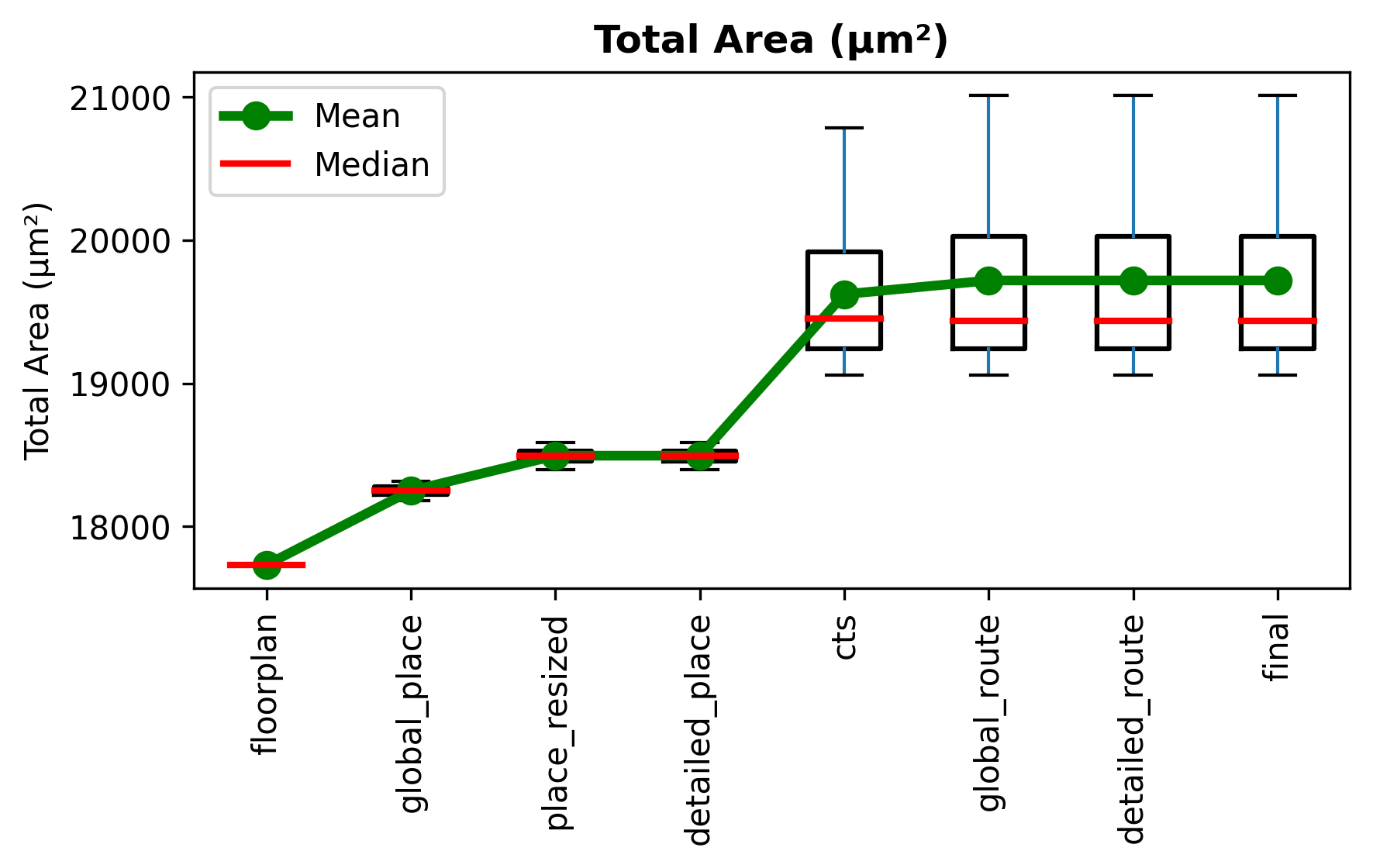}
        \caption{}
    \end{subfigure}
    \begin{subfigure}{0.49\columnwidth}
        \centering
        \includegraphics[width=0.8\linewidth]{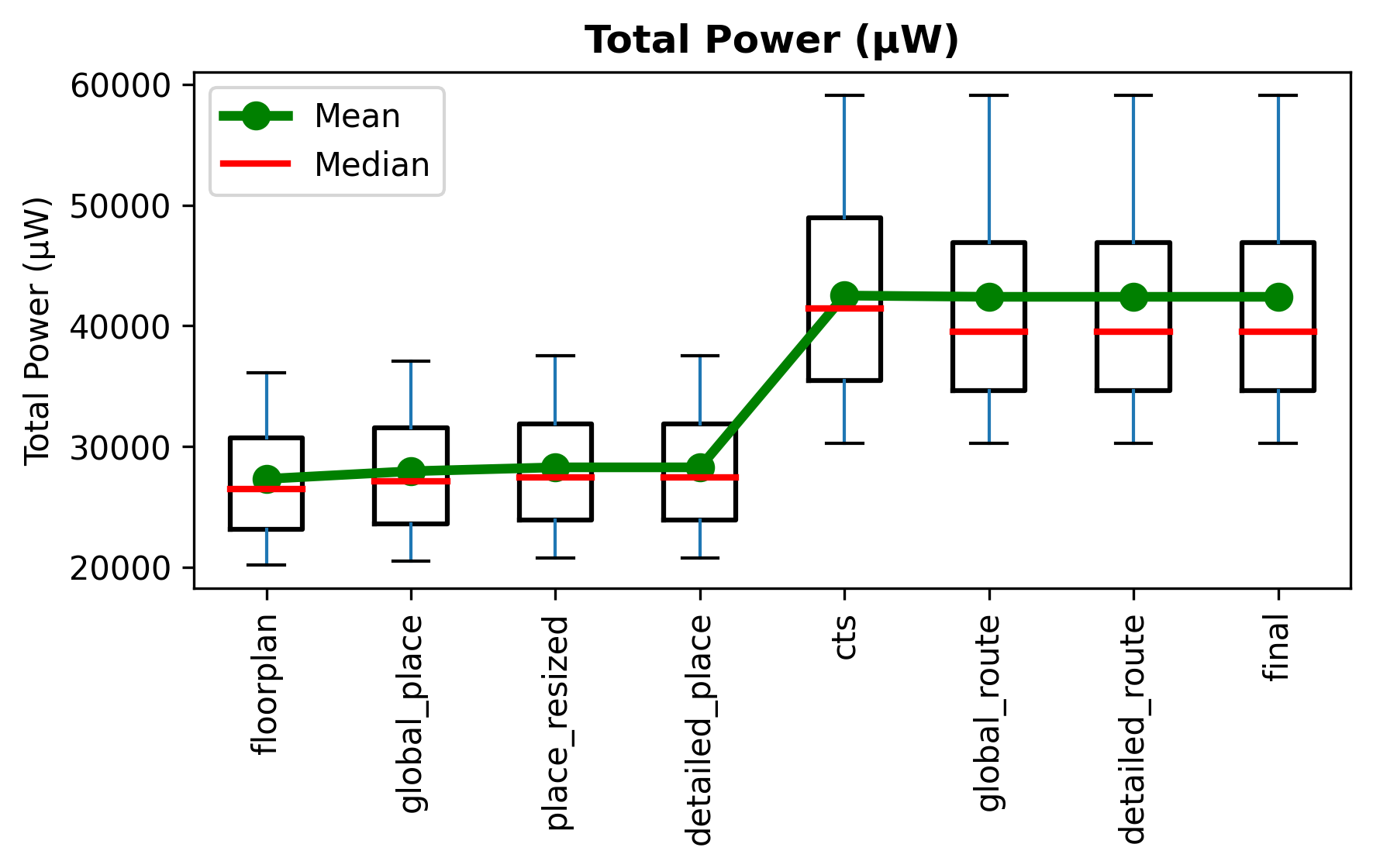}
        \caption{}
    \end{subfigure}
    \begin{subfigure}{0.49\columnwidth}
        \centering
        \includegraphics[width=0.8\linewidth]{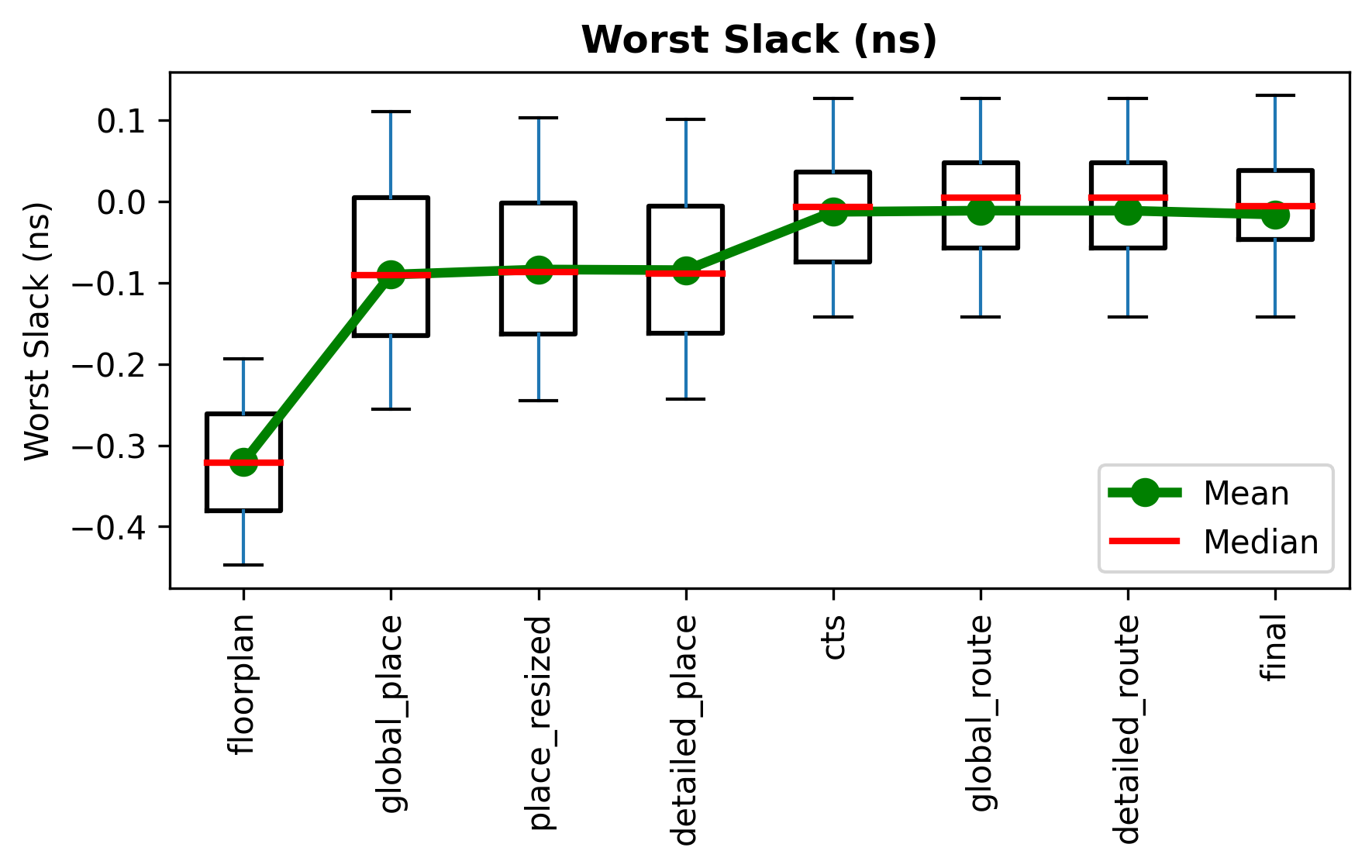}
        \caption{}
    \end{subfigure}
    \begin{subfigure}{0.49\columnwidth}
        \centering
        \includegraphics[width=0.8\linewidth]{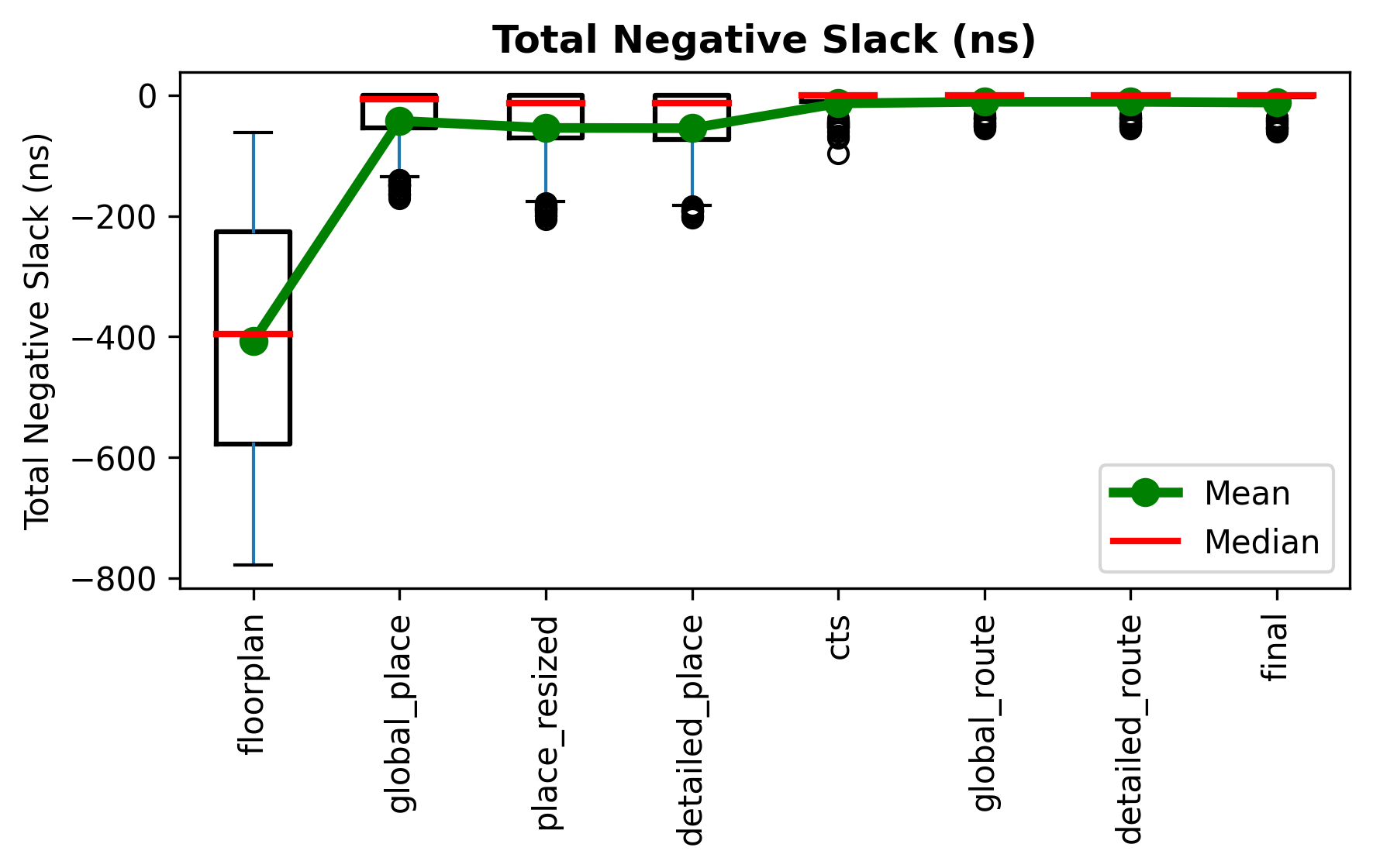}
        \caption{}
    \end{subfigure}
    \begin{subfigure}{0.49\columnwidth}
        \centering
        \includegraphics[width=0.8\linewidth]{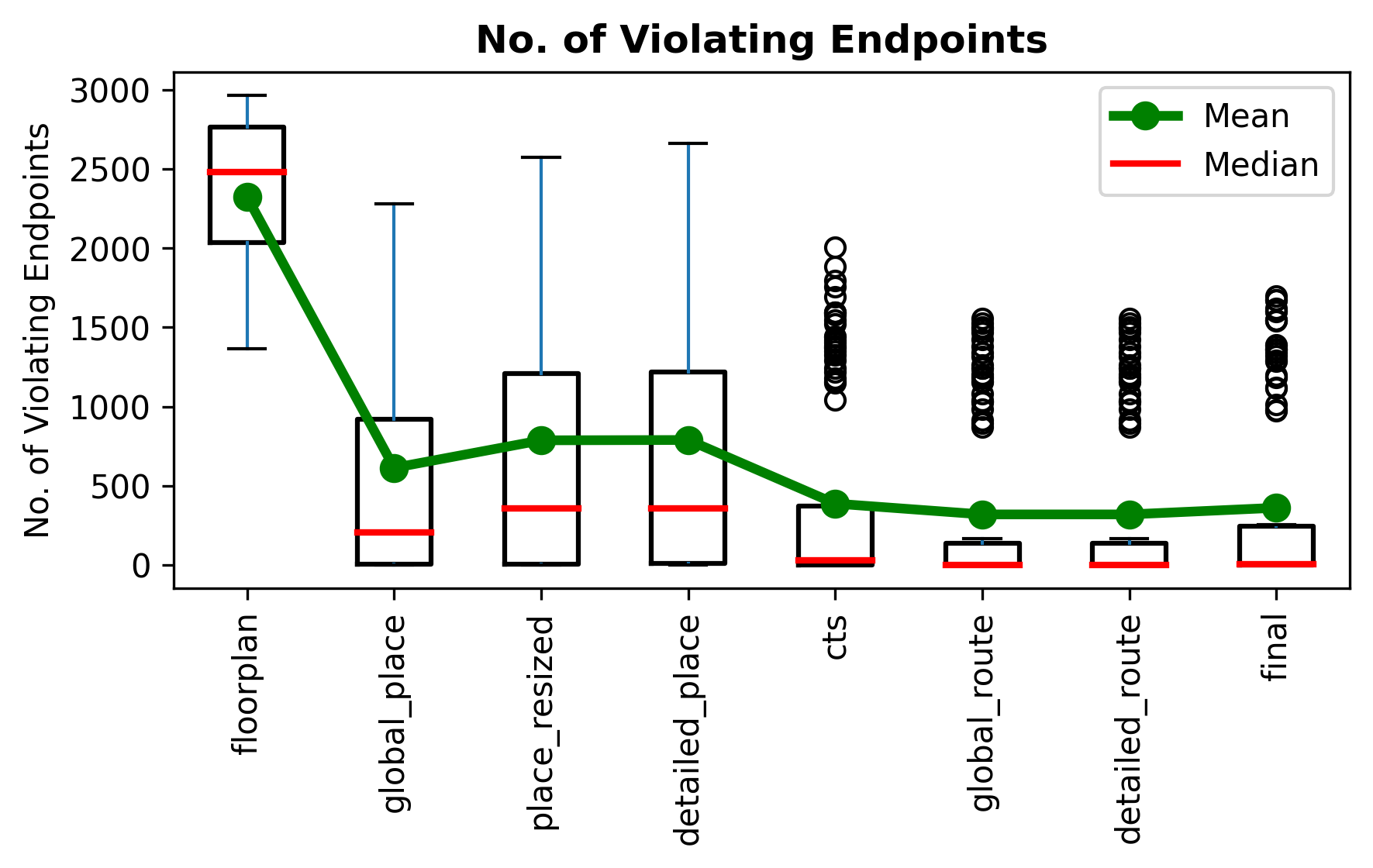}
        \caption{}
    \end{subfigure}
    \begin{subfigure}{0.49\columnwidth}
        \centering
        \includegraphics[width=0.8\linewidth]{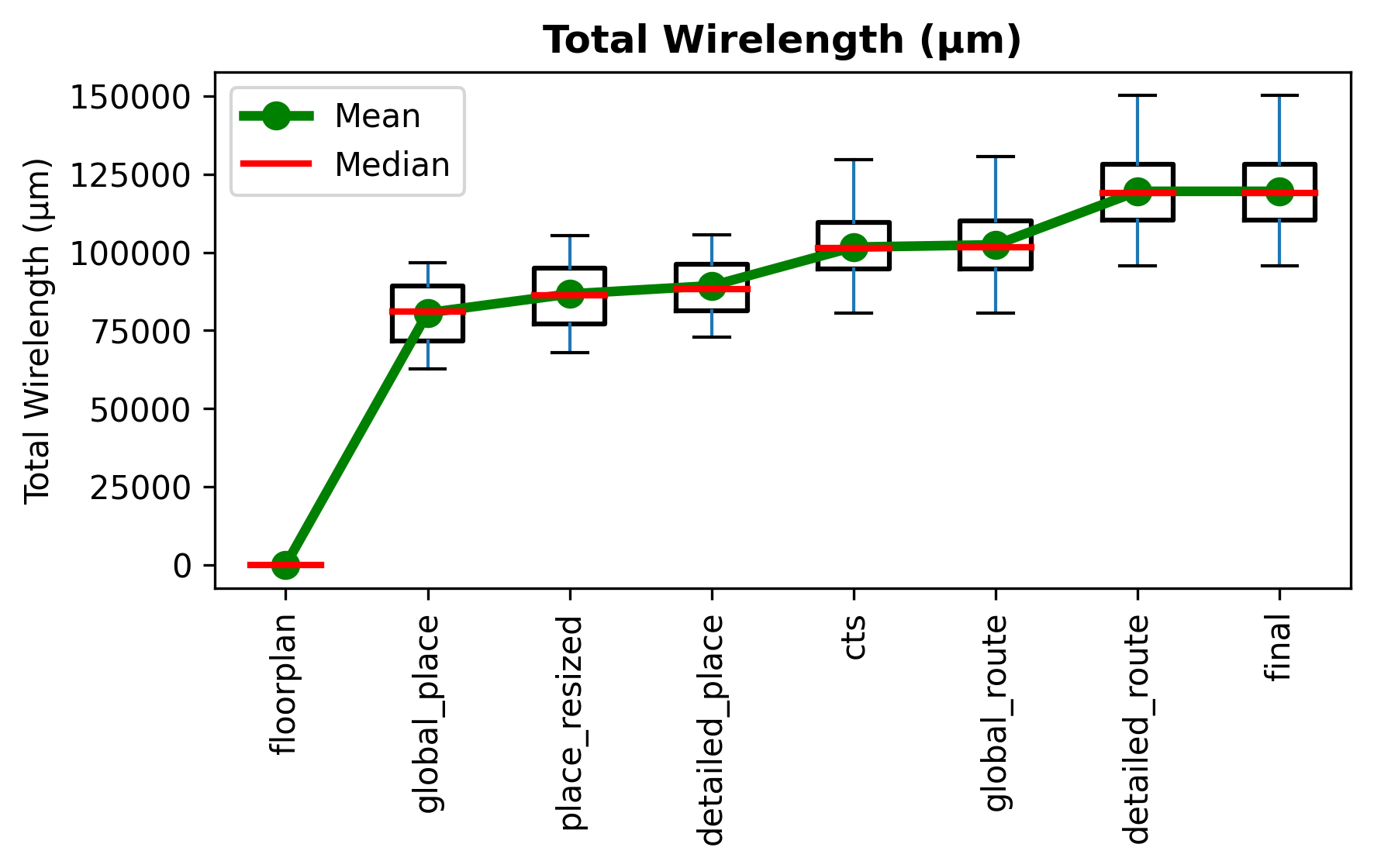}
        \caption{}
    \end{subfigure}

    \caption{Change in the distribution of (a) total area, (b) total power, (c) worst slack, (d) total negative slack, (e) number of violating endpoints, and (f) total wirelength across physical design stages for the \textit{ac97\_ctrl} circuit implemented in the NG45 technology node.}
    \label{fig:stage_analysis}
\end{figure}

The change in quality metrics for the \textit{ac97\_ctrl} circuit across stages of the physical design flow is shown in Fig.~\ref{fig:stage_analysis}. The distributions of total area, total power, worst slack, total negative slack, number of violating endpoints, and net length for each design stage is analyzed.
Trend lines indicate the mean value of each metric at each stage of the flow.
The results indicate that quality metrics change as the circuit progresses from floorplanning through placement, clock tree synthesis, and routing.
Total area and total power exhibit incremental changes during early design stages, followed by more significant changes during clock tree synthesis and routing due to the insertion of clock buffers and additional physical resources (metals, gates, and gate sizes).
Timing-related metrics, including worst slack and total negative slack, demonstrate design stage-dependent behavior.
Significant improvements typically occur after optimizing placement and clock tree synthesis, followed by stable or minor degradation in timing parameters during detailed routing.
Across all evaluated metrics, variability increases during later design stages, reflecting the cumulative effects of physical optimization and the influence of multiple design constraints.
The distributions indicate that each physical design stage produces distinct and measurable changes in circuit quality. The observed behavior motivates the inclusion of data from intermediate design stages of the physical design flow.

% \vspace{-0.05in}

\section{EDA-schema for Machine Learning}
\label{sec:machine_learning}

The primary aim of this work is to provide a foundational dataset for machine learning applications.
In Section~\ref{subsec:feature_set_analysis}, the datasets and feature sets utilized in prior research efforts are analyzed, assessing the effectiveness of the proposed data model in achieving interoperability and comparability.
A baseline analysis of three prediction tasks that utilize metrics provided by EDA-Schema-V2 and the open datasets are described in Section~\ref{subsec:baseline_analysis}.

% \subsection{Feature Set Analysis}
% \label{subsec:feature_set_analysis}
% In this section, prior research across a broad range of machine learning tasks in electronic design automation is analyzed to evaluate the interoperability and expressiveness of the proposed data model.
% A comprehensive overview of eight predictive EDA tasks—spanning QoR, timing, parasitic, net length, IR-drop, power, and routability predictions—along with representative design optimization and design generation studies, is summarized in Table~\ref{tab:feature_analysis}.
% Each work is annotated with its learning objective, data modality, feature representation, and network architecture, where \textbf{bold} features are natively available in EDA-Schema-V2 and \textit{italicized} features are directly derivable from its structural and spatial entities.

% The bold features listed in Table~\ref{tab:feature_analysis} are directly available within EDA-schema for ML tasks.
% In addition, the italicized features represent attributes that are not directly included in EDA-schema, but are derived using the provided structures and features, demonstrating the ability to expand the schema.
% For example, in \cite{agnesina2020vlsi}, the number of nets with a fanout of $\leq$ 10 is not explicitly listed in EDA-schema.
% A value, however, is calculated by using the net fanout feature.
% Similarly, metrics like the distance between a driver cell and target cells as well as pin and net density are computed using the structural and spatial features of the netlist and net graphs in EDA-schema.

\subsection{Feature Coverage Analysis}
\label{subsec:feature_set_analysis}

Machine learning methods are increasingly applied to a wide range of electronic design automation (EDA) problems, including power prediction, wirelength estimation, timing prediction, parasitic parameter prediction, routability analysis, and electromigration (EM) or IR-drop estimation.
Prior studies utilize a diverse set of modeling approaches, including convolutional neural networks (CNNs) \cite{ahn2024dtoc, xie2018routenet, almeida2025eh}, graph neural networks (GNNs) \cite{lu2022driving, du2024powpredict, xie2021net2, shrestha2025differentiable, shrestha2022graph, shrestha2023graph, jakang2025}, encoder–decoder models \cite{li2024lightweight, chhabria2022encoder, chen2024global, yu2025lared}, and hybrid multimodal architectures that combine graph and image representations \cite{cao2025optimization}.
Each modeling approach operates on different stages of the physical design flow and relies on structural, spatial, and analytic features extracted from circuit representations.
Compatibility between prior modeling approaches and the proposed schema is evaluated through a feature coverage analysis of representative prior work. The selected studies span six predictive tasks in digital physical design: power prediction, wirelength prediction, timing prediction, parasitic impedance prediction, routability prediction, and EM or IR-drop analysis.
Features used across the predictive tasks, together with the associated design stage, modeling technique, toolchain assumptions, and required inputs, are listed in Table~\ref{tab:feature_analysis}.
Features directly provided by the schema are classified as \textit{Available Features}.
% Available features include gate-level attributes such as cell function, drive strength, and power or leakage metrics; net-level properties include resistance, capacitance, and connectivity; timing-related quantities include slack, arrival time, and slew; and spatial layout properties include cell placement maps, routing maps, and power-delivery network representations.
Each attribute corresponds to values stored within the structural, metric, and spatial entities defined in EDA-Schema-V2.
Additional features frequently used in prior work are classified as \textit{Derivable Features}.
Derivable features are not available explicitly but are computed from the structural graphs and spatial layout representations contained in EDA-Schema-V2.
Examples include fan-in and fan-out statistics, cell and pin density maps, congestion indicators such as RUDY scores, Manhattan distances between cells, and neighborhood statistics used in graph-based models.
As the schema stores complete netlist graphs, placement coordinates, routing geometries, and spatial layout images, derived quantities are computed consistently across different predictive tasks.

The analysis of feature coverage indicates that most features used in existing machine learning pipelines map naturally to the graph-based structural entities, spatial image entities, and metric entities defined in EDA-Schema-V2.
Structural entities, spatial image entities, and metric entities provide multimodal representations of circuit designs by combining structural graph information, spatial layout images, and stage-level analytic metrics.
The proposed representation, therefore, supports a wide range of model architectures reported in prior work, including GNN-based models that utilize circuit topology for prediction of structural-based tasks, CNN-based models for spatial congestion and routability analysis, and hybrid graph–image architectures for timing and parasitic impedance estimation.
Certain features, including number of fanouts, pin or net density measures, and distance-based attributes, are not stored explicitly in EDA-Schema-V2.
Such quantities are derived from the netlist graphs and spatial layout representations contained in the schema.
The proposed data model, therefore, satisfies the feature requirements of the representative studies listed in Table~\ref{tab:feature_analysis} while supporting consistent derivation of features and reuse of data and representations across predictive tasks.

% Please add the following required packages to your document preamble:
% \usepackage{multirow}
\begin{landscape}
% \begin{table*}[!h]
    \begin{center}
    % \captionof{table}{Features used in prior machine learning based digital circuit design automation for timing prediction, wirelength prediction, parasitic impedance prediction, power prediction, routability prediction, and EM/IR analysis.}
    % \captionof{table}{Features used in prior machine learning based digital circuit design automation for power prediction, wirelength prediction, timing prediction, parasitic impedance prediction, routability prediction, and analysis of EM/IR. Features are categorized as \textit{Available} when directly provided in EDA-Schema-V2, \textit{Derivable Features} when obtained from structural or spatial representations, and \textit{Other Features/Contraints} when reported in prior work but are not presently available in the schema.}
    % \vspace{-0.05in}
    % \resizebox{\columnwidth}{!}{
    \setlength{\tabcolsep}{2.5pt}
    \fontsize{5.75}{7.5}\selectfont{
        \begin{longtable}{|l|l|l|l|l|l|ll|l|l|}
        \caption{(page 1/2) Features used in prior machine learning based digital circuit design automation for power prediction, wirelength prediction, timing prediction, parasitic impedance prediction, routability prediction, and analysis of EM/IR. Features are categorized as \textit{Available} when directly provided in EDA-Schema-V2, \textit{Derivable Features} when obtained from structural or spatial representations, and \textit{Other Features/Contraints} when reported in prior work but are not presently available in the schema.}
        \label{tab:feature_analysis}\\
        \hline
        \textbf{\begin{tabular}[c]{@{}l@{}}Prediction\\ Task\end{tabular}}                         & \textbf{Ref}                                       & \textbf{\begin{tabular}[c]{@{}l@{}}Problem\\ Type\end{tabular}}       & \textbf{\begin{tabular}[c]{@{}l@{}}Metric to\\ Predict\end{tabular}}                                      & \textbf{\begin{tabular}[c]{@{}l@{}}Initial \\ Stage\end{tabular}}                   & \textbf{\begin{tabular}[c]{@{}l@{}}Final\\ Stage\end{tabular}}                  & \multicolumn{1}{l|}{\textbf{Toolset}}                                                                           & \textbf{PDK}                                                        & \textbf{Features}                                                                                                                                                                                                                                                                                                                                                                                               & \textbf{\begin{tabular}[c]{@{}l@{}}Network\\ Architecture\end{tabular}}               \\ \hline
        \multirow{2}{*}{\textbf{\begin{tabular}[c]{@{}l@{}}\\\\\\\\Power\\ Prediction\end{tabular}}}       & \cite{lu2022driving}              & Regression                                                            & total power                                                                                               & \begin{tabular}[c]{@{}l@{}}Post-placement, CTS,\\ initial routing\end{tabular}      & \begin{tabular}[c]{@{}l@{}}Post final\\ routing\end{tabular}                    & \multicolumn{1}{l|}{Synopsys ICC2}                                                                              & \begin{tabular}[c]{@{}l@{}}TSMC\\ 28nm\end{tabular}                 & \begin{tabular}[c]{@{}l@{}}Available Features:\\ - Standard Cell: switching power, internal power, leakage power\\ - Pin: slew, delay, slack\end{tabular}                                                                                                                                                                                                                                                       & GNN + LSTM                                                                            \\ \cline{2-10} 
                                                                                                   & \cite{du2024powpredict}           & Regression                                                            & \begin{tabular}[c]{@{}l@{}}Total power,\\ internal power,\\ switching power,\\ leakage power\end{tabular} & Post-placement                                                                      & Post-routing                                                                    & \multicolumn{2}{l|}{CircuitNet 14 nm dataset}                                                                                                                                         & \begin{tabular}[c]{@{}l@{}}Available Features:\\ - Gate: drive strength, is sequential, number of fanout, x/y co-ordinates\\ - Pin: capacitance, rise/fall slew\\ - Net: HPWL, x/y co-ordinates\\ - Routability Metrics: RUDY\\ \\ Derivable Features: cell density\\ \\ Other Features: macro regions, non-linear power model (NLPM) LUT data\end{tabular}                                                     & GNN                                                                                   \\ \hline
        \multirow{2}{*}{\textbf{\begin{tabular}[c]{@{}l@{}}\\Wirelength\\ Prediction\end{tabular}}}  & \cite{xie2021net2}                & Regression                                                            & net length                                                                                                & Pre-synthesis                                                                       & Post-floorplan                                                                  & \multicolumn{1}{l|}{\begin{tabular}[c]{@{}l@{}}Synopsys\\ DC Compiler,\\ Cadence\\ Innovus\end{tabular}}        & \begin{tabular}[c]{@{}l@{}}Nangate\\ 45nm\end{tabular}              & \begin{tabular}[c]{@{}l@{}}Available Features:\\ - Net: number of fan-in and number of fan-out\\ - Gate: area\\ \\ Derivable Features: cluster-based edge features from hMETIS \cite{karypis1997multilevel}\end{tabular}                                                                                                                                                                       & GAT                                                                                   \\ \cline{2-10} 
                                                                                                   & \cite{shrestha2025differentiable} & Regression                                                            & net length                                                                                                & Post-placement                                                                      & Post-routing                                                                    & \multicolumn{1}{l|}{\begin{tabular}[c]{@{}l@{}}Synopsys\\ DC Compiler,\\ IC Compiler\end{tabular}}              & \begin{tabular}[c]{@{}l@{}}commercial\\ 28nm\end{tabular}           & \begin{tabular}[c]{@{}l@{}}Available Features:\\ - Net: number of fan-in, number of fan-out, x/y co-ordinates\\ - Gate: number of fan-in, number of fan-out, width, x/y co-ordinates\end{tabular}                                                                                                                                                                                                               & GNN                                                                                   \\ \hline
        \multirow{3}{*}{\textbf{\begin{tabular}[c]{@{}l@{}}\\\\\\\\\\\\\\Timing\\ Prediction\end{tabular}}}      & \cite{shrestha2022graph}          & Regression                                                            & gate arc delay                                                                                            & \begin{tabular}[c]{@{}l@{}}Post floorplan,\\ placement, CTS\end{tabular}            & Post-routing                                                                    & \multicolumn{1}{l|}{\begin{tabular}[c]{@{}l@{}}Synopsys\\ DC Compiler,\\ IC Compiler,\\ PrimeTime\end{tabular}} & \begin{tabular}[c]{@{}l@{}}commercial\\ 65nm\end{tabular}           & \begin{tabular}[c]{@{}l@{}}Available Features:\\ - Constraints: clock period, aspect ratio, utilization\\ - Gate: functionality\\ - Net: number of fanout, capacitance\\ - Gate Arc: logic level, initial phase delay, arrival time\\ \\ Other Constraints: max skew, max fanout, max clock network capacitance, max latency\end{tabular}                                                                       & GNN                                                                                   \\ \cline{2-10} 
                                                                                                   & \cite{ahn2024dtoc}                & Regression                                                            & \begin{tabular}[c]{@{}l@{}}arc delay,\\ arc output slew\end{tabular}                                      & Post-placement                                                                      & Post-routing                                                                    & \multicolumn{1}{l|}{\begin{tabular}[c]{@{}l@{}}Synopsys\\ ICC2\end{tabular}}                                    & \multicolumn{1}{l|}{\begin{tabular}[c]{@{}l@{}}Nangate\\ 15nm\end{tabular}}                                                                    & \begin{tabular}[c]{@{}l@{}}Available Features\\ - Net: fanout, length, resistance, capacitance\\ - Timing Path: arc delay, output slew, rise/fall\\ - RoutabilityMetrics: net RUDY, long range net RUDY\\ \\ Derivable Features:\\ - cell RUDY map, net density map, source/sink location map, pin capacitance location map\end{tabular}                                                                        & CNN                                                                                   \\ \cline{2-10} 
                                                                                                   & \cite{cao2025optimization}        & Regression                                                            & \begin{tabular}[c]{@{}l@{}}arc delay,\\ endpoint\\ arrival time\end{tabular}                              & Post-placement                                                                      & Post-routing                                                                    & \multicolumn{1}{l|}{\begin{tabular}[c]{@{}l@{}}Synopsys\\ IC Compiler,\\ PrimeTime\end{tabular}}                & \begin{tabular}[c]{@{}l@{}}TSMC\\ 22nm\end{tabular}                 & \begin{tabular}[c]{@{}l@{}}Available Features:\\ - Standard Cell: function, drive strength, number of fanins, number of fanouts\\ - Pin: is output pin?, x/y co-ordinates, rise/fall capacitance, rise/fall slew, rise/fall slack\\ - Cell arc: delay, slew\\ - Net: length\\ - threshold voltage type is uniform (RVT-TT) across the whole dataset\\ \\ Other Features: is cell sizable (boolean)\end{tabular} & \begin{tabular}[c]{@{}l@{}}Heterogeneous\\ GAT + U-net\end{tabular}                   \\ \hline
\endfirsthead
        \newpage
        \caption{(page 2/2) Features used in prior machine learning based digital circuit design automation for power prediction, wirelength prediction, timing prediction, parasitic impedance prediction, routability prediction, and analysis of EM/IR. Features are categorized as \textit{Available} when directly provided in EDA-Schema-V2, \textit{Derivable Features} when obtained from structural or spatial representations, and \textit{Other Features/Contraints} when reported in prior work but are not presently available in the schema.}\\
        \hline
        \textbf{\begin{tabular}[c]{@{}l@{}}Prediction\\ Task\end{tabular}}                         & \textbf{Ref}                                       & \textbf{\begin{tabular}[c]{@{}l@{}}Problem\\ Type\end{tabular}}       & \textbf{\begin{tabular}[c]{@{}l@{}}Metric to\\ Predict\end{tabular}}                                      & \textbf{\begin{tabular}[c]{@{}l@{}}Initial \\ Stage\end{tabular}}                   & \textbf{\begin{tabular}[c]{@{}l@{}}Final\\ Stage\end{tabular}}                  & \multicolumn{1}{l|}{\textbf{Toolset}}                                                                           & \textbf{PDK}                                                        & \textbf{Features}                                                                                                                                                                                                                                                                                                                                                                                               & \textbf{\begin{tabular}[c]{@{}l@{}}Network\\ Architecture\end{tabular}}               \\ \hline
        \multirow{2}{*}{\textbf{\begin{tabular}[c]{@{}l@{}}\\Parasitic\\ Prediction\end{tabular}}}   & \cite{shrestha2023graph}          & Regression                                                            & net capacitance                                                                                           & Post-placement                                                                      & Post-routing                                                                    & \multicolumn{1}{l|}{\begin{tabular}[c]{@{}l@{}}Synopsys DC/\\ IC Compiler,\\ PrimeTime\end{tabular}}            & \begin{tabular}[c]{@{}l@{}}commercial\\ 65nm\end{tabular}           & \begin{tabular}[c]{@{}l@{}}Available Features:\\ - Net: capacitance, length, x/y co-ordinates\\ \\ Derivable Features: pin density, net density\end{tabular}                                                                                                                                                                                                                                                    & Spatial GCN                                                                           \\ \cline{2-10} 
                                                                                                   & \cite{jakang2025}                 & Regression                                                            & \begin{tabular}[c]{@{}l@{}}net resistance,\\ net capacitance\end{tabular}                                 & Post-placement                                                                      & Post-routing                                                                    & \multicolumn{1}{l|}{\begin{tabular}[c]{@{}l@{}}Synopsys DC/\\ IC Compiler,\\ PrimeTime\end{tabular}}            & ASAP 7nm                                                            & \begin{tabular}[c]{@{}l@{}}Available Features:\\ - Pin: x/y co-ordinates, rise/fall capacitance\\ - Net: bounding box, area, length, degree (number of fanins + fanouts)\end{tabular}                                                                                                                                                                                                                           & \begin{tabular}[c]{@{}l@{}}Hybrid HGNN +\\ Graph Transformer\end{tabular}             \\ \hline
        \multirow{3}{*}{\textbf{\begin{tabular}[c]{@{}l@{}}\\\\\\\\\\Routability\\ Prediction\end{tabular}}} & \cite{xie2018routenet}            & \begin{tabular}[c]{@{}l@{}}Regression /\\ Classification\end{tabular} & \begin{tabular}[c]{@{}l@{}}\#DRV and\\ congestion\\ hotspot\end{tabular}                                  & \begin{tabular}[c]{@{}l@{}}Post–global placement,\\ detailed placement\end{tabular} & \begin{tabular}[c]{@{}l@{}}Post–global\\ routing\end{tabular}                   & \multicolumn{1}{l|}{\begin{tabular}[c]{@{}l@{}}Cadence\\ Encounter\end{tabular}}                                & NA                                                                  & \begin{tabular}[c]{@{}l@{}}Available Features:\\ - Routability Metrics: long-range RUDY, short-range RUDY, RUDY-pins\\ \\ Derivable Features: cell density, pin density\\ \\ Other Features:\\ - Macro: macro region map, macro pin density (per metal layer)\\ - Track Route / Global Route congestion map\end{tabular}                                                                                        & \begin{tabular}[c]{@{}l@{}}CNN /\\ Spatial CNN\end{tabular}                           \\ \cline{2-10} 
                                                                                                   & \cite{li2024lightweight}          & \begin{tabular}[c]{@{}l@{}}Regression /\\ Classification\end{tabular} & \begin{tabular}[c]{@{}l@{}}Tile-level routing\\ overflow,\\ DRC hotspots\end{tabular}                     & \begin{tabular}[c]{@{}l@{}}Post–global placement,\\ detailed placement\end{tabular} & \begin{tabular}[c]{@{}l@{}}Post–global\\ routing,\\ Post–placement\end{tabular} & \multicolumn{2}{l|}{CircuitNet 28 nm dataset}                                                                                                                                         & \begin{tabular}[c]{@{}l@{}}Available Features:\\ - Routability Metrics: RUDY-nets, RUDY-pins\\ \\ Derivable Features: cell density, pin density\\ \\ Other Features:\\ - Macro: macro region map, macro pin density (per metal layer)\\ - Congestion map (global routing overflow)\end{tabular}                                                                                                                 & \begin{tabular}[c]{@{}l@{}}Inception-Boosted\\ U-Net\end{tabular}                     \\ \cline{2-10} 
                                                                                                   & \cite{almeida2025eh}              & Classification                                                        & DRV                                                                                                       & \begin{tabular}[c]{@{}l@{}}Initial detail routing\\ iteration\end{tabular}          & \begin{tabular}[c]{@{}l@{}}Post-global\\ routing\end{tabular}                   & \multicolumn{1}{l|}{\begin{tabular}[c]{@{}l@{}}OpenROAD,\\ Cadence\\ Innovus\end{tabular}}                      & \begin{tabular}[c]{@{}l@{}}Nangate\\ 45nm\end{tabular}              & \begin{tabular}[c]{@{}l@{}}Derivable Features: pin density, pin neighborhood density\\ \\ Other Features: vertical/horizontal routing overflow\end{tabular}                                                                                                                                                                                                                                                     & CNN                                                                                   \\ \hline
        \multirow{3}{*}{\textbf{\begin{tabular}[c]{@{}l@{}}\\\\\\EM/IR\\ Prediction\end{tabular}}}       & \cite{chhabria2022encoder}        & Generation                                                            & \begin{tabular}[c]{@{}l@{}}IR drop map\\ EM hotspot map\end{tabular}                                      & \begin{tabular}[c]{@{}l@{}}Post-routing/\\ Pre-IR drop\\ analysis\end{tabular}      & \begin{tabular}[c]{@{}l@{}}Post-routing/\\ Post-IR drop\\ analysis\end{tabular} & \multicolumn{1}{l|}{\begin{tabular}[c]{@{}l@{}}OpeNPDN,\\ PDNSim\end{tabular}}                                  & \begin{tabular}[c]{@{}l@{}}commercial\\ 12 nm\\ FinFET\end{tabular} & \begin{tabular}[c]{@{}l@{}}Available Features:\\ - Power Delivery Network: power bump pattern\\ \\ Derivable Features: Power distributions map, PDN density map\end{tabular}                                                                                                                                                                                                                                    & \begin{tabular}[c]{@{}l@{}}Encoder-Decoder\\ (U-Net,\\ LSTM,\\ 3D U-Net)\end{tabular} \\ \cline{2-10} 
                                                                                                   & \cite{chen2024global}             & Generation                                                            & IR drop map                                                                                               & \begin{tabular}[c]{@{}l@{}}Post-routing/\\ Pre-IR drop\\ analysis\end{tabular}      & \begin{tabular}[c]{@{}l@{}}Post-routing/\\ Post-IR drop\\ analysis\end{tabular} & \multicolumn{1}{l|}{\begin{tabular}[c]{@{}l@{}}OpenROAD/\\ ICCAD 2023\\ dataset\end{tabular}}                   & \multicolumn{1}{l|}{\begin{tabular}[c]{@{}l@{}}Nangate\\ 45nm\end{tabular}}                                                           & \begin{tabular}[c]{@{}l@{}}Derivable Features: PDN density map, resistance map, effective distance map,\\ current map, shortest path resistance map, shortest path voltage map\end{tabular}                                                                                                                                                                                                                     & \begin{tabular}[c]{@{}l@{}}Encoder-Decoder\\ (U-Net with\\ Inception)\end{tabular}    \\ \cline{2-10} 
                                                                                                   & \cite{yu2025lared}                & Generation                                                            & IR drop map                                                                                               & \begin{tabular}[c]{@{}l@{}}Post-routing/\\ Pre-IR drop\\ analysis\end{tabular}      & \begin{tabular}[c]{@{}l@{}}Post-routing/\\ Post-IR drop\\ analysis\end{tabular} & \multicolumn{2}{l|}{CircuitNet 14 nm dataset}                                                                                                                                         & \begin{tabular}[c]{@{}l@{}}Derivable Features: power map, effective resistance to power/ground net,\\ minimum path resistance to power/ground net\end{tabular}                                                                                                                                                                                                                                                  & \begin{tabular}[c]{@{}l@{}}Rebuilder\\ Encoder-Decoder\\ (LaRED)\end{tabular}         \\ \hline
        \end{longtable}
    }
    \end{center}
% \end{table*}
\end{landscape}

Despite promising open and reproducible results, benchmarking across prior work remains challenging. The primary limitation arises from the reliance on commercial EDA tools and proprietary process design kits (PDKs), together with the use of closed or internal benchmark datasets.
Many studies rely on industrial designs without releasing the associated netlists, constraints, or technology information.
Other studies depend on commercial signoff tools provided by vendors including Synopsys and Cadence \cite{lu2022driving, xie2021net2, shrestha2025differentiable, shrestha2022graph, ahn2024dtoc, cao2025optimization, shrestha2023graph, jakang2025, xie2018routenet, almeida2025eh}, or on proprietary foundry PDKs \cite{lu2022driving,  du2024powpredict, shrestha2025differentiable, shrestha2022graph, cao2025optimization, shrestha2023graph, li2024lightweight, chhabria2022encoder, yu2025lared} including from GlobalFoundries, Intel, Samsung, and TSMC for feature extraction and label generation.
Commercial environments used to generate intermediate representations and ground-truth metrics are, therefore, legally constrained and/or difficult to reproduce.
In several cases, studies refer only to a “commercial” design flow without specifying the configuration of the toolchain or the PDK.
Confidentiality requirements often restrict the release of such details; however, the absence of toolchain and dataset information limits reproducibility. Training data, intermediate representations, and extracted features are frequently unavailable, preventing independent validation and fair comparison across machine learning based methods and algorithms. The proposed schema and dataset instead provide standardized and open representations of circuit structure, layout, and results from analysis, which enables reproducible experimentation and consistent benchmarking of machine learning approaches for digital physical design.

\subsection{Baseline Analysis}
\label{subsec:baseline_analysis}

Baseline analysis is performed across multiple quality-of-results (QoR) metrics using EDA-Schema-V2.
Prediction is considered for circuit-level metrics including \textit{total area}, \textit{total power}, and \textit{total wirelength}, routing metrics including \textit{per-net interconnect length}, timing-related metrics including \textit{worst arrival time}, \textit{worst slack}, and \textit{total negative slack (TNS)}, and path-level timing metrics including \textit{timing path arrival time} and \textit{timing path slack}.
In addition, timing arc quantities including \textit{net arc delay}, \textit{cell arc delay}, and \textit{cell arc slew} are evaluated to capture delay and transition time behavior at the detail of individual timing arcs.
Given the results of metrics evaluated in the initial stages of the physical design flow (floorplan, global\_place, place\_resize, detailed\_place, cts, and global\_route) and the final-stage (\textit{detailed\_route}), values reported by the design tools at earlier stages are treated as baseline estimates for the given metrics evaluated in the final-stage.
Baseline error is evaluated by comparing stage-level estimates to final-stage values across the dataset.
Half-perimeter wirelength (HPWL) prior to global routing is treated as the baseline estimate for both total circuit wirelength and per-net interconnect length.
For metrics evaluated on timing-paths, the paths from the initial and final stages are matched using the startpoint name, endpoint name, and type of timing check (setup or hold).
For metrics evaluated on timing arcs, cell arcs and net arcs are matched using associated pin pairs along the timing path.
For metrics evaluated on interconnect, nets are matched using the net name. Paths, arcs, and nets that are added, removed, or structurally modified across stages are excluded from the comparison.
Machine learning models developed to address the aforementioned prediction problems are expected to outperform the baseline estimates produced by EDA tools.

\begin{figure}[!h]
    \vspace{-0.1in}
    \begin{center}
        \includegraphics[width=0.975\columnwidth]{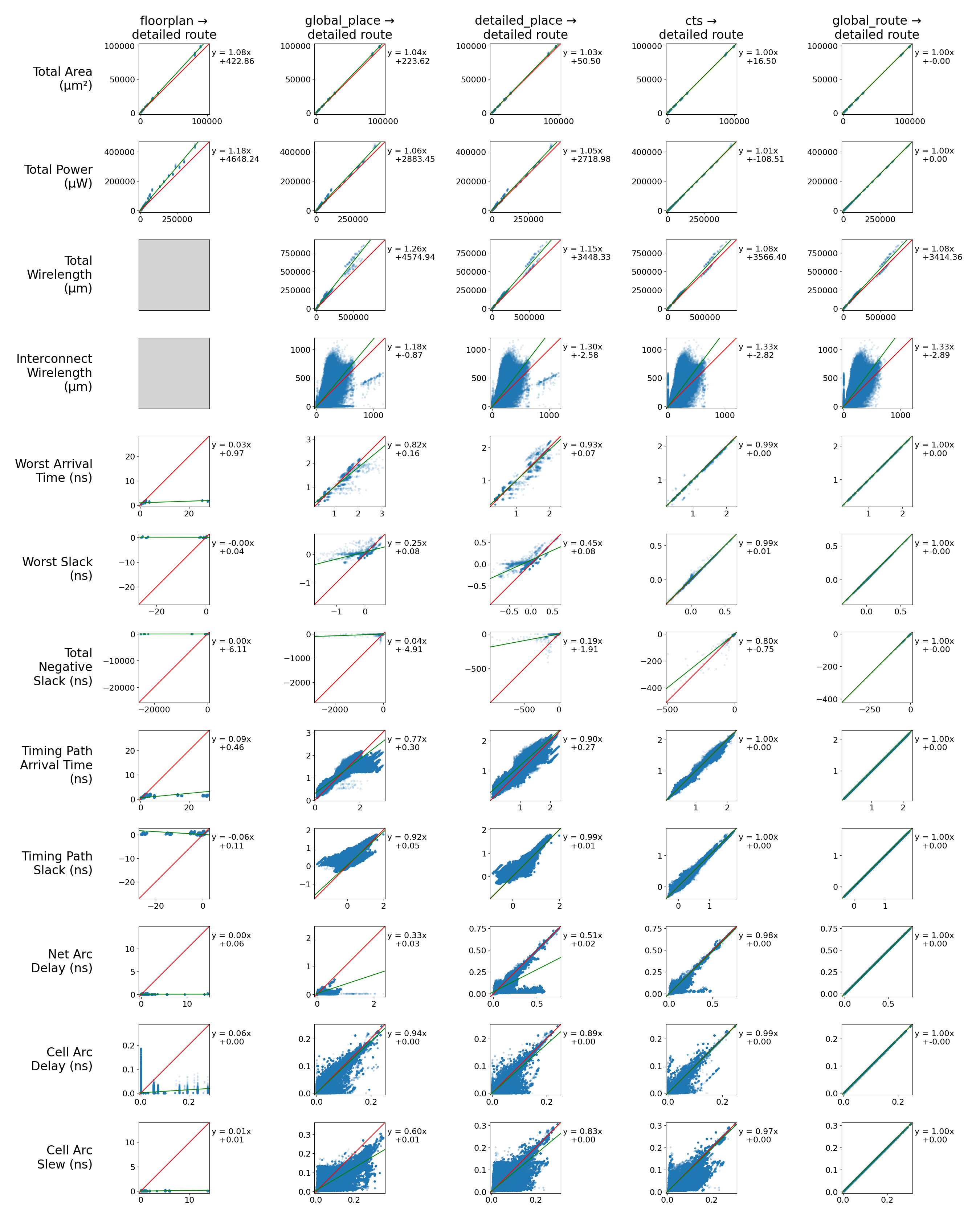}
    \end{center}
    \caption{Inter-design stage correlation of QoR and timing metrics across all benchmark circuits in the NG45 technology node. Columns compare intermediate design stages with the final stage, with intermediate-stage values on the x-axis and final-stage values on the y-axis. Rows correspond to metrics including area, power, wirelength, timing, and arc delays. Scatter plots aggregate all circuits and operating points; the red line denotes the (y=x) reference and the green line represents a least-squares fit.
    }
    \label{fig:ng45_stage_analysis}
    \vspace{-0.2in}
\end{figure}

The correlation between values from the initial-stage and final-stage of evaluated QoR metrics for all circuits in the NG45 dataset is shown in Fig.~\ref{fig:ng45_stage_analysis}.
The diagonal $x=y$ line denotes perfect prediction, where the baseline exactly matches the final metric.
Points above the diagonal indicate \textit{underestimation}, where the baseline value is smaller than the final actual value, while points below the diagonal indicate \textit{overestimation}, where the baseline value is larger that in the final actual value.
Across most metrics, agreement with the final-stage values improves as the circuit progresses through later stages of the physical design flow.
Early-stage estimates frequently \textit{underestimate} total power, arrival time, and routing-related metrics as parasitic effects, buffering, and detailed routing structures are not yet captured.
As the design progresses through detailed placement, clock tree synthesis, and routing, additional physical effects are applied and accounted for, resulting in closer alignment with the QoR values reported from the final stage of the physical design flow.
Several metric-dependent trends are evident. Circuit-level metrics such as total area and total power exhibit strong linear correlation with final-stage values even at early stages, indicating that coarse physical characteristics are largely determined early in the design flow.
In contrast, wirelength-related metrics exhibit systematic underestimation at early design stages, with slopes significantly greater than unity, which reflect the absence of effects due to detailed routing and parasitic impedance. Timing metrics exhibit more complex behavior: arrival time exhibits moderate correlation beginning at placement and improves across later stages, whereas slack-based metrics cluster near zero with weak correlation in early design stages. Similarly, path-level and arc-level timing quantities exhibit substantial dispersion early in the design flow and only converge toward final values after clock tree synthesis and routing. The described observations demonstrate that prediction accuracy depends strongly on both the evaluated metric and the design stage, and motivate the need for the evaluation of metrics that remain stable across design stages.

The accuracy of baseline predictions is first evaluated using standard regression metrics including Mean Absolute Error (MAE), Mean Absolute Percentage Error (MAPE), and the coefficient of determination ($R^2$), as defined by
\vspace{-0.1in}
\begin{equation}
MAE = \frac{1}{n}\sum_{i=1}^{n} |M_i - M_{b,i}|\text{,}
\end{equation}
\begin{equation}
MAPE = \frac{100\%}{n}\sum_{i=1}^{n}\left|1 - \frac{M_{b,i}}{M_i}\right|\text{, and}
\end{equation}
\begin{equation}
R^2 = 1 - \frac{\sum_{i=1}^{n}(M_i - M_{b,i})^2}{\sum_{i=1}^{n}(M_i - \bar{M})^2}
\end{equation}
\noindent where $M_i$ denotes the value of metric $M$ after the \textit{final} design stage, $M_{b,i}$ is the baseline estimate obtained from an intermediate stage, $\bar{M}$ is the mean of the final-stage metric values, and $n$ is the number of samples in the dataset.
The MAE, MAPE, and R$^2$ provide a first-order characterization of the accuracy of the prediction and are primarily applicable to circuit-level features including area, power, total wirelength, and interconnect length.
However, evaluating early-stage estimates of physical design metrics presents several additional challenges.

Metrics such as per-net interconnect length and per timing path arrival time exhibit long-tailed behavior.
For example, many nets are short while only a small number are very long in length, and most paths provide moderate arrival times while a small fraction arrive very late.
In practice, the few long interconnects are more significant than the large number of short interconnects, and similarly, the relatively small number of late-arriving timing paths are more critical than the many early-arriving paths, since long wires and late arrivals disproportionately impact delay, congestion, and timing risk.
To characterize the robustness of a prediction in long-tailed distributions, the \(95^{th}\) percentile absolute error is reported in addition to mean-based statistics for metrics evaluated for wirelength and arrival time.
In addition, to directly evaluate the accuracy of a prediction on the most critical interconnects and timing paths, error metrics are also computed on the longest \(5\%\) of nets and the slowest \(5\%\) of timing paths.
The additional analysis narrows the prediction of performance parameters on the design instances that most strongly effect circuit timing and physical design quality.

Timing-related metrics including worst slack and total negative slack assume both positive and negative values and frequently cluster near zero as designs approach timing closure.
In such cases, percentage-based metrics including MAPE become unstable and produce misleading values.
Similarly, $R^2$ becomes unreliable when the variance of the values from the final-stage is extremely small.
As indicated by the results shown in Fig.~\ref{fig:ng45_stage_analysis}, timing-related metrics evaluated at earlier design stages often cluster tightly near zero relative to values from the final design stage, resulting in very limited spread and unstable estimates of variance. 
Consequently, MAPE is not reported for slack-based metrics, and $R^2$ is omitted for all timing-based metrics.

Slack-based metrics including worst slack and total negative slack are inherently relative quantities, defined by the difference between arrival time and required time.
As a result, absolute error metrics alone do not fully characterize the quality of the prediction.
In timing analysis, the direction of the prediction error is also important. 
In general, prediction behavior is ranked as follows: perfect prediction is ideal, conservative predictions that slightly overestimate delay or underestimate slack are preferred as such predictions are less likely to hide timing violations, and optimistic predictions that underestimate delay or overestimate slack are least desirable as such predictions may incorrectly indicate timing closure.

To analyze the directional bias of a prediction, overestimation and underestimation are separated as distinct error metrics. Let $M_{b,i}$ denote the baseline estimate and $M_i$ the metric value of the final design stage for sample $i$. Overestimation occurs when $M_{b,i} > M_i$, while underestimation occurs when $M_{b,i} < M_i$.
Directional bias is quantified using the Mean Positive Error (MPE) and Mean Negative Error (MNE), which are given by, respectively,
% \vspace{-0.1in}
\begin{equation}
MPE = \frac{1}{n_p}\sum_{M_{b,i} > M_i} (M_{b,i} - M_i)\text{, and}
\end{equation}
\begin{equation}
MNE = \frac{1}{n_n}\sum_{M_{b,i} < M_i} (M_i - M_{b,i})\text{,}
\end{equation}
\noindent where $n_p$ and $n_n$ denote the number of overestimated samples and underestimated samples, respectively. The metrics separately quantify pessimistic and optimistic prediction behavior in the baseline estimates.

For arc-level timing metrics including \textit{net arc delay}, \textit{cell arc delay}, and \textit{cell arc slew}, similar limitations are considered when using percentage-based or correlation-based metrics.
Arc delays and slews at early stages of the design flow are often small valued and exhibit weak correlation with the final-stage values due to incomplete placement, incomplete buffering, and inaccurate estimation of parasitic impedance.
As a result, many samples contain values that are close to zero, which results in percentage-based metrics, including MAPE, that evaluate the excessively large or unstable values.
In addition, arc delay and slew values at early design stages are often close to zero, resulting in limited variance and a coefficient of determination score ($R^2$) that is unreliable and highly sensitive to minor deviations in predicted values.
Under such considerations, $R^2$ produces large negative values that do not meaningfully reflect the performance of the prediction.
To avoid reporting numerically unstable results, extreme values are thresholded in the reported tables. MAPE values exceeding 10000\% are denoted using >10000\%, and very negative $R^2$ values are denoted using <-1.
Evaluation, therefore, focuses primarily on measures that utilize absolute error, which provide more stable and interpretable comparisons across design stages.

Finally, to measure the ability of baseline estimates to correctly identify timing violations, the True Positive Rate (TPR) and True Negative Rate (TNR) are used as given by, respectively,
% \vspace{-0.1in}
\begin{equation}
TPR = \frac{TP}{TP + FN}\text{, and}
\end{equation}
\begin{equation}
TNR = \frac{TN}{TN + FP}\text{,}
\end{equation}
\noindent where $TP$ and $TN$ represent correctly predicted violating and non-violating instances, respectively, and $FP$ and $FN$ denote false positives and false negatives.

% Please add the following required packages to your document preamble:
% \usepackage{multirow}
\begin{landscape}
    \begin{center}
        \captionof{table}{Averaged baseline results of evaluated metrics between design stages for the NG45, SKY130, IHP130, and ASAP7 datasets. The error obtained when using the initial-stage tool estimate to predict the final post-routing QoR metric is listed in each column. Metrics include Mean Absolute Error (MAE), Mean Absolute Percentage Error (MAPE),  coefficient of determination ($R^2$), and timing-specific metrics of Mean Positive Error (MPE), Mean Negative Error (MNE), True Positive Rate (TPR), and True Negative Rate (TNR).}
        \label{tab:baseline}
        \resizebox{\linewidth}{!}
        {
            \setlength{\tabcolsep}{2pt}
            \begin{tabular}{l|l|rrrr|rrrr|rrrr|rrrr|rrrr|}
            \cline{2-22}
                                                                                                                                                      & \multicolumn{1}{c|}{}                                   & \multicolumn{4}{c|}{\textbf{floorplan to  detailed route}}                                                                                                                        & \multicolumn{4}{c|}{\textbf{global place to detailed route}}                                                                                                               & \multicolumn{4}{c|}{\textbf{detailed place to detailed route}}                                                                                                             & \multicolumn{4}{c|}{\textbf{CTS to detailed route}}                                                                                                                        & \multicolumn{4}{c|}{\textbf{global route to detailed route}}                                                                                             \\ \cline{3-22} 
            \multirow{-2}{*}{}                                                                                                                        & \multicolumn{1}{c|}{\multirow{-2}{*}{\textbf{Metrics}}} & \multicolumn{1}{c|}{\textbf{NG45}}           & \multicolumn{1}{c|}{\textbf{SKY130}}         & \multicolumn{1}{c|}{\textbf{IPH130}}         & \multicolumn{1}{c|}{\textbf{ASAP7}}  & \multicolumn{1}{c|}{\textbf{NG45}}         & \multicolumn{1}{c|}{\textbf{SKY130}}       & \multicolumn{1}{c|}{\textbf{IPH130}}       & \multicolumn{1}{c|}{\textbf{ASAP7}} & \multicolumn{1}{c|}{\textbf{NG45}}         & \multicolumn{1}{c|}{\textbf{SKY130}}       & \multicolumn{1}{c|}{\textbf{IPH130}}       & \multicolumn{1}{c|}{\textbf{ASAP7}} & \multicolumn{1}{c|}{\textbf{NG45}}         & \multicolumn{1}{c|}{\textbf{SKY130}}       & \multicolumn{1}{c|}{\textbf{IPH130}}       & \multicolumn{1}{c|}{\textbf{ASAP7}} & \multicolumn{1}{c|}{\textbf{NG45}}  & \multicolumn{1}{c|}{\textbf{SKY130}}  & \multicolumn{1}{c|}{\textbf{IPH130}} & \multicolumn{1}{c|}{\textbf{ASAP7}} \\ \hline
            \multicolumn{1}{|l|}{}                                                                                                                    & \textbf{MAE}                                            & \multicolumn{1}{r|}{1,781.97}                & \multicolumn{1}{r|}{18,567.03}               & \multicolumn{1}{r|}{48,738.62}               & 225.06                               & \multicolumn{1}{r|}{871.07}                & \multicolumn{1}{r|}{8,385.61}              & \multicolumn{1}{r|}{35,724.45}             & 73.79                               & \multicolumn{1}{r|}{571.13}                & \multicolumn{1}{r|}{7,505.03}              & \multicolumn{1}{r|}{19,487.21}             & 47.23                               & \multicolumn{1}{r|}{49.02}                 & \multicolumn{1}{r|}{854.81}                & \multicolumn{1}{r|}{2,339.67}              & 1.98                                & \multicolumn{1}{r|}{0.00}           & \multicolumn{1}{r|}{0.00}             & \multicolumn{1}{r|}{0.00}            & 0.00                                \\ \cline{2-22} 
            \multicolumn{1}{|l|}{}                                                                                                                    & \textbf{MAPE}                                           & \multicolumn{1}{r|}{12.43 \%}                & \multicolumn{1}{r|}{22.51 \%}                & \multicolumn{1}{r|}{20.74 \%}                & 15.35 \%                             & \multicolumn{1}{r|}{7.07 \%}               & \multicolumn{1}{r|}{11.71 \%}              & \multicolumn{1}{r|}{13.60 \%}              & 7.50 \%                             & \multicolumn{1}{r|}{4.14 \%}               & \multicolumn{1}{r|}{9.27 \%}               & \multicolumn{1}{r|}{5.56 \%}               & 3.79 \%                             & \multicolumn{1}{r|}{0.49 \%}               & \multicolumn{1}{r|}{1.22 \%}               & \multicolumn{1}{r|}{0.72 \%}               & 0.24 \%                             & \multicolumn{1}{r|}{0.00 \%}        & \multicolumn{1}{r|}{0.00 \%}          & \multicolumn{1}{r|}{0.00 \%}         & 0.00 \%                             \\ \cline{2-22} 
            \multicolumn{1}{|l|}{\multirow{-3}{*}{\textbf{\begin{tabular}[c]{@{}l@{}}Total\\ Area (\textmu m$^2$)\end{tabular}}}}      & \textbf{R$^2$}                                          & \multicolumn{1}{r|}{0.989}                   & \multicolumn{1}{r|}{0.957}                   & \multicolumn{1}{r|}{0.912}                   & 0.969                                & \multicolumn{1}{r|}{0.997}                 & \multicolumn{1}{r|}{0.991}                 & \multicolumn{1}{r|}{0.945}                 & 0.997                               & \multicolumn{1}{r|}{0.998}                 & \multicolumn{1}{r|}{0.991}                 & \multicolumn{1}{r|}{0.979}                 & 0.998                               & \multicolumn{1}{r|}{1.000}                 & \multicolumn{1}{r|}{1.000}                 & \multicolumn{1}{r|}{1.000}                 & 1.000                               & \multicolumn{1}{r|}{1.000}          & \multicolumn{1}{r|}{1.000}            & \multicolumn{1}{r|}{1.000}           & 1.000                               \\ \hline
            \multicolumn{1}{|l|}{}                                                                                                                    & \textbf{MAE}                                            & \multicolumn{1}{r|}{11,875.86}               & \multicolumn{1}{r|}{46,947.57}               & \multicolumn{1}{r|}{17,699.64}               & 6,071.61                             & \multicolumn{1}{r|}{5,670.15}              & \multicolumn{1}{r|}{19,274.83}             & \multicolumn{1}{r|}{16,651.34}             & 2,266.72                            & \multicolumn{1}{r|}{5,043.04}              & \multicolumn{1}{r|}{19,382.95}             & \multicolumn{1}{r|}{15,644.47}             & 2,177.82                            & \multicolumn{1}{r|}{280.22}                & \multicolumn{1}{r|}{2,270.66}              & \multicolumn{1}{r|}{123.65}                & 55.15                               & \multicolumn{1}{r|}{0.00}           & \multicolumn{1}{r|}{0.00}             & \multicolumn{1}{r|}{0.00}            & 0.00                                \\ \cline{2-22} 
            \multicolumn{1}{|l|}{}                                                                                                                    & \textbf{MAPE}                                           & \multicolumn{1}{r|}{27.89 \%}                & \multicolumn{1}{r|}{50.85 \%}                & \multicolumn{1}{r|}{23.71 \%}                & 33.47 \%                             & \multicolumn{1}{r|}{16.70 \%}              & \multicolumn{1}{r|}{31.36 \%}              & \multicolumn{1}{r|}{18.87 \%}              & 19.23 \%                            & \multicolumn{1}{r|}{15.37 \%}              & \multicolumn{1}{r|}{30.00 \%}              & \multicolumn{1}{r|}{16.83 \%}              & 17.57 \%                            & \multicolumn{1}{r|}{0.57 \%}               & \multicolumn{1}{r|}{1.67 \%}               & \multicolumn{1}{r|}{0.24 \%}               & 0.29 \%                             & \multicolumn{1}{r|}{0.00 \%}        & \multicolumn{1}{r|}{0.00 \%}          & \multicolumn{1}{r|}{0.00 \%}         & 0.00 \%                             \\ \cline{2-22} 
            \multicolumn{1}{|l|}{\multirow{-3}{*}{\textbf{\begin{tabular}[c]{@{}l@{}}Total\\ Power (\textmu W)\end{tabular}}}}         & \textbf{R$^2$}                                          & \multicolumn{1}{r|}{0.948}                   & \multicolumn{1}{r|}{0.648}                   & \multicolumn{1}{r|}{0.807}                   & 0.823                                & \multicolumn{1}{r|}{0.987}                 & \multicolumn{1}{r|}{0.946}                 & \multicolumn{1}{r|}{0.781}                 & 0.976                               & \multicolumn{1}{r|}{0.989}                 & \multicolumn{1}{r|}{0.944}                 & \multicolumn{1}{r|}{0.798}                 & 0.977                               & \multicolumn{1}{r|}{1.000}                 & \multicolumn{1}{r|}{0.997}                 & \multicolumn{1}{r|}{1.000}                 & 1.000                               & \multicolumn{1}{r|}{1.000}          & \multicolumn{1}{r|}{1.000}            & \multicolumn{1}{r|}{1.000}           & 1.000                               \\ \hline
            \multicolumn{1}{|l|}{}                                                                                                                    & \textbf{MAE}                                            & \multicolumn{4}{c|}{\cellcolor[HTML]{EFEFEF}}                                                                                                                                     & \multicolumn{1}{r|}{30,529.92}             & \multicolumn{1}{r|}{83,777.43}             & \multicolumn{1}{r|}{117,233.52}            & 6,460.04                            & \multicolumn{1}{r|}{20,115.77}             & \multicolumn{1}{r|}{61,602.30}             & \multicolumn{1}{r|}{57,977.64}             & 3,479.88                            & \multicolumn{1}{r|}{13,920.52}             & \multicolumn{1}{r|}{40,751.29}             & \multicolumn{1}{r|}{26,046.32}             & 2,562.35                            & \multicolumn{1}{r|}{13,698.67}      & \multicolumn{1}{r|}{38,941.02}        & \multicolumn{1}{r|}{25,473.38}       & 2,491.87                            \\ \cline{2-2} \cline{7-22} 
            \multicolumn{1}{|l|}{}                                                                                                                    & \textbf{MAPE}                                           & \multicolumn{4}{c|}{\cellcolor[HTML]{EFEFEF}}                                                                                                                                     & \multicolumn{1}{r|}{29.21 \%}              & \multicolumn{1}{r|}{34.24 \%}              & \multicolumn{1}{r|}{24.43 \%}              & 17.17 \%                            & \multicolumn{1}{r|}{19.61 \%}              & \multicolumn{1}{r|}{24.30 \%}              & \multicolumn{1}{r|}{11.49 \%}              & 7.05 \%                             & \multicolumn{1}{r|}{13.16 \%}              & \multicolumn{1}{r|}{16.18 \%}              & \multicolumn{1}{r|}{5.51 \%}               & 3.40 \%                             & \multicolumn{1}{r|}{12.53 \%}       & \multicolumn{1}{r|}{14.86 \%}         & \multicolumn{1}{r|}{5.33 \%}         & 3.28 \%                             \\ \cline{2-2} \cline{7-22} 
            \multicolumn{1}{|l|}{\multirow{-3}{*}{\textbf{\begin{tabular}[c]{@{}l@{}}Total\\ wirelength\\ (\textmu m)\end{tabular}}}}  & \textbf{R$^2$}                                          & \multicolumn{4}{c|}{\cellcolor[HTML]{EFEFEF}}                                                                                                                                     & \multicolumn{1}{r|}{0.916}                 & \multicolumn{1}{r|}{0.902}                 & \multicolumn{1}{r|}{0.928}                 & 0.987                               & \multicolumn{1}{r|}{0.959}                 & \multicolumn{1}{r|}{0.939}                 & \multicolumn{1}{r|}{0.977}                 & 0.995                               & \multicolumn{1}{r|}{0.980}                 & \multicolumn{1}{r|}{0.975}                 & \multicolumn{1}{r|}{0.996}                 & 0.997                               & \multicolumn{1}{r|}{0.981}          & \multicolumn{1}{r|}{0.977}            & \multicolumn{1}{r|}{0.996}           & 0.997                               \\ \cline{1-2} \cline{7-22} 
            \multicolumn{1}{|l|}{}                                                                                                                    & \textbf{MAE}                                            & \multicolumn{4}{c|}{\cellcolor[HTML]{EFEFEF}}                                                                                                                                     & \multicolumn{1}{r|}{4.83}                  & \multicolumn{1}{r|}{12.41}                 & \multicolumn{1}{r|}{13.19}                 & 1.45                                & \multicolumn{1}{r|}{4.13}                  & \multicolumn{1}{r|}{11.8}                  & \multicolumn{1}{r|}{11.09}                 & 1.41                                & \multicolumn{1}{r|}{4.19}                  & \multicolumn{1}{r|}{12.01}                 & \multicolumn{1}{r|}{11.25}                 & 1.44                                & \multicolumn{1}{r|}{4.19}           & \multicolumn{1}{r|}{11.84}            & \multicolumn{1}{r|}{11.17}           & 1.43                                \\ \cline{2-2} \cline{7-22} 
            \multicolumn{1}{|l|}{}                                                                                                                    & \textbf{MAPE}                                           & \multicolumn{4}{c|}{\cellcolor[HTML]{EFEFEF}}                                                                                                                                     & \multicolumn{1}{r|}{87.59 \%}              & \multicolumn{1}{r|}{94.04 \%}              & \multicolumn{1}{r|}{72.60 \%}              & 113.14 \%                           & \multicolumn{1}{r|}{81.33 \%}              & \multicolumn{1}{r|}{86.04 \%}              & \multicolumn{1}{r|}{73.29 \%}              & 125.55 \%                           & \multicolumn{1}{r|}{78.68 \%}              & \multicolumn{1}{r|}{72.85 \%}              & \multicolumn{1}{r|}{71.34 \%}              & 123.19 \%                           & \multicolumn{1}{r|}{78.20 \%}       & \multicolumn{1}{r|}{69.82 \%}         & \multicolumn{1}{r|}{70.16 \%}        & 122.69 \%                           \\ \cline{2-2} \cline{7-22} 
            \multicolumn{1}{|l|}{}                                                                                                                    & \textbf{R$^2$}                                          & \multicolumn{4}{c|}{\cellcolor[HTML]{EFEFEF}}                                                                                                                                     & \multicolumn{1}{r|}{0.735}                 & \multicolumn{1}{r|}{0.751}                 & \multicolumn{1}{r|}{0.89}                  & 0.886                               & \multicolumn{1}{r|}{0.77}                  & \multicolumn{1}{r|}{0.773}                 & \multicolumn{1}{r|}{0.916}                 & 0.892                               & \multicolumn{1}{r|}{0.777}                 & \multicolumn{1}{r|}{0.791}                 & \multicolumn{1}{r|}{0.915}                 & 0.893                               & \multicolumn{1}{r|}{0.779}          & \multicolumn{1}{r|}{0.798}            & \multicolumn{1}{r|}{0.918}           & 0.895                               \\ \cline{2-2} \cline{7-22} 
            \multicolumn{1}{|l|}{}                                                                                                                    & \textbf{MAE P95}                                        & \multicolumn{4}{c|}{\cellcolor[HTML]{EFEFEF}}                                                                                                                                     & \multicolumn{1}{r|}{12.72}                 & \multicolumn{1}{r|}{38.53}                 & \multicolumn{1}{r|}{53.88}                 & 5.57                                & \multicolumn{1}{r|}{10.32}                 & \multicolumn{1}{r|}{39.23}                 & \multicolumn{1}{r|}{46.46}                 & 5.44                                & \multicolumn{1}{r|}{10.82}                 & \multicolumn{1}{r|}{41.92}                 & \multicolumn{1}{r|}{46.28}                 & 5.85                                & \multicolumn{1}{r|}{10.79}          & \multicolumn{1}{r|}{41.09}            & \multicolumn{1}{r|}{46.08}           & 5.83                                \\ \cline{2-2} \cline{7-22} 
            \multicolumn{1}{|l|}{}                                                                                                                    & \textbf{MAPE P95}                                       & \multicolumn{4}{c|}{\cellcolor[HTML]{EFEFEF}}                                                                                                                                     & \multicolumn{1}{r|}{230.82 \%}             & \multicolumn{1}{r|}{293.66 \%}             & \multicolumn{1}{r|}{339.20 \%}             & 410.10 \%                           & \multicolumn{1}{r|}{186.85 \%}             & \multicolumn{1}{r|}{247.83 \%}             & \multicolumn{1}{r|}{326.04 \%}             & 403.03 \%                           & \multicolumn{1}{r|}{181.48 \%}             & \multicolumn{1}{r|}{238.43 \%}             & \multicolumn{1}{r|}{291.33 \%}             & 402.78 \%                           & \multicolumn{1}{r|}{180.77 \%}      & \multicolumn{1}{r|}{234.56 \%}        & \multicolumn{1}{r|}{291.33 \%}       & 402.78 \%                           \\ \cline{2-2} \cline{7-22} 
            \multicolumn{1}{|l|}{}                                                                                                                    & \textbf{MAE TOP5}                                       & \multicolumn{4}{c|}{\cellcolor[HTML]{EFEFEF}}                                                                                                                                     & \multicolumn{1}{r|}{44.73}                 & \multicolumn{1}{r|}{116.02}                & \multicolumn{1}{r|}{86.57}                 & 11.67                               & \multicolumn{1}{r|}{46.91}                 & \multicolumn{1}{r|}{117.87}                & \multicolumn{1}{r|}{79.59}                 & 11.37                               & \multicolumn{1}{r|}{47.4}                  & \multicolumn{1}{r|}{121.41}                & \multicolumn{1}{r|}{80.45}                 & 11.65                               & \multicolumn{1}{r|}{47.61}          & \multicolumn{1}{r|}{121.04}           & \multicolumn{1}{r|}{80.05}           & 11.61                               \\ \cline{2-2} \cline{7-22} 
            \multicolumn{1}{|l|}{\multirow{-7}{*}{\textbf{\begin{tabular}[c]{@{}l@{}}Interconnect\\ length (\textmu m)\end{tabular}}}} & \textbf{MAPE TOP5}                                      & \multicolumn{4}{c|}{\multirow{-10}{*}{\cellcolor[HTML]{EFEFEF}\begin{tabular}[c]{@{}c@{}}Estimated wirelength is not available\\ as cells have not been placed yet\end{tabular}}} & \multicolumn{1}{r|}{24.45 \%}              & \multicolumn{1}{r|}{25.39 \%}              & \multicolumn{1}{r|}{19.20 \%}              & 18.59 \%                            & \multicolumn{1}{r|}{23.99 \%}              & \multicolumn{1}{r|}{24.99 \%}              & \multicolumn{1}{r|}{18.51 \%}              & 18.19 \%                            & \multicolumn{1}{r|}{24.18 \%}              & \multicolumn{1}{r|}{25.04 \%}              & \multicolumn{1}{r|}{18.56 \%}              & 19.03 \%                            & \multicolumn{1}{r|}{24.12 \%}       & \multicolumn{1}{r|}{24.74 \%}         & \multicolumn{1}{r|}{18.49 \%}        & 18.90 \%                            \\ \hline
            \multicolumn{1}{|l|}{}                                                                                                                    & \textbf{MAE}                                            & \multicolumn{1}{r|}{3.10}                    & \multicolumn{1}{r|}{10.84}                   & \multicolumn{1}{r|}{8.55}                    & 0.09                                 & \multicolumn{1}{r|}{0.16}                  & \multicolumn{1}{r|}{0.77}                  & \multicolumn{1}{r|}{0.41}                  & 0.06                                & \multicolumn{1}{r|}{0.13}                  & \multicolumn{1}{r|}{1.21}                  & \multicolumn{1}{r|}{0.34}                  & 0.06                                & \multicolumn{1}{r|}{0.01}                  & \multicolumn{1}{r|}{0.14}                  & \multicolumn{1}{r|}{0.06}                  & 0.01                                & \multicolumn{1}{r|}{0.00}           & \multicolumn{1}{r|}{0.00}             & \multicolumn{1}{r|}{0.00}            & 0.00                                \\ \cline{2-22} 
            \multicolumn{1}{|l|}{\multirow{-2}{*}{\textbf{\begin{tabular}[c]{@{}l@{}}Worst Arrival\\ Time (ns)\end{tabular}}}}                        & \textbf{MAPE}                                           & \multicolumn{1}{r|}{192.92 \%}               & \multicolumn{1}{r|}{183.71 \%}               & \multicolumn{1}{r|}{331.31 \%}               & 15.78 \%                             & \multicolumn{1}{r|}{16.19 \%}              & \multicolumn{1}{r|}{21.48 \%}              & \multicolumn{1}{r|}{16.98 \%}              & 11.17 \%                            & \multicolumn{1}{r|}{13.81 \%}              & \multicolumn{1}{r|}{33.30 \%}              & \multicolumn{1}{r|}{15.82 \%}              & 10.44 \%                            & \multicolumn{1}{r|}{1.60 \%}               & \multicolumn{1}{r|}{4.69 \%}               & \multicolumn{1}{r|}{2.32 \%}               & 2.14 \%                             & \multicolumn{1}{r|}{0.00 \%}        & \multicolumn{1}{r|}{0.00 \%}          & \multicolumn{1}{r|}{0.00 \%}         & 0.00 \%                             \\ \hline
            \multicolumn{1}{|l|}{}                                                                                                                    & \textbf{MAE}                                            & \multicolumn{1}{r|}{3.18}                    & \multicolumn{1}{r|}{11.28}                   & \multicolumn{1}{r|}{8.83}                    & 0.03                                 & \multicolumn{1}{r|}{0.21}                  & \multicolumn{1}{r|}{1.15}                  & \multicolumn{1}{r|}{0.27}                  & 0.05                                & \multicolumn{1}{r|}{0.15}                  & \multicolumn{1}{r|}{1.62}                  & \multicolumn{1}{r|}{0.31}                  & 0.05                                & \multicolumn{1}{r|}{0.01}                  & \multicolumn{1}{r|}{0.09}                  & \multicolumn{1}{r|}{0.04}                  & 0.01                                & \multicolumn{1}{r|}{0.00}           & \multicolumn{1}{r|}{0.00}             & \multicolumn{1}{r|}{0.00}            & 0.00                                \\ \cline{2-22} 
            \multicolumn{1}{|l|}{}                                                                                                                    & \textbf{MPE}                                            & \multicolumn{1}{r|}{0.06}                    & \multicolumn{1}{r|}{0.30}                    & \multicolumn{1}{r|}{0.30}                    & 0.04                                 & \multicolumn{1}{r|}{0.09}                  & \multicolumn{1}{r|}{0.27}                  & \multicolumn{1}{r|}{0.27}                  & 0.04                                & \multicolumn{1}{r|}{0.04}                  & \multicolumn{1}{r|}{0.17}                  & \multicolumn{1}{r|}{0.20}                  & 0.03                                & \multicolumn{1}{r|}{0.01}                  & \multicolumn{1}{r|}{0.03}                  & \multicolumn{1}{r|}{0.04}                  & 0.00                                & \multicolumn{4}{c|}{\cellcolor[HTML]{EFEFEF}}                                                                                                            \\ \cline{2-18}
            \multicolumn{1}{|l|}{}                                                                                                                    & \textbf{MNE}                                            & \multicolumn{1}{r|}{3.68}                    & \multicolumn{1}{r|}{12.15}                   & \multicolumn{1}{r|}{9.35}                    & 0.02                                 & \multicolumn{1}{r|}{0.24}                  & \multicolumn{1}{r|}{1.26}                  & \multicolumn{1}{r|}{0.26}                  & 0.05                                & \multicolumn{1}{r|}{0.17}                  & \multicolumn{1}{r|}{1.68}                  & \multicolumn{1}{r|}{0.36}                  & 0.06                                & \multicolumn{1}{r|}{0.02}                  & \multicolumn{1}{r|}{0.12}                  & \multicolumn{1}{r|}{0.13}                  & 0.01                                & \multicolumn{4}{c|}{\multirow{-2}{*}{\cellcolor[HTML]{EFEFEF}\begin{tabular}[c]{@{}c@{}}No positive or negative error\\ $n_p$ = $n_n$ = 0\end{tabular}}} \\ \cline{2-22} 
            \multicolumn{1}{|l|}{}                                                                                                                    & \textbf{TPR}                                            & \multicolumn{1}{r|}{95.67 \%}                & \multicolumn{1}{r|}{96.32 \%}                & \multicolumn{1}{r|}{94.52 \%}                & 67.94 \%                             & \multicolumn{1}{r|}{95.67 \%}              & \multicolumn{1}{r|}{96.21 \%}              & \multicolumn{1}{r|}{64.38 \%}              & 93.08 \%                            & \multicolumn{1}{r|}{95.67 \%}              & \multicolumn{1}{r|}{98.53 \%}              & \multicolumn{1}{r|}{58.90 \%}              & 99.45 \%                            & \multicolumn{1}{r|}{100.00 \%}             & \multicolumn{1}{r|}{99.47 \%}              & \multicolumn{1}{r|}{96.58 \%}              & 100.00 \%                           & \multicolumn{1}{r|}{100.00 \%}      & \multicolumn{1}{r|}{100.00 \%}        & \multicolumn{1}{r|}{100.00 \%}       & 100.00 \%                           \\ \cline{2-22} 
            \multicolumn{1}{|l|}{\multirow{-5}{*}{\textbf{\begin{tabular}[c]{@{}l@{}}Worst\\ Slack (ns)\end{tabular}}}}                               & \textbf{TNR}                                            & \multicolumn{1}{r|}{28.70 \%}                & \multicolumn{1}{r|}{15.76 \%}                & \multicolumn{1}{r|}{28.11 \%}                & 97.70 \%                             & \multicolumn{1}{r|}{57.56 \%}              & \multicolumn{1}{r|}{21.39 \%}              & \multicolumn{1}{r|}{95.54 \%}              & 63.75 \%                            & \multicolumn{1}{r|}{56.26 \%}              & \multicolumn{1}{r|}{8.39 \%}               & \multicolumn{1}{r|}{91.58 \%}              & 58.79 \%                            & \multicolumn{1}{r|}{93.24 \%}              & \multicolumn{1}{r|}{60.59 \%}              & \multicolumn{1}{r|}{99.22 \%}              & 91.53 \%                            & \multicolumn{1}{r|}{100.00 \%}      & \multicolumn{1}{r|}{100.00 \%}        & \multicolumn{1}{r|}{100.00 \%}       & 100.00 \%                           \\ \hline
            \multicolumn{1}{|l|}{}                                                                                                                    & \textbf{MAE}                                            & \multicolumn{1}{r|}{1,764.18}                & \multicolumn{1}{r|}{17,768.52}               & \multicolumn{1}{r|}{3,755.75}                & 1.09                                 & \multicolumn{1}{r|}{57.74}                 & \multicolumn{1}{r|}{727.04}                & \multicolumn{1}{r|}{0.57}                  & 5.65                                & \multicolumn{1}{r|}{22.45}                 & \multicolumn{1}{r|}{959.12}                & \multicolumn{1}{r|}{1.05}                  & 5.75                                & \multicolumn{1}{r|}{2.11}                  & \multicolumn{1}{r|}{70.22}                 & \multicolumn{1}{r|}{0.04}                  & 1.01                                & \multicolumn{1}{r|}{0.00}           & \multicolumn{1}{r|}{0.00}             & \multicolumn{1}{r|}{0.00}            & 0.00                                \\ \cline{2-22} 
            \multicolumn{1}{|l|}{}                                                                                                                    & \textbf{MPE}                                            & \multicolumn{1}{r|}{1.26}                    & \multicolumn{1}{r|}{4.74}                    & \multicolumn{1}{r|}{2.50}                    & 3.92                                 & \multicolumn{1}{r|}{24.60}                 & \multicolumn{1}{r|}{5.30}                  & \multicolumn{1}{r|}{1.82}                  & 3.76                                & \multicolumn{1}{r|}{23.22}                 & \multicolumn{1}{r|}{2.74}                  & \multicolumn{1}{r|}{1.88}                  & 4.68                                & \multicolumn{1}{r|}{4.81}                  & \multicolumn{1}{r|}{22.51}                 & \multicolumn{1}{r|}{0.17}                  & 1.56                                & \multicolumn{4}{c|}{\cellcolor[HTML]{EFEFEF}}                                                                                                            \\ \cline{2-18}
            \multicolumn{1}{|l|}{\multirow{-3}{*}{\textbf{\begin{tabular}[c]{@{}l@{}}Total\\ Negative\\ Slack (ns)\end{tabular}}}}                    & \textbf{MNE}                                            & \multicolumn{1}{r|}{2,296.57}                & \multicolumn{1}{r|}{20,176.54}               & \multicolumn{1}{r|}{5,201.66}                & 2.10                                 & \multicolumn{1}{r|}{101.63}                & \multicolumn{1}{r|}{849.90}                & \multicolumn{1}{r|}{7.15}                  & 11.49                               & \multicolumn{1}{r|}{37.22}                 & \multicolumn{1}{r|}{1,031.87}              & \multicolumn{1}{r|}{9.00}                  & 10.54                               & \multicolumn{1}{r|}{6.34}                  & \multicolumn{1}{r|}{123.39}                & \multicolumn{1}{r|}{0.76}                  & 3.78                                & \multicolumn{4}{c|}{\multirow{-2}{*}{\cellcolor[HTML]{EFEFEF}\begin{tabular}[c]{@{}c@{}}No positive or negative error\\ $n_p$ = $n_n$ = 0\end{tabular}}} \\ \hline
            \multicolumn{1}{|l|}{}                                                                                                                    & \textbf{MAE}                                            & \multicolumn{1}{r|}{0.4756}                  & \multicolumn{1}{r|}{2.9814}                  & \multicolumn{1}{r|}{1.7195}                  & 0.1836                               & \multicolumn{1}{r|}{0.2313}                & \multicolumn{1}{r|}{0.9552}                & \multicolumn{1}{r|}{0.6682}                & 0.1509                              & \multicolumn{1}{r|}{0.2398}                & \multicolumn{1}{r|}{0.9151}                & \multicolumn{1}{r|}{0.6392}                & 0.1462                              & \multicolumn{1}{r|}{0.0022}                & \multicolumn{1}{r|}{0.0319}                & \multicolumn{1}{r|}{0.0041}                & 0.0012                              & \multicolumn{1}{r|}{0.0000}         & \multicolumn{1}{r|}{0.0000}           & \multicolumn{1}{r|}{0.0000}          & 0.0000                              \\ \cline{2-22} 
            \multicolumn{1}{|l|}{}                                                                                                                    & \textbf{MAPE}                                           & \multicolumn{1}{r|}{75.89 \%}                & \multicolumn{1}{r|}{92.31 \%}                & \multicolumn{1}{r|}{95.11 \%}                & 51.92 \%                             & \multicolumn{1}{r|}{46.25 \%}              & \multicolumn{1}{r|}{39.61 \%}              & \multicolumn{1}{r|}{43.00 \%}              & 42.55 \%                            & \multicolumn{1}{r|}{45.63 \%}              & \multicolumn{1}{r|}{37.87 \%}              & \multicolumn{1}{r|}{41.34 \%}              & 41.05 \%                            & \multicolumn{1}{r|}{0.27 \%}               & \multicolumn{1}{r|}{0.80 \%}               & \multicolumn{1}{r|}{0.18 \%}               & 0.22 \%                             & \multicolumn{1}{r|}{0.00 \%}        & \multicolumn{1}{r|}{0.00 \%}          & \multicolumn{1}{r|}{0.00 \%}         & 0.00 \%                             \\ \cline{2-22} 
            \multicolumn{1}{|l|}{}                                                                                                                    & \textbf{MAE P95}                                        & \multicolumn{1}{r|}{1.0541}                  & \multicolumn{1}{r|}{7.1590}                  & \multicolumn{1}{r|}{4.6540}                  & 0.3296                               & \multicolumn{1}{r|}{0.3600}                & \multicolumn{1}{r|}{1.5150}                & \multicolumn{1}{r|}{0.8949}                & 0.2385                              & \multicolumn{1}{r|}{0.3570}                & \multicolumn{1}{r|}{1.5320}                & \multicolumn{1}{r|}{0.8899}                & 0.2387                              & \multicolumn{1}{r|}{0.0107}                & \multicolumn{1}{r|}{0.1980}                & \multicolumn{1}{r|}{0.0220}                & 0.0036                              & \multicolumn{1}{r|}{0.0000}         & \multicolumn{1}{r|}{0.0000}           & \multicolumn{1}{r|}{0.0000}          & 0.0000                              \\ \cline{2-22} 
            \multicolumn{1}{|l|}{}                                                                                                                    & \textbf{MAPE P95}                                       & \multicolumn{1}{r|}{133.31 \%}               & \multicolumn{1}{r|}{169.45 \%}               & \multicolumn{1}{r|}{241.72 \%}               & 74.00 \%                             & \multicolumn{1}{r|}{69.54 \%}              & \multicolumn{1}{r|}{64.36 \%}              & \multicolumn{1}{r|}{65.89 \%}              & 70.09 \%                            & \multicolumn{1}{r|}{69.32 \%}              & \multicolumn{1}{r|}{64.39 \%}              & \multicolumn{1}{r|}{65.77 \%}              & 70.06 \%                            & \multicolumn{1}{r|}{1.27 \%}               & \multicolumn{1}{r|}{4.81 \%}               & \multicolumn{1}{r|}{0.98 \%}               & 0.68 \%                             & \multicolumn{1}{r|}{0.00 \%}        & \multicolumn{1}{r|}{0.00 \%}          & \multicolumn{1}{r|}{0.00 \%}         & 0.00 \%                             \\ \cline{2-22} 
            \multicolumn{1}{|l|}{}                                                                                                                    & \textbf{MAE TOP5}                                       & \multicolumn{1}{r|}{2.6279}                  & \multicolumn{1}{r|}{10.1956}                 & \multicolumn{1}{r|}{1.8427}                  & 0.3048                               & \multicolumn{1}{r|}{0.2398}                & \multicolumn{1}{r|}{0.9467}                & \multicolumn{1}{r|}{0.7733}                & 0.1986                              & \multicolumn{1}{r|}{0.1971}                & \multicolumn{1}{r|}{1.0037}                & \multicolumn{1}{r|}{0.7550}                & 0.1975                              & \multicolumn{1}{r|}{0.0107}                & \multicolumn{1}{r|}{0.1110}                & \multicolumn{1}{r|}{0.0246}                & 0.0037                              & \multicolumn{1}{r|}{0.0000}         & \multicolumn{1}{r|}{0.0000}           & \multicolumn{1}{r|}{0.0000}          & 0.0000                              \\ \cline{2-22} 
            \multicolumn{1}{|l|}{\multirow{-6}{*}{\textbf{\begin{tabular}[c]{@{}l@{}}Timing\\ Path\\ Arrival\\ Time (ns)\end{tabular}}}}              & \textbf{MAPE TOP5}                                      & \multicolumn{1}{r|}{208.58 \%}               & \multicolumn{1}{r|}{164.89 \%}               & \multicolumn{1}{r|}{58.23 \%}                & 36.41 \%                             & \multicolumn{1}{r|}{15.32 \%}              & \multicolumn{1}{r|}{14.73 \%}              & \multicolumn{1}{r|}{24.60 \%}              & 23.24 \%                            & \multicolumn{1}{r|}{12.57 \%}              & \multicolumn{1}{r|}{15.69 \%}              & \multicolumn{1}{r|}{23.97 \%}              & 23.01 \%                            & \multicolumn{1}{r|}{0.70 \%}               & \multicolumn{1}{r|}{1.68 \%}               & \multicolumn{1}{r|}{0.73 \%}               & 0.42 \%                             & \multicolumn{1}{r|}{0.00 \%}        & \multicolumn{1}{r|}{0.00 \%}          & \multicolumn{1}{r|}{0.00 \%}         & 0.00 \%                             \\ \hline
            \multicolumn{1}{|l|}{}                                                                                                                    & \textbf{MAE}                                            & \multicolumn{1}{r|}{0.2450}                  & \multicolumn{1}{r|}{2.6909}                  & \multicolumn{1}{r|}{1.8879}                  & 0.0400                               & \multicolumn{1}{r|}{0.0532}                & \multicolumn{1}{r|}{0.2179}                & \multicolumn{1}{r|}{0.0736}                & 0.0166                              & \multicolumn{1}{r|}{0.0280}                & \multicolumn{1}{r|}{0.2873}                & \multicolumn{1}{r|}{0.0975}                & 0.0203                              & \multicolumn{1}{r|}{0.0023}                & \multicolumn{1}{r|}{0.0316}                & \multicolumn{1}{r|}{0.0040}                & 0.0012                              & \multicolumn{1}{r|}{0.0000}         & \multicolumn{1}{r|}{0.0000}           & \multicolumn{1}{r|}{0.0000}          & 0.0000                              \\ \cline{2-22} 
            \multicolumn{1}{|l|}{}                                                                                                                    & \textbf{MPE}                                            & \multicolumn{1}{r|}{0.0361}                  & \multicolumn{1}{r|}{1.7111}                  & \multicolumn{1}{r|}{0.9992}                  & 0.0788                               & \multicolumn{1}{r|}{0.0282}                & \multicolumn{1}{r|}{0.0851}                & \multicolumn{1}{r|}{0.0556}                & 0.0150                              & \multicolumn{1}{r|}{0.0299}                & \multicolumn{1}{r|}{0.1210}                & \multicolumn{1}{r|}{0.0741}                & 0.0167                              & \multicolumn{1}{r|}{0.0125}                & \multicolumn{1}{r|}{0.0351}                & \multicolumn{1}{r|}{0.0159}                & 0.0080                              & \multicolumn{4}{c|}{\cellcolor[HTML]{EFEFEF}}                                                                                                            \\ \cline{2-18}
            \multicolumn{1}{|l|}{}                                                                                                                    & \textbf{MNE}                                            & \multicolumn{1}{r|}{0.2749}                  & \multicolumn{1}{r|}{2.8765}                  & \multicolumn{1}{r|}{2.8290}                  & 0.0222                               & \multicolumn{1}{r|}{0.0641}                & \multicolumn{1}{r|}{0.4140}                & \multicolumn{1}{r|}{0.1082}                & 0.0203                              & \multicolumn{1}{r|}{0.0273}                & \multicolumn{1}{r|}{0.5783}                & \multicolumn{1}{r|}{0.1349}                & 0.0296                              & \multicolumn{1}{r|}{0.0217}                & \multicolumn{1}{r|}{0.1342}                & \multicolumn{1}{r|}{0.0167}                & 0.0149                              & \multicolumn{4}{c|}{\multirow{-2}{*}{\cellcolor[HTML]{EFEFEF}\begin{tabular}[c]{@{}c@{}}No positive or negative error\\ $n_p$ = $n_n$ = 0\end{tabular}}} \\ \cline{2-22} 
            \multicolumn{1}{|l|}{}                                                                                                                    & \textbf{TPR}                                            & \multicolumn{1}{r|}{98.01 \%}                & \multicolumn{1}{r|}{96.65 \%}                & \multicolumn{1}{r|}{74.89 \%}                & 70.21 \%                             & \multicolumn{1}{r|}{96.50 \%}              & \multicolumn{1}{r|}{98.01 \%}              & \multicolumn{1}{r|}{74.19 \%}              & 83.93 \%                            & \multicolumn{1}{r|}{94.90 \%}              & \multicolumn{1}{r|}{97.61 \%}              & \multicolumn{1}{r|}{73.80 \%}              & 85.13 \%                            & \multicolumn{1}{r|}{93.04 \%}              & \multicolumn{1}{r|}{95.99 \%}              & \multicolumn{1}{r|}{84.56 \%}              & 95.92 \%                            & \multicolumn{1}{r|}{100.00 \%}      & \multicolumn{1}{r|}{100.00 \%}        & \multicolumn{1}{r|}{100.00 \%}       & 100.00 \%                           \\ \cline{2-22} 
            \multicolumn{1}{|l|}{\multirow{-5}{*}{\textbf{\begin{tabular}[c]{@{}l@{}}Timing\\ Path\\ Slack (ns)\end{tabular}}}}                       & \textbf{TNR}                                            & \multicolumn{1}{r|}{88.69 \%}                & \multicolumn{1}{r|}{77.61 \%}                & \multicolumn{1}{r|}{93.99 \%}                & 99.96 \%                             & \multicolumn{1}{r|}{96.00 \%}              & \multicolumn{1}{r|}{90.69 \%}              & \multicolumn{1}{r|}{99.93 \%}              & 98.69 \%                            & \multicolumn{1}{r|}{97.65 \%}              & \multicolumn{1}{r|}{89.43 \%}              & \multicolumn{1}{r|}{99.92 \%}              & 98.56 \%                            & \multicolumn{1}{r|}{99.63 \%}              & \multicolumn{1}{r|}{98.34 \%}              & \multicolumn{1}{r|}{99.99 \%}              & 99.73 \%                            & \multicolumn{1}{r|}{100.00 \%}      & \multicolumn{1}{r|}{100.00 \%}        & \multicolumn{1}{r|}{100.00 \%}       & 100.00 \%                           \\ \hline
            \multicolumn{1}{|l|}{}                                                                                                                    & \textbf{MAE}                                            & \multicolumn{1}{r|}{0.1128}                  & \multicolumn{1}{r|}{0.7713}                  & \multicolumn{1}{r|}{0.4259}                  & 0.0105                               & \multicolumn{1}{r|}{0.0200}                & \multicolumn{1}{r|}{0.0631}                & \multicolumn{1}{r|}{0.0306}                & 0.0066                              & \multicolumn{1}{r|}{0.0122}                & \multicolumn{1}{r|}{0.0466}                & \multicolumn{1}{r|}{0.0165}                & 0.0039                              & \multicolumn{1}{r|}{0.0005}                & \multicolumn{1}{r|}{0.0046}                & \multicolumn{1}{r|}{0.0004}                & 0.0002                              & \multicolumn{1}{r|}{0.0000}         & \multicolumn{1}{r|}{0.0000}           & \multicolumn{1}{r|}{0.0000}          & 0.0000                              \\ \cline{2-22} 
            \multicolumn{1}{|l|}{}                                                                                                                    & \textbf{MAPE}                                           & \multicolumn{1}{r|}{\textgreater 10000 \%}   & \multicolumn{1}{r|}{403.06 \%}               & \multicolumn{1}{r|}{547.23 \%}               & 39.38 \%                             & \multicolumn{1}{r|}{\textgreater 10000 \%} & \multicolumn{1}{r|}{28.21 \%}              & \multicolumn{1}{r|}{31.14 \%}              & 20.69 \%                            & \multicolumn{1}{r|}{\textgreater 10000 \%} & \multicolumn{1}{r|}{19.99 \%}              & \multicolumn{1}{r|}{8.91 \%}               & 11.16 \%                            & \multicolumn{1}{r|}{\textgreater 10000 \%} & \multicolumn{1}{r|}{2.36 \%}               & \multicolumn{1}{r|}{0.32 \%}               & 0.95 \%                             & \multicolumn{1}{r|}{0.00 \%}        & \multicolumn{1}{r|}{0.00 \%}          & \multicolumn{1}{r|}{0.00 \%}         & 0.00 \%                             \\ \cline{2-22} 
            \multicolumn{1}{|l|}{\multirow{-3}{*}{\textbf{\begin{tabular}[c]{@{}l@{}}Net Arc\\ Delay (ns)\end{tabular}}}}                             & \textbf{R$^2$}                                          & \multicolumn{1}{r|}{\textless -1}            & \multicolumn{1}{r|}{\textless -1}            & \multicolumn{1}{r|}{\textless -1}            & \textless -1                         & \multicolumn{1}{r|}{\textless -1}          & \multicolumn{1}{r|}{-0.029}                & \multicolumn{1}{r|}{\textless -1}          & -0.402                              & \multicolumn{1}{r|}{-0.024}                & \multicolumn{1}{r|}{0.748}                 & \multicolumn{1}{r|}{0.876}                 & 0.777                               & \multicolumn{1}{r|}{0.976}                 & \multicolumn{1}{r|}{0.981}                 & \multicolumn{1}{r|}{0.999}                 & 0.979                               & \multicolumn{1}{r|}{1.000}          & \multicolumn{1}{r|}{1.000}            & \multicolumn{1}{r|}{1.000}           & 1.000                               \\ \hline
            \multicolumn{1}{|l|}{}                                                                                                                    & \textbf{MAE}                                            & \multicolumn{1}{r|}{0.0011}                  & \multicolumn{1}{r|}{0.0031}                  & \multicolumn{1}{r|}{0.0003}                  & 0.0018                               & \multicolumn{1}{r|}{0.0008}                & \multicolumn{1}{r|}{0.0016}                & \multicolumn{1}{r|}{0.0002}                & 0.0003                              & \multicolumn{1}{r|}{0.0009}                & \multicolumn{1}{r|}{0.0013}                & \multicolumn{1}{r|}{0.0001}                & 0.0002                              & \multicolumn{1}{r|}{0.0001}                & \multicolumn{1}{r|}{0.0003}                & \multicolumn{1}{r|}{0.0000}                & 0.0000                              & \multicolumn{1}{r|}{0.0000}         & \multicolumn{1}{r|}{0.0000}           & \multicolumn{1}{r|}{0.0000}          & 0.0000                              \\ \cline{2-22} 
            \multicolumn{1}{|l|}{}                                                                                                                    & \textbf{MAPE}                                           & \multicolumn{1}{r|}{\textgreater 10000 \%}   & \multicolumn{1}{r|}{\textgreater 10000 \%}   & \multicolumn{1}{r|}{\textgreater 10000 \%}   & \textgreater 10000 \%                & \multicolumn{1}{r|}{\textgreater 10000 \%} & \multicolumn{1}{r|}{\textgreater 10000 \%} & \multicolumn{1}{r|}{\textgreater 10000 \%} & \textgreater 10000 \%               & \multicolumn{1}{r|}{\textgreater 10000 \%} & \multicolumn{1}{r|}{\textgreater 10000 \%} & \multicolumn{1}{r|}{\textgreater 10000 \%} & \textgreater 10000 \%               & \multicolumn{1}{r|}{\textgreater 10000 \%} & \multicolumn{1}{r|}{\textgreater 10000 \%} & \multicolumn{1}{r|}{\textgreater 10000 \%} & \textgreater 10000 \%               & \multicolumn{1}{r|}{0.00 \%}        & \multicolumn{1}{r|}{0.00 \%}          & \multicolumn{1}{r|}{0.00 \%}         & 0.00 \%                             \\ \cline{2-22} 
            \multicolumn{1}{|l|}{\multirow{-3}{*}{\textbf{\begin{tabular}[c]{@{}l@{}}Cell Arc\\ Delay (ns)\end{tabular}}}}                            & \textbf{R$^2$}                                          & \multicolumn{1}{r|}{-0.308}                  & \multicolumn{1}{r|}{\textless -1}            & \multicolumn{1}{r|}{\textless -1}            & -0.141                               & \multicolumn{1}{r|}{0.928}                 & \multicolumn{1}{r|}{0.873}                 & \multicolumn{1}{r|}{0.641}                 & 0.851                               & \multicolumn{1}{r|}{0.875}                 & \multicolumn{1}{r|}{0.894}                 & \multicolumn{1}{r|}{0.657}                 & 0.925                               & \multicolumn{1}{r|}{0.987}                 & \multicolumn{1}{r|}{0.974}                 & \multicolumn{1}{r|}{0.975}                 & 0.952                               & \multicolumn{1}{r|}{1.000}          & \multicolumn{1}{r|}{1.000}            & \multicolumn{1}{r|}{1.000}           & 1.000                               \\ \hline
            \multicolumn{1}{|l|}{}                                                                                                                    & \textbf{MAE}                                            & \multicolumn{1}{r|}{0.0761}                  & \multicolumn{1}{r|}{0.8919}                  & \multicolumn{1}{r|}{0.4735}                  & 0.0152                               & \multicolumn{1}{r|}{0.0082}                & \multicolumn{1}{r|}{0.0415}                & \multicolumn{1}{r|}{0.0059}                & 0.0026                              & \multicolumn{1}{r|}{0.0028}                & \multicolumn{1}{r|}{0.0392}                & \multicolumn{1}{r|}{0.0011}                & 0.0017                              & \multicolumn{1}{r|}{0.0003}                & \multicolumn{1}{r|}{0.0054}                & \multicolumn{1}{r|}{0.0004}                & 0.0003                              & \multicolumn{1}{r|}{0.0000}         & \multicolumn{1}{r|}{0.0000}           & \multicolumn{1}{r|}{0.0000}          & 0.0000                              \\ \cline{2-22} 
            \multicolumn{1}{|l|}{}                                                                                                                    & \textbf{MAPE}                                           & \multicolumn{1}{r|}{183.03 \%}               & \multicolumn{1}{r|}{462.31 \%}               & \multicolumn{1}{r|}{1134.69 \%}              & 34.33 \%                             & \multicolumn{1}{r|}{27.44 \%}              & \multicolumn{1}{r|}{37.69 \%}              & \multicolumn{1}{r|}{11.05 \%}              & 7.44 \%                             & \multicolumn{1}{r|}{13.43 \%}              & \multicolumn{1}{r|}{35.60 \%}              & \multicolumn{1}{r|}{2.05 \%}               & 5.89 \%                             & \multicolumn{1}{r|}{1.27 \%}               & \multicolumn{1}{r|}{4.66 \%}               & \multicolumn{1}{r|}{0.53 \%}               & 1.01 \%                             & \multicolumn{1}{r|}{0.00 \%}        & \multicolumn{1}{r|}{0.00 \%}          & \multicolumn{1}{r|}{0.00 \%}         & 0.00 \%                             \\ \cline{2-22} 
            \multicolumn{1}{|l|}{\multirow{-3}{*}{\textbf{\begin{tabular}[c]{@{}l@{}}Cell Arc\\ Slew (ns)\end{tabular}}}}                             & \textbf{R$^2$}                                          & \multicolumn{1}{r|}{\textless -1}            & \multicolumn{1}{r|}{\textless -1}            & \multicolumn{1}{r|}{\textless -1}            & \textless -1                         & \multicolumn{1}{r|}{0.332}                 & \multicolumn{1}{r|}{0.414}                 & \multicolumn{1}{r|}{0.965}                 & 0.878                               & \multicolumn{1}{r|}{0.786}                 & \multicolumn{1}{r|}{0.427}                 & \multicolumn{1}{r|}{0.987}                 & 0.911                               & \multicolumn{1}{r|}{0.973}                 & \multicolumn{1}{r|}{0.938}                 & \multicolumn{1}{r|}{0.994}                 & 0.964                               & \multicolumn{1}{r|}{1.000}          & \multicolumn{1}{r|}{1.000}            & \multicolumn{1}{r|}{1.000}           & 1.000                               \\ \hline
            \end{tabular}
    }
    \end{center}

    % \hl{add more baseline evaluations on buffer insertion/removal, resizing, cloning}
% \vspace{3in}

\end{landscape}

The resulting baseline values across all metrics, design stages, and technology nodes are listed in Table~\ref{tab:baseline}.
The baseline values establish a standardized point of comparison for future research efforts that utilize the EDA-Schema-V2 dataset.
Machine learning methods and other predictive techniques targeting the listed tasks will benefit from evaluation with the provided baselines and are expected to demonstrate improved prediction accuracy across the reported metrics.
Consequently, Table~\ref{tab:baseline} serves as a reference benchmark leaderboard for early-stage QoR prediction problems in physical design and provides a consistent target for the evaluation of subsequent work.

\section{Conclusion}
\label{sec:conclusions}

This paper presents EDA-Schema-V2, a standardized multimodal schema for machine-learning applications in digital physical design. The schema organizes artifacts generated throughout the RTL-to-GDSII design flow within a unified entity–relationship structure that captures both structural and spatial circuit characteristics. Graph representations model logical and physical connectivity among design elements, while image-based representations capture spatial layout characteristics and analyzed circuit properties are provided as heatmaps. By relying on widely adopted electronic design automation (EDA) interchange formats and technology layer conventions, the schema remains compatible with both open-source and commercial physical design tool flows.

A large-scale dataset is generated using the OpenROAD tool chain together with the four open process design kits (SkyWater 130\,nm, Nangate 45\,nm, ASAP7 7\,nm, and IHP SG13G2 130\,nm). The dataset is generated from IWLS'05 benchmark circuits through systematic sweeps of key physical design parameters. The resulting dataset contains approximately 7{,}800 physical design instances derived from 18 benchmark circuits and spans more than 275 million gates, 75 million nets, and over 36 million extracted timing paths. Stage-resolved information across the physical design flow from synthesis through detailed routing is captured together with multimodal representations that include circuit graphs, spatial layout images, and quality-of-results (QoR) metrics.

A baseline analysis is provided to quantify the accuracy of predicted final post-routing QoR metrics during execution of intermediate stages of the physical design flow and across different process design kits (PDKs). The evaluated baseline metrics establish a standardized reference for early-stage QoR prediction and provide a consistent benchmark for evaluating future machine learning algorithms and methods. To support reproducibility and broader community adoption, the EDA-Schema-V2 specification, dataset, and supporting tooling are publicly released through an open repository available at \href{https://github.com/drexel-ice/EDA-schema}{https://github.com/drexel-ice/EDA-schema}. EDA-Schema-V2 together with the accompanying dataset improves reproducibility, comparability, and accessibility in machine-learning-for-EDA research while providing a scalable foundation for future data-driven digital design automation.

\bibliographystyle{myIEEE}
\bibliography{references}

\appendix

\newpage

\section{Appendix}
\label{sec:appendix}

The appendix provides additional analyses that support the characterization of the dataset and evaluation of baseline metrics presented in the paper.
Three complementary aspects of the dataset are examined across the SKY130, IHP130, and ASAP7 (NG45 provided in paper) technology nodes.
First, the sensitivity of quality-of-results (QoR) metrics to variations in physical design parameters is analyzed in Section~\ref{sec:appendix_param_sensitivity}.
Second, the evolution of QoR metrics across stages of the physical design flow are evaluated in Section~\ref{sec:appendix_inter_stage_analysis}.
Third, correlations between intermediate-stage estimates and final detailed-routing results are analyzed in Section~\ref{sec:appendix_baseline_analysis}.

\subsection{Parameter Sensitivity Analysis}
\label{sec:appendix_param_sensitivity}

\begin{figure}[!h]
\centering
\includegraphics[width=0.975\columnwidth]{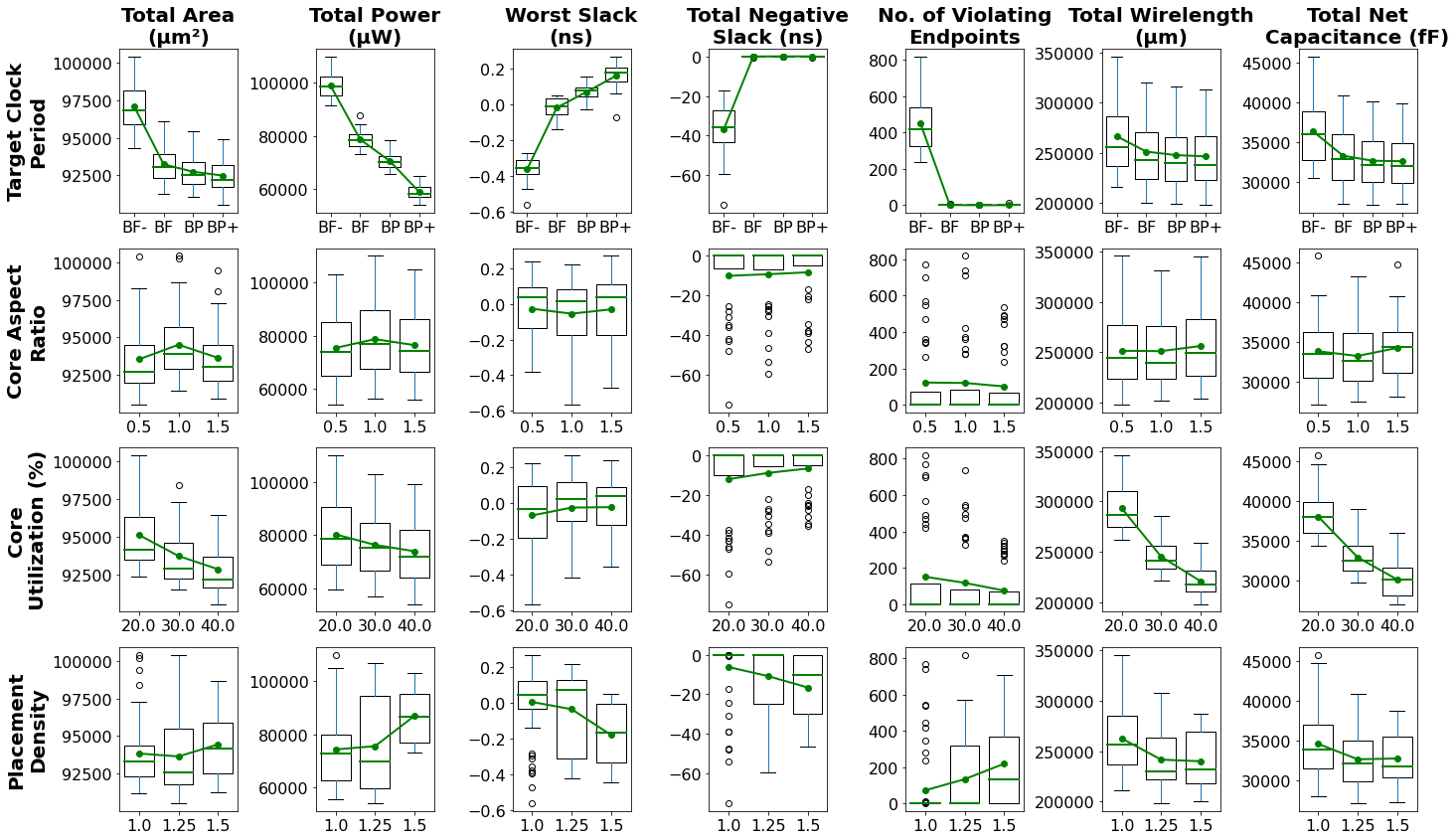}
\caption{Distribution of final-stage QoR metrics for total area, total power, worst-case slack, total negative slack, number of slack violating endpoints, total wirelength, and total net capacitance for the \textit{ac97\_ctrl} circuit with variations in clock period, core aspect ratio, core utilization, and placement density using the SKY130 technology node.}
\label{fig:sky130_parameter_sensitivity_analysis}
\vspace{-0.1in}
\end{figure}

\begin{figure}[!h]
\centering
\includegraphics[width=0.975\columnwidth]{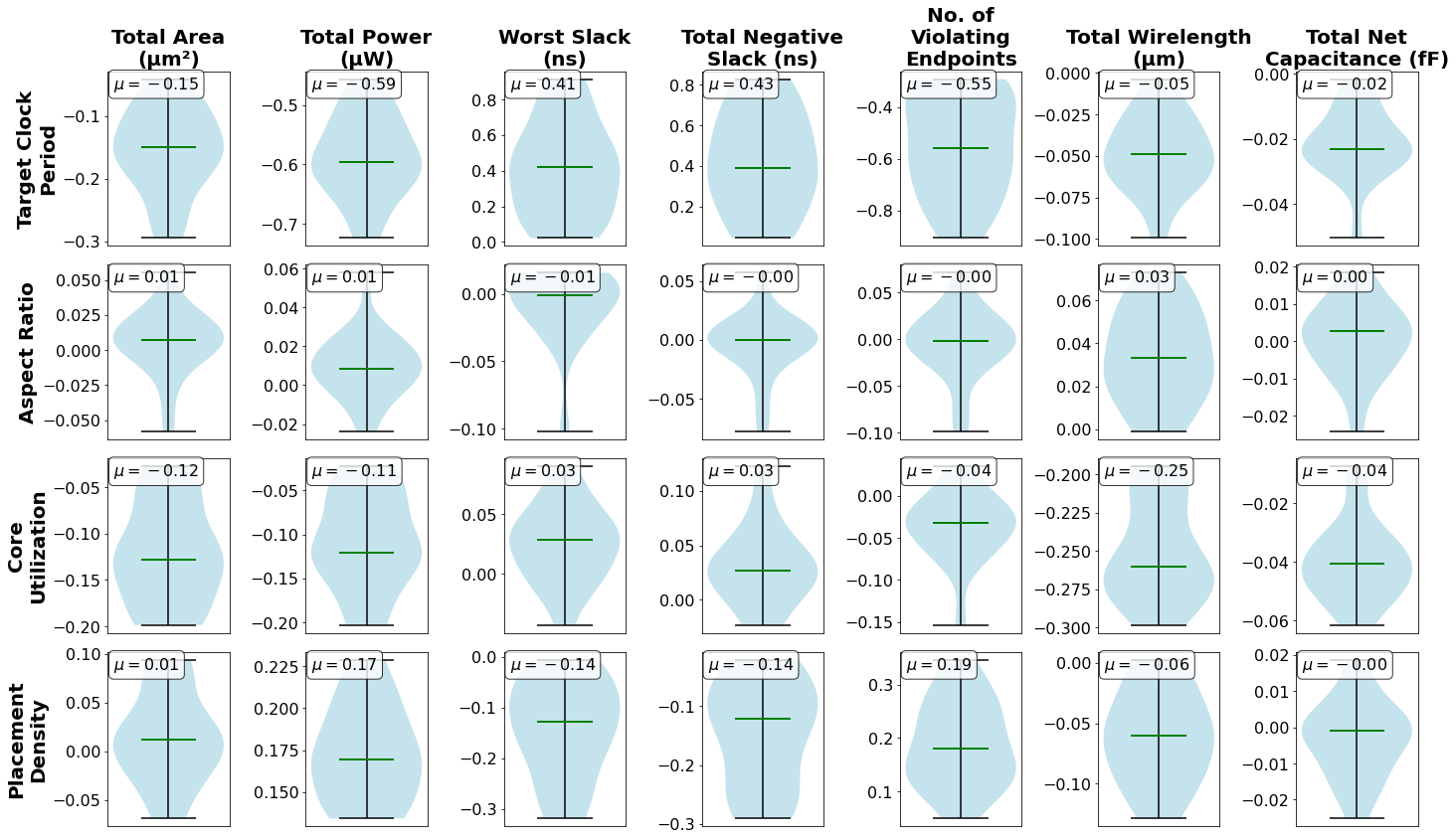}
\caption{Distribution of Pearson correlation coefficients between QoR metrics and design parameters across all circuits in the SKY130 dataset. The correlation coefficient are shown with violin plots for each parameter-metric pair.}
\label{fig:sky130_parameter_correlation_violins}
\end{figure}

\begin{figure}[!h]
\centering
\includegraphics[width=0.975\columnwidth]{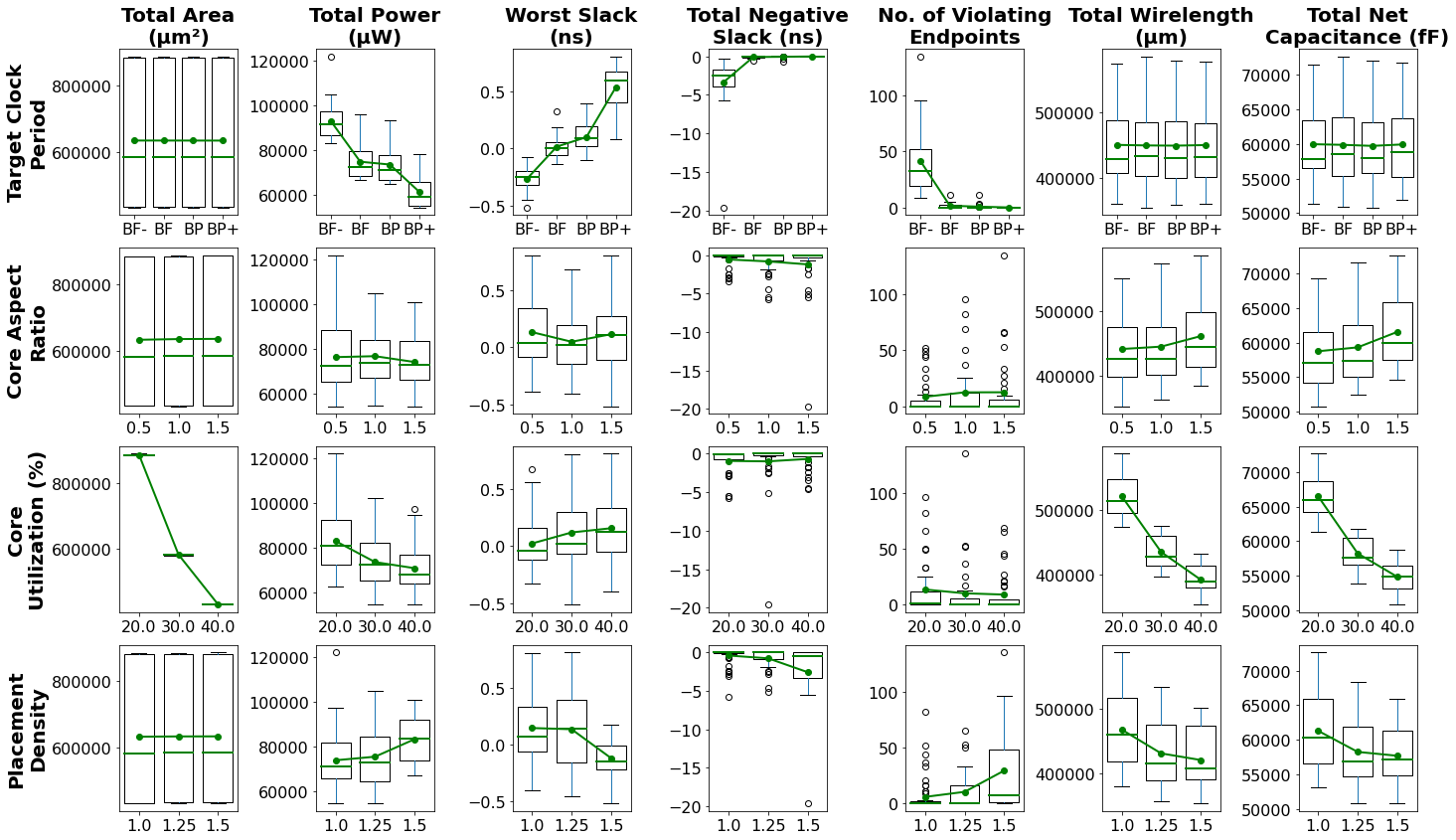}
\caption{Distribution of final-stage QoR metrics for total area, total power, worst-case slack, total negative slack, number of slack violating endpoints, total wirelength, and total net capacitance for the \textit{ac97\_ctrl} circuit with variations in clock period, core aspect ratio, core utilization, and placement density using the IHP130 technology node.}
\label{fig:ihp130_parameter_sensitivity_analysis}
\end{figure}

\begin{figure}[!h]
\centering
\includegraphics[width=0.975\columnwidth]{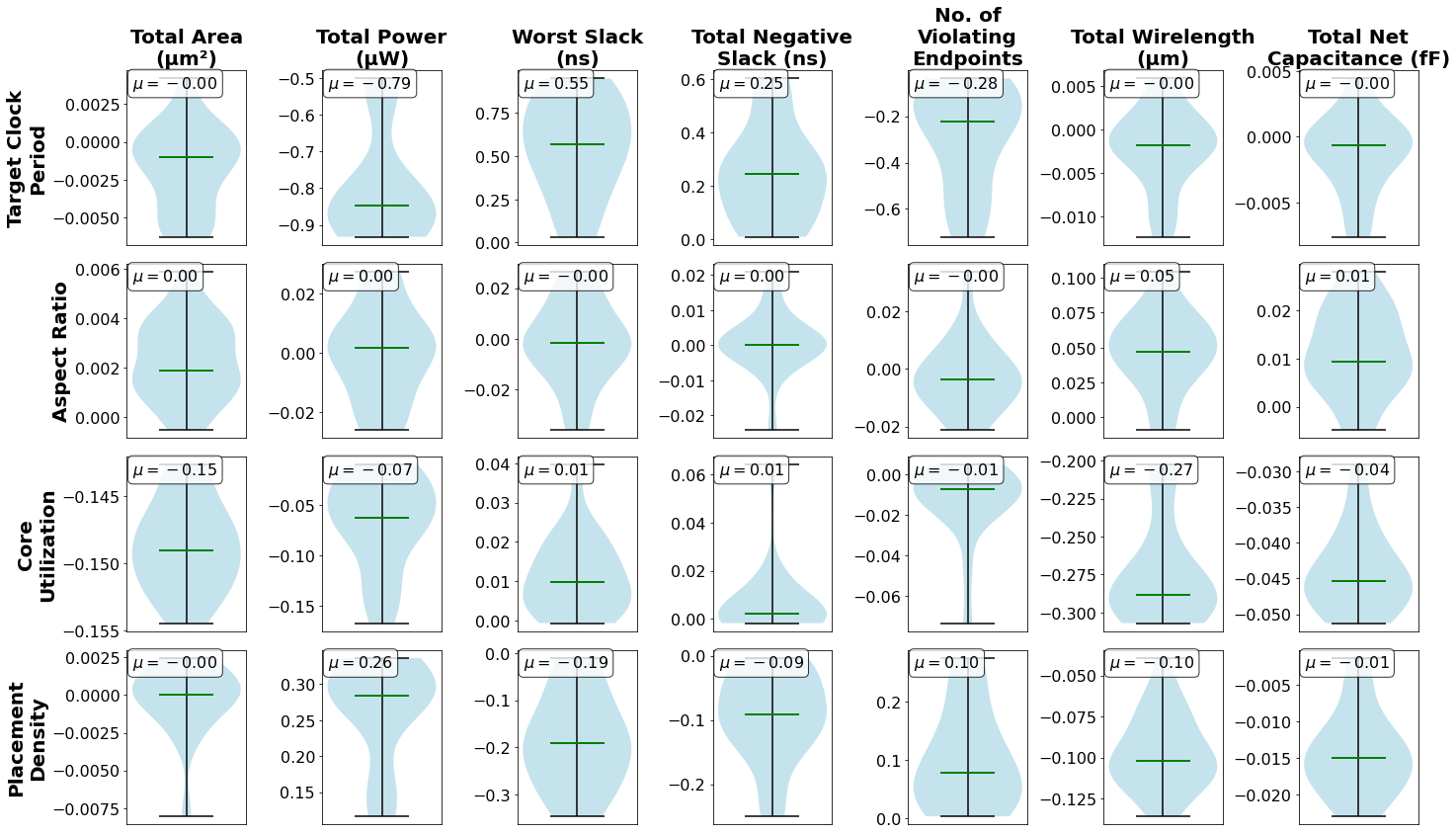}
\caption{Distribution of Pearson correlation coefficients between QoR metrics and design parameters across all circuits in the IHP130 dataset. The correlation coefficient are shown with violin plots for each parameter-metric pair.}
\label{fig:ihp130_parameter_correlation_violins}
\end{figure}

\begin{figure}[!h]
\centering
\includegraphics[width=0.975\columnwidth]{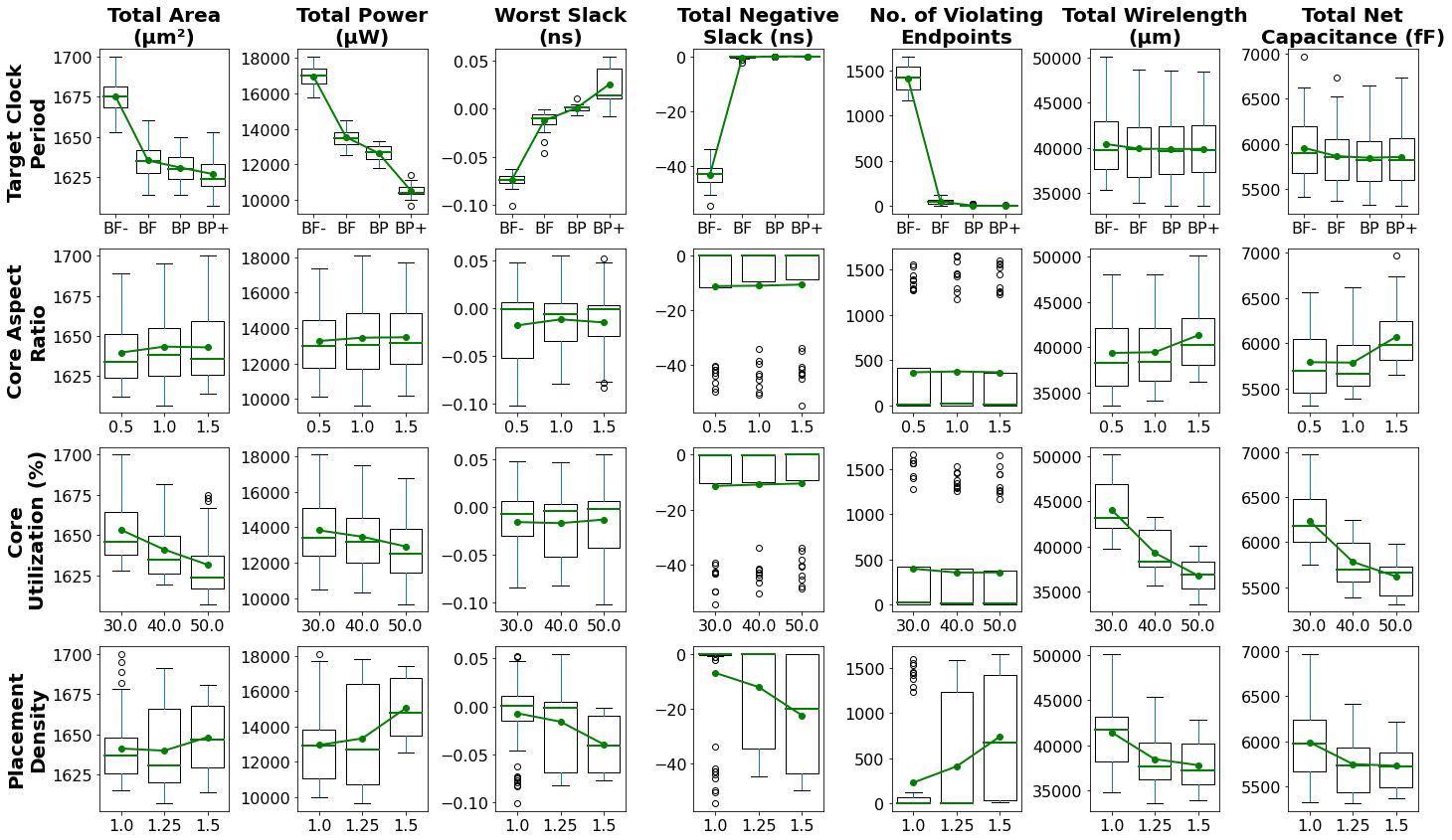}
\caption{Distribution of final-stage QoR metrics for total area, total power, worst-case slack, total negative slack, number of slack violating endpoints, total wirelength, and total net capacitance for the \textit{ac97\_ctrl} circuit with variations in clock period, core aspect ratio, core utilization, and placement density using the ASAP7 technology node.}
\label{fig:asap7_parameter_sensitivity_analysis}
\vspace{-0.1in}
\end{figure}

\begin{figure}[!h]
\centering
\includegraphics[width=0.975\columnwidth]{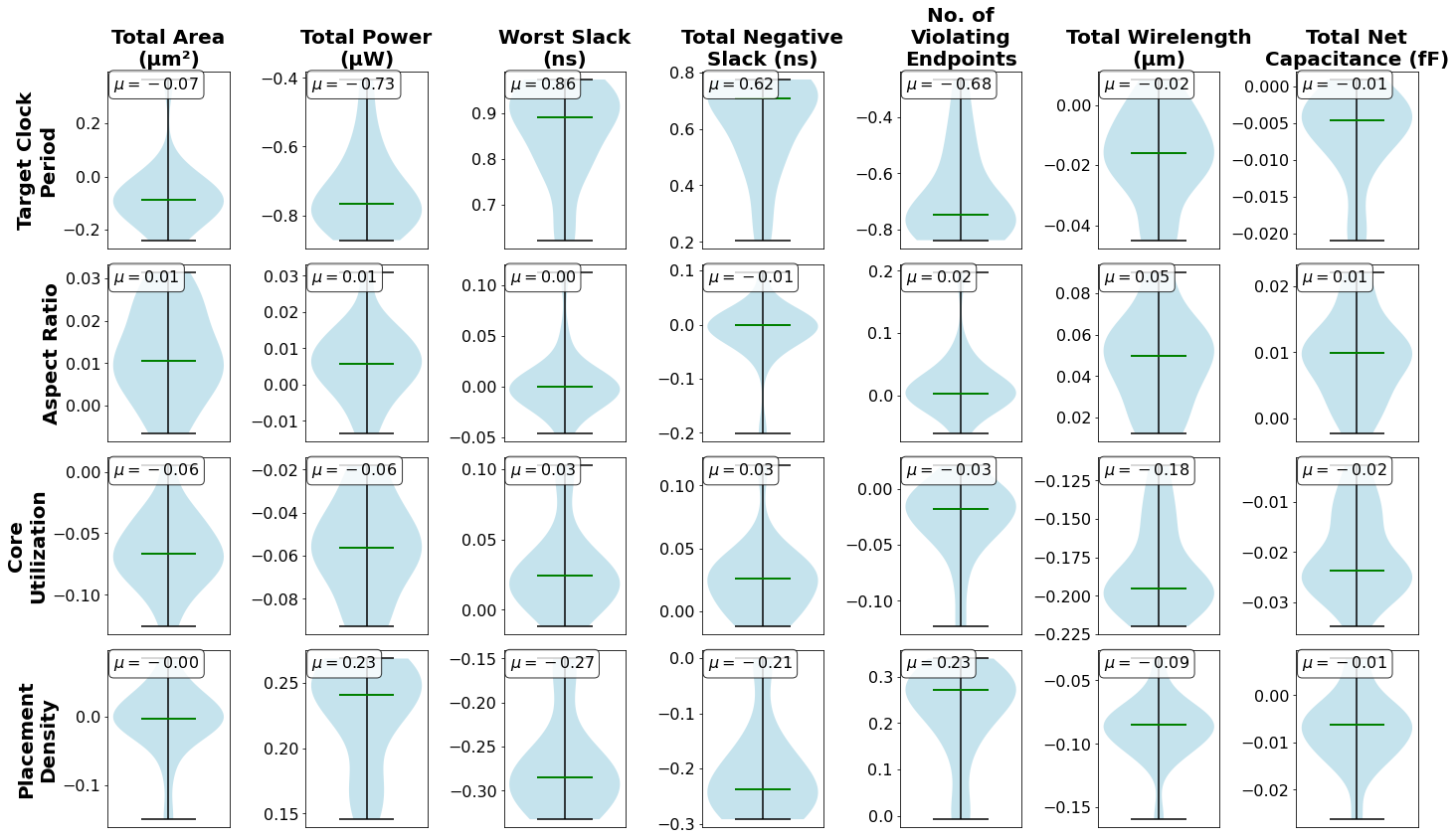}
\caption{Distribution of Pearson correlation coefficients between QoR metrics and design parameters across all circuits in the ASAP7 dataset. The correlation coefficient are shown with violin plots for each parameter-metric pair.}
\label{fig:asap7_parameter_correlation_violins}
\end{figure}

The results of the parameter sensitivity analysis for the SKY130, IHP130, and ASAP7 technology nodes are provided in Figures~\ref{fig:sky130_parameter_sensitivity_analysis} and \ref{fig:sky130_parameter_correlation_violins}, Figures~\ref{fig:ihp130_parameter_sensitivity_analysis} and \ref{fig:ihp130_parameter_correlation_violins}, and Figures~\ref{fig:asap7_parameter_sensitivity_analysis} and \ref{fig:asap7_parameter_correlation_violins}, respectively.
% The results from analysis of parameter sensitivity for the SKY130 technology node are provided in Figures~\ref{fig:sky130_parameter_sensitivity_analysis} and \ref{fig:sky130_parameter_correlation_violins}, while the corresponding analysis for the IHP130 technology is provided through Figures~\ref{fig:ihp130_parameter_sensitivity_analysis} and \ref{fig:ihp130_parameter_correlation_violins}.
% Similarly, the results from the analysis of parameter sensitivity for the ASAP7 technology node are shown in  Figures~\ref{fig:asap7_parameter_sensitivity_analysis} and 
% \ref{fig:asap7_parameter_correlation_violins}.
The statistical distribution of the final-stage QoR metrics due to sensitivity to design parameters is analyzed for total area, total power, worst-case slack, total negative slack, number of slack violating endpoints, total wirelength, and total net capacitance for the \textit{ac97\_ctrl} circuit with variations in clock period, core aspect ratio, core utilization, and placement density.
Plots of the Pearson correlation coefficients between QoR metrics and design parameters across all benchmark circuits of each dataset are provided for SKY130, IHP130, and ASAP7 technology nodes in Figures~\ref{fig:sky130_parameter_correlation_violins}, \ref{fig:ihp130_parameter_correlation_violins}, and \ref{fig:asap7_parameter_correlation_violins}, respectively.

~\newpage
\subsection{Inter-Design Stage Analysis}
\label{sec:appendix_inter_stage_analysis}
\begin{figure}[!h]
    \centering
    \begin{subfigure}{0.32\columnwidth}
        \includegraphics[width=0.9\linewidth]{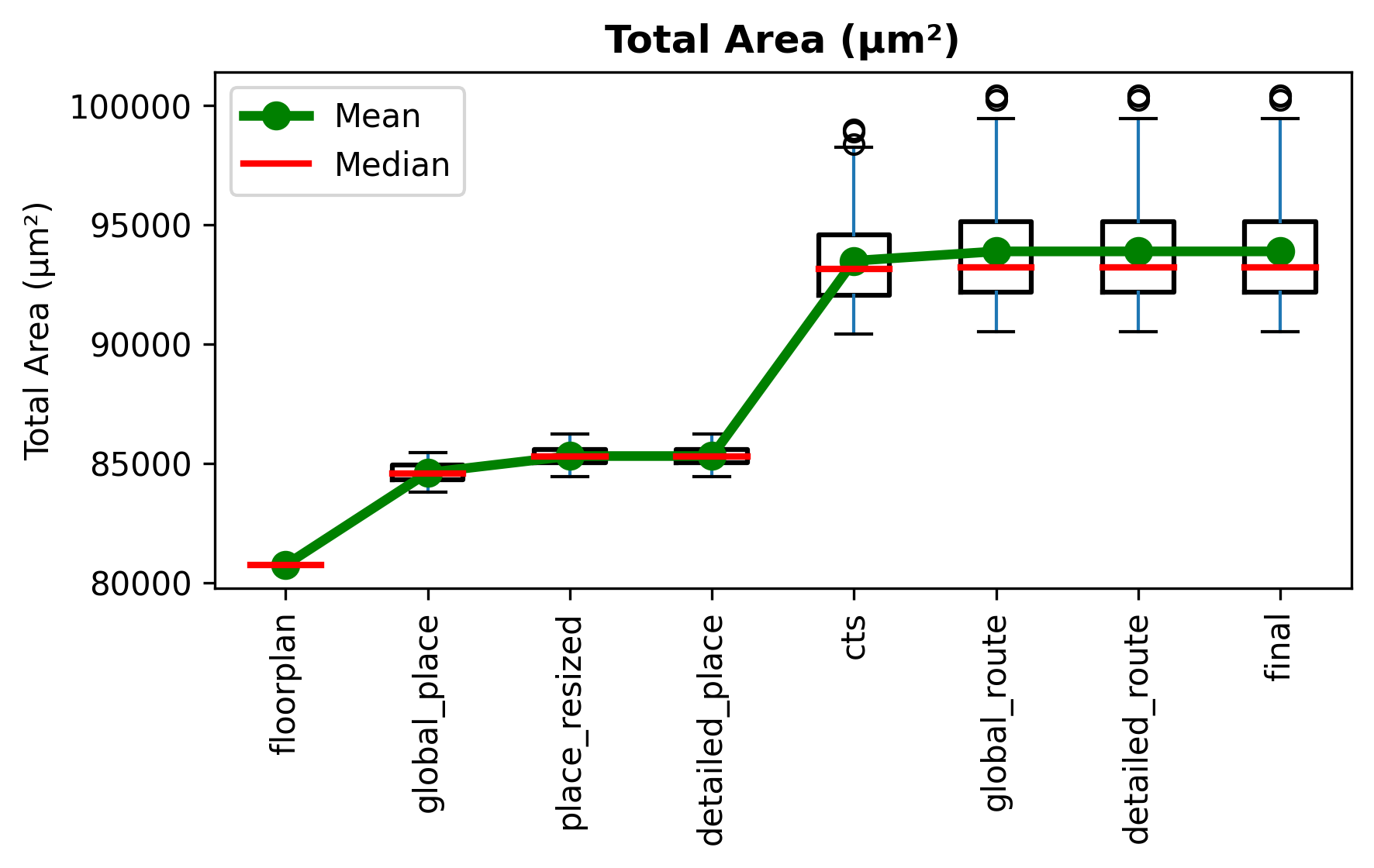}
        \caption{}
    \end{subfigure}
    \begin{subfigure}{0.32\columnwidth}
        \includegraphics[width=0.9\linewidth]{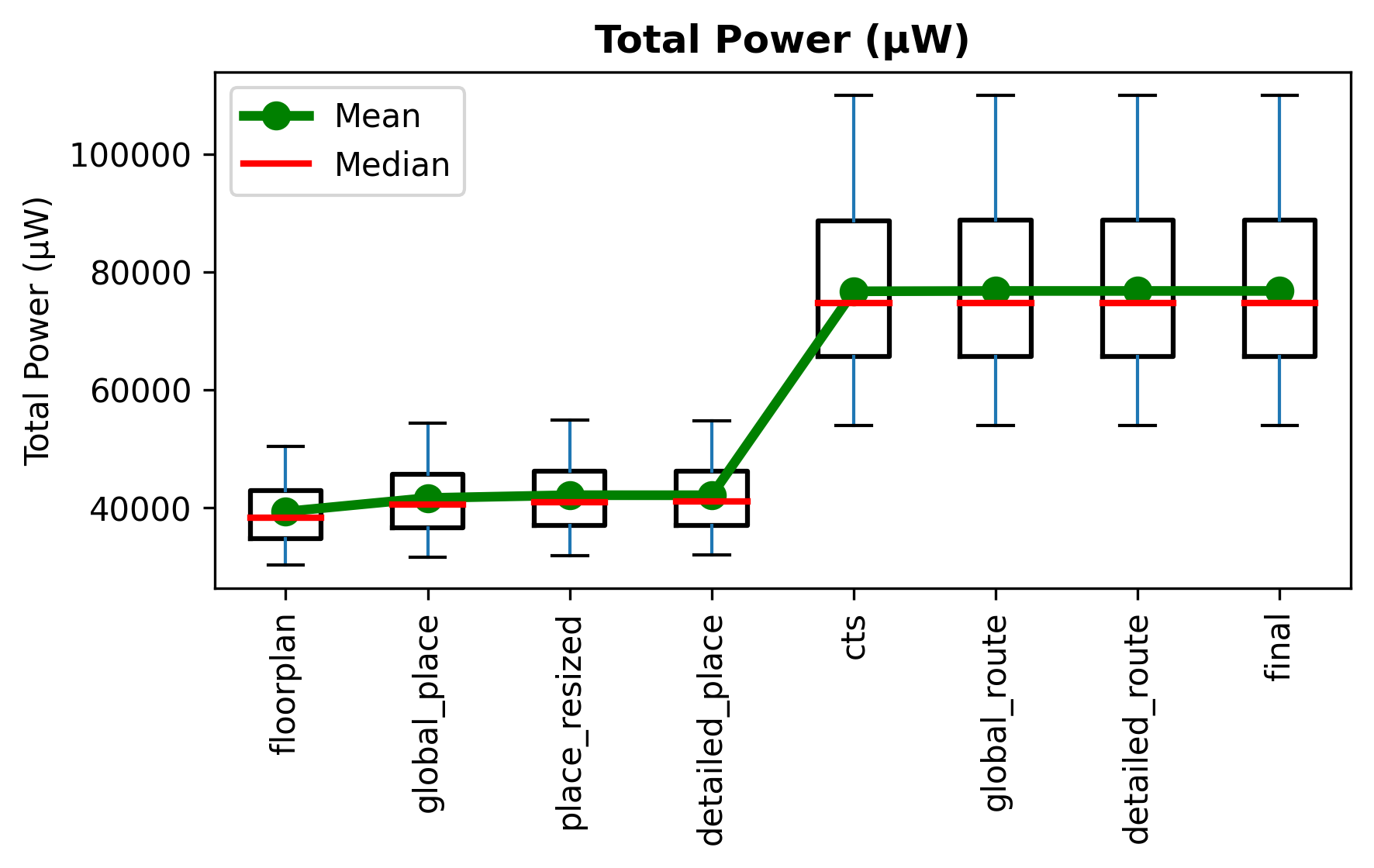}
        \caption{}
    \end{subfigure}
    \begin{subfigure}{0.32\columnwidth}
        \includegraphics[width=0.9\linewidth]{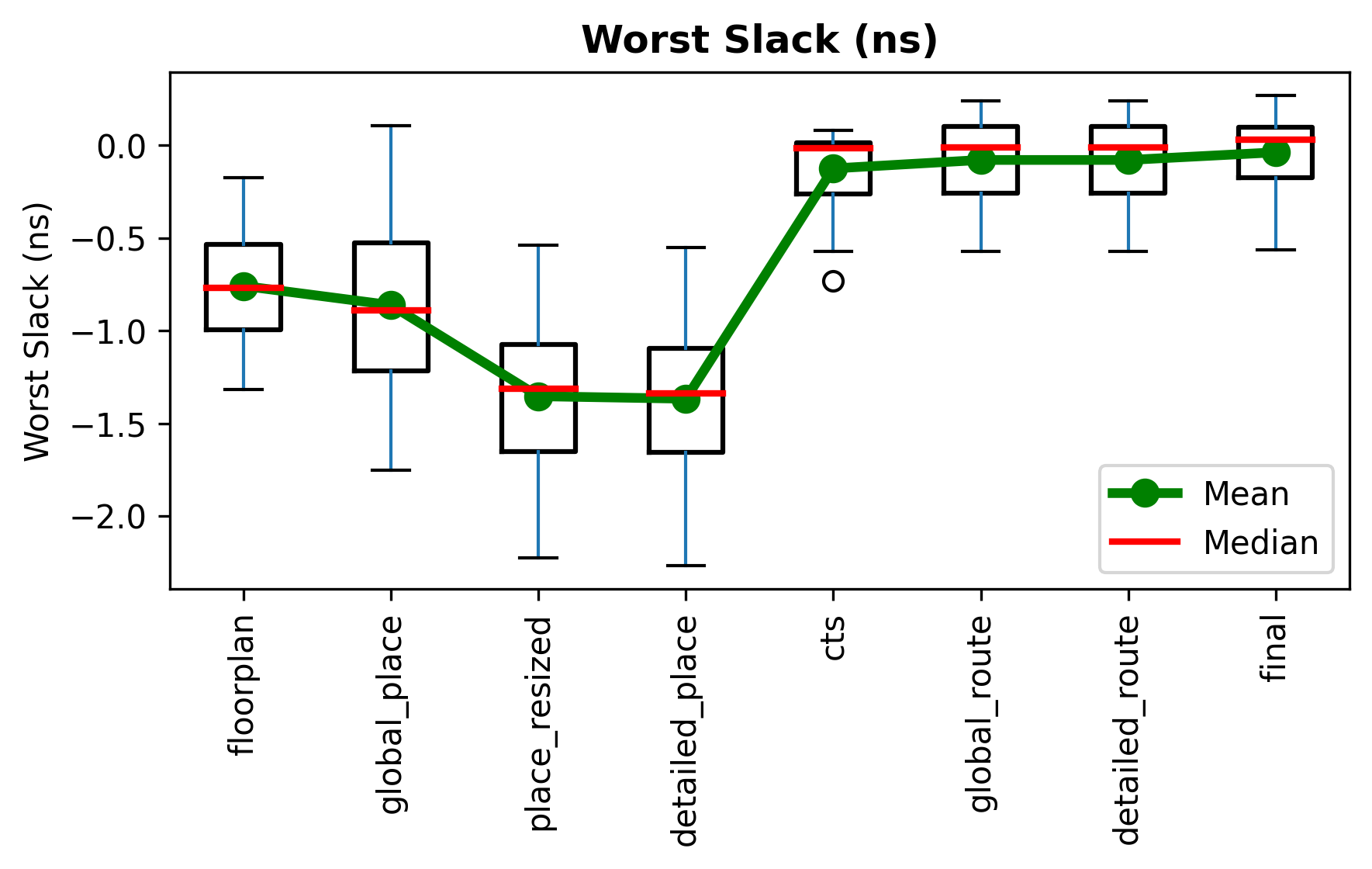}
        \caption{}
    \end{subfigure}

    \begin{subfigure}{0.32\columnwidth}
        \includegraphics[width=0.9\linewidth]{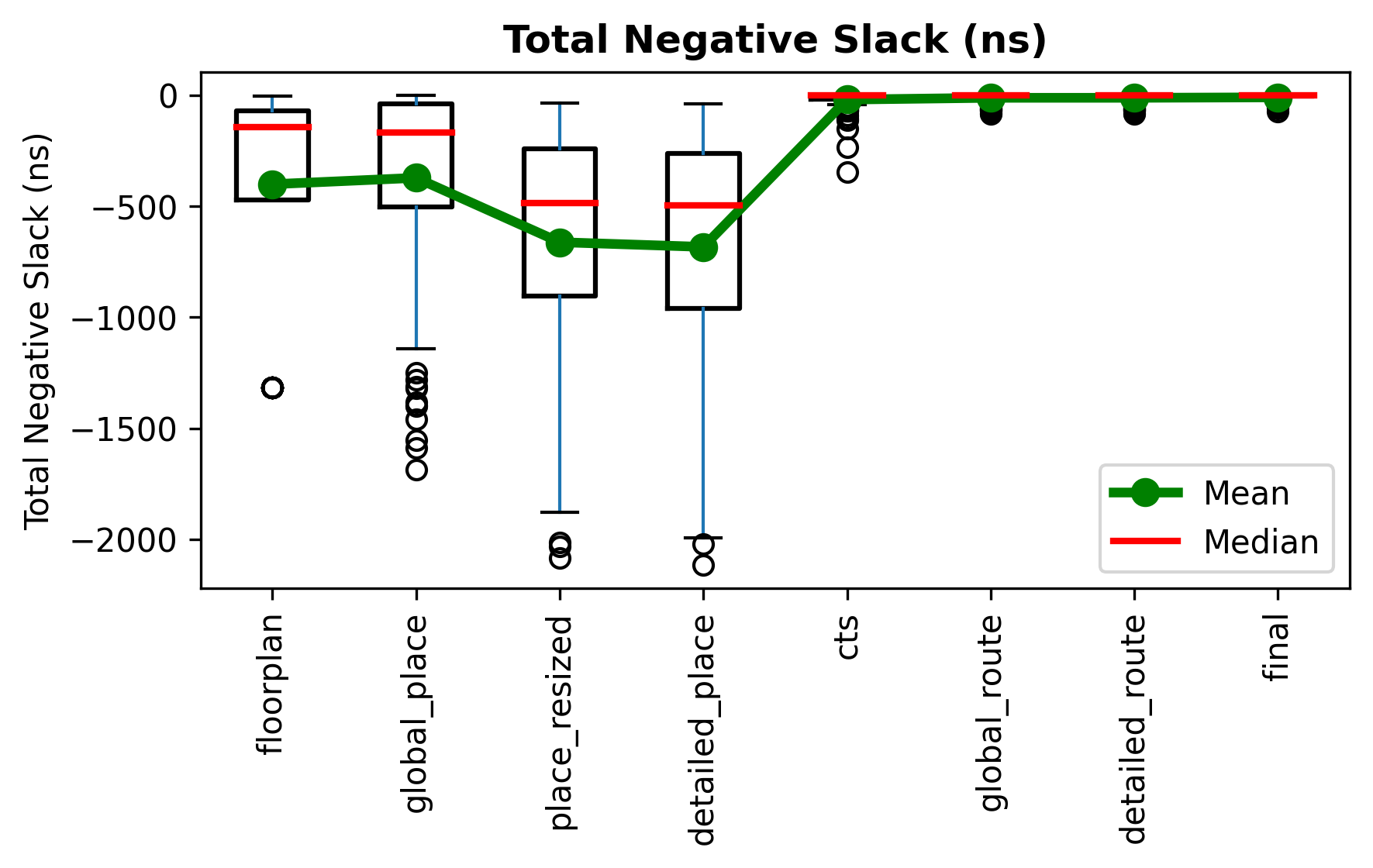}
        \caption{}
    \end{subfigure}
    \begin{subfigure}{0.32\columnwidth}
        \includegraphics[width=0.9\linewidth]{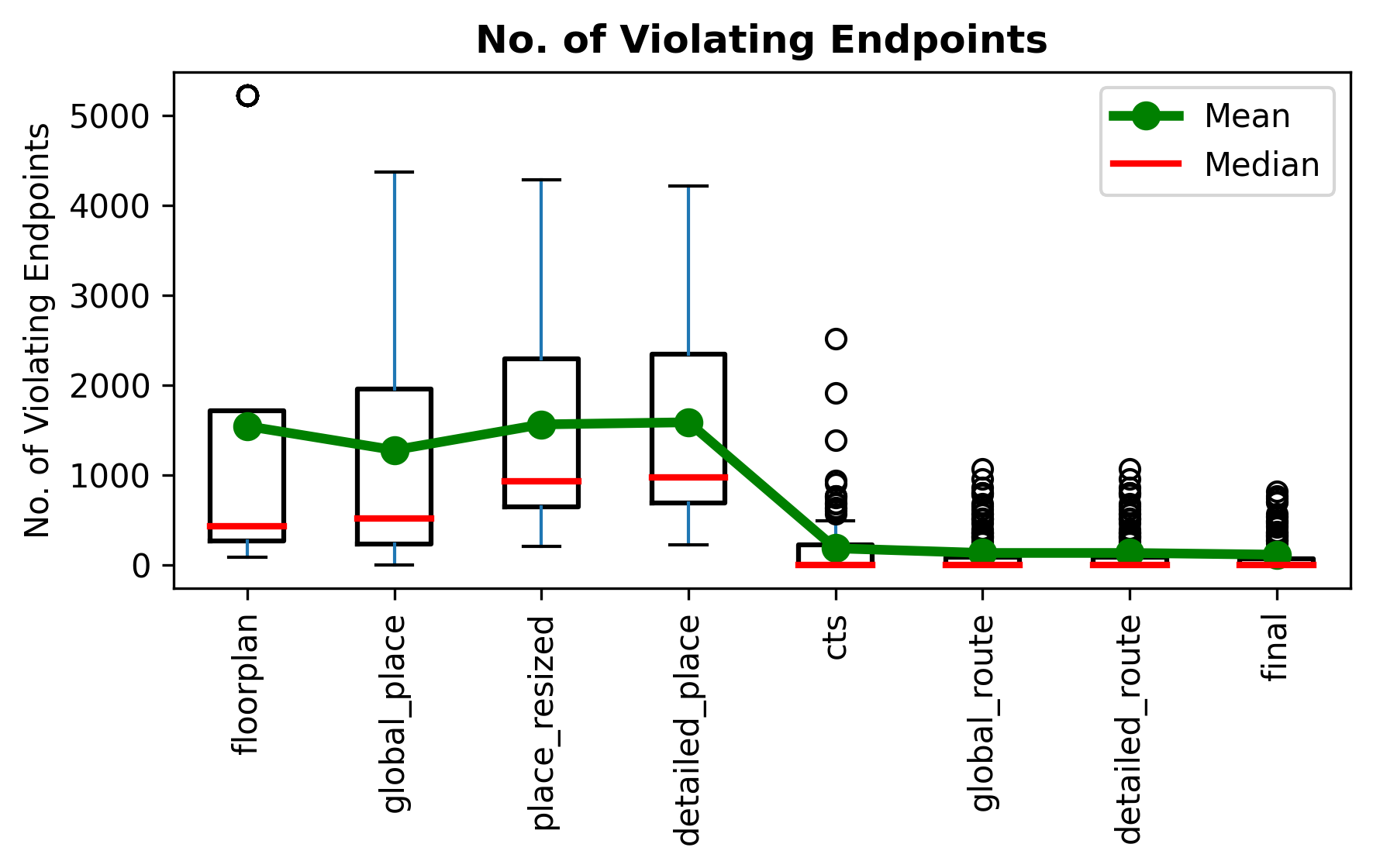}
        \caption{}
    \end{subfigure}
    \begin{subfigure}{0.32\columnwidth}
        \includegraphics[width=0.9\linewidth]{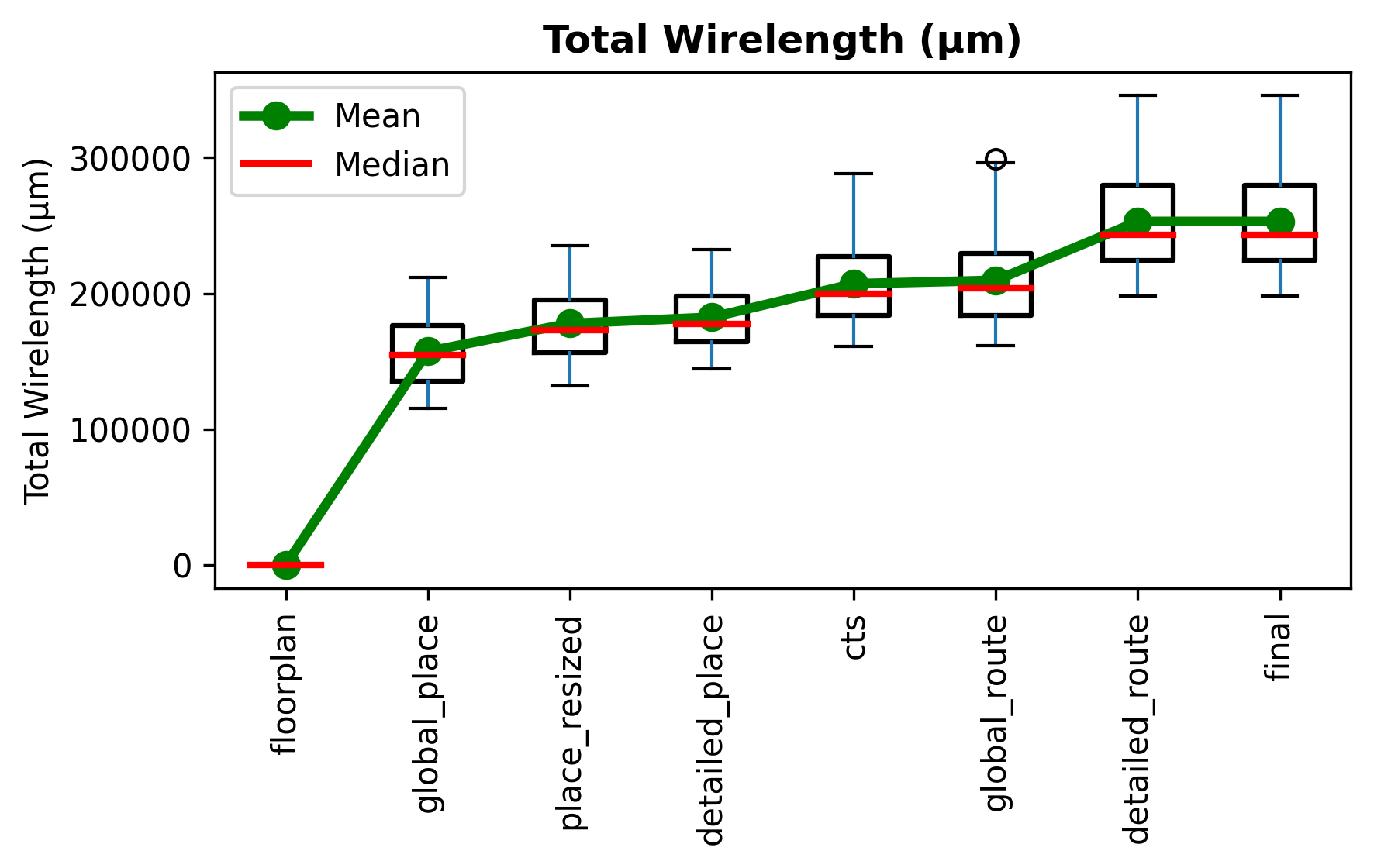}
        \caption{}
    \end{subfigure}
    \vspace{-0.1in}

    \caption{Change in the distribution of (a) total area, (b) total power, (c) worst slack, (d) total negative slack, (e) number of violating endpoints, and (f) total wirelength across physical design stages for the \textit{ac97\_ctrl} circuit implemented in the SKY130 technology node.}
    \label{fig:sky130_stage_analysis}
    \vspace{-0.2in}
\end{figure}

\begin{figure}[!h]
    \centering
    \begin{subfigure}{0.32\columnwidth}
        \includegraphics[width=0.9\linewidth]{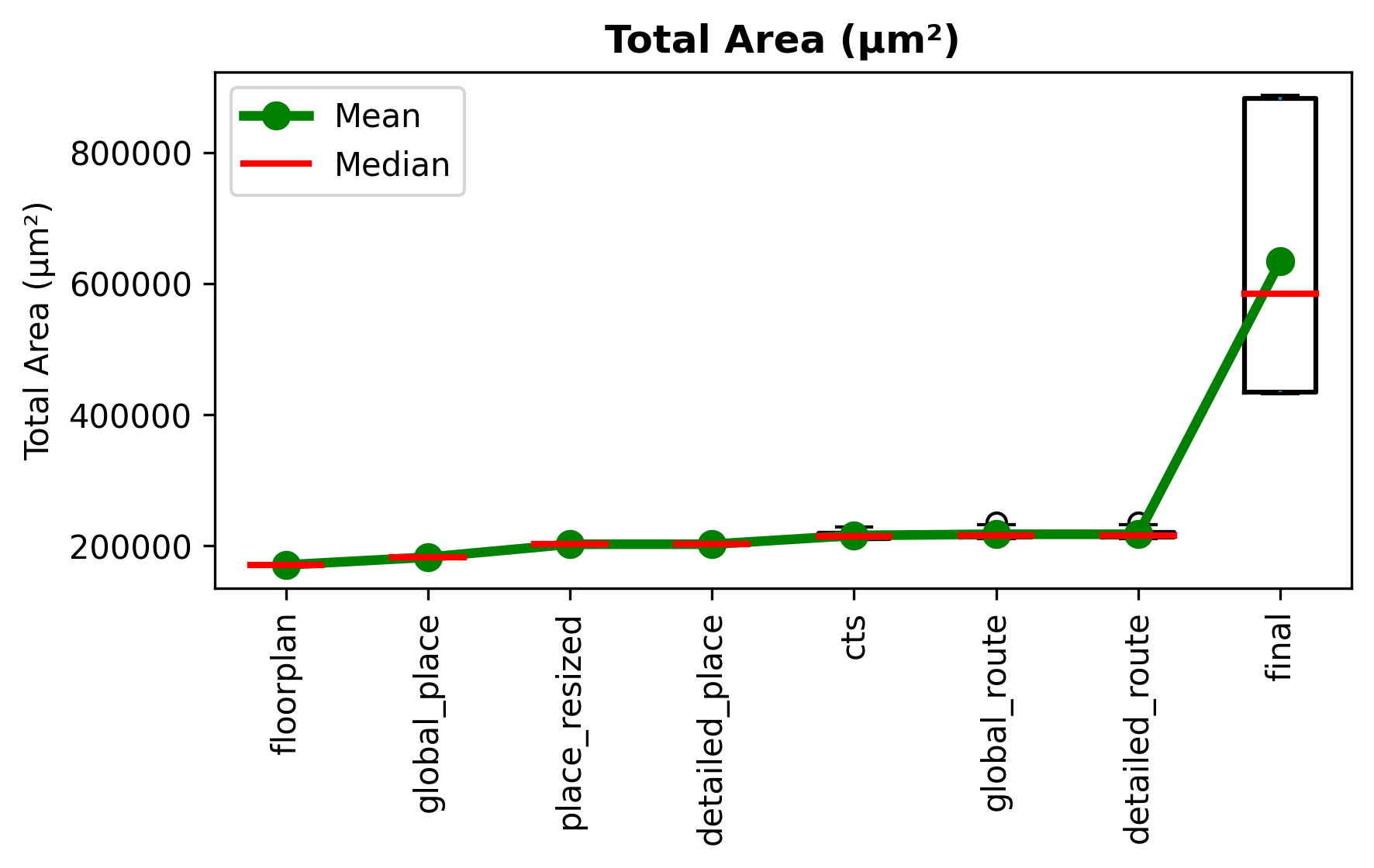}
        \caption{}
    \end{subfigure}
    \begin{subfigure}{0.32\columnwidth}
        \includegraphics[width=0.9\linewidth]{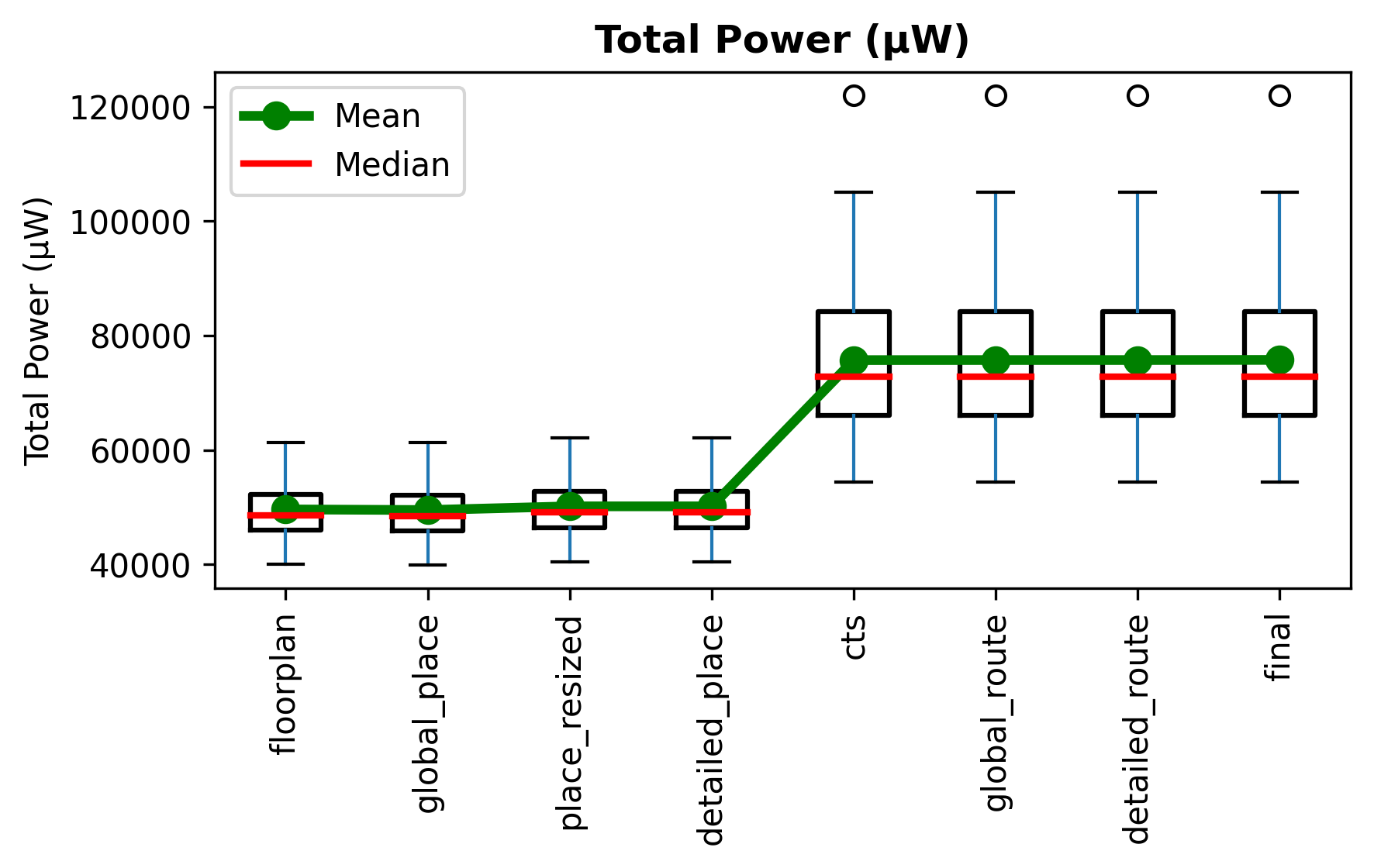}
        \caption{}
    \end{subfigure}
    \begin{subfigure}{0.32\columnwidth}
        \includegraphics[width=0.9\linewidth]{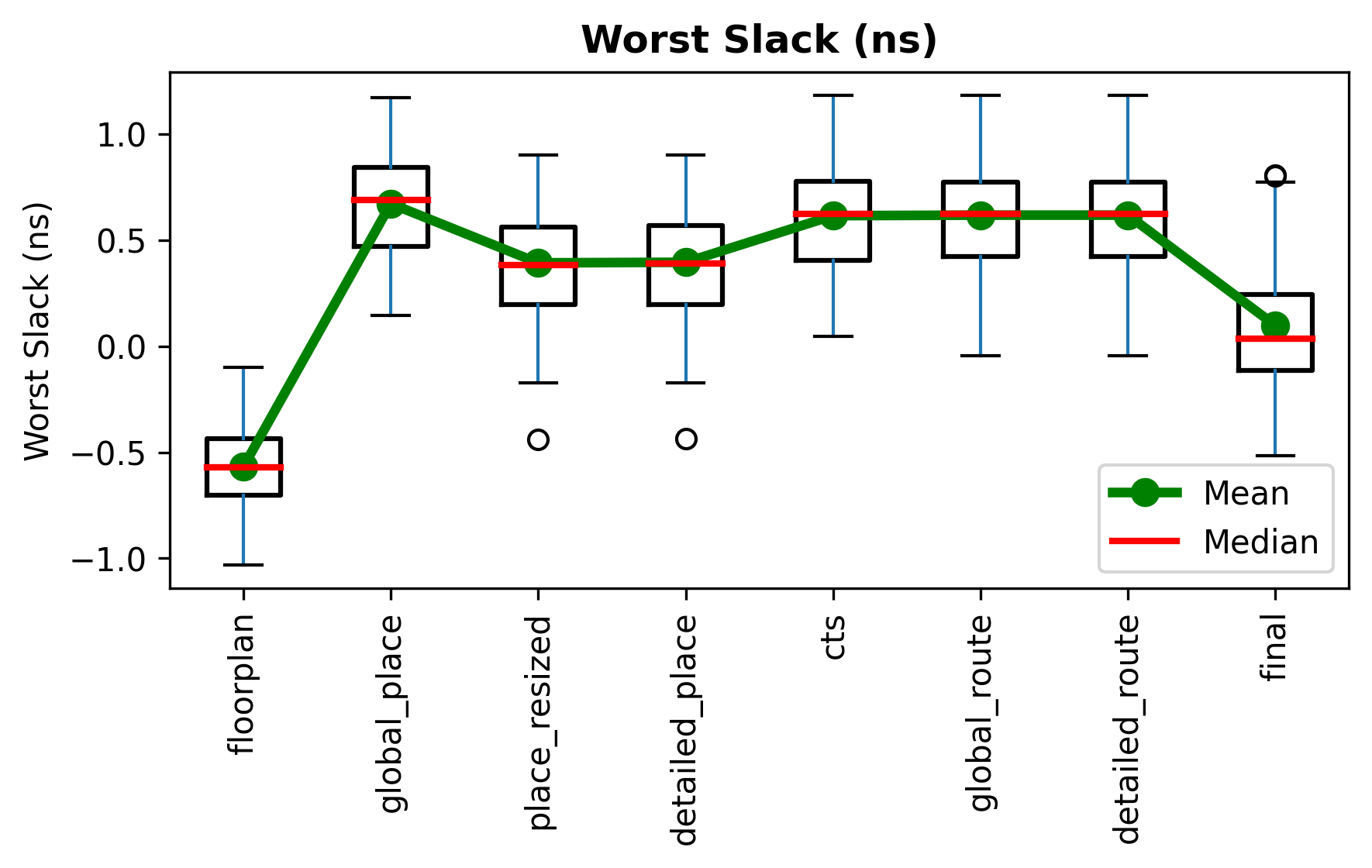}
        \caption{}
    \end{subfigure}

    \begin{subfigure}{0.32\columnwidth}
        \includegraphics[width=0.9\linewidth]{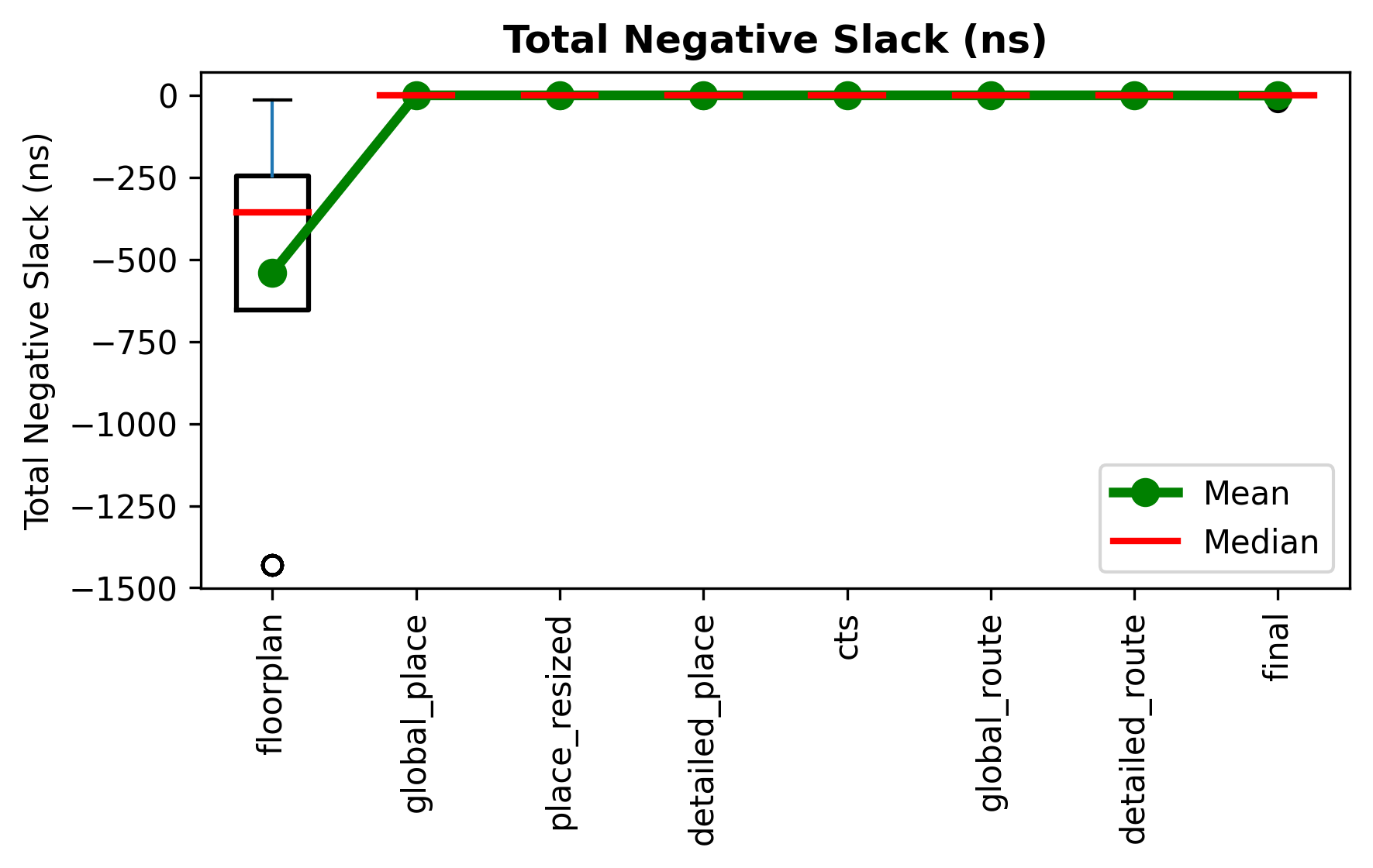}
        \caption{}
    \end{subfigure}
    \begin{subfigure}{0.32\columnwidth}
        \includegraphics[width=0.9\linewidth]{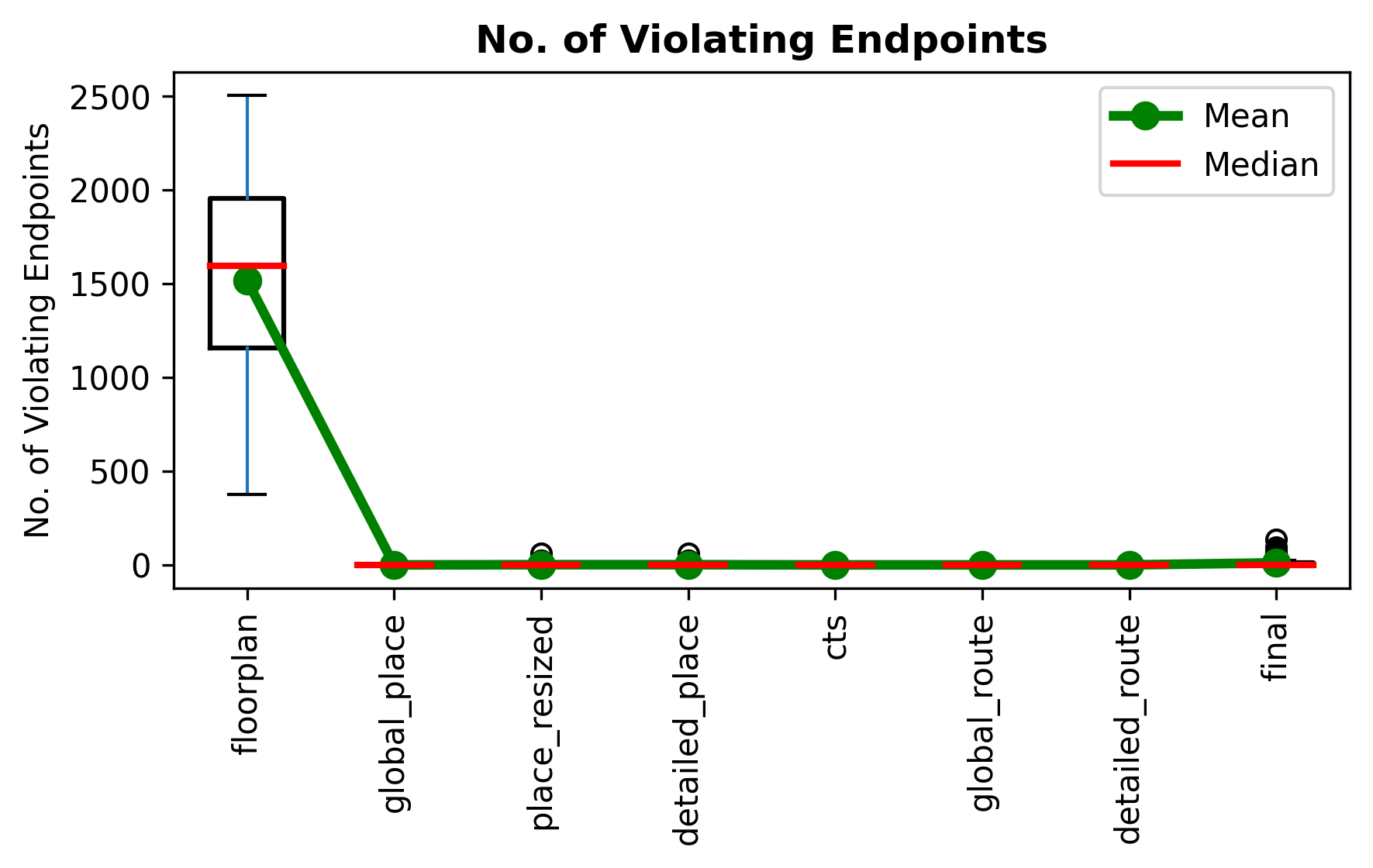}
        \caption{}
    \end{subfigure}
    \begin{subfigure}{0.32\columnwidth}
        \includegraphics[width=0.9\linewidth]{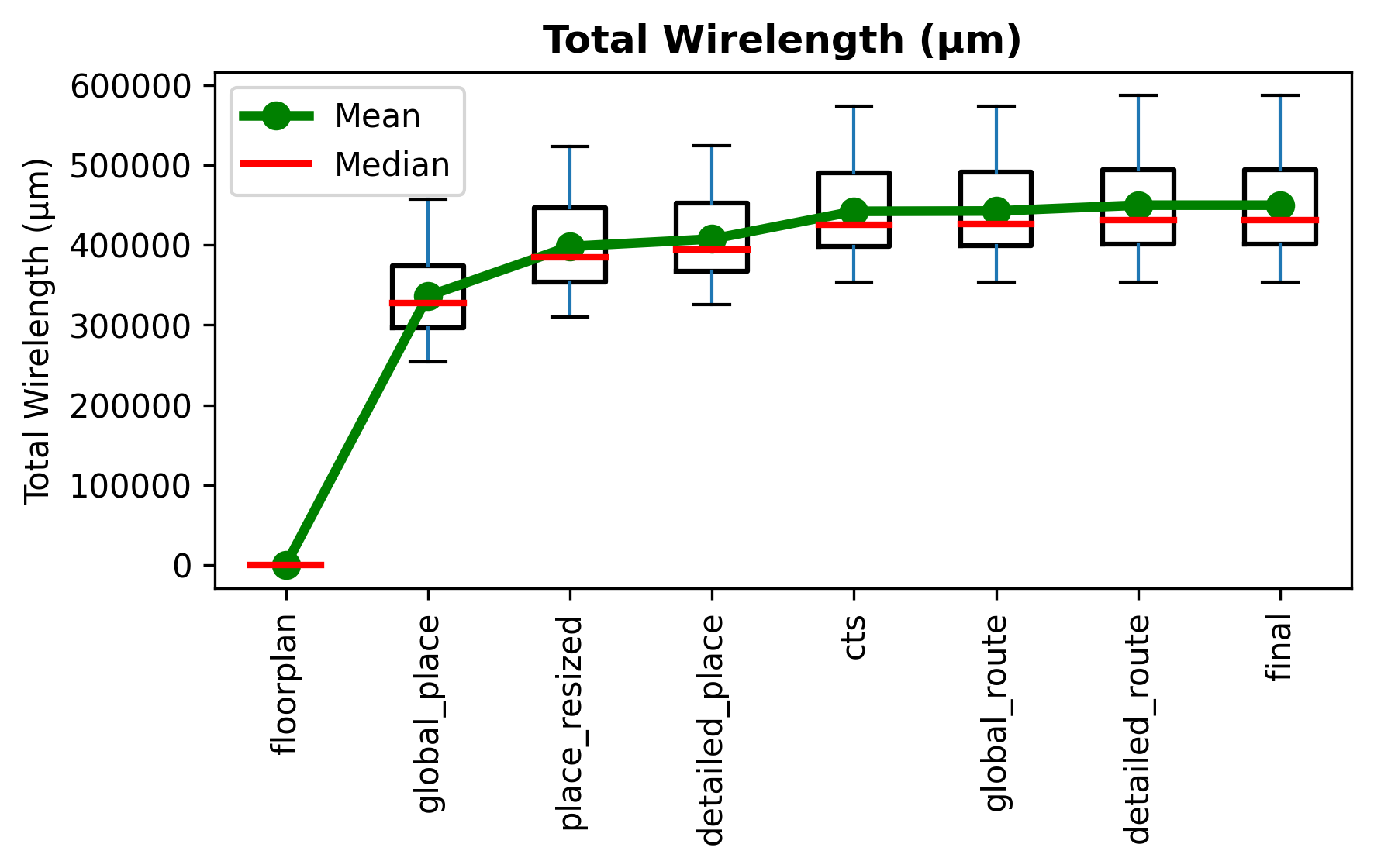}
        \caption{}
    \end{subfigure}

    \caption{Change in the distribution of (a) total area, (b) total power, (c) worst slack, (d) total negative slack, (e) number of violating endpoints, and (f) total wirelength across physical design stages for the \textit{ac97\_ctrl} circuit implemented in the IHP130 technology node.}
    \label{fig:ihp130_stage_analysis}
\end{figure}

\begin{figure}[!h]
    \centering
    \begin{subfigure}{0.32\columnwidth}
        \includegraphics[width=0.9\linewidth]{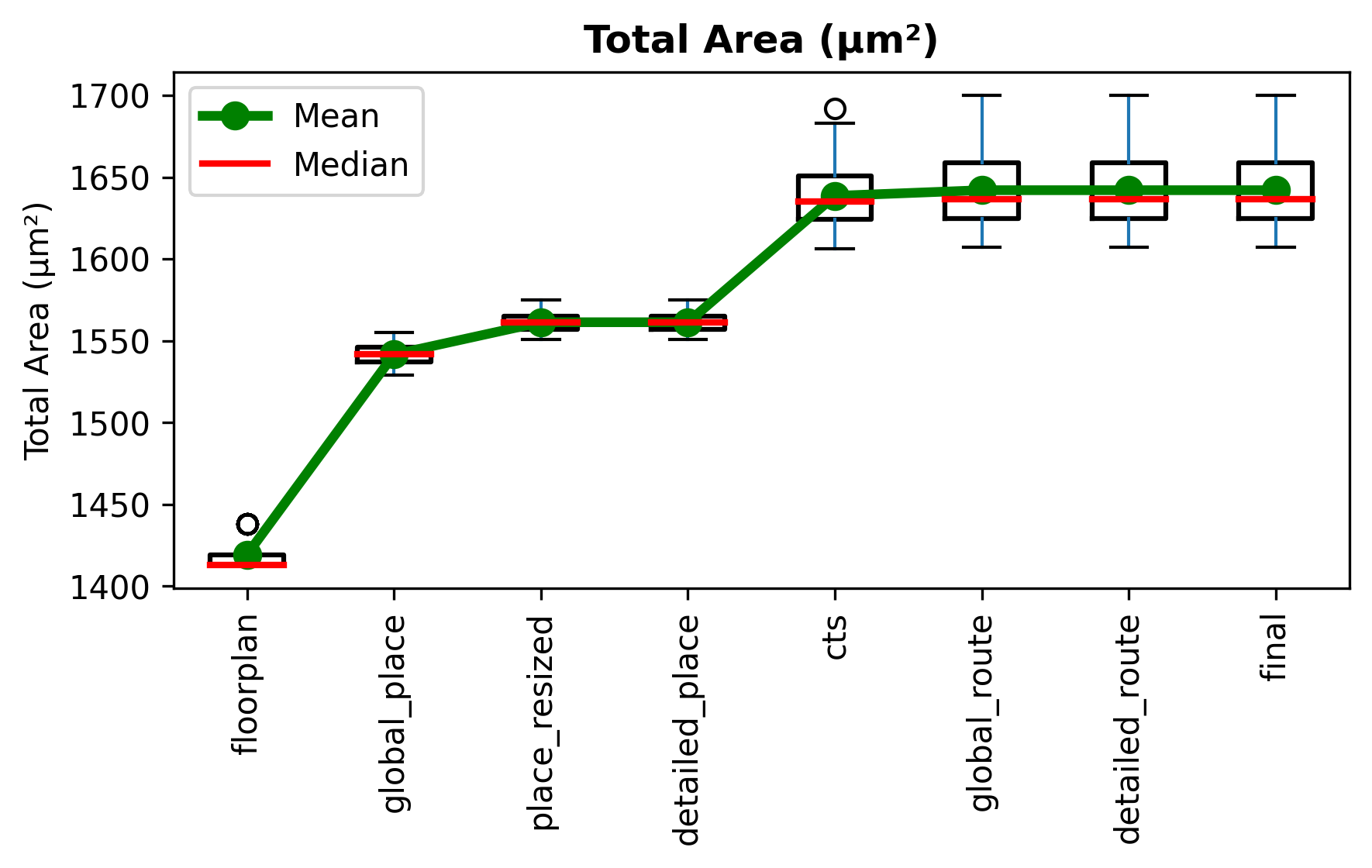}
        \caption{}
    \end{subfigure}
    \begin{subfigure}{0.32\columnwidth}
        \includegraphics[width=0.9\linewidth]{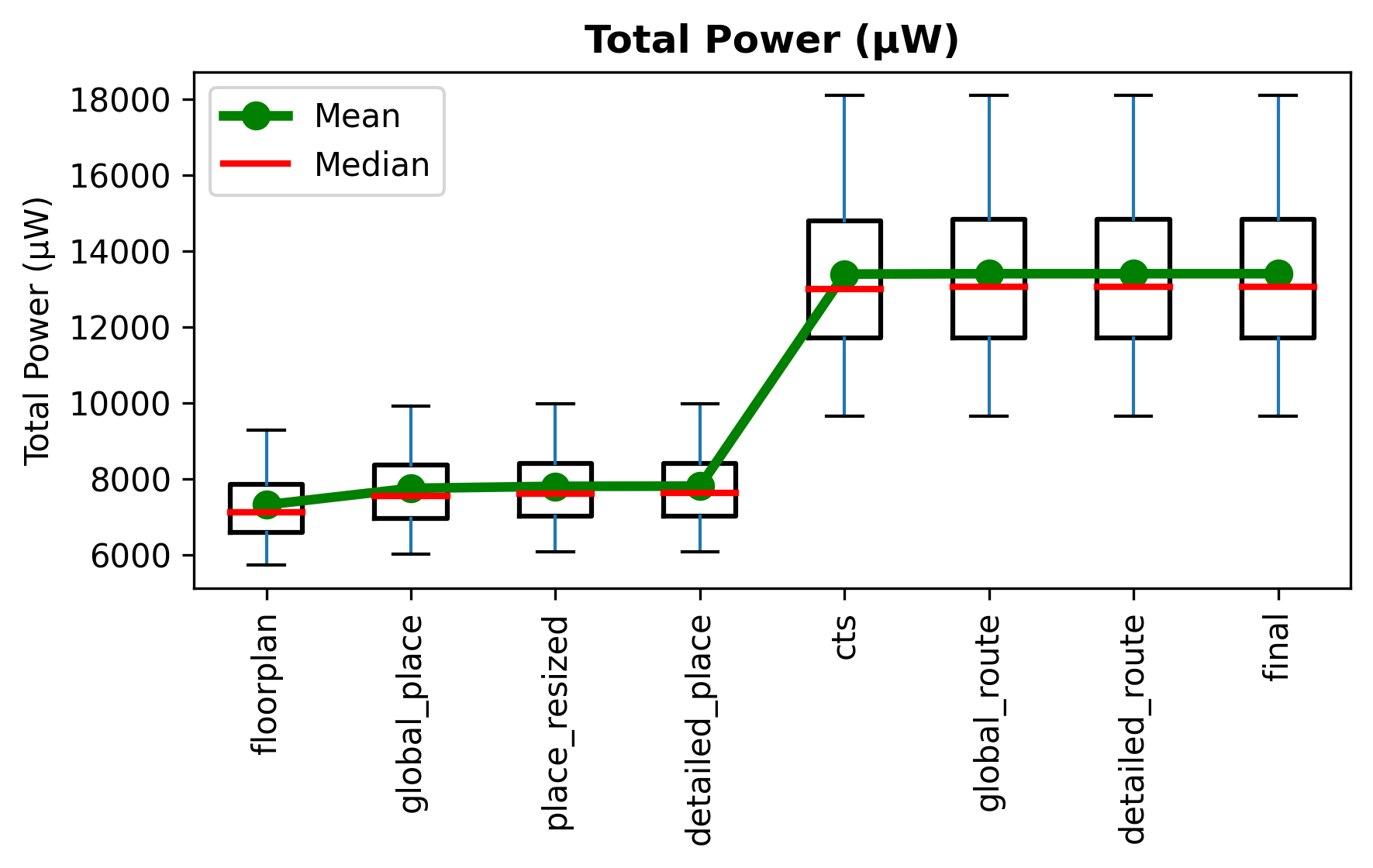}
        \caption{}
    \end{subfigure}
    \begin{subfigure}{0.32\columnwidth}
        \includegraphics[width=0.9\linewidth]{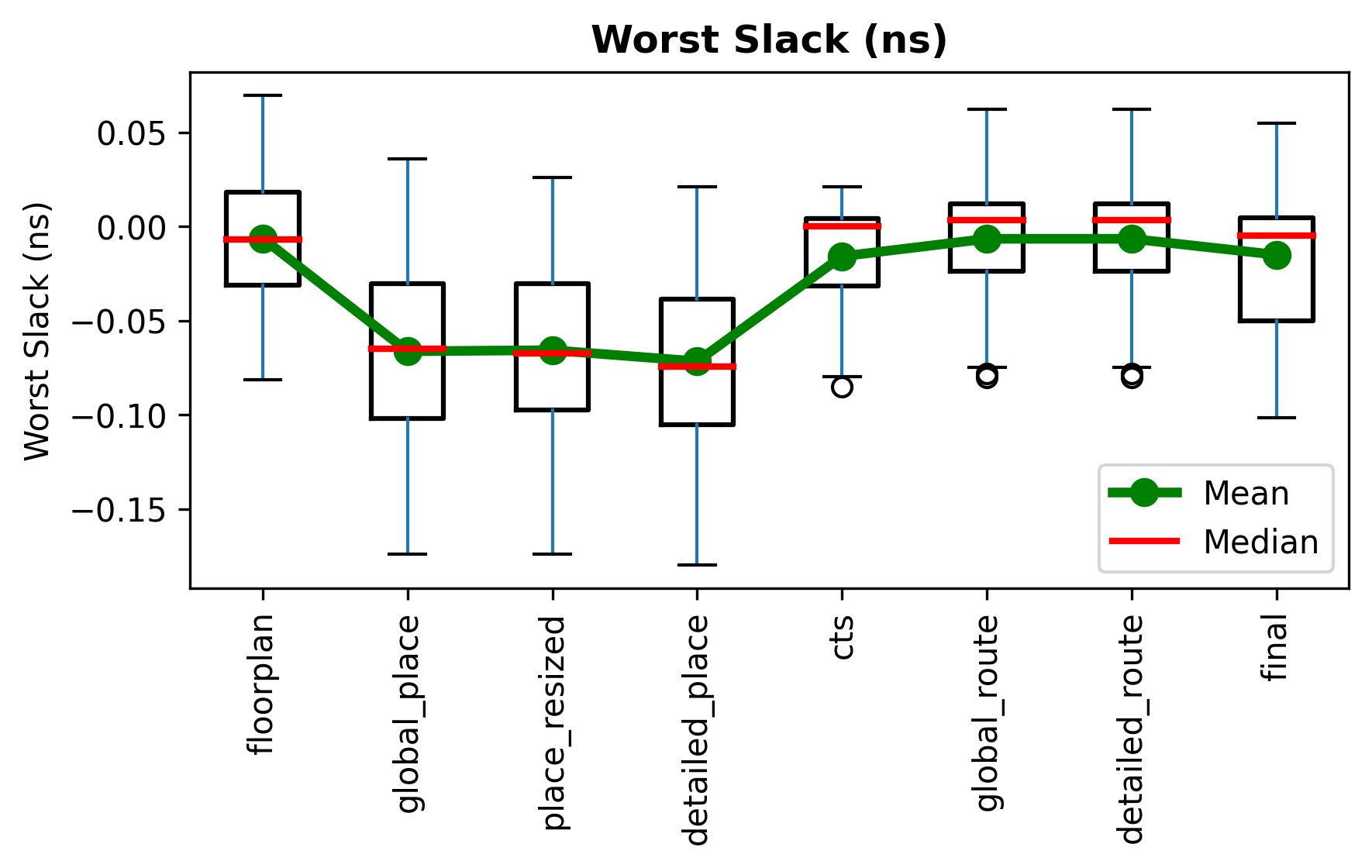}
        \caption{}
    \end{subfigure}

    \begin{subfigure}{0.32\columnwidth}
        \includegraphics[width=0.9\linewidth]{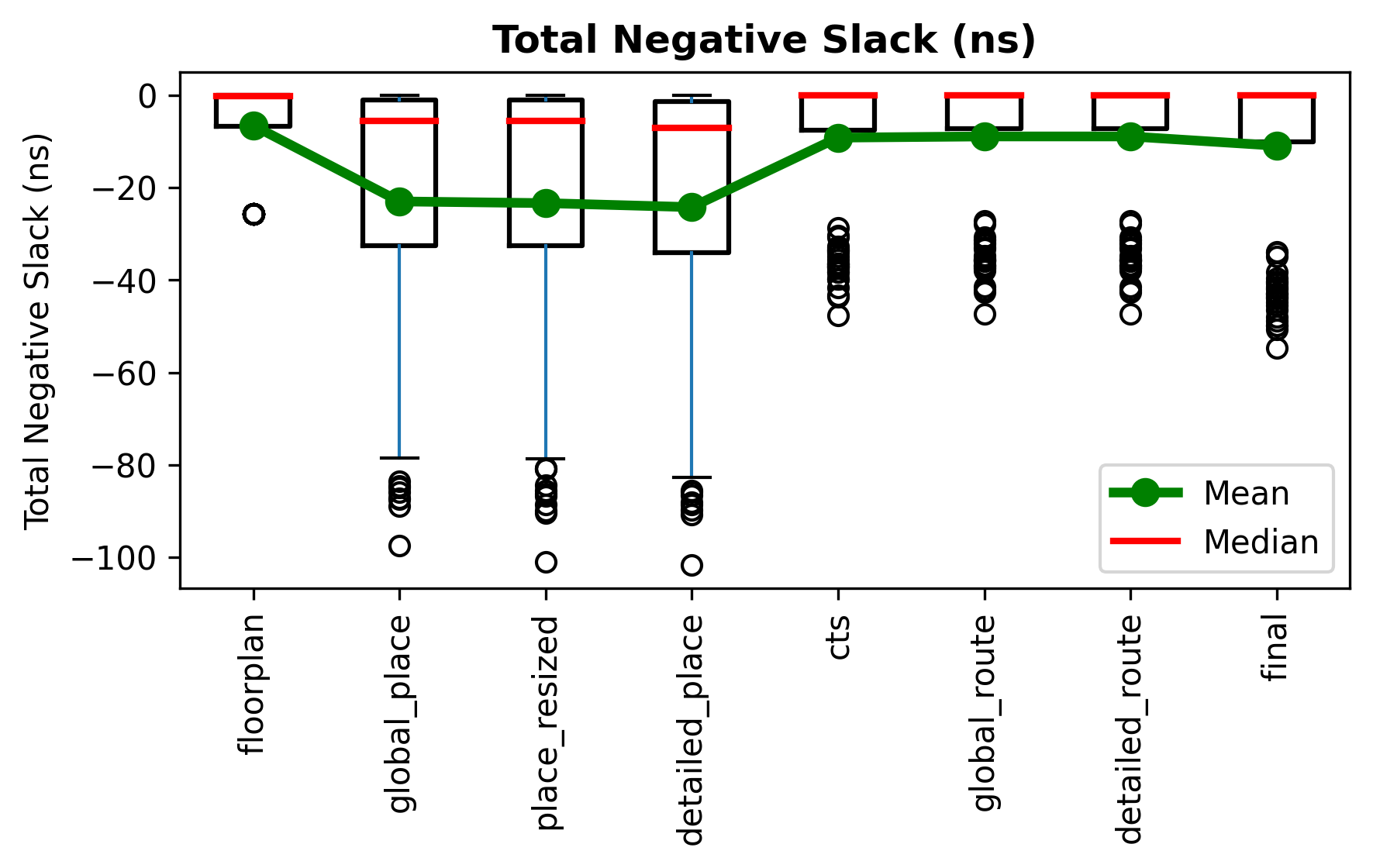}
        \caption{}
    \end{subfigure}
    \begin{subfigure}{0.32\columnwidth}
        \includegraphics[width=0.9\linewidth]{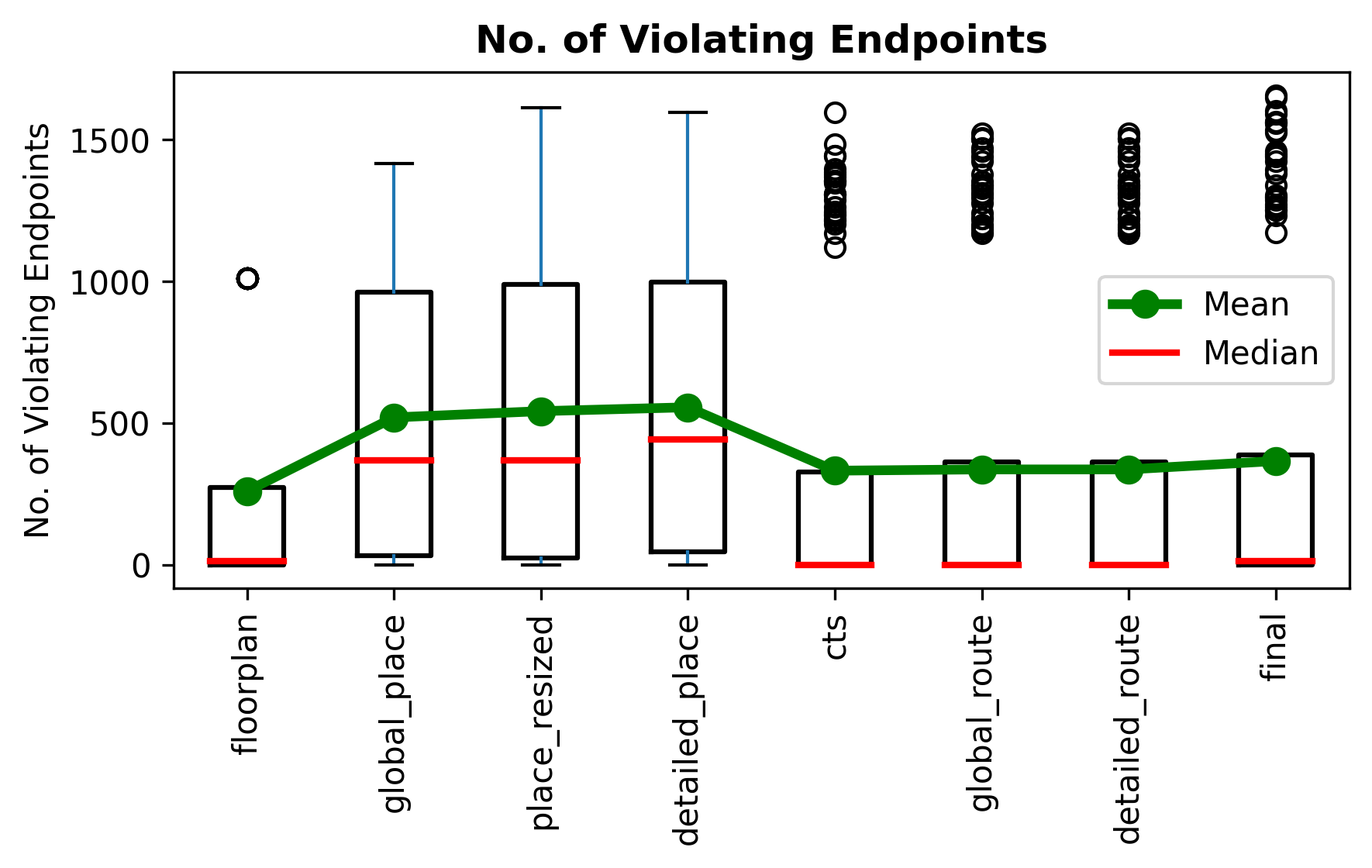}
        \caption{}
    \end{subfigure}
    \begin{subfigure}{0.32\columnwidth}
        \includegraphics[width=0.9\linewidth]{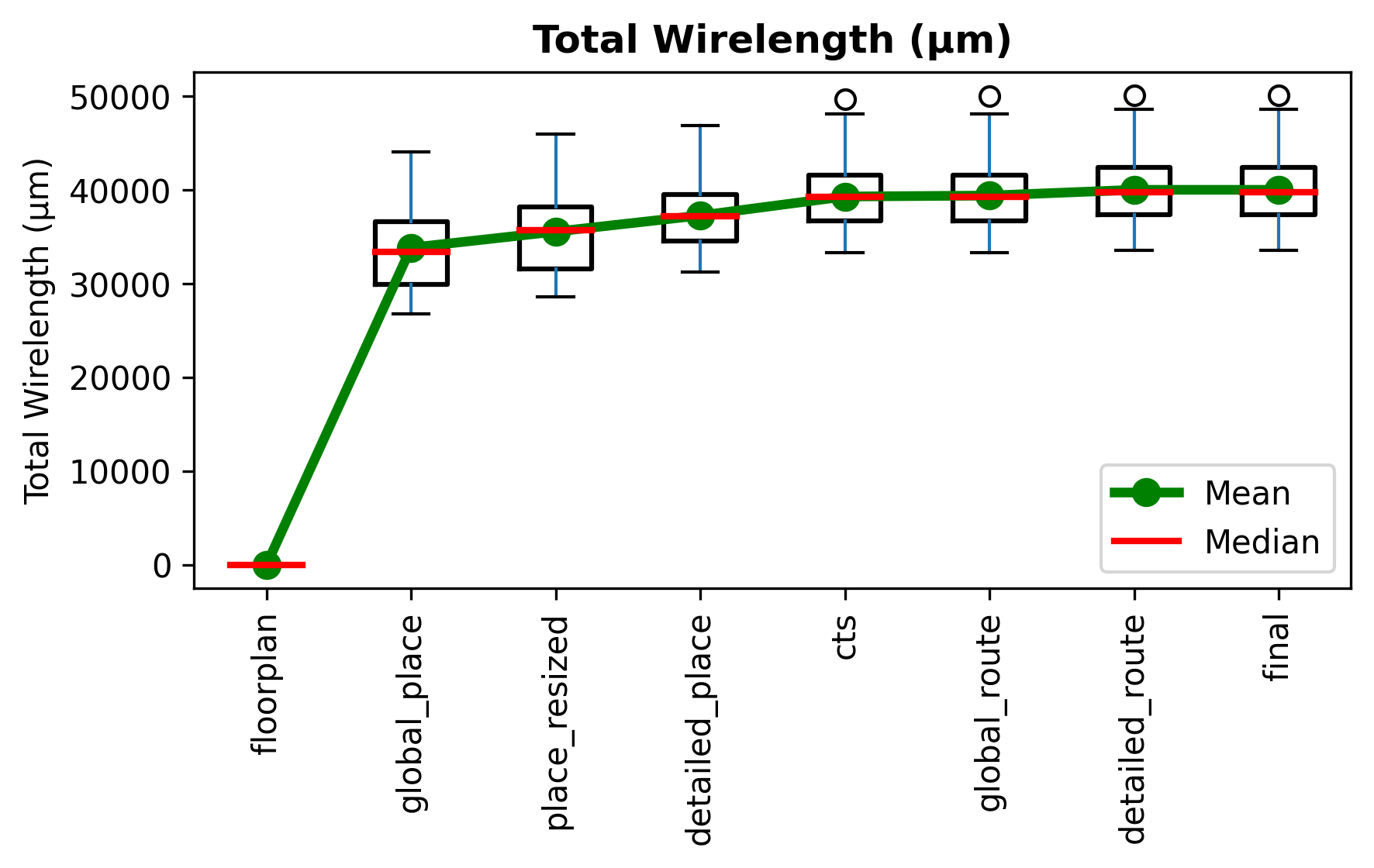}
        \caption{}
    \end{subfigure}

    \caption{Change in the distribution of (a) total area, (b) total power, (c) worst slack, (d) total negative slack, (e) number of violating endpoints, and (f) total wirelength across physical design stages for the \textit{ac97\_ctrl} circuit implemented in the ASAP7 technology node.}
    \label{fig:asap7_stage_analysis}
\end{figure}

Results from an inter-design stage analysis of the \textit{ac97\_ctrl} circuit across the SKY130, IHP130, and ASAP7 technology nodes are presented in Figures~\ref{fig:sky130_stage_analysis}, \ref{fig:ihp130_stage_analysis}, and \ref{fig:asap7_stage_analysis}, respectively.
The evolution of QoR metrics, including total area, total power, and worst-case arrival time, through the various physical design stages are illustrated in the figure, which provides insight into changes in circuit characteristics with progression through the physical design flow.

\subsection{Baseline Analysis}
\label{sec:appendix_baseline_analysis}

\begin{figure}[!h]
\vspace{-0.1in}
\centering
\includegraphics[width=0.975\columnwidth]{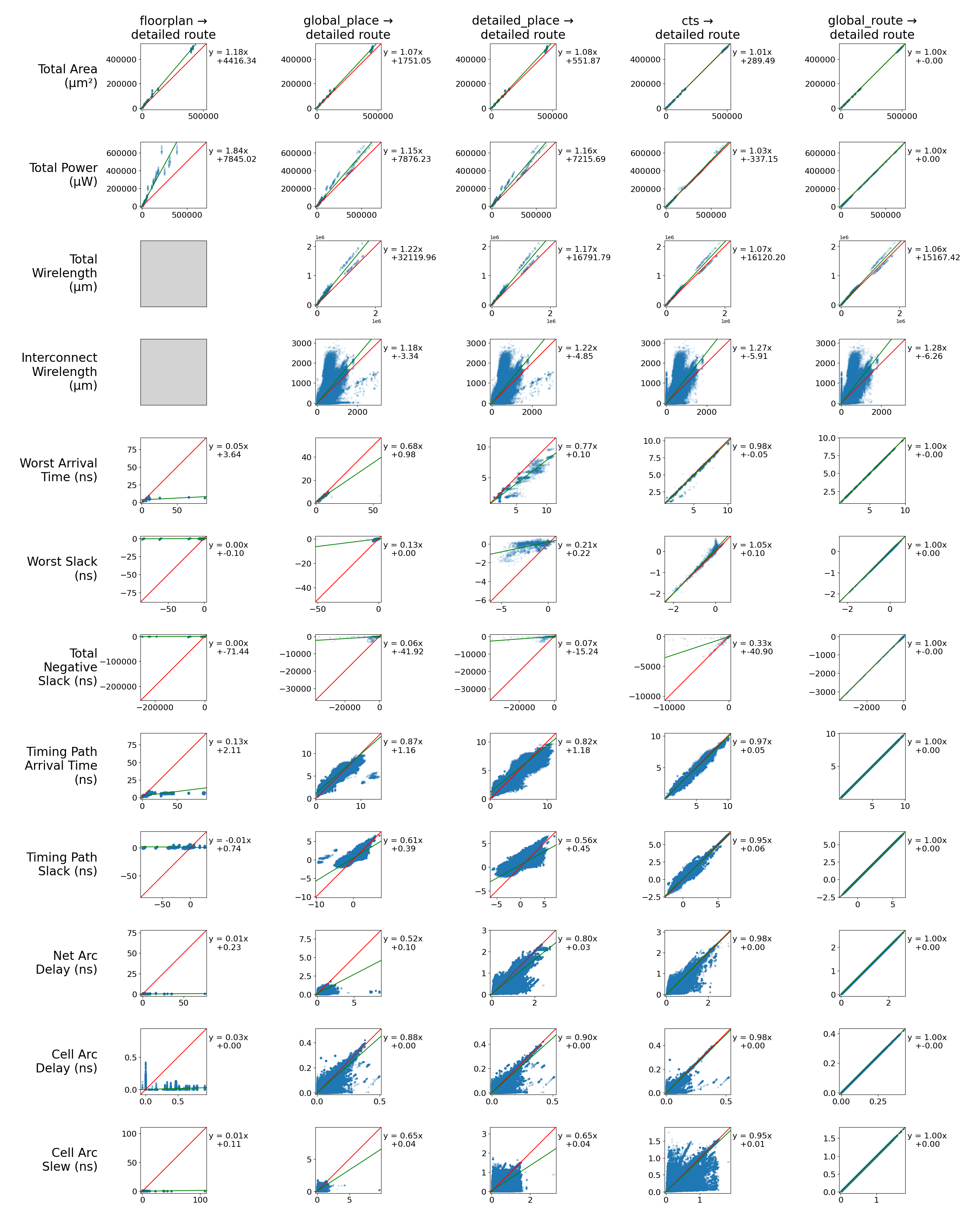}
\caption{Inter-design stage correlation of QoR and timing metrics across all benchmark circuits in the SKY130 technology node. Columns compare intermediate design stages with the final stage, with intermediate-stage values on the x-axis and final-stage values on the y-axis. Rows correspond to metrics including area, power, wirelength, timing, and arc delays. Scatter plots aggregate all circuits and operating points; the red line denotes the (y=x) reference and the green line represents a least-squares fit.}
\label{fig:sky130_baseline}
\end{figure}

\begin{figure}[!h]
\vspace{-0.1in}
\centering
\includegraphics[width=0.975\columnwidth]{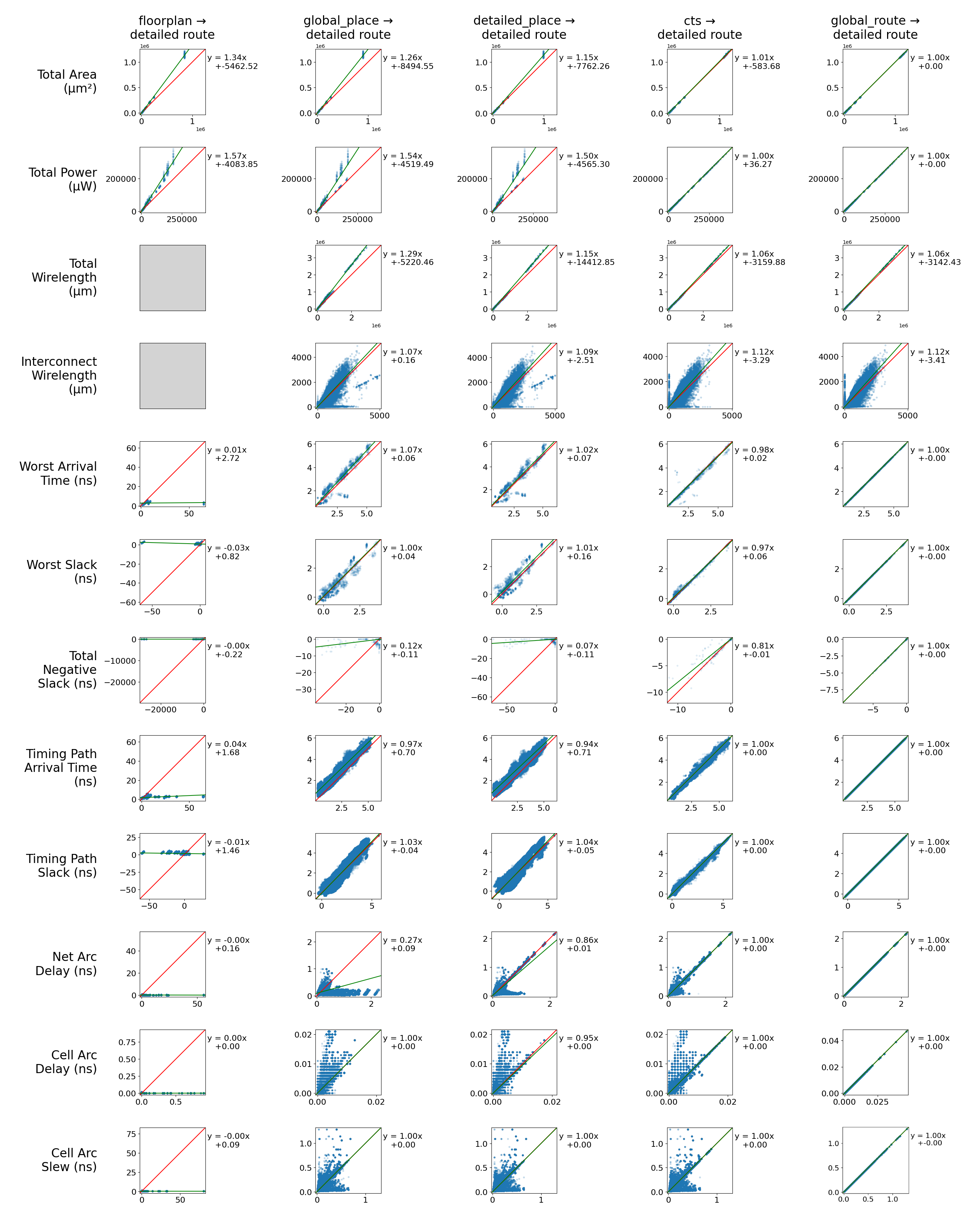}
\caption{Inter-design stage correlation of QoR and timing metrics across all benchmark circuits in the IHP130 technology node. Columns compare intermediate design stages with the final stage, with intermediate-stage values on the x-axis and final-stage values on the y-axis. Rows correspond to metrics including area, power, wirelength, timing, and arc delays. Scatter plots aggregate all circuits and operating points; the red line denotes the (y=x) reference and the green line represents a least-squares fit.}
\label{fig:ihp130_baseline}
\end{figure}

\begin{figure}[!h]
\vspace{-0.1in}
\centering
\includegraphics[width=0.975\columnwidth]{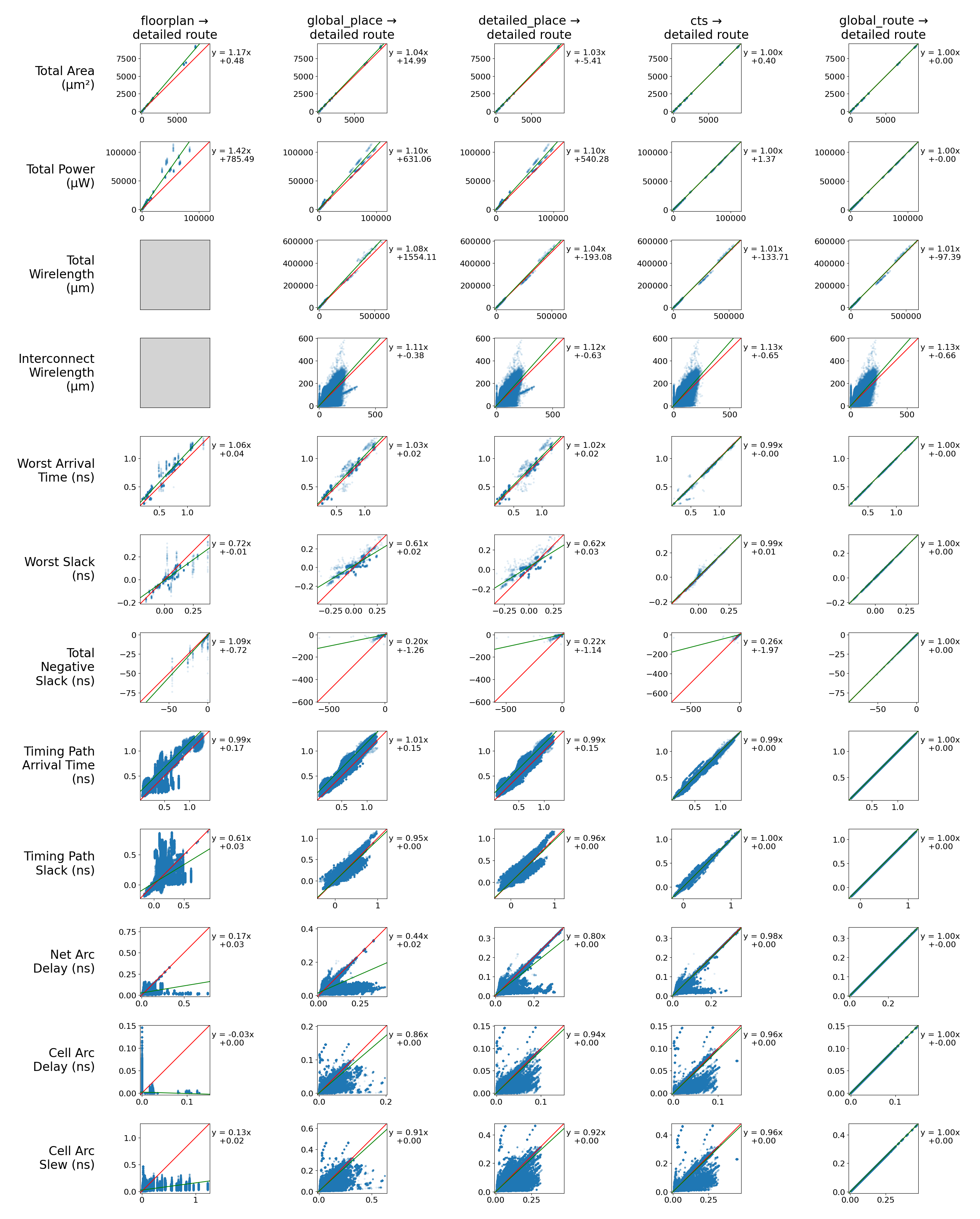}
\caption{Inter-design stage correlation of QoR and timing metrics across all benchmark circuits in the ASAP7 technology node. Columns compare intermediate design stages with the final stage, with intermediate-stage values on the x-axis and final-stage values on the y-axis. Rows correspond to metrics including area, power, wirelength, timing, and arc delays. Scatter plots aggregate all circuits and operating points; the red line denotes the (y=x) reference and the green line represents a least-squares fit.}
\label{fig:asap7_baseline}
\end{figure}

The correlation between intermediate-stage QoR estimates and the final detailed-routing results across all benchmark circuits for the SKY130, IHP130, and ASAP7 technology nodes are shown in Figures~\ref{fig:sky130_baseline}, \ref{fig:ihp130_baseline}, and \ref{fig:asap7_baseline}, respectively.
The comparison of baselines highlights the change in key design metrics across stages of the design flow and evaluates the predictive consistency of intermediate-stage estimates relative to metric scores from the final design stage.

\end{document}